\documentclass[aps,prx,notitlepage,onecolumn,longbibliography,superscriptaddress]{revtex4-1}

\usepackage[colorlinks=true,citecolor=blue]{hyperref}
\usepackage{amsmath,latexsym,amssymb,bm,color,graphicx,multirow,array,wrapfig,scalerel,datetime,enumerate}
\usepackage[section]{placeins}
\usepackage{makecell}

\newcommand{\UseFigurePdfFiles}{}
\newcounter{Num}
\newcommand{\FigName}{figures/Fig_}
\newcommand{\NewFigName}[1]{\renewcommand{\FigName}{#1}\tikzsetexternalprefix{\FigName}\setcounter{Num}{0}}

\usepackage{subcaption}
\usepackage{ragged2e}
\DeclareCaptionJustification{justified}{\justifying}
\newcommand{\dd}{\mathrm{d}}
\newcommand{\al}{\alpha}
\newcommand{\be}{\beta}

\newcommand{\g}{\gamma}
\newcommand{\s}{\sigma}
\newcommand{\om}{\omega}

\newcommand{\ra}{\rightarrow}
\newcommand{\emp}{\,\!}
\newcommand{\Smile}{\!\!\smile\!}
\newcommand{\Z}{\mathbb{Z}}
\newcommand{\Zf}{\mathbb{Z}_2^f}

\newcommand{\cT}{\mathcal T}

\newcommand{\tcP}{\tilde{\mathcal P}}
\newcommand{\timesw}{\times_{\om_2}}
\newcommand{\floor}[1]{\left\lfloor #1 \right\rfloor}
\newcommand{\eq}[1]{Eq.~(\ref{#1})}
\newcommand{\eqs}[2]{Eqs.~(\ref{#1}) and (\ref{#2})}
\newcommand{\fig}[1]{Fig.~\ref{#1}}
\newcommand{\Ref}[1]{Ref.~\onlinecite{#1}}
\newcommand{\tab}[1]{Table~\ref{#1}}
\renewcommand{\v}[1]{\boldsymbol{#1}}
\graphicspath{{figures/}}

\usepackage{tikz,ifthen}
\usetikzlibrary{arrows}
\usetikzlibrary{external}
\usetikzlibrary{matrix}
\usetikzlibrary{positioning}
\usetikzlibrary{calc}
\usetikzlibrary{decorations.markings}
\usetikzlibrary{shapes.geometric}
\tikzset{->-/.style={decoration={
  markings,
  mark=at position #1 with {\arrow{>}}},postaction={decorate}}
}
\tikzset{-<-/.style={decoration={
  markings,
  mark=at position #1 with {\arrow{<}}},postaction={decorate}}
}
\newcommand{\Ellipse}[5]{
\coordinate (A) at (#2);
\coordinate (B) at (#3);
\fill [#1] ($(A)!.5!(B)$) let \p1 = ($(B)-(A)$) in ellipse [x radius={#4/2*veclen(\x1,\y1)}, y radius={#5/2*veclen(\x1,\y1)}, rotate={asin(\y1* (-sign(\x1)*sign(\x1)+sign(\x1)+1) /veclen(\x1,\y1))}];
}
\newcommand{\Overline}[4]{
\draw[draw=white,double=white,double distance=.8pt,ultra thick]($(#1)!.1!(#2)$)--($(#1)!.9!(#2)$);
\draw[->-=#3,thick,#4] (#1)--(#2);
}
\newcommand{\OverlineBend}[5]{
\draw[draw=white,double=white,double distance=.8pt,ultra thick,bend #4=5]($(#1)!.05!(#2)$)to($(#1)!.92!(#2)$);
\draw[->-=#3,thick,bend #4=5,#5] (#1) to (#2);
}
\newcommand{\OverlineHex}[4]{
\draw[draw=white,double=#3,double distance=.8pt,ultra thick]($(#1)!.1!(#2)$)--($(#1)!.9!(#2)$);
\draw[thick,#3] ($(#1)!0!(#2)$)--($(#1)!.2!(#2)$);
\draw[thick,#3] ($(#1)!.8!(#2)$)--($(#1)!1!(#2)$);
\draw[->-=.5,thick,#3] ($(#1)!{#4}!(#2)$)--($(#1)!{#4+.01}!(#2)$);
}
\newcommand{\midpt}[4]{
\coordinate (Ax) at (#3);
\coordinate (Bx) at (#4);
\coordinate (ABx) at ($(Ax)!.5!(Bx)$);
\coordinate (#1) at($(ABx)!#2!270:(Ax)$);
}
\newcommand{\tikzfig}[2]{
\stepcounter{Num}
\vcenter{\hbox{
\ifdefined\UseFigurePdfFiles
\IfFileExists{\FigName\arabic{Num}.pdf}{\includegraphics[]{\FigName\arabic{Num}.pdf}}{
\fi
\tikzsetnextfilename{\arabic{Num}}
\begin{tikzpicture}[#1]
#2;
\end{tikzpicture}
\ifdefined\UseFigurePdfFiles
}
\fi
}}
}

\begin{document}

\title{Construction and classification of symmetry protected topological phases in interacting fermion systems}

\author{Qing-Rui Wang}
\affiliation{Department of Physics, The Chinese University of Hong Kong, Shatin, New Territories, Hong Kong, China}

\author{Zheng-Cheng Gu}
\affiliation{Department of Physics, The Chinese University of Hong Kong, Shatin, New Territories, Hong Kong, China}


\begin{abstract}
The classification and lattice model construction of symmetry protected topological (SPT) phases in interacting fermion systems are very interesting but challenging. In this paper, we give a systematic fixed point wave function construction of fermionic SPT (FSPT) states for generic fermionic symmetry group $G_f=\mathbb{Z}_2^f \times_{\omega_2} G_b$ which is a central extension of bosonic symmetry group $G_b$ (may contain time reversal symmetry) by the fermion parity symmetry group $\mathbb{Z}_2^f = \{1,P_f\}$. 
Our construction is based on the concept of equivalence class of finite depth fermionic symmetric local unitary (FSLU) transformations and decorating symmetry domain wall picture, subjected to certain obstructions. We will also discuss the systematical construction and classification of boundary anomalous SPT (ASPT) states which leads to a trivialization of the corresponding bulk FSPT states.  
Thus, we conjecture that the obstruction-free and trivialization-free constructions naturally lead to a classification of FSPT phases.
Each fixed-point wave function admits an exactly solvable commuting-projector Hamiltonian.
We believe that our classification scheme can be generalized to point/space group symmetry as well as continuum Lie group symmetry.
\end{abstract}

\maketitle

\tableofcontents

\section{Introduction}
\subsection{The goal of this paper}
Topological phases of quantum matter have become a fascinating subject in the past three decades. The concept of long range entanglement and equivalence class of finite depth local unitary (LU) transformation \cite{Chen2010} provided us a paradigm towards classifying and systematically constructing these intriguing quantum states. It was realized that the patterns of long-range entanglement are the essential data to characterize various topological phases of quantum matter.    

In recent years, the research on the interplay between topology and symmetry also has achieved a lot of fruitful results. The concept of equivalence class of finite depth symmetric LU (SLU) transformations suggests that in the presence of global symmetry, even short-range entangled (SRE) states still can belong to many different phases if they do not break any symmetry of
the system! (It is well known that the traditional Landau's symmetry
breaking states are characterized by different broken symmetries.)
Thus, these new SRE states of quantum matter
are named as symmetry protected topological (SPT) phases \cite{gu09,chenScience2012,chen13}. Topological insulators (TIs) \cite{hasan10,qi11} are the simplest examples of SPT phases, which are protected by time-reversal and charge-conservation symmetries. 

By definition, all SPT phases can be adiabatically connected to a trivial disorder phase (e.g., a product state or an atomic insulator) in the absence of global symmetry. In \Ref{gu09}, it was first pointed out that the well-known spin-1 Haldane chain \cite{haldane83} is actually an SPT phase which can be adiabatically connected to a trivial disorder phase in the absence of any symmetry. 
Thus, SPT phases can always be constructed by applying LU transformations onto a trivial product state. Such a special property makes it possible to systematically construct and classify SPT phases for interacting systems. For example, Refs.~\onlinecite{chenScience2012,chen13} introduced a systematic way of constructing fixed-point partition functions and exactly solvable lattice models for interacting bosonic systems using group cohomology theory, and it has been believed that such a construction is fairly complete for bosonic SPT (BSPT) phases protected by unitary symmetry up to 3D. Physically, the corresponding fixed-point ground-state wave functions of such a construction can be regarded as a superposition of fluctuation symmetry domain walls. Later, it was pointed out that by further decorating the $E_8$ state onto the symmetry domain wall \cite{wen15}, the fluctuation symmetry domain wall picture can actually describe all BSPT phases, which are believed to be classified by cobordism theory \cite{kapustin14a,kapustin14}. 
In section \ref{sec:SLU}, we will review how to use the equivalence class of finite depth SLU transformation  approach and fluctuation symmetry domain wall picture to classify and construct all BSPT phases with unitary symmetries up to 3D.

Although the SRE SPT phases seems to be not as interesting as long-range entangled (LRE) topological phases due to the absence of bulk fractionalized excitations, the concept of ``gauging'' the global symmetry of SPT phases establishes a direct mapping from SPT phases to intrinsic topological phases. In fact, it has been shown that different BSPT phases protected by a unitary symmetry group $G$ can be characterized by different types of braiding statistics of $G$-flux in 2D and different types of the so-called three loop braiding statistics of flux lines in 3D \cite{levin12,cheng2014,wanggu16,threeloop,ran14,wangcj15,wangj15,Juven2015,lin15,juven16,Juven2018,WCWG2018}.
Very recently, it has been further conjectured that all topological phases in 3D interacting systems can actually be realized by ``gauging'' certain SPT phases \cite{Lan1,Lan2}.  

Moreover, the classification of SPT phases in interacting systems turns out to be a one to one correspondence with the classification of global anomalies on the boundary \cite{witten15}. For example,    
anomalous surface topological order has been proposed as another very powerful way to identify and characterize different 3D SPT phases in interacting systems \cite{vishwanath13,wangc13,chen14,wangPRX2016, bonderson13,wangc13b,fidkowski13,chen14a,chong14,metlitski14,metlitski15,Fidkowski2018}.
In high energy physics, it is well known that global anomalies can be characterized and classified by cobordism/spin cobordism for interacting boson/fermion systems, thus it is not a surprise that the classification of SPT phase are closely related to cobordism/spin cobordism theory \cite{kapustin14,juven16,Juven2018}.

Despite the fact that great success has been made on the construction and classification of SPT phases in interacting boson systems and free fermion systems, understandings of SPT phases in interacting fermion systems are still very limited, especially on the construction of microscopic models.
Previously, a lot of efforts have been made on the reduction of the free-fermion classifications \cite{ff1,ff2,ff3,ff4} under the effect of interactions \cite{fidkowski10,fidkowski11,qi13,yao13,ryu12,gu14b,you2014,morimoto15}. On the other hand, stacking BSPT states onto free-fermion SPT states is another obvious way to generate some new SPT phases \cite{wangc-science,chong14}. 
Apparently, these two approaches will miss those fermionic SPT (FSPT) phases that can neither be realized in free-fermion systems nor in interacting bosonic systems \cite{wanggu16,neupert14}. Moreover, it has been further shown that certain BSPT phases become ``trivial'' (adiabatically connected to a product state) \cite{GuWen2014,chong14} when embedded into interacting fermion systems. Therefore, a systematical understanding for the classification and construction of SPT phases in interacting fermion systems is very desired.   

Very recently, based on the concept of equivalence class of finite depth fermionic SLU (FSLU) transformation and decorated symmetry domain wall picture, a breakthrough has been made on the full construction and classification of FSPT states with a total symmetry $G_f=G_b\times \mathbb Z_2^f$ (where $G_b$ is the bosonic unitary symmetry and $\mathbb Z_2^f$ is the fermion parity conservation symmetry) \cite{WangGu2017}. The fixed-point wave functions generated by FSLU transformations can be realized by exactly solvable lattice models and the resulting classification results all agree with previous studies in 1D and 2D using other methods \cite{chen11b,fidkowski11,GuWen2014,cheng15,wangcj16,Gaiotto2016}.
Most surprisingly, such a completely different physical approach precisely matches the potential global anomaly for interacting fermion systems classified by spin cobordism theory \cite{kapustin14,freed14,gaiotto16,freed16,morgan16,Kapustin2017,Morgan2018,Kapustin2018}.

It turns out that the mathematical objects that classify 1D FSPT phases with a total symmetry $G_f=G_b\times \mathbb Z_2^f$ can be summarized as two cohomology groups of the symmetry group $G_b$: $H^1(G_b,\mathbb Z_2 )$ and $H^2(G_b, U(1))$, which correspond to the complex fermion decoration on $G_b$ symmetry domain walls and classification of 1D BSPT phases. 

The mathematical objects that classify 2D FSPT phases with a total symmetry $G_f=G_b\times \mathbb Z_2^f$ are slightly complicated and can be summarized as three cohomology groups of the symmetry group $G_b$ \cite{cheng15,gaiotto16}: $H^1 (G_b,\mathbb Z_2 )$, $BH^2(G_b,\mathbb Z_2)$, and $H^3(G_b, U(1))$. $H^1 (G_b,\mathbb Z_2 )$ corresponds to the Majorana chain decoration on $G_b$ symmetry domain walls. Naively, one may expect that the complex fermion decorations on the intersection point of $G_b$ symmetry domain walls should be described by the data
$H^2(G_b,\mathbb Z_2)$. However, it turns out that such a decoration scheme will suffer from obstructions, and only the subgroup $BH^2(G_b,\mathbb Z_2)$ classifies valid and inequivalent 2D FSPT phases. More precisely, $BH^2(G_b,\mathbb Z_2)$ is defined by $n_{2} \in H^2(G_b,\mathbb Z_2)$ that satisfy $Sq^2(n_{2})=0$ in
$H^{4}(G_b,U(1))$, where $Sq^2$ is the Steenrod square operation, $Sq^2:
H^{d}(G_b,\mathbb Z_2)\rightarrow H^{d+2}(G_b,\mathbb Z_2)$ \cite{Steenrod1947}. Finally, $H^3(G_b,U(1))$ is the well-known classification of BSPT phases.

Similarly, The mathematical objects that classify 3D FSPT phases with a total symmetry $G_f=G_b\times \mathbb Z_2^f$ can also be summarized as three cohomology groups of the symmetry group $G_b$ \cite{WangGu2017,Kapustin2017}: $\tilde{B}H^2 (G_b,\mathbb Z_2 )$, $BH^3(G_b,\mathbb Z_2)$, and $H^4_{\rm rigid}(G_b, U(1))$. 
As a subgroup of $H^2 (G_b,\mathbb Z_2 )$, $\tilde{B}H^2 (G_b,\mathbb Z_2 )$ corresponds to the Majorana chain decoration on the intersection lines of $G_b$ symmetry domain walls subject to a much more subtle and complicated objections related to discrete spin structure. Again, as a subgroup of $H^3 (G_b,\mathbb Z_2 )$, $BH^3(G_b,\mathbb Z_2)$ corresponds to the complex fermion decoration on the intersection points of $G_b$ symmetry domain walls. And it is formed by elements $n_{3} \in H^3(G_b,\mathbb Z_2)$ that satisfy $Sq^2(n_{3})=0$ in
$H^{4}(G_b,U(1))$. Finally, $H^4_{\rm rigid}(G_b, U(1))\equiv H^4(G_b,U(1))/\Gamma^4$ corresponds to stable BSPT phases when embedded into interacting fermion systems. We note that $\Gamma^4$ is a normal subgroup of $H^4(G_b,U(1))$ generated by $(-)^{Sq^2(n_2)}$, where $n_2 \in H^2 (G_b , \mathbb Z_2 )$ and $(-)^{Sq^2(n_2)}$ are viewed as elements of $H^4(G_b, U(1))$. Physically, $\Gamma^4$ corresponds to those trivialized BSPT phases when embedded into interacting fermion systems.

In this paper, we aim to generalize the above constructions and classifications of FSPT phases to generic fermionic symmetry group $G_f=\Zf\timesw G_b$, which is a central extension of bosonic symmetry group $G_b$ (may contain time reversal symmetry) by the fermion parity symmetry group $\Zf=\{1,P_f\}$. We will show that the equivalence class of finite depth FSLU transformation and decorated symmetry domain wall picture still apply for generic cases, subjected to much more complicated obstruction conditions. Moreover, we will also clarify the physical meaning of obstruction by introducing the notion of anomalous SPT (ASPT) states \cite{WQG2018}, that is, a new kind of SPT states that can only be realized on the boundary of certain SPT states in one dimension higher. In the meanwhile, this also implies that the corresponding bulk SPT states are actually trivialized. Finally, we will show that if $G_b$ is time reversal symmetry, an additional layer of $p+ip$ topological superconducting state decoration on the symmetry domain wall will lead to new FSPT states, which is the analogy of decorating the $E_8$ state onto the symmetry domain wall for BSPT phases with time reversal symmetry \cite{wen15}.

\subsection{Some generalities of fermionic symmetry groups}
\label{sec:G}

For a fermionic system with total symmetry group $G_f$, there is always a subgroup: the fermion parity symmetry group $\Zf=\{1,P_f=(-1)^F\}$, where $F$ is the total fermion number operator. The subgroup $\Zf$ is in the center of $G_f$, because all physical symmetries should not change the fermion parity of the state, i.e., commute with $P_f$. Therefore, we can construct a quotient group $G_b=G_f/\Zf$, which we will call bosonic symmetry group.

Conversely, for a given bosonic symmetry group $G_b$, there are many different fermionic symmetry group $G_f$ which is the central extension of $G_b$ by $\Zf$. We have the following short exact sequence:
\begin{align}\label{seq}
1 \ra \Zf \ra G_f \ra G_b \ra 1.
\end{align}
Different extensions $G_f$ are specified by 2-cocycles $\om_2\in H^2(G_b,\Z_2=\{0,1\})$. 
This is the reason why we denote $G_f$ as $\Zf\timesw G_b$.
The group element $g_f$ of $G_f$ has the form $g_f=(P_f^{n(g)}, g_b)\in \Zf\times G_b$, with $n(g)=0,1$. We may also simply denote it as $g_f=P_f^{n(g)}g_b$. And the multiplication rule in $G_f$ is given by
\begin{align}\label{multi}
g_f \cdot h_f = \left(P_f^{n(g)},g_b\right)\cdot\left(P_f^{n(h)},h_b\right) := \left(P_f^{n(g)+n(h)+\om_2(g_b,h_b)}, g_b h_b\right),
\end{align}
where we have $P_f^{n(g)+n(h)+\om_2(g_b,h_b)}\in\Zf$ and $g_b h_b\in G_b$. The associativity condition of $g\cdot h\cdot k$ ($g,h,k\in G_b$) gives rise to the cocycle equation for $\om_2$:
\begin{align}\label{dw}
(\dd \om_2)(g,h,k) := \om_2(h,k)+ \om_2(gh,k) + \om_2(g,hk) + \om_2(g,h) = 0 \quad (\text{mod 2}).
\end{align}
We will omit the subscript of $g_b$ and use merely $g$ to denote the group element of $G_b$ henceforth. One can also show that adding coboundaries to $\om_2$ will give rise to isomorphic $G_f$. Therefore, $\om_2$ is an element in $H^2(G_b,\Z_2)$ and classifies the central extension of $G_b$ by $\Zf$. Note that there is another constraint for $\om_2$ as $\om_2(e,g) = \om_2(g,e) = 0$ ($e$ is the identity element of $G_b$).

Another ingredient of the symmetry group is associated with time reversal symmetry which is antiunitary. We can use a function $s_1$ with
\begin{align}
s_1(g)=
\begin{cases}
0, & g \text{ is unitary}\\
1, & g \text{ is antiunitary}
\end{cases}
\end{align}
to indicate whether $g\in G_b$ is antiunitary or not. The function $s_1$ is a group homomorphism from $G_b$ to $\Z_2$ because of the property
\begin{align}
(\dd s_1)(g,h) := s_1(h) + s_1(gh) + s_1(g) = 0\quad (\text{mod 2}).
\end{align}
So $s_1$ can also be viewed as a 1-cocycle in $H^1(G_b,\Z_2)$.

Let us consider some examples. The superconductor with time reversal symmetry $T^2=P_f$ ($T^2=-1$ when acting on single fermion states) has bosonic symmetry group $G_b=\Z_2^T=\{e,T\}$ and fermionic symmetry group $G_f=\Z_4^{Tf}=\Zf\timesw\Z_2^T$. In terms of our language, the 2-cocycle $\om_2$ and 1-cocycle $s_1$ have nonzero values $\om_2(T,T)=1$ and $s_1(T)=1$. They are nontrivial cocycles in $H^2(\Z_2^T,\Z_2)=\Z_2$ and $H^1(\Z_2^T,\Z_2)=\Z_2$, respectively. By choosing different $\om_2$ and $s_1$, we have three other fermionic symmetry groups $G_f$: $\Zf\times\Z_2$ (trivial $\om_2$ and trivial $s_1$), $\Z_4^f=\Zf\timesw\Z_2$ (nontrivial $\om_2$ and trivial $s_1$) and $\Zf\times\Z_2^T$ (trivial $\om_2$ and nontrivial $s_1$). We will calculate the classifications of FSPT phases with these four fermionic symmetry groups in Appendix~\ref{sec:E:Z2}.

\subsection{Summary of main results}

\subsubsection{Summary of data and equations}

\begin{table}[h]
\centering
\begin{tabular}{|c|c|c|c|c|c|c|}
\hline
data $\backslash$ dim 
& \multicolumn{2}{c|}{1D} & \multicolumn{2}{c|}{2D} & \multicolumn{2}{c|}{3D} \\
\hline
\hline
$C^1(G_b,\cdot)$ & $n_1$ & complex fermion & $n_1$ & Kitaev chain & $n_1$ & $p+ip$ superconductor \\
\hline
$C^2(G_b,\cdot)$  & $\nu_2$ & phase factor & $n_2$ & complex fermion & $n_2$ & Kitaev chain \\
\hline
$C^3(G_b,\cdot)$  & - & - & $\nu_3$ & phase factor & $n_3$ & complex fermion \\
\hline
$C^4(G_b,\cdot)$  & - & - & - & - & $\nu_4$ & phase factor \\
\hline
\end{tabular}
\caption{Layers of classification data. The cochains $n_d \in C^d(G_b,\Z_2)$, $n_{d-1} \in C^{d-1}(G_b,\Z_2)$ and $n_{d-2} \in C^{d-2}(G_b,\Z_T)$ describe the decorations of 0D complex fermions, 1D Kitaev chains, and 2D $p+ip$ superconductors (SC) in the $d$-spacial dimension model respectively. And $\nu_{d+1} \in C^{d+1}(G_b,U(1)_T)$ is the bosonic $U(1)$ phase factor in the wave function, which is related to the group cohomology classification of BSPT phases in $d$-spacial dimension. There is also an $n_0$ data in each dimension, if we want to classify fermionic invertible topological orders. We omit these $n_0$ states in this paper, for they do not need any $G_b$ symmetry protection. But they are important in the trivialization group $\Gamma^2$ for $\nu_2$ or $n_2$.
}
\label{tab:data}
\end{table}

\begin{table}[h]
\centering
\begin{tabular}{|c|l|l|l|l|c|}
\hline
layers $\backslash$ dim & 
\multicolumn{1}{c|}{1D} & \multicolumn{1}{c|}{2D} & \multicolumn{1}{c|}{3D} & \multicolumn{1}{c|}{physical meanings} \\
\hline
\hline
$p+ip$ SC & 
\multicolumn{1}{c|}{-} & \multicolumn{1}{c|}{-}
&  $\dd n_1=0$ &  \multicolumn{1}{c|}{no chiral Majorana mode} \\
\hline
Kitaev chain 
& \multicolumn{1}{c|}{-}
&  $\dd n_1=0
$ &  $\dd n_2=(\omega_2 + s_1\Smile n_1)\Smile n_1$ &  \multicolumn{1}{c|}{no free Majorana fermion} \\
\hline
complex fermion 
& $\dd n_1=0
$ &  $\dd n_2=(\omega_2+ s_1\Smile n_1)\Smile n_1$ &  $\dd n_3=(\omega_2 + n_2)\Smile n_2 + s_1\Smile(n_2\Smile_1n_2)$ &  \multicolumn{1}{c|}{fermion parity conservation} \\
\hline
 phase factor 
&  $\dd\nu_2=(-)^{\omega_2\smile n_1}$ &  $\dd\nu_3=\mathcal O_4[n_2]$ &  $\dd\nu_4=\mathcal O_5[n_3]$ & \multicolumn{1}{c|}{twisted cocycle equation} \\
\hline
\end{tabular}
\caption{Consistency equations and their physical meanings for each layers. The physical meanings of the twisted equations (of the form $\dd n_i=\mathcal O_{i+1}$ or $\dd \nu_i=\mathcal O_{i+1}$) are given in the last column of the table. The explicit expressions of $\mathcal O_4[n_2]$ and $\mathcal O_5[n_3]$ are given in Eqs.~(\ref{2D:O4}) and (\ref{3D:O5}), respectively. 
}
\label{tab:eq}
\end{table}

As discussed above, to specify the total symmetry group $G_f$ of a fermionic system, we have an 1-cocycles $s_1\in H^1(G_b,\Z_2)$ which is related to time reversal symmetry and a 2-cocycle $\omega_2\in H^2(G_b,\Z_2)$ which tells us how $G_b$ is extended by $\Zf$. They satisfy the (mod 2) cocycle equations:
\begin{align}
\dd s_1 &= 0,\\
\dd \omega_2 &= 0.
\end{align}

Given the input information of the total symmetry group $G_f$ (i.e., $G_b$ with $s_1$ and $\om_2$), we summarize the classification data, symmetry conditions, consistency equations and extra coboundary (states trivialized by ASPT state in one lower dimensions) for FSPT states in different physical dimensions in Eqs.~(\ref{1D:data})-(\ref{3D:Gamma}) (see also \tab{tab:data} for the classification data, and \tab{tab:eq} for the physical meanings of the consistency equations). 

We note that the cochains $n_d \in C^d(G_b,\Z_2)$, $n_{d-1} \in C^{d-1}(G_b,\Z_2)$ and $n_{d-2} \in C^{d-2}(G_b,\Z_T)$ describe the decorations of 0D complex fermions, 1D Kitaev chains, and 2D $p+ip$ superconductors (SC) in the $d$-spacial dimension model respectively. In 1D, it is only possible to decorate complex fermion onto the $G_b$ symmetry domain wall and the constraint $\dd n_1=0$ is nothing but the fermion parity conservation requirement for a valid FSLU transformation. In 2D, it is possible to decorate both Majoran chain onto the $G_b$ symmetry domain wall and compelx fermion onto the intersection point of $G_b$ symmetry domain walls. In order to construct FSPT states, we must decorate closed Majoran chain onto the $G_b$ symmetry domain wall and this implies $\dd n_1=0$. Again, fermion parity conservation of FSLU transformation requires that
$\dd n_2=(\omega_2 + s_1\Smile n_1)\Smile n_1$. In 3D, it is even possible to decorate 2D $p+ip$ SC state onto the $G_b$ symmetry domain wall if $G_b$ contains anti-unitary symmetry. However, in order to construct such FSPT states, we must require that there is no chiral Majorana mode on the intersection lines of $G_b$ symmetry domain walls. Furthermore, $\dd n_2=(\omega_2+ s_1\Smile n_1)\Smile n_1$ corresponds to the absence of free Majorana fermion on the intersection points of $G_b$ symmetry domain walls and $\dd n_3=(\omega_2 + n_2)\Smile n_2 + s_1\Smile(n_2\Smile_1n_2)$ again corresponds to fermion parity conservation of FSLU transformation.
Finally, the bosonic $U(1)$-valued phase factor $\nu_{d+1} \in C^{d+1}(G_b,U(1)_T)$ must satisfy the so-called twisted cocycle condition $\dd \nu_i=\mathcal O_{i+1}$, which are generated by fixed point conditions of FSPT wave functions.
We note that the bosonic layer data $\nu_{d+1}$ without superscript always means the inhomogeneous cochain in the twisted cocycle equation. The homogeneous cochain is obtained by a symmetry action and may have additional sign factors. There is also a symmetry action on the first term of the coboundary definition in $\dd \nu_{d+1}$. Because time reversal symmetry has nontrivial actions on both $\Z_T$ and $U(1)_T$, there is an exponent $1-2s_1(g_0)=\pm 1$ for the first term of $\dd \nu_{d+1}$.

Based on the above decoration construction, we can obtain the FSPT classifications by solving the consistency equations layer by layer as shown in \tab{tab:eq}. The solutions of these equations can be used to construct FSPT states. And the final classifications are obtained from these data by quotient some subgroups. We note that $B^i$ are the coboundary subgroups defined for the corresponding cochain groups $C^i$ in the usual sense. The trivialization subgroups $\Gamma^i$ of the classification data correspond to the states that are trivialized by boundary ASPT states. In $d$ spacial dimensions, the $U(1)$ factor $\nu_{d+1}$ in $\Gamma^{d+1}$ corresponds to BSPT state trivialized by fermions \cite{GuWen2014}. The complex fermion decoration data $n_d$ in the next layer $\Gamma^{d}$ is trivialized by boundary ASPT states with Kitaev chains \cite{WQG2018}. And the Kitaev chain decoration data $n_{d-1}$ in $\Gamma^{d-1}$ is trivialized by boundary ASPT states with 2D $p+ip$ chiral superconductors.

A subtle trivialization subgroup is $\Gamma^4$ which trivializes some 3D BSPT states in $H^4(G_b,U(1)_T)$ [see \eq{3D:Gamma}]. Depending on whether the corresponding 2D ASPT state has $p+ip$ superconductor components or not, $\Gamma^4$ can be divided into two parts: $\Gamma^4 = \Gamma_{n_0=0}^4 \cup \Gamma_{n_0\ne 0}^4$. The first one $\Gamma_{n_0=0}^4$ is related to the ASPT state with boundary Majorana chain $n_1$ and complex fermion $n_2$ decorations [see the last line of \eq{3D:Gamma} for the expression]. In this subgroup, the 2D ASPT state satisfies $\dd n_2=\mathcal O_3[n_1]$ in \eq{2D:eq}, and the 3D BSPT with 4-cocycle $\mathcal O_4[n_2]$ in \eq{2D:O4} becomes trivial 3D FSPT state. The second part $\Gamma_{n_0\ne 0}^4$ is related to $n_0\ne 0$ layers of $p+ip$ superconductors as 2D ASPT states. By gauging fermion parity, one can derive a complicated expression for $\Gamma_{n_0\ne 0}^4$ \footnote{Chenjie Wang, private communication.}. To the best of our knowledge, so far there is no known example of $G_f$ corresponding to a nontrivial solution of $\Gamma_{n_0\ne 0}^4$. Therefore, it is possible that $\Gamma_{n_0\ne 0}^4$ is always trivial for realistic physical systems, and we will study the full derivation of $\Gamma_{n_0\ne 0}^4$ elsewhere.	

Our FSPT classification results in different spacial dimensions are summarized below.

1D: $(n_1,\nu_2)$
\begin{align}
&\begin{cases}\label{1D:data}
n_1 \in H^1(G_b,\Z_2)
,\\
\nu_2 \in C^2(G_b,U(1)_T)/B^2(G_b,U(1)_T)/\Gamma^2.
\end{cases}\\
&\begin{cases}\label{1D:symm}
n_1(gg_0,gg_1) = n_1(g_0,g_1) = n_1(g_0^{-1}g_1),\\
\nu_2(g,ga,gab) = \emp^g\nu_2(a,b) = \nu_2(a,b)^{1-2s_1(g)} \cdot (-1)^{(\omega_2\smile n_1)(g,a,b)}.
\end{cases}\\
&\begin{cases}\label{1D:eq}
\dd n_1 = 0
,\\
\dd \nu_2 = (-1)^{\omega_2\smile n_1}.\\
\end{cases}\\
&\quad\Gamma^2 = \{(-1)^{\omega_2} \in H^2(G_b,U(1)_T) \}.\label{1D:Gamma}
\end{align}

2D: $(n_1,n_2,\nu_3)$
\begin{align}
&\begin{cases}\label{2D:data}
n_1 \in H^1(G_b,\Z_2)
,\\
n_2 \in C^2(G_b,\Z_2)/B^2(G_b,\Z_2)/\Gamma^2,\\
\nu_3 \in C^3(G_b,U(1)_T)/B^3(G_b,U(1)_T)/\Gamma^3.
\end{cases}\\
&\begin{cases}\label{2D:symm}
n_1(gg_0,gg_1)=n_1(g_0,g_1)=n_1(g_0^{-1}g_1),\\
n_2(gg_0,gg_1,gg_2)=n_2(g_0,g_1,g_2)=n_2(g_0^{-1}g_1,g_1^{-1}g_2),\\
\nu_3(g,ga,gab,gabc)=\emp^g\nu_3(a,b,c)=\nu_3(a,b,c)^{1-2s_1(g)} \cdot \mathcal O_4^\mathrm{symm}(g,ga,gab,gabc)
\quad\text{[see \eq{2D:Osymm}]}.
\end{cases}\\
&\begin{cases}\label{2D:eq}
\dd n_1 = 0
,\\
\dd n_2 = \omega_2\smile n_1 + s_1\smile n_1\smile n_1,\\
\dd \nu_3 = \mathcal O_4[n_2] \quad\text{[see \eq{2D:O4}]}.
\end{cases}\\
&\begin{cases}\label{2D:Gamma}
\Gamma^2 = \{\omega_2 \in H^2(G_b,\Z_2) \},\\
\Gamma^3 = \{(-1)^{\omega_2\smile n_1} \in H^3(G_b,U(1)_T) | n_1\in H^1(G_b,\Z_2)\}.
\end{cases}
\end{align}

3D: $(n_1,n_2,n_3,\nu_4)$
\begin{align}
&\begin{cases}\label{3D:data}
n_1 \in H^1(G_b,\Z_T),\\
n_2 \in C^2(G_b,\Z_2)/B^2(G_b,\Z_2)/\Gamma^2,\\
n_3 \in C^3(G_b,\Z_2)/B^3(G_b,\Z_2)/\Gamma^3,\\
\nu_4 \in C^4(G_b,U(1)_T)/B^4(G_b,U(1)_T)/\Gamma^4.
\end{cases}\\
&\begin{cases}\label{3D:symm}
n_1(g,ga)=\emp^gn_1(e,a)=\emp^{g}n_1(a)=(-1)^{s_1(g)}n_1(a),\\
n_2(gg_0,gg_1,gg_2)=n_2(g_0,g_1,g_2)=n_2(g_0^{-1}g_1,g_1^{-1}g_2),\\
n_3(gg_0,gg_1,gg_2,gg_3)=n_3(g_0,g_1,g_2,g_3)=n_3(g_0^{-1}g_1,g_1^{-1}g_2,g_2^{-1}g_3),\\
\nu_4(g,ga,gab,gabc,gabcd)=\emp^g\nu_4(a,b,c,d)=\nu_4(a,b,c,d)^{1-2s_1(g)} \cdot \mathcal O_5^\mathrm{symm}(g,ga,gab,gabc,gabcd)
\quad\text{[see \eq{3D:Osymm}]}.
\end{cases}\\
&\begin{cases}\label{3D:eq}
\dd n_1=0,\\
\dd n_2 = \omega_2\smile n_1 + s_1\smile n_1\smile n_1,\\
\dd n_3 = \omega_2\smile n_2 + n_2\smile n_2 + s_1\smile(n_2\smile_1n_2),\\
\dd \nu_4 = \mathcal O_5[n_3] \quad\text{[see \eq{3D:O5}]}.
\end{cases}\\
&\begin{cases}\label{3D:Gamma}
\Gamma^2 = \{\omega_2\smile n_0 \in H^2(G_b,\Z_2) | n_0\in H^0(G_b,\Z_T) \},\\
\Gamma^3 = \{\omega_2\smile n_1+s_1\smile n_1\smile n_1 + (\om_2\smile_1 \om_2) \floor{n_0/2} \in H^3(G_b,\Z_2) | n_1\in H^1(G_b,\Z_2), n_0\in H^0(G_b,\Z_T) \},\\
\Gamma^4 = \{ \mathcal O_4[n_2] \text{ [see \eq{2D:O4}]} \in H^4(G_b,U(1)_T) | n_2 \text{ satisfying \eq{2D:dn2} for some }n_1\in H^1(G_b,\Z_2) \} \cup \Gamma^4_{n_0\ne 0}.
\end{cases}
\end{align}

\subsubsection{Summary of classification examples}

Using the above data, we calculate the classifications for FSPT phases with several simple symmetry groups. They are summarized in \tab{tab:examples}. Some of the derivations are given in Appendix~\ref{sec:E}. In particular, we calculate the classifications for 2D FSPT phases with arbitrary unitary finite Abelian group $G_f$ in Appendix~\ref{sec:E:Ab}. Our results are exactly the same as that in \Ref{wanggu16}, which uses a totally different approach by investigating the braiding statistics of the gauge flux. The calculations for 3D FSPT phases with arbitrary unitary finite Abelian group $G_f$ are given in \Ref{ZWWG2019}. The results are also consistent with 3D loop braiding statistics approaches.
 We calculate the classifications of FSPT phases for the four fermionic symmetry groups with $G_b\cong \Z_2$ in Appendix~\ref{sec:E:Z2}. They are also consistent with previously known results. As an example of non-Abelian $G_f$, we calculate the FSPT phases with quaternion group symmetry $G_f=Q_8^f$ in Appendix~\ref{sec:E:Q8}.

\begin{table}[ht]
\centering
\begin{tabular}{ |c|c|c|c| }
\hline
$G_f\ \backslash\ \mathrm{dim}$ 
& 1 & 2 & 3\\
\hline
\hline
$\Z_2^f\times\Z_2$ 
& $\Z_2$ & $\Z_8$ & $\Z_1$ \\
\hline
$\Z_2^f\times\Z_{2k+1}$ 
& $\Z_1$ & $\Z_{2k+1}$ & $\Z_1$ \\
\hline
$\Z_2^f\times\Z_{2k}$ 
& $\Z_2$ &
$\begin{cases}
\Z_{4k}\times\Z_2,\ \ k\text{ even}\\
\Z_{8k},\quad\quad\quad k\text{ odd}
\end{cases}$
& $\Z_1$ \\
\hline
$\Z_2^f\times\Z_{2}\times \Z_2$ 
& $\Z_{2}^3$ & $\Z_8^2 \times \Z_4$ & $\Z_{2}^2 $   \\
\hline
$\Z_2^f\times\Z_{2}\times \Z_4$ 
& $\Z_{2}^3$ & $\Z_8^2 \times \Z_2^3$ & $\Z_4 \times \Z_2$   \\
\hline
$\Z_2^f\times\Z_{2}\times \Z_8$ 
& $\Z_{2}^3$ & $ \Z_{16} \times \Z_{8}\times \Z_{2}^3 $ & $\Z_{8} \times \Z_{2}$   \\
\hline
$\Z_2^f\times\Z_{4}\times \Z_4$ 
& $\Z_2^2 \times \Z_4$ & $\Z_{8}^2 \times \Z_{4} \times \Z_{2}^3 $ & $\Z_{4}^2 \times \Z_{2}$   \\
\hline
$\Z_2^f\times\Z_{4}\times \Z_8$ 
& $\Z_2^2 \times \Z_4$ & $\Z_{16} \times \Z_{8} \times \Z_{4} \times \Z_{2}^3 $ & $\Z_{8} \times \Z_{4} \times \Z_{2}$   \\
\hline
$\Z_2^f\times\Z_{2}\times \Z_2\times \Z_2$ 
& $\Z_2^4$ & $\Z_{8}^3 \times \Z_{4}^3 \times \Z_{2} $ & $\Z_{2}^8$   \\
\hline
$\Z_2^f\times\Z_{2}\times \Z_2\times \Z_4$ 
& $\Z_2^4$ & $\Z_{8}^3 \times \Z_{4} \times \Z_{2}^6 $ & $\Z_{4}^3 \times \Z_{2}^5$   \\
\hline
$\Z_2^f\times\Z_{2}\times \Z_4\times \Z_4$ 
& $\Z_2^4$ & $\Z_{8}^3 \times \Z_{4} \times \Z_{2}^8 $ & $\Z_{4}^4 \times \Z_{2}^6$   \\
\hline
$\Z_{2k}^f(k=2^n\ge 2)$ 
& $\Z_1$ & $\Z_{k/2}$ & $\Z_1$ \\
\hline
$\Z_4^f\times\Z_2$ 
& $\Z_2$ & $\Z_4$ & $\Z_2$ \\
\hline
$\Z_4^f\times\Z_4$ 
& $\Z_4$ & $\Z_8\times \Z_2$ & $\Z_2$ \\
\hline
$\Z_8^f\times\Z_2$ 
& $\Z_2$ & $\Z_4\times \Z_2$ & $\Z_4$ \\
\hline
$\Z_4^f\times\Z_2\times\Z_2$ 
& $\Z_2^3$ & $\Z_4^2\times \Z_2^2$ & $\Z_2^5$ \\
\hline
$\Z_4^f\times\Z_2\times\Z_4$ 
& $\Z_4\times\Z_2^2$ & $\Z_8\times\Z_4\times \Z_2^3$ & $\Z_4\times \Z_2^4$ \\
\hline
$\Z_4^f\times\Z_4\times\Z_4$ 
& $\Z_4^3$ & $\Z_8^2\times\Z_4^2\times \Z_2^2$ & $\Z_4^2\times \Z_2^4$ \\
\hline
$\Z_2^f\times\Z_2^T$ 
& $\Z_4$ & $\Z_1$ & $\Z_1$ \\
\hline
$\Z_4^{Tf}=\Z_2^f\timesw\Z_2^T$ 
& $\Z_2$ & $\Z_2$ & $\Z_{16}$ \\
\hline
$Q_8^{f}=\Z_2^f\timesw(\Z_2\times\Z_2)$ 
& $\Z_1$ & $\Z_2$ & $\Z_1$ \\
\hline
\end{tabular}
\caption{Classifications of FSPT phases with some simple fermionic symmetry group $G_f$ in different spacial dimensions. Invertible topological orders protected by $\Zf$ only are not included in this table.}
\label{tab:examples}
\end{table}

\subsection{Organization of the paper}

The rest of the paper is organized as follows. In section~\ref{sec:SLU}, we review the key concept of SLU transformations. Using this approach, we show the classifications of BSPT phases in various dimensions. In section~\ref{sec:FSLU}, we summarize the procedures of constructing FSPT states. The definition of FSLU transformations is given in section~\ref{sec:FSLU_}. All layers of degrees of freedom and their symmetry transformation rules are summarized in section~\ref{sec:FSLU:dof}. In section~\ref{sec:FSLU:symm}, we discuss briefly the two essential requirements of the FSLU transformations: the coherence equations and the symmetry conditions. Using the outlined procedure, the details of the classifications of 1D, 2D and 3D FSPT phases are given in sections~\ref{sec:1D}, \ref{sec:2D}, and \ref{sec:3D}, respectively. In each dimension, we first give the symmetric decoration procedures. Then the $F$ move (FSLU transformation) and its coherence equation are given explicitly. As the final step in classifying FSPT phases in each dimension, we discuss some new coboundaries associated with ASPT states in one lower dimension. We summarize this work in section~\ref{sec:con}.

In Appendix~\ref{sec:0D}, we show the classification of the simplest 0D FSPT phases. In Appendix~\ref{sec:moves}, we list all possible 2D and 3D moves that admit a branching structure. In Appendix~\ref{sec:Kast}, the (local) Kasteleyn orientations for 2D and 3D lattices are discussed briefly. In Appendix~\ref{App:obs}, we discuss the Bockstein homomorphism mapping a $\Z_2$-valued cocycle to a $\Z_T$-valued cocycle. It is useful in checking whether the obstruction function $(-1)^{f_k}$, where $f_k$ is a $\Z_2$-valued cocycle, is a $U(1)_T$-valued coboundary or not. The detail calculations of FSPT phases for some simple groups are given in Appendix~\ref{sec:E}. Some of the results are already summarized in \tab{tab:examples}.

\section{SLU transformation and classification of BSPT phases}
\label{sec:SLU}
\NewFigName{figures/Fig_BSPT_}

\subsection{SLU transformation and BSPT phases} 
From the definition of SPT states, it is easy to see that (in the absence of global symmetry):
\begin{equation}
  |\text{SPT}\rangle  = U^M_{circ} |\text{Trivial}\rangle \label{LU}
\end{equation} 
Namely, an SPT state can be connected to a trivial state (e.g., a product state) vial LU transformation (in the absence of global symmetry). Clearly, Eq.(\ref{LU}) implies that the support space \footnote{Considering the
entanglement density matrix $\rho_A$ for a SPT state in region $A$, $\rho_A$ may act on a subspace of the Hilbert space in region A, and the subspace is called the support space $\tilde{V}_A$ of region $A$} of any SPT state in a region must be one dimensional. This is simply because a trivial state (e.g., a product state) has a one dimensional support space, and any SPT state will become a product state via a proper local basis change (induced by a LU transformation). 

In the presence of global symmetry, we can further introduce the notion of symmetric local unitary (SLU) transformations classifying SPT phases in interacting bosonic systems. By SLU transformation, we mean the corresponding piecewise LU operator is invariant under symmetry $G$. More precisely, we have
$ U_{pwl}= \prod_{\v i} e^{-i  H_b(g_{\v i0},g_{\v i1},g_{\v i2},\cdots)}\equiv \prod_{\v i}U(g_{\v i0},g_{\v i1},g_{\v i2},\cdots)$ and $U(gg_{\v i0},gg_{\v i2},gg_{\v i3},\cdots)=U(g_{\v i0},g_{\v i1},g_{\v i2},\cdots)$ for any $g \in G$. (We note that here we choose the group element basis $g_{\v i0},g_{\v i1},g_{\v i2},\cdots$ to represent bosonic symmetric unitary operator acting on a region labeled by $\v i$.)    
However, we need to enforce the SLU transformations to be one dimensional (when acting on the support space $\rho_A$ for any region $A$), and we call them invertible SLU transformations. Thus, we claim that SPT phases in interacting bosonic systems can be classified by equivalence class of invertible SLU transformations.

SPT phases are also referred as invertible (non-chiral) topological phases. It turns out that the novel concept of invertible SLU transformation even allows us to construct very general fixed-point SPT states.
All of these fixed-point wave functions admit exactly solvable parent Hamiltonians consisting of commuting projectors on an arbitrary triangulation with an arbitrary branching structure. 

\subsection{Fixed-point wave function and classification for BSPT phases in 1D}

As a warm up, let us begin with fixed-point wave function in 1D and use SLU transformation to derive the well known classification results of 1D BSPT phases. Without loss of generality, here we assume that every (locally ordered) vertex $i$ of the 1D lattice has bosonic degrees of freedom labeled by a group element $g_i\in G$.

Our 1D fixed-point state is a superposition of those basis states with all possible 1D graph with a branching structure. 

\begin{align}
|\Psi\rangle 
=\sum_{\text{all conf.}} \Psi\left(
\tikzfig{scale=.6}{
\pgfmathsetmacro{\N}{5};
\pgfmathsetmacro{\Nm}{4};
\foreach \i in {0,1,...,\N} {
	\coordinate (\i) at (\i,0);
}
\foreach \i in {0,1,...,\Nm} {
	\pgfmathsetmacro{\c}{\i+1};
	\draw[->-=.5,line width=.5] (\i)--(\c);
	\node[above,scale=.9] at (\i) {$g_{\i}$};
}
\foreach \i in {0,1,...,\N} {
	\node[circle,draw=black,fill=black,scale=.15] at (\i) {};
}
\node[above,scale=.9] at (5,0) {$g_5$};
\node[below] at (5,0) {};
}
\right) \big|
\vcenter{\hbox{\includegraphics[scale=1]{\FigName\arabic{Num}.pdf}}}
\big\rangle.
\end{align}

In the following, we will derive the rules of wave function renormalization generated by SLU transformations for the above wave function and show how to construct all BSPT states in 1D.
To obtain a fixed-point wave function, we need to understand the changes of the wave function under renormalization. In 1D, renormalization can be understood as removing some degrees of freedom by reducing the number of vertices. The basic renormalization process is known as (2-1) Pachner move of triangulation of 1D manifold.  

To be more precise, the (2-1) move is an SLU transformation between two different 1D graphs:
\begin{align}
\label{eq:1Dmove}
\Psi\left(
\vcenter{\hbox{\includegraphics[scale=0.25]{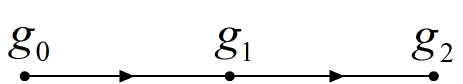}}}
\right)=
\frac{1}{|G|^{1/2}}\nu_2(g_0,g_1,g_2)
\Psi\left(
\vcenter{\hbox{\includegraphics[scale=0.25]{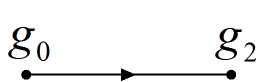}}}
\right),
\end{align}
 
We note that the $|G|$ is the order of the group $G$ and we introduce the normalization factor $1/|G|^{1/2}$ in the above expression due to the change of vertex number. 
Here, $\nu_2(g_0,g_1,g_2)$ is a $U(1)$-valued function with variables $g_i\in G$. Since we are constructing symmetric state, $\nu_2$ should be symmetric under the action of $G$ with $\nu_2(gg_0,gg_1,gg_2)=\nu_2(g_0,g_1,g_2)$ (We note that $\nu_2(gg_0,gg_1,gg_2)=\nu_2^*(g_0,g_1,g_2)$ if $g$ is anti-unitary).    

Since we are constructing fixed-point wave function, it should be invariant under renormalization. For instance, we can use two different sequences of the above (2-1) moves \eq{eq:1Dmove} to connect a fixed initial state and a fixed final state. Different approaches should give rise to the same wave function. These constraints give us the consistent equations for $\nu_2$.

The simplest example is the following two paths between two fixed states:
\begin{align}
\label{eq:1Dmove1}
\Psi\left(
\vcenter{\hbox{\includegraphics[scale=0.25]{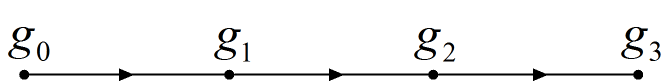}}}
\right) 
&=
\frac{1}{|G|^{1/2}}\nu_2(g_1,g_2,g_3)\ 
\Psi\left(
\vcenter{\hbox{\includegraphics[scale=0.25]{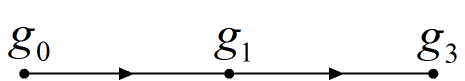}}}
\right) \nonumber \\
&=
\frac{1}{|G|}\nu_2(g_1,g_2,g_3) \nu_2(g_0,g_1,g_3) \ 
\Psi\left(
\vcenter{\hbox{\includegraphics[scale=0.25]{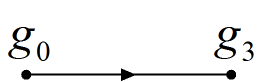}}}
\right),
\end{align}
 
\begin{align}
\label{eq:1Dmove2}
\Psi\left(
\vcenter{\hbox{\includegraphics[scale=0.25]{1Dc.png}}}
\right) 
&=
\frac{1}{|G|^{1/2}}\nu_2(g_0,g_1,g_2)\ 
\Psi\left(
\vcenter{\hbox{\includegraphics[scale=0.25]{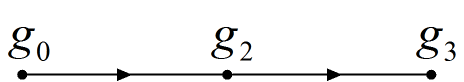}}}
\right) \nonumber \\
&=
\frac{1}{|G|}\nu_2(g_0,g_1,g_2) \nu_2(g_0,g_2,g_3) \ 
\Psi\left(
\vcenter{\hbox{\includegraphics[scale=0.25]{1Df.png}}}
\right).
\end{align}
The constraint is that the products of $F$ moves for the above two processes equal to each other:
\begin{align}
\label{eq:1DRG}
\nu_2(g_0,g_1,g_3) \nu_2(g_1,g_2,g_3)=\nu_2(g_0,g_2,g_3) \nu_2(g_0,g_1,g_2).
\end{align}
The above equation implies:
\begin{align}
\dd \nu_2(g_0,g_1,g_2,g_3) = \frac{\nu_2(g_1,g_2,g_3) \nu_2(g_0,g_1,g_3)}{\nu_2(g_0,g_2,g_3) \nu_2(g_0,g_1,g_2)} =1.
\end{align}
The above equation is exactly the same as the cocycle equation of group cohomology theory, and it means $\nu_2$ should be a $U(1)_T$-valued 2-cocycle.  

Using an SLU transformation, we can further redefine the basis state $|\{g_l\}\rangle$ as
\begin{align}
|\{g_l\}\rangle' = U_{\mu_1,m_0} |\{g_l\}\rangle 
= \prod_{\langle ij \rangle} \mu_1(g_i,g_j) |\{g_l\}\rangle,
\end{align}
In the new basis, one find that the phase factor in \eq{eq:1Dmove} becomes
\begin{align}
\nu_2'(g_0,g_1,g_2) \equiv \nu_2(g_0,g_1,g_2) \frac{\mu_1(g_1,g_2)\mu_1(g_0,g_1)}{\mu_1(g_0,g_2)}.
\end{align}
Since our gapped phases are defined by SLU transformations, $\nu_2'$ and $\nu_2$ belong to the same phase. In general, the elements $\nu_2$ in the same group cohomology class in $H^2(G, U(1)_T)$ correspond to the same 1D BSPT phase. 

SLU transformations not only give rise to the local rules of constructing fixed-point wave functions, but also give rise to commuting projector parent Hamiltonian for these fixed-point wave functions.

For example, in 1D, the parent Hamiltonian can be expressed as $H=-\sum_i H_i$ where the matrix element of $H_i$ are defined as:
\begin{align}
\label{eq:1DH}
\left\langle\Psi\left(
\vcenter{\hbox{\includegraphics[scale=0.25]{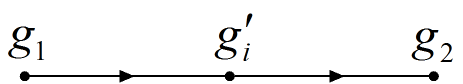}}}
\right) \right |H_i
\left|\Psi\left(
\vcenter{\hbox{\includegraphics[scale=0.25]{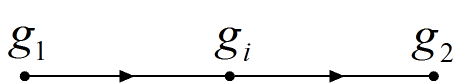}}}
\right)\right \rangle
\end{align}
which only acts the states on site $i$ and its neighboring sites. However, $H_i$ will not alter the states on neighboring sites of $i$.

The above amplitude can be computed by SLU transformations by considering the following moves for a three site patch: 
\begin{align}
\label{eq:1DH_}
\Psi\left(
\vcenter{\hbox{\includegraphics[scale=0.25]{1DH1.png}}}
\right)=&
\frac{1}{|G|^{1/2}}\nu_2(g_1,g_i,g_2)
\Psi\left(
\vcenter{\hbox{\includegraphics[scale=0.25]{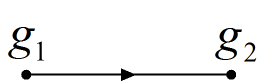}}}
\right)\nonumber\\
=&\frac{1}{|G|}\sum_{g_i^\prime}\frac{\nu_2(g_1,g_i,g_2)}{\nu_2(g_1,g_i^\prime,g_2)}
\Psi\left(
\vcenter{\hbox{\includegraphics[scale=0.25]{1DH2.png}}}
\right),
\end{align}
which implies that:
\begin{align}
\langle g_i^\prime,g_1g_2 |H_i|g_i,g_1g_2\rangle=\frac{1}{|G|}\frac{\nu_2(g_1,g_i,g_2)}{\nu_2(g_1,g_i^\prime,g_2)}=\frac{1}{|G|}\frac{\nu_2(g_1,g_i,g_i^\prime)}{\nu_2(g_i,g_i^\prime,g_2)}
\end{align}
where we use the 2-cocycle condition of $\nu_2$ in the last step. Equivalently, we can just define the action of $H_i$ on site $i$ and its neighboring sites $1,2$ as:
\begin{align}
H_i|g_i,g_1g_2\rangle=\frac{1}{|G|}\sum_{g_i^\prime}\frac{\nu_2(g_1,g_i,g_i^\prime)}{\nu_2(g_i,g_i^\prime,g_2)}|g_i^\prime,g_1g_2\rangle
\end{align}

\subsection{Fixed-point wave function and classification for BSPT phases in 2D}

The fixed-point wave functions for BSPT phases in 2D are similar to the 1D case. We can again use the group element basis to construct the local Hilbert space on each vertex of arbitrary triangulation. 

\begin{align}
|\Psi\rangle = \sum_{\text{all conf.}} \Psi\left(
\tikzfig{scale=1.5}{
\coordinate (0) at (0,0);
\coordinate (1) at (1.5,0);
\coordinate (2) at (1.5,1);
\coordinate (3) at (0,1);
\coordinate (4) at (.4,.3);
\coordinate (5) at (.6,.7);
\coordinate (6) at (1,.4);
\coordinate (7) at (0,.5);
\coordinate (8) at (.6,0);
\coordinate (9) at (1.1,0);
\coordinate (11) at (.55,1);
\coordinate (12) at (1.2,1);
\coordinate (13) at (1.5,.5);
\draw[->-=.5] (0)--(8);
\draw[->-=.5] (8)--(9);
\draw[->-=.5] (9)--(1);
\draw[->-=.5] (1)--(13);
\draw[->-=.5] (13)--(2);
\draw[->-=.5] (2)--(12);
\draw[->-=.5] (12)--(11);
\draw[->-=.5] (11)--(3);
\draw[->-=.5] (3)--(7);
\draw[->-=.5] (0)--(7);
\draw[->-=.5] (0)--(4);
\draw[->-=.5] (7)--(4);
\draw[->-=.5] (7)--(5);
\draw[->-=.5] (3)--(5);
\draw[->-=.5] (4)--(5);
\draw[->-=.5] (5)--(6);
\draw[->-=.5] (8)--(4);
\draw[->-=.5] (9)--(6);
\draw[->-=.5] (11)--(5);
\draw[->-=.5] (12)--(5);
\draw[->-=.5] (12)--(6);
\draw[->-=.3] (13)--(6);
\draw[->-=.5] (1)--(6);
\draw[-<-=.5] (12)--(13);
\draw[->-=.7] (4)--(6);
\draw[->-=.5] (8)--(6);
\node[below left,scale=.8] at (0) {$g_0$};
\node[below right,scale=.8] at (1) {$g_3$};
\node[above right,scale=.8] at (2) {$g_5$};
\node[above left,scale=.8] at (3) {$g_8$};
\node[right,scale=.8] at (4) {$g_{10}$};
\node[right,scale=.8] at (5) {$g_{11}$};
\node[right,scale=.8] at (6) {$g_{12}$};
\node[left,scale=.8] at (7) {$g_9$};
\node[below,scale=.8] at (8) {$g_1$};
\node[below,scale=.8] at (9) {$g_2$};
\node[above,scale=.8] at (11) {$g_{7}$};
\node[above,scale=.8] at (12) {$g_{6}$};
\node[right,scale=.8] at (13) {$g_{4}$};
}
\right) \stretchleftright{\Bigg|}{\ 
\vcenter{\hbox{\includegraphics[scale=1]{\FigName\arabic{Num}.pdf}}}
\ }{\Big\rangle}.
\end{align}

We assume that the triangulation admits a branching structure that can be labeled by a set of local arrows on all links (edges) with no oriented loop for any triangle. Mathematically, the branching structure can be regarded as a discrete version of a $\rm spin^c$ structure and can be consistently defined on arbitrary triangulation of orientable manifolds. The basic renormalization process is known as (2-2) and (2-0)/(0-2) Pachner move of triangulation of 2D manifold. Moreover, according to the definition of BSPT phases, we also require that the support space of SLU transformations to be one-dimensional, such that it can adiabatically connect to a product state in the absence of global symmetry. Below, we will discuss physically consistent conditions for those SLU transformations generating fixed point wave functions. 

An example of (2-2) move (now we call it the standard (2-2) move, which is the analogy of $F$ move in a unitary fusion category theory) is presented as follows:
\begin{align}
\label{eq:2Dmove}
\Psi\left(
\vcenter{\hbox{\includegraphics[scale=0.25]{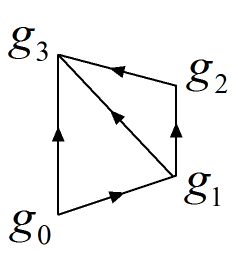}}}
\right)= 
\nu_3(g_0,g_1,g_2,g_3)
\quad
\Psi\left(
\vcenter{\hbox{\includegraphics[scale=0.25]{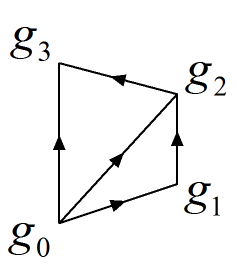}}}
\right),
\end{align}
Here, $\nu_3(g_0,g_1,g_2,g_3)$ is a $U(1)$-valued 3-cochain that is symmetric under $g$ action $\nu_3(gg_0,gg_1,gg_2,gg_3)=\nu_3(g_0,g_1,g_2,g_3)$ (Again, $\nu_3(gg_0,gg_1,gg_2,gg_3)=\nu_3^*(g_0,g_1,g_2,g_3)$ if $g$ is anti-unitary.) 

Apart from the (2-2) move, there is another (2-0) move that can change the total number of vertices for triangulations.  
\begin{align}
\label{eq:2D20}
\Psi\left(
\vcenter{\hbox{\includegraphics[scale=0.2]{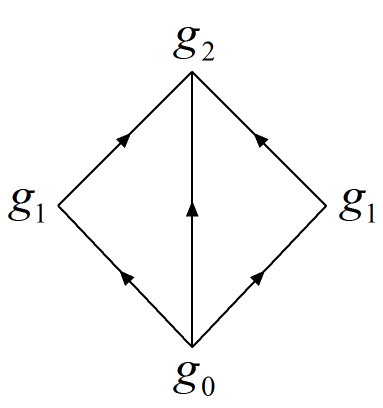}}}
\right)=\frac{1}{|G|^{1/2}}  
\Psi\left(
\vcenter{\hbox{\includegraphics[scale=0.2]{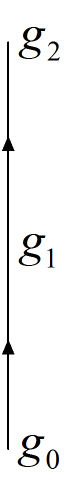}}}
\right),
\end{align}
We also add a normalization factor $|G|^{-1/2}$ in the front of the (2-0) move operator, for the vertex number is reduced by one from the left state to the right state. \footnote{In principle, we can also add an arbitrary phase factor into the above move. However, since such a phase factor must be a symmetric $U(1)$-valued function of group elements $g_0,g_1,g_2$, it can always be removed by the basis redefinition \eq{eq:2Dnu3}.}

It is easy to check that the other (2-2) moves with different branching structure, e.g, the analogy of $H$-move, can always be derived by the standard (2-2)-move and (2-0)/(0-2) move. Considering the SLU transformation for the following patch:
\begin{align}
\label{eq:2DH0}
\Psi\left(
\vcenter{\hbox{\includegraphics[scale=0.25]{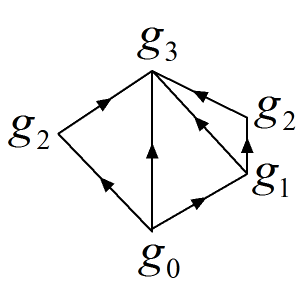}}}
\right)&= 
\nu_3(g_0,g_1,g_2,g_3)
\quad
\Psi\left(
\vcenter{\hbox{\includegraphics[scale=0.25]{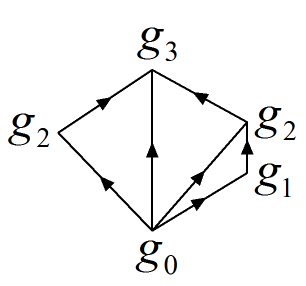}}}
\right) 
=
\frac{1}{|G|^{1/2}}\nu_3(g_0,g_1,g_2,g_3)
\quad
\Psi\left(
\vcenter{\hbox{\includegraphics[scale=0.25]{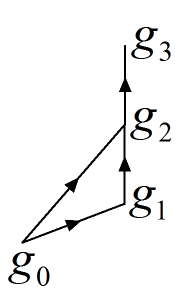}}}
\right) \nonumber\\
&= \nu_3(g_0,g_1,g_2,g_3)
\quad
\Psi\left(
\vcenter{\hbox{\includegraphics[scale=0.25]{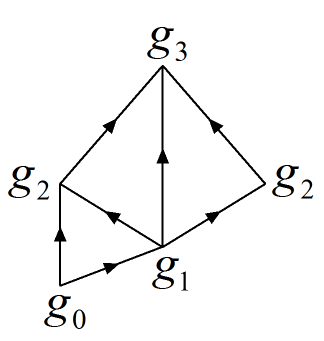}}}
\right),
\end{align}
Thus, we conclude that 
\begin{align}
\label{eq:2DH}
\Psi\left(
\vcenter{\hbox{\includegraphics[scale=0.25]{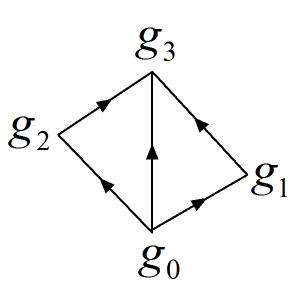}}}
\right)=
\nu_3(g_0,g_1,g_2,g_3)
\quad
\Psi\left(
\vcenter{\hbox{\includegraphics[scale=0.25]{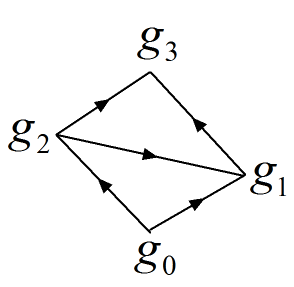}}}
\right).
\end{align}
The rest (2-2) moves are the analogies of dual F-move and dual-H move, they can also be derived from the basis (2-2) move and (2-0)/(0-2) move. For example, let us assume:
\begin{align}
\label{eq:2Ddua}
\Psi\left(
\vcenter{\hbox{\includegraphics[scale=0.25]{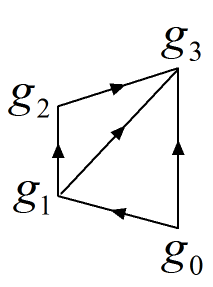}}}
\right)=
\bar\nu_3(g_0,g_1,g_2,g_3)
\Psi\left(
\vcenter{\hbox{\includegraphics[scale=0.25]{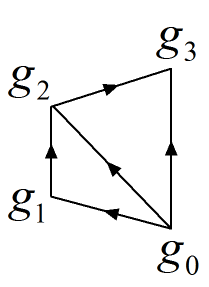}}}
\right).
\end{align}
where $\bar\nu_3(g_0,g_1,g_2,g_3)$ is another $U(1)$-valued function which is different from $\nu_3(g_0,g_1,g_2,g_3)$.
Considering the SLU transformation on the following patch:
\begin{align}
\label{eq:2Ddual0}
\Psi\left(
\vcenter{\hbox{\includegraphics[scale=0.25]{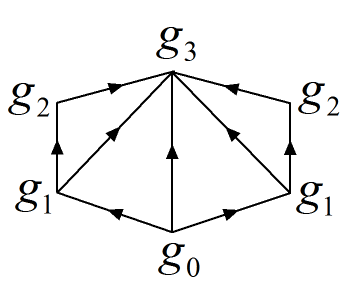}}}
\right)=&
\nu_3(g_0,g_1,g_2,g_3)\bar\nu_3(g_0,g_1,g_2,g_3)
\Psi\left(
\vcenter{\hbox{\includegraphics[scale=0.25]{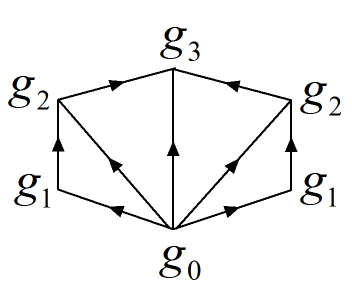}}}
\right)\nonumber\\
=&
\frac{1}{|G|}\nu_3(g_0,g_1,g_2,g_3)\bar\nu_3(g_0,g_1,g_2,g_3)
\Psi\left(
\vcenter{\hbox{\includegraphics[scale=0.25]{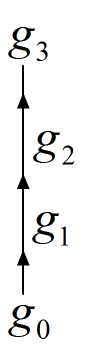}}}
\right).
\end{align}
On the other hand, by applying the (2-0) move directly, we have
\begin{align}
\label{eq:2Ddual0_}
\Psi\left(
\vcenter{\hbox{\includegraphics[scale=0.25]{2Dg1.png}}}
\right)= \frac{1}{|G|} 
\Psi\left(
\vcenter{\hbox{\includegraphics[scale=0.25]{2Dg.png}}}
\right).
\end{align}
Thus, we conclude that $\bar\nu_3(g_0,g_1,g_2,g_3)\nu_3(g_0,g_1,g_2,g_3)=1$ or $\bar\nu_3(g_0,g_1,g_2,g_3)=\nu_3^{-1}(g_0,g_1,g_2,g_3)$.

Moreover, the combination of (2-2) move and (2-0) move will further allow us to define a new set of renormalization move which reduces the number of vertices, namely, the (3-1) move. For example, considering the SLU transformation for the following patch, 
\begin{align}
\label{eq:2DH0_}
\Psi\left(
\vcenter{\hbox{\includegraphics[scale=0.25]{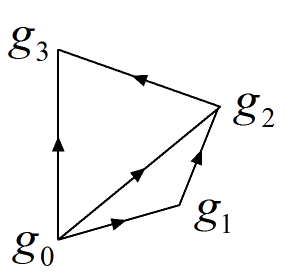}}}
\right)=&
\nu_3^{-1}(g_0,g_1,g_2,g_3)
\quad
\Psi\left(
\vcenter{\hbox{\includegraphics[scale=0.25]{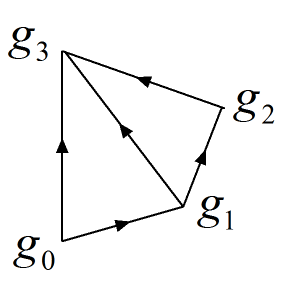}}}
\right) 
= \nu_3^{-1}(g_0,g_1,g_2,g_3)
\quad
\Psi\left(
\vcenter{\hbox{\includegraphics[scale=0.25]{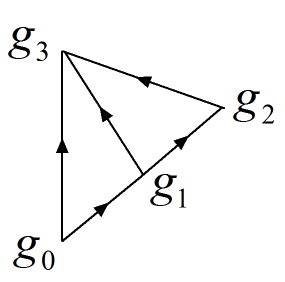}}}
\right) \nonumber \\
=& \nu_3^{-1}(g_0,g_1,g_2,g_3)
\quad
\Psi\left(
\vcenter{\hbox{\includegraphics[scale=0.25]{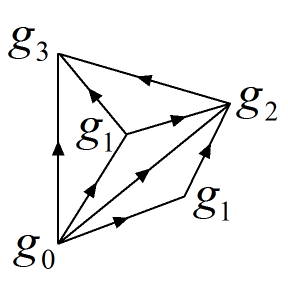}}}
\right),
\end{align}

In Fig. \ref{fig:2Dmove} of the Appendix, we list all possible (2-2) and (3-1) moves that are consistent with a branching structure.

\begin{figure}
\centering
\includegraphics[scale=0.45]{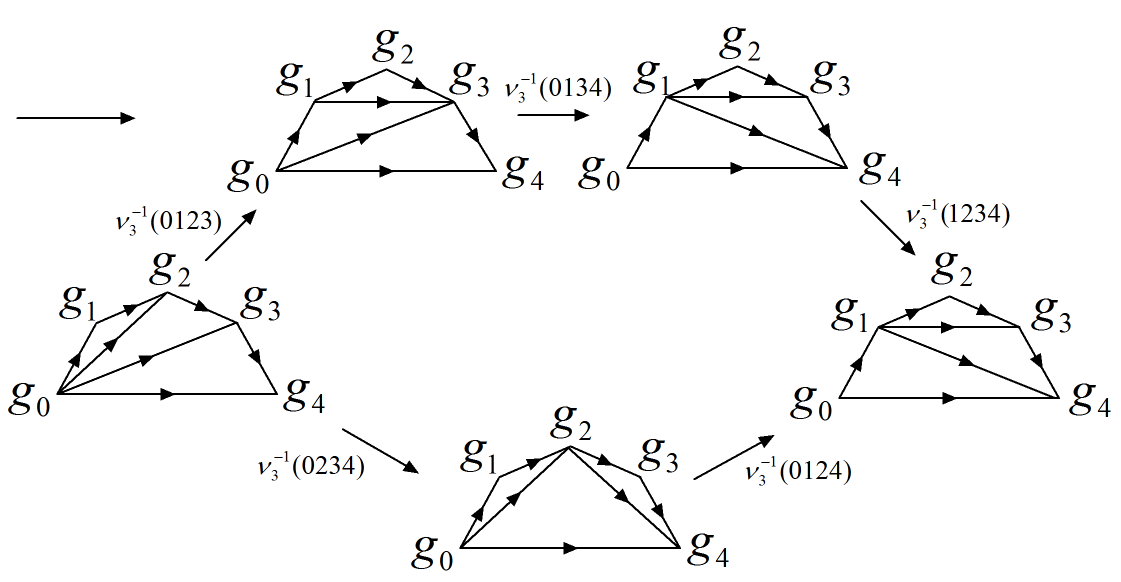}
\caption{The self consistent equation on a big patch. Mathematically it is known as the 3-cocycle equation.}
\label{fig:Pontagon}
\end{figure}

In the above, we discuss the SLU moves. The most important one is the standard (2-2) move in \eq{eq:2Dmove}. Similar to the 1D case, if we apply the (2-2) move for bigger patch as seen in Fig. \ref{fig:Pontagon}, we can derive the consistent conditions for $\nu_3$ describing fixed point wave functions:  
\begin{align}
(\dd\nu_3)(g_0,g_1,g_2,g_3,g_4) \equiv \frac{\nu_3(g_1,g_2,g_3,g_4) \nu_3(g_0,g_1,g_3,g_4) \nu_3(g_0,g_1,g_2,g_3)}{\nu_3(g_0,g_2,g_3,g_4) \nu_3(g_0,g_1,g_2,g_4)}=1.
\end{align}
Mathematically, this equation is known as the 3-cocycle equation.

Similar to the 1D case, we can use SLU to redefine the basis state $|\{g_l\}\rangle$ as
\begin{align}\label{eq:2Dbasis}
|\{g_l\}\rangle' = U_{\mu_2} |\{g_l\}\rangle = \prod_{\langle ijk \rangle} \mu_2(g_i,g_j,g_k)^{s_{\langle ijk \rangle}}   |\{g_l\}\rangle,
\end{align}
where $s_{\langle ijk \rangle}=\pm 1$ denotes the orientation of the triangle $\langle ijk \rangle$. One finds that the phase factor in \eq{eq:2Dmove} becomes
\begin{align}\label{eq:2Dnu3}
\nu_3'(g_0,g_1,g_2,g_3) = \nu_3(g_0,g_1,g_2,g_3) \frac{\mu_2(g_1,g_2,g_3)\mu_2(g_0,g_1,g_3)}{\mu_2(g_0,g_2,g_3)\mu_2(g_0,g_1,g_2)},
\end{align}
So the elements $\nu_3$ in the same group cohomology class in $H^3(G, U(1)_T)$ correspond to the same 2D BSPT phase.

Again, similar to the 1D case, the 2D SLU transformations can also be used to construct commuting projector parent Hamiltonian for these fixed point wave functions. The Hamiltonian term is a sequence of moves that change the group element of one vertex from $g_\ast$ to another $g_\ast'$. For example, we can consider the following moves for a triangular lattice:
\begin{align}
\tikzfig{scale=1}{
\coordinate (5) at ({0*60}:1);
\coordinate (4) at ($({1*60}:1)+(.1,0)$);
\coordinate (2) at ($({2*60}:1)+(.1,0)$);
\coordinate (0) at ({3*60}:1);
\coordinate (1) at ($({4*60}:1)-(.1,0)$);
\coordinate (3) at ($({5*60}:1)-(.1,0)$);;
\draw[->-=.5,black,line width=.6](0)--(1);
\draw[->-=.5,black,line width=.6](1)--(3);
\draw[->-=.5,black,line width=.6](3)--(5);
\draw[->-=.5,black,line width=.6](4)--(5);
\draw[->-=.5,black,line width=.6](2)--(4);
\draw[->-=.5,black,line width=.6](0)--(2);
\draw[-<-=.5,black,line width=.6](0,0)--(0);
\draw[-<-=.5,black,line width=.6](0,0)--(1);
\draw[-<-=.5,black,line width=.6](0,0)--(2);
\draw[->-=.5,black,line width=.6](0,0)--(3);
\draw[->-=.5,black,line width=.6](0,0)--(4);
\draw[->-=.5,black,line width=.6](0,0)--(5);
\node[left,scale=.8]at (0) {$g_0$};
\node[left,scale=.8]at (1) {$g_1$};
\node[left,scale=.8]at (2) {$g_2$};
\node[right,scale=.8]at (3) {$g_3$};
\node[right,scale=.8]at (4) {$g_4$};
\node[right,scale=.8]at (5) {$g_5$};
\node[xshift=9,yshift=-4,scale=.8]at (0,0) {$g_\ast$};
}
\!\!\!\rightarrow\!\!\!\!
\tikzfig{scale=1}{
\coordinate (5) at ({0*60}:1);
\coordinate (4) at ($({1*60}:1)+(.1,0)$);
\coordinate (2) at ($({2*60}:1)+(.1,0)$);
\coordinate (0) at ({3*60}:1);
\coordinate (1) at ($({4*60}:1)-(.1,0)$);
\coordinate (3) at ($({5*60}:1)-(.1,0)$);;
\draw[->-=.5,black,line width=.6](0)--(1);
\draw[->-=.5,black,line width=.6](1)--(3);
\draw[->-=.5,black,line width=.6](3)--(5);
\draw[->-=.5,black,line width=.6](4)--(5);
\draw[->-=.5,black,line width=.6](2)--(4);
\draw[->-=.5,black,line width=.6](0)--(2);
%
\draw[-<-=.5,black,line width=.6](0,0)--(1);
\draw[-<-=.5,black,line width=.6](0,0)--(2);
\draw[->-=.5,black,line width=.6](0,0)--(5);
\draw[->-=.5,black,line width=.6](1)--(2);
\draw[->-=.5,black,line width=.6](2)--(5);
\draw[->-=.5,black,line width=.6](1)--(5);
\node[left,scale=.8]at (0) {$g_0$};
\node[left,scale=.8]at (1) {$g_1$};
\node[left,scale=.8]at (2) {$g_2$};
\node[right,scale=.8]at (3) {$g_3$};
\node[right,scale=.8]at (4) {$g_4$};
\node[right,scale=.8]at (5) {$g_5$};
\node[xshift=5,yshift=-4,scale=.8]at (0,0) {$g_\ast$};
}
\!\!\!\rightarrow\!\!\!\!
\tikzfig{scale=1}{
\coordinate (5) at ({0*60}:1);
\coordinate (4) at ($({1*60}:1)+(.1,0)$);
\coordinate (2) at ($({2*60}:1)+(.1,0)$);
\coordinate (0) at ({3*60}:1);
\coordinate (1) at ($({4*60}:1)-(.1,0)$);
\coordinate (3) at ($({5*60}:1)-(.1,0)$);;
\draw[->-=.5,black,line width=.6](0)--(1);
\draw[->-=.5,black,line width=.6](1)--(3);
\draw[->-=.5,black,line width=.6](3)--(5);
\draw[->-=.5,black,line width=.6](4)--(5);
\draw[->-=.5,black,line width=.6](2)--(4);
\draw[->-=.5,black,line width=.6](0)--(2);
%
%
\draw[->-=.5,black,line width=.6](1)--(2);
\draw[->-=.5,black,line width=.6](2)--(5);
\draw[->-=.5,black,line width=.6](1)--(5);
\node[left,scale=.8]at (0) {$g_0$};
\node[left,scale=.8]at (1) {$g_1$};
\node[left,scale=.8]at (2) {$g_2$};
\node[right,scale=.8]at (3) {$g_3$};
\node[right,scale=.8]at (4) {$g_4$};
\node[right,scale=.8]at (5) {$g_5$};
}
\!\!\!\rightarrow\!\!\!\!
\tikzfig{scale=1}{
\coordinate (5) at ({0*60}:1);
\coordinate (4) at ($({1*60}:1)+(.1,0)$);
\coordinate (2) at ($({2*60}:1)+(.1,0)$);
\coordinate (0) at ({3*60}:1);
\coordinate (1) at ($({4*60}:1)-(.1,0)$);
\coordinate (3) at ($({5*60}:1)-(.1,0)$);;
\draw[->-=.5,black,line width=.6](0)--(1);
\draw[->-=.5,black,line width=.6](1)--(3);
\draw[->-=.5,black,line width=.6](3)--(5);
\draw[->-=.5,black,line width=.6](4)--(5);
\draw[->-=.5,black,line width=.6](2)--(4);
\draw[->-=.5,black,line width=.6](0)--(2);
%
\draw[-<-=.5,black,line width=.6](0,0)--(1);
\draw[-<-=.5,black,line width=.6](0,0)--(2);
\draw[->-=.5,black,line width=.6](0,0)--(5);
\draw[->-=.5,black,line width=.6](1)--(2);
\draw[->-=.5,black,line width=.6](2)--(5);
\draw[->-=.5,black,line width=.6](1)--(5);
\node[left,scale=.8]at (0) {$g_0$};
\node[left,scale=.8]at (1) {$g_1$};
\node[left,scale=.8]at (2) {$g_2$};
\node[right,scale=.8]at (3) {$g_3$};
\node[right,scale=.8]at (4) {$g_4$};
\node[right,scale=.8]at (5) {$g_5$};
\node[xshift=5,yshift=-5,scale=.8]at (0,0) {$g_\ast'$};
}
\!\!\!\rightarrow\!\!\!\!
\tikzfig{scale=1}{
\coordinate (5) at ({0*60}:1);
\coordinate (4) at ($({1*60}:1)+(.1,0)$);
\coordinate (2) at ($({2*60}:1)+(.1,0)$);
\coordinate (0) at ({3*60}:1);
\coordinate (1) at ($({4*60}:1)-(.1,0)$);
\coordinate (3) at ($({5*60}:1)-(.1,0)$);;
\draw[->-=.5,black,line width=.6](0)--(1);
\draw[->-=.5,black,line width=.6](1)--(3);
\draw[->-=.5,black,line width=.6](3)--(5);
\draw[->-=.5,black,line width=.6](4)--(5);
\draw[->-=.5,black,line width=.6](2)--(4);
\draw[->-=.5,black,line width=.6](0)--(2);
\draw[-<-=.5,black,line width=.6](0,0)--(0);
\draw[-<-=.5,black,line width=.6](0,0)--(1);
\draw[-<-=.5,black,line width=.6](0,0)--(2);
\draw[->-=.5,black,line width=.6](0,0)--(3);
\draw[->-=.5,black,line width=.6](0,0)--(4);
\draw[->-=.5,black,line width=.6](0,0)--(5);
\node[left,scale=.8]at (0) {$g_0$};
\node[left,scale=.8]at (1) {$g_1$};
\node[left,scale=.8]at (2) {$g_2$};
\node[right,scale=.8]at (3) {$g_3$};
\node[right,scale=.8]at (4) {$g_4$};
\node[right,scale=.8]at (5) {$g_5$};
\node[xshift=9,yshift=-5,scale=.8]at (0,0) {$g_\ast'$};
}.
\end{align}
We have shifted the lattice a little, such that the branching structure is induced by a time direction from left to right.
The first step of the above figures is a combination of three (2-2) moves. The second step is a (3-1) move that removes the vertex with group label $g_\ast$ at the center. The third step is a (1-3) move that create a vertex with group label $g_\ast'$ at the center. And the last step is a combination of three (2-2) moves that change the lattice to the original shape. Since our wave function is at the fixed-point, the terms for different vertices commute with each other.

\subsection{Fixed-point wave function and classification for BSPT phases in 3D}

The fixed-point wave functions for BSPT phases in 3D are similar to the 1D and 2D cases. We can again use the group element basis to construct the local Hilbert space on each vertex of arbitrary triangulation. The basic renormalization process is known as (2-3) and (2-0) Pachner move of triangulation of 3D manifold.

\begin{align}
|\Psi\rangle = \sum_{\text{all conf.}} \Psi\left( 
\tikzfig{scale=2}{
\pgfmathsetmacro{\x}{.2};
\pgfmathsetmacro{\y}{.3};
\draw[->-=.5,dashed] (0,0)--(\x,1+\y);
\draw[-<-=.5,dashed] (\x,1+\y)--(1+\x,\y);
\draw[->-=.5,dashed] (0,0)--(1+\x,\y);
\draw[->-=.7] (0,1)--(1,1);
\draw[->-=.5] (1,1)--(1+\x,1+\y);
\draw[->-=.5] (1+\x,1+\y)--(\x,1+\y);
\draw[->-=.5] (0,1)--(\x,1+\y);
\draw[->-=.5] (0,0)--(1,0);
\draw[->-=.7] (1,0)--(1,1);
\draw[->-=.5] (1,0)--(1+\x,\y);
\draw[->-=.5] (0,0)--(1,1);
\draw[->-=.5] (0,0)--(0,1);
\draw[->-=.5] (1+\x,\y)--(1,1);
\draw[->-=.5] (1,1)--(\x,1+\y);
\draw[->-=.5] (1+\x,\y)--(1+\x,1+\y);
\draw[->-=.7,dashed] (0,0)--(\x,\y);
\draw[->-=.5,dashed] (1+\x,\y)--(\x,\y);
\draw[->-=.5,dashed] (\x,\y)--(\x,1+\y);
\node [below left,scale=.8] at (0,0) {$g_0$};
\node [below right,scale=.8] at (1,0) {$g_1$};
\node [right,scale=.8] at (1+\x,\y) {$g_2$};
\node [left,scale=.8] at (\x,\y) {$g_3$};
\node [left,scale=.8] at (0,1) {$g_4$};
\node [right,scale=.8] at (1,1) {$g_5$};
\node [above right,scale=.8] at (1+\x,1+\y) {$g_6$};
\node [above left,scale=.8] at (\x,1+\y) {$g_7$};
}
\right) \stretchleftright{\Bigg|\ }{
\vcenter{\hbox{\includegraphics[scale=1]{\FigName\arabic{Num}.pdf}}}
}{\Big\rangle}.
\end{align}

An example of $2-3$ move (now we call it the standard $2-3$ move) is presented as follows:
\begin{align}
\label{eq:3Dmove}
\Psi\left(
\vcenter{\hbox{\includegraphics[scale=0.3]{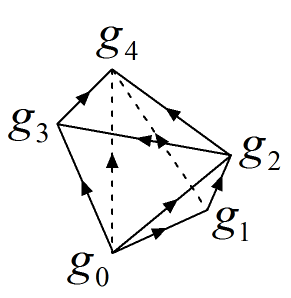}}}
\right)= 
\nu_4(g_0,g_1,g_2,g_3,g_4)
\quad
\Psi\left(
\vcenter{\hbox{\includegraphics[scale=0.3]{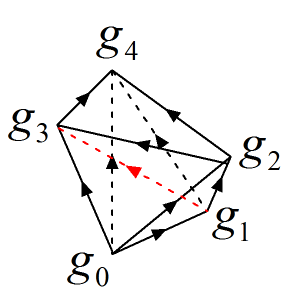}}}
\right),
\end{align}
Here, $\nu_4(g_0,g_1,g_2,g_3,g_4)$ is a $U(1)$-valued 4-cochain that is symmetric under $g$ action $\nu_4(gg_0,gg_1,gg_2,gg_3,gg_4)=\nu_4(g_0,g_1,g_2,g_3,g_4)$ ($\nu_4(gg_0,gg_1,gg_2,gg_3,gg_4)=\nu_4^*(g_0,g_1,g_2,g_3,g_4)$ if $g$ is anti-unitary) 

Again, apart from the (2-3) move, there are two (2-0) moves consisting with the branching structure that can change the total number of vertices for triangulations.  
\begin{align}
\label{eq:3D20a}
\Psi\left(
\vcenter{\hbox{\includegraphics[scale=0.3]{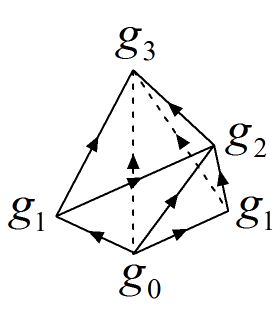}}}
\right)=\frac{1}{|G|^{1/2}}  
\Psi\left(
\vcenter{\hbox{\includegraphics[scale=0.3]{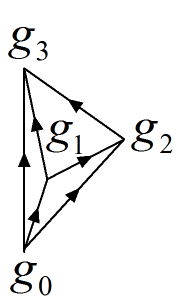}}}
\right),
\end{align}
and
\begin{align}
\label{eq:3D20b}
\Psi\left(
\vcenter{\hbox{\includegraphics[scale=0.3]{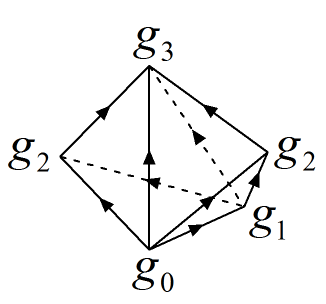}}}
\right)=\frac{1}{|G|^{1/2}}  
\Psi\left(
\vcenter{\hbox{\includegraphics[scale=0.3]{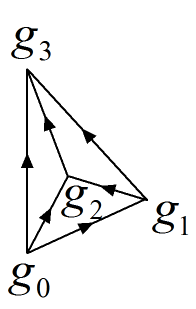}}}
\right),
\end{align}

Again, we add a normalization factor $|G|^{-1/2}$ in the front of the (2-0) move operator, for the vertex number is reduced by one from the left state to the right state. \footnote{In principle, we can also add an arbitrary phase factor into the above move. However, since such a phase factor must be a $U(1)$-valued function of group elements $g_0,g_1,g_2,g_3$, it can always be removed by the basis redefinition \eq{eq:3Dnu4}.}



It is easy to check that other (2-3) move with different branching structure can always be generated by the standard by the standard (2-3) move and (2-0)/(0-2) move. Moreover, the combination of (2-3) move and (2-0) move will further allow us to define a new set of renormalization move which reduces the number of vertices, namely, (3-1) move. In Figs.~\ref{fig:3Dmovea} and \ref{fig:3Dmoveb} of the Appendix, we list all possible (2-3) and (4-1) moves that admit a branching structure.

In the above, we discuss the SLU moves. The most important one is the standard $(2-3)$ move in \eq{eq:2Dmove}. Similar to the 1D and 2D cases, if we apply the (2-3) move for bigger patch, we can derive the consistent conditions for $\nu_4$ describing fixed point wave functions:  
\begin{align}
(\dd\nu_4)(g_0,g_1,g_2,g_3,g_4,g_5) \equiv \frac{\nu_4(g_1,g_2,g_3,g_4,g_5) \nu_4(g_0,g_1,g_3,g_4,g_5) \nu_4(g_0,g_1,g_2,g_3,g_5)}{\nu_4(g_0,g_2,g_3,g_4,g_5) \nu_4(g_0,g_1,g_2,g_4,g_5)\nu_4(g_0,g_1,g_3,g_2,g_4)}=1.
\end{align}

Similar to the 1D and 2D cases, we can use SLU to redefine the basis state $|\{g_l\}\rangle$ as
\begin{align}\label{eq:3Dbasis}
|\{g_l\}\rangle' = U_{\mu_3} |\{g_l\}\rangle = \prod_{\langle ijk \rangle} \mu_2(g_i,g_j,g_k)^{s_{\langle ijk \rangle}}   |\{g_l\}\rangle,
\end{align}
one finds that the phase factor in \eq{eq:2Dmove} becomes
\begin{align}\label{eq:3Dnu4}
\nu_4'(g_0,g_1,g_2,g_3,g_4)=\nu_4(g_0,g_1,g_2,g_3,g_4) \frac{\mu_3(g_1,g_2,g_3,g_4)\mu_3(g_0,g_1,g_3,g_4)\mu_3(g_0,g_1,g_2,g_3)}{\mu_3(g_0,g_2,g_3,g_4)\mu_3(g_0,g_1,g_2,g_4)},
\end{align}
So the elements $\nu_4$ in the same group cohomology class in $H^4(G, U(1)_T)$ correspond to the same 3D BSPT phase.

We can also use the above moves to construct a 3D commuting projector parent Hamiltonian. Each term of the Hamiltonian is a sequence of 3D moves that changes the label of a vertex from $g_\ast$ to $g_\ast'$. All the terms commute with each other because the wave function is at the fixed-point.

Finally, we notice that for anti-unitary symmetry, e.g, time reversal symmetry, the above construction and classification scheme is not complete. It has been pointed out \cite{wen15} that the decoration of $E_8$ state on the $G$-symmetry domain walls will give rise to new BSPT states beyond group cohomology classification. Apparently, the data $H^1(G,\mathbb Z_T)$ classifies such a decorating pattern and the corresponding additional BSPT states. Since $H^1(G,\mathbb Z_T)$ is trivial for all unitary symmetry group $G$ and $H^1(\mathbb Z_2^T,\mathbb Z_T)=\mathbb Z_2$ for the (anti-unitary) time-reversal symmetry, we understand why the beyond group cohomology BSPT phases only arise for anti-unitary symmetry. Thus, we conclude that the two cohomology groups of the symmetry group $G$: $H^1(G,\mathbb Z_T)$ and $H^4(G, U(1)_T)$ give rise to a complete classification of BSPT phases in 3D.  

\section{FSLU transformation and FSPT phases}
\label{sec:FSLU}
\NewFigName{figures/Fig_FSLU_}

\subsection{Fermionic symmetric local unitary transformations}
\label{sec:FSLU_}

In Ref.~\onlinecite{Gu2015}, it was shown that fermionic local unitary (FLU) transformations can be used to define and classify intrinsic topological phases for interacting fermion systems. The Fock space structure and fermion parity conservation symmetry of fermion systems can be naturally encoded into FLU transformations.   It is well-known that the finite-time FLU evolution is closely
related to fermionic quantum circuits with finite
depth, which is defined through piecewise FLU operators.  A
piecewise FLU operator has the form
$ U_{pwl}= \prod_{\v i} e^{-i  H_f(\v i)}\equiv \prod_{\v i}U(\v i)$,
where $ H_f(\v i) $ is a fermionic Hermitian operator and $U(\v i)$ is the corresponding fermionic unitary operator defined in Fock space that preserve fermion parity (e.g., contains even number of fermion creation and annihilation operators) and act on a region labeled by $\v
i$. Note that regions labeled by different $\v i$'s are not
overlapping. We further require that the size of each region is less
than some finite number $l$. The unitary operator $U_{pwl}$
defined in this way is called a piecewise fermionic local
unitary operator with range $l$.  A fermion quantum
circuit with depth $M$ is given by the product of $M$
piecewise fermionic local unitary operators:
$U^M_{circ}= U_{pwl}^{(1)} U_{pwl}^{(2)} \cdots
U_{pwl}^{(M)}$.
It is believed that any FLU evolution can be
simulated with a constant depth fermionic quantum circuit and
vice versa. Therefore, the equivalence relation between gapped states in interacting fermion systems can be rewritten in terms of constant
depth fermionic quantum circuits:
\begin{equation}
|\psi(1)\rangle  \sim |\psi(0)\rangle  \text{ iff } |\psi(1)\rangle  = U^M_{circ} |\psi(0)\rangle 
\end{equation}
Thus, we can use the term FLU transformation to refer to both FLU evolution and constant depth fermionic quantum circuit. From the definition of FSPT state, it is easy to see that (in the absence of global symmetry):
\begin{equation}
  |\text{FSPT}\rangle  = U^M_{circ} |\text{Trivial}\rangle \label{FLU}
\end{equation} 
Namely, an FSPT state can be connected to a trivial state (e.g., a product state) vial FLU transformation (in the absence of global symmetry). Similar to the BSPT case, Eq.(\ref{FLU}) implies that the support space of any FSPT in region $A$ must be one dimensional. This is simply because a trivial state (e.g., a product state) has a one dimensional support space, and any FSPT state will become a product state via a proper local basis change (induced by a FLU transformation). 

In the presence of global symmetry, we can further introduce the notion of invertable fermionic symmetric local unitary (FSLU) transformations to define and classify FSPT phases in interacting fermion systems. By FSLU transformation, we mean that the corresponding piecewise FLU operator is invariant under total symmetry group $G_f$.

\subsection{Layers of degrees of freedom}
\label{sec:FSLU:dof}

The are at most four layers of degrees of freedom in total in our fixed-point wave function of FSPT state (up to four spacetime dimensions). The bosonic states are always at the vertices. And the fermionic degrees of freedom (complex fermions, Majorana fermions and 2D $p+ip$ chiral superconductors) are decorated on the intersecting sub-manifold of the bosonic state. In summary, the degrees of freedom of our FSPT states are \footnote{We assume $G_b$ to be a finite group in this paper.}
\begin{itemize}
\item
$|G_b|$ level bosonic (spin) state $|g_i\rangle$ ($g_i\in G_b$) on each vertex $i$.
\item
$|G_b|$ species of complex fermions $c_{ij...k}^\s$ ($\s\in G_b$) on each codimension-0 simplex $\langle ij...k\rangle$.
\item
$|G_b|$ species of Majorana fermions $\g_{ij...k,A}^\s$ and $\g_{ij...k,B}^\s$ ($\s\in G_b$), which come from complex fermions $a_{ij...k}^\s=(\g_{ij...k,A}^\s+i\g_{ij...k,B}^\s)/2$, on the two sides of each codimension-1 simplex $\langle ij...k\rangle$.
\item
$|G_b|$ species of 2D $p+ip$ chiral superconductors (may have several copies) on the dual surface of each codimension-2 simplex. The chiral Majorana modes along the edge of the dual surface are labelled by $\psi_{ij...k,L}^\s$ or $\psi_{ij...k,R}^\s$ depending on the chirality.
\end{itemize}
The above degrees of freedom has different symmetry transformation rules. The
symmetry transformation of $G_b$ on bosonic state is the same as that in the BSPT states ($g\in G_b$):
\begin{align}
U(g) |g_i\rangle &= |gg_i\rangle,
\end{align}
For complex fermions, we choose the symmetry transformations under $G_b$ to be
\begin{align}\label{symm:c}
U(g) c_{ij...k}^\s U(g)^\dagger &= (-1)^{\om_2(g,\s)} c_{ij...k}^{g\s}.
\end{align}
The symmetry transformation rules of Majorana fermions $\g_{ij...k,A/B}^\s$ are induced by the transformation of complex fermion $a_{ij...k}^\s$:
\begin{align}
U(g) a_{ij...k}^\s U(g)^\dagger &= (-1)^{\om_2(g,\s)} a_{ij...k}^{g\s},\\
U(g) \g_{ij...k,A}^\s U(g)^\dagger &= (-1)^{\om_2(g,\s)} \g_{ij...k,A}^{g\s},\\
U(g) \g_{ij...k,B}^\s U(g)^\dagger &= (-1)^{\om_2(g,\s)+s_1(g)} \g_{ij...k,B}^{g\s},
\end{align}
And the symmetry transformations of chiral Majorana modes on the boundary of decorated $p+ip$ superconductors are chosen to be
\begin{align}
U(g) \psi_{ij...k,R}^\s U(g)^\dagger &= (-1)^{\om_2(g,\s)} \psi_{ij...k,g(R)}^{g\s},\\
U(g) \psi_{ij...k,L}^\s U(g)^\dagger &= (-1)^{\om_2(g,\s)+s_1(g)} \psi_{ij...k,g(L)}^{g\s}.
\end{align}
We will discuss more about why we choose the transformation rules for Majorana modes in section~\ref{sec:3D:pip_symm}.

In this way, the $G_b$ species of fermions span a space that support a projective representation of $G_b$ with coefficient $P_f^{\om_2}$:
\begin{align}\label{U_proj}
U(g)U(h) = P_f^{\om_2(g,h)} U(gh),
\end{align}
with $P_f=-1$ when acting on fermion parity odd states. We note that the projective representation $U$ of $G_b$ is equivalent to a linear representation $\tilde U$ of $G_f$ by
\begin{align}\label{UU}
\tilde U(P_f^n g) := P_f^n U(g).
\end{align}
One can check directly that $\tilde U$ is indeed a genuine linear representation of $G_f$:
\begin{align}
\tilde U \left( P_f^m g \right) \tilde U \left( P_f^n h \right) = \tilde U \left( P_f^m g \cdot P_f^n h \right),
\end{align}
where the dot product in $G_f$ is defined in \eq{multi}.

In the previous constructions of FSPT state for $G_f=\Zf\times G_b$, we put only one species of fermions on each simplex \cite{GuWen2014,WangGu2017}. They transform trivially under the action of $G_b$ (and $G_f$). To construct FSPT state for $G_f=\Zf\timesw G_b$, we need projective representation of $G_b$ with coefficient $\om_2$ to make it a linear representation of $G_f$. We have chosen the canonical $|G_b|$-dimensional projective representation \eq{symm:c}, which can be constructed for arbitrary finite symmetry group $G_b$. Although there are $|G_b|$ species fermions $c_{ij...k}^\s$ ($\s\in G_b$) on each simplex $\langle ij...k\rangle$, (at most) only one of them $c_{ij...k}^{g_i}$ will be decorated or in the occupied state, and all others fermion species $c_{ij...k}^\s$ ($\s\neq g_i$) are in the vacuum states. In such a way, the FSPT constructions for symmetry group $G_f=\Zf\times G_b$ can be generalized to the case of $G_f=\Zf\timesw G_b$.

\subsection{Symmetry conditions and consistency equations}
\label{sec:FSLU:symm}

Since we are constructing FSPT states, the $F$ moves should be compatible with the symmetry action $U(g)$ defined in Section~\ref{sec:FSLU:dof}. To be more precise, let us consider the following two dimensional commuting digram:
\begin{align}\label{FUg}
\tikzfig{}{
\node (p1) at (0,0) {$|\Psi_\cT(\{g_i\})\rangle$};
\node [below=of p1] (pp1) {$|\Psi_\cT(\{gg_i\})\rangle$};
\node [right=of pp1,xshift=25pt] (pp2) {$|\Psi_{\cT'}(\{gg_i\})\rangle$};
\node [above=of pp2] (p2) {$|\Psi_{\cT'}(\{g_i\})\rangle$};
\draw[<-] (p1)--(p2) node [midway,above] {$F(\{g_i\})$};
\draw[<-] (pp1)--(pp2) node [midway,above] {$F(\{gg_i\})$};
\draw[->] (p1)--(pp1) node [midway,left] {$U(g)$};
\draw[->] (p2)--(pp2) node [midway,right] {$U(g)$};
\node [above=of p1,yshift=-15pt] (p1_up) {$\vdots$};
\draw[->] (p1_up)--(p1) node [] {};
\node [above=of p2,yshift=-15pt] (p2_up) {$\vdots$};
\draw[->] (p2_up)--(p2) node [] {};
\node [below=of pp1,yshift=15pt] (pp1_down) {$\vdots$};
\draw[<-] (pp1_down)--(pp1) node [] {};
\node [below=of pp2,yshift=15pt] (pp2_down) {$\vdots$};
\draw[<-] (pp2_down)--(pp2) node [] {};
\node [left=of p1] (p1_left) {$\cdots$};
\draw[->] (p1)--(p1_left) node [] {};
\node [right=of p2] (p2_right) {$\cdots$};
\draw[<-] (p2)--(p2_right) node [] {};
\node [left=of pp1,xshift=2.5pt] (pp1_left) {$\cdots$};
\draw[->] (pp1)--(pp1_left) node [] {};
\node [right=of pp2,xshift=-2.5pt] (pp2_right) {$\cdots$};
\draw[<-] (pp2)--(pp2_right) node [] {};
}
\end{align}
where the horizontal $F$ move changes the triangulations of the spacial manifold, and the vertical symmetry action $U(g)$ changes the bosonic degrees of freedom from $\{g_i\}$ to $\{gg_i\}$ (similar for the fermionic ones). The outside parts ``...'' correspond to other triangulations of the spacial manifold and other symmetry actions on the states. This two dimensional diagram should commute for arbitrary horizontal Pachner move of arbitrary triangulations and vertical symmetry action with arbitrary $g\in G_b$. The requirement of symmetric fixed-point wave function implies the following conditions:
\begin{enumerate}
\item
The diagram \eq{FUg} commutes, i.e.,
\begin{align}\label{Fg=gF}
F(\{gg_i\})=U(g)F(\{g_i\})U(g)^\dagger, \quad \forall g\in G_b.
\end{align}
\item
The vertical direction of diagram \eq{FUg} should form a projective representation of $G_b$ with coefficient $\om_2$, i.e., $U(g)U(h)=P_f^{\om_2(g,h)}U(gh)$ when acting on these states. This projective representation of $G_b$ induces a linear representation of $G_f$ by \eq{UU}.
\item
The horizontal direction of diagram \eq{FUg} should satisfy some coherence equation, which is known as super (fermionic) pentagon equation in 2D. In the FSPT setting, it is a twisted cocycle equation.
\end{enumerate}

For the triangulations of $d$-dimensional space manifold, the Pachner move involves $d+2$ vertices. So the basic $F$ move can be denoted by $F(g_0,g_1,...,g_{d+1})$. With the help of \eq{Fg=gF}, we can obtain the generic $F$ move
\begin{align}
F(g_0,g_1,...,g_{d+1}) = \emp^{g_0}F(e,g_0^{-1}g_1,...,g_0^{-1}g_{d+1}) = U(g_0) F(e,g_0^{-1}g_1,...,g_0^{-1}g_{d+1}) U(g_0)^\dagger,
\end{align}
provided that we have defined the \emph{standard} $F$ symbol with the first argument being the identity element $e\in G_b$. Using this definition of $F$ move, \eq{Fg=gF} is automatically satisfied. This is because of the following commuting diagram
\begin{align}
\tikzfig{scale=1}{
\node (up) at (3,1.5) {$F(e,g_0^{-1}g_1,...,g_0^{-1}g_{d+1})$};
\node (0) at (0,0) {$F(g_0,g_1,...,g_{d+1})$};
\node (1) at (6,0) {$F(gg_0,gg_1,...,gg_{d+1})$};
\draw[->] (up)--(0) node [midway,left,yshift=3] {$U(g_0)$};
\draw[->] (up)--(1) node [midway,right,,yshift=3] {$U(gg_0)$};
\draw[->,dashed] (0)--(1) node [midway,yshift=7] {$U(g)$};
}.
\end{align}
We can deduce the dashed arrow $U(g)$ from solid arrows $U(g_0)$ and $U(gg_0)$, due to $U(g)U(h) = P_f^{\om_2(g,h)} U(gh)$ and the fermion parity even property of the $F$ operators.

%


\section{Fixed-point wave function and classification of FSPT states in 1D}
\label{sec:1D}
\NewFigName{figures/Fig_1D_}

In this section, we give the explicit constructions and classifications of 1D FSPT states. The fixed-point wave functions are obtained by decorating complex fermions to BSPT states consistently. Formally, the wave function is a superposition of all basis states $|\{g_i\}\rangle$ with legitimate decorations:
\begin{align}
\label{eq:1Dwf}
|\Psi\rangle = \sum_{\{g_i\}} \Psi(\{g_i\}) \ |\{g_i\}\rangle
=\sum_{\text{all conf.}} \Psi\left(
\tikzfig{scale=.6}{
\pgfmathsetmacro{\N}{5};
\pgfmathsetmacro{\Nm}{4};
\foreach \i in {0,1,...,\N} {
	\coordinate (\i) at (\i,0);
}
\foreach \i in {0,1,...,\Nm} {
	\pgfmathsetmacro{\c}{\i+1};
	\draw[line width=.5] (\i)--(\c);
}
\foreach \i in {0,1,...,\N} {
	\node[circle,draw=black,fill=black,scale=.1] at (\i) {};
}
\foreach \i in {2,4} {
	\pgfmathsetmacro{\c}{\i+1};
	\node[circle,shading=ball,scale=.4] at ($(\i)!.5!(\c)$) {};
}
\foreach \i in {0,1,3} {
	\pgfmathsetmacro{\c}{\i+1};
	\node[circle,draw=blue,fill=white,scale=.35] at ($(\i)!.5!(\c)$) {};
}
}
\right) |
\vcenter{\hbox{\includegraphics[scale=1]{\FigName\arabic{Num}.pdf}}}
\rangle.
\end{align}
The basis state $|\{g_i\}\rangle$ is a state (with vertex $i$ labelled by $g_i\in G_b$) decorated by complex fermions $c_{ij}^\s$ at link $\langle ij\rangle$. The constructed fixed-point wave function $|\Psi\rangle$ should be both symmetric and topological (invariant under retriangulations of the lattice). As will shown below, these constraints would give us the consistency conditions for the 1D FSPT classifications summarized in Eqs.~(\ref{1D:data}) - (\ref{1D:Gamma}).

We note that the 1D Kitaev chain is a fermionic invertible topological order. Since it does not need any bosonic symmetry protection ($\Zf$ can not be broken), we do not consider it as FSPT state. The Kitaev chain layer is useful when considering it as the ASPT on the boundary of a 2D FSPT state. The 2D classification data $n_2$ will be trivialized.

This section will be organized as follows. The two layers of degrees of freedom (bosonic states and complex fermions) are introduced in section~\ref{sec:1D:dof}. In section~\ref{sec:1D:docoration}, we propose the procedures of symmetrically decorating complex and Majorana fermions to BSPT states. Then the construction and the consistency equations of the FSLU transformations are discussed in sections~\ref{sec:1D:F} and \ref{sec:1D:dnu}, respectively.

\subsection{Two layers of degrees of freedom}
\label{sec:1D:dof}

The basic idea to construct FSPT states is to decorate complex fermions to the BSPT states. Therefore, there are two layers of degrees of freedom, including the bosonic ones, in the 1D lattice model:
\begin{itemize}
\item
$|G_b|$ level bosonic (spin) state $|g_i\rangle$ ($g_i\in G_b$) on each vertex $i$.
\item
$|G_b|$ species of complex fermions $c_{ij}^\s$ ($\s\in G_b$) at the center of each link $\langle ij\rangle$.
\end{itemize}
These degrees of freedom are summarized in one unit cell in the following figure:
\begin{align}
\tikzfig{scale=1}{
\draw[->-=.8,line width=1](0,0)--(1,0);
\node[below] at (0,0) {$g_i$};
\node[below] at (1,0) {$g_j$};
\draw[line width=1] (0,-.017)--(0,.1);
\draw[line width=1] (1,-.017)--(1,.1);
\node[circle,shading=ball,scale=.6] at (.5,0) {};
\node[blue,below,scale=.9] at (.5,0) {$c_{ij}^\s$};
}
(g_i,g_j,\s\in G_b).
\end{align}
Here, we choose the link direction from vertex $i$ to vertex $j$ (from left to right). The vertices are labelled by $g_i$ and $g_j$ which are elements of $G_b$. Blue ball is the decorated complex fermion $c_{ij}^\s$ ($\s\in G_b$) at the center of link $\langle ij\rangle$.

The symmetry transformations of these degrees of freedom under $G_b$ are the same as the discussions in section~\ref{sec:FSLU:dof}. To be more specific to the 1D case, we summarize them as ($g,g_i,\s\in G_b$):
\begin{align}
\label{1D:symm:gi}
U(g) |g_i\rangle &= |gg_i\rangle,\\
\label{1D:symm:c}
U(g) c_{ij}^\s U(g)^\dagger &= (-1)^{\om_2(g,\s)} c_{ij}^{g\s}.
\end{align}
While the bosonic degrees of freedom on each vertex form a linear representation of $G_b$, the complex fermions form projective representation of $G_b$ with coefficient $(-1)^{\om_2}$. In this way, they all transform linearly under the action of $G_f$ defined by \eq{UU}.

Although there are $|G_b|$ species of complex fermions in the Hilbert space of the system, we will see later that (at most) only one of them is decorated nontrivially in the fixed-point wave function. If we consider the case of $\om_2=0$ (i.e., $G_f=\Zf\times G_b$), the symmetry transformation rules are independent of group element label $\s$ of the fermions [see \eq{1D:symm:c}]. Therefore, we can reduce these $|G_b|$ species of fermions to only one species without group element label. The resulting states are exactly the ones studied in Refs.~\onlinecite{GuWen2014,WangGu2017}.

\subsection{Decoration of complex fermions}
\label{sec:1D:docoration}

In the group cohomology theory of BSPT phases \cite{chen13}, the fixed-point wave functions are constructed as superpositions of all basis state $|\{g_i\}\rangle$. The coefficients in front of these basis states are $U(1)$-valued cocycles. To construct FSPT states, we have introduced the fermionic degrees of freedom associated to the basis states in the previous section. In the following, we would discuss the detailed procedures of systematically decorating complex fermions. These decorations should be designed to be symmetric under symmetry actions.

The complex fermion decoration is specified by a $\Z_2$-valued 1-cochain $n_1\in C^1(G_b,\Z_2)$, which is the first classification data for 1D FSPT phases. If $n_1(g_i,g_j)=0$, all the modes of complex fermions $c_{ij}^\s$ ($\s\in G_b$) at link $\langle ij\rangle$ are unoccupied (shown by blue circles in figures). On the other hand, if $n_1(g_i,g_j)=1$, exactly one complex fermion $c_{ij}^{g_i}$ will be decorated at the center of the oriented link $\langle ij\rangle$ (shown by filled blue balls in figures). And all other complex fermions $c_{ij}^\s$ ($\s\neq g_i$) are still in vacuum states.

The above complex fermion decoration rule is $G_b$-symmetric. Under a $U(g)$-action, the vertex labels of link $\langle ij\rangle$ become $gg_i$ and $gg_j$. According to the decoration rule, the decorated complex fermion [if $n_1(gg_i,gg_j) = n_1(g_i,g_j) = 1$] should be $c_{ij}^{gg_i}$, which is exactly the complex fermion $c_{ij}^{g_i}$ by a $U(g)$-action.

\subsection{$F$ moves}
\label{sec:1D:F}

For a fixed triangulation of spacial manifold, we can decorate complex fermions symmetrically as discussed above. However, we want to construct fixed-point wave functions that are invariant under retriangulation of the space. To connect different triangulations, there are FSLU transformations for each Pachner move. For the 1D lattice, there is essentially only one Pachner move given by
\begin{align}
\label{1D:Fmove}
\Psi\left(
\!\!\!\!
\tikzfig{scale=1}{
\coordinate (0) at (0,0);
\coordinate (1) at (1,0);
\node[below] at (0) {$g_0$};
\node[below] at (1) {$g_2$};
\draw[line width=1] (0,-.017)--(0,.1);
\draw[line width=1] (1,-.017)--(1,.1);
\draw[->-=.8,line width=1] (0)--(1);
\node[circle,draw=blue,fill=white,scale=.55] at ($(0)!.5!(1)$) {};
\node[blue,below,scale=1,scale=.9] at ($(0)!.5!(1)$) {$c_{02}^{g_0}$};
\pgfmathsetmacro{\dx}{.15};
\pgfmathsetmacro{\dy}{.2};
\node[]at(0,\dy){};
}
\!\!\!
\right)
\quad = \quad
F(g_0,g_1,g_2)
\quad
\Psi\left(
\!\!\!\!
\tikzfig{scale=1}{
\coordinate (0) at (0,0);
\coordinate (1) at (1,0);
\coordinate (2) at (2,0);
\node[below] at (0) {$g_0$};
\node[below] at (1) {$g_1$};
\node[below] at (2) {$g_2$};
\draw[line width=1] (0,-.017)--(0,.1);
\draw[line width=1] (1,-.017)--(1,.1);
\draw[line width=1] (2,-.017)--(2,.1);
\draw[->-=.8,line width=1] (0)--(1);
\draw[->-=.8,line width=1] (1)--(2);
\node[circle,draw=blue,fill=white,scale=.55] at ($(0)!.5!(1)$) {};
\node[circle,draw=blue,fill=white,scale=.55] at ($(1)!.5!(2)$) {};
\node[blue,below,scale=1,scale=.9] at ($(0)!.5!(1)$) {$c_{01}^{g_0}$};
\node[blue,below,scale=1,scale=.9] at ($(1)!.5!(2)$) {$c_{12}^{g_1}$};
\pgfmathsetmacro{\dx}{.15};
\pgfmathsetmacro{\dy}{.2};
\node[]at(0,\dy){};
}
\!\!\!
\right),
\end{align}
where the FSLU $F$ operator is defined as
\begin{align}
\label{1D:F}
F(g_0,g_1,g_2) &= 
|G_b|^{1/2}
\nu_2(g_0,g_1,g_2)
\big(c^{g_0}_{02}\big)^{\dagger n_1(g_0,g_2)}
\big(c^{g_0}_{01}\big)^{n_1(g_0,g_1)}
\big(c^{g_1}_{12}\big)^{n_1(g_1,g_2)}
.
\end{align}

In the above expression of $F$ symbol, $|G_b|^{1/2}$ is the normalization factor because the number of the lattice sites is reduced by one. $\nu_2(g_0,g_1,g_2)$ is a $U(1)$ phase factor depending on three group elements of $G_b$. For BSPT states, the $F$ operator has only these two bosonic factors. For FSPT states, however, there are complex fermion term of the form $c^\dagger c c$. The complex fermion term annihilate the possibly decorated (depending on $n_1$) complex fermions $c^{g_1}_{12}$ and $c^{g_0}_{01}$ on the two links in the right-hand-side state, and create a new complex fermion $c_{02}^{g_0}$ at the center of link $\langle 02\rangle$ in the left-hand-side state.

As discussed above, the $F$ move should be a FSLU operator. Therefore, it should be both fermion parity even and symmetric under $G_b$-action. These two conditions give us several consistency equations for the classification data $n_1$ and $\nu_2$.

\subsubsection{Fermion parity conservation}

Since the complex fermions are decorated according to $n_1(g_i,g_j)$, the complex fermion parity change of the $F$ move \eq{1D:Fmove} is
\begin{align}
\Delta P_f^c (012) = (-1)^{n_1(g_0,g_2) + n_1(g_0,g_1) + n_1(g_1,g_2)} = (-1)^{\dd n_1(g_0,g_1,g_2)}.
\end{align}
As a result, the conservation of fermion parity under the $F$ move enforces the condition
\begin{align}\label{1D:dn1}
\dd n_1=0\quad \text{(mod 2)},
\end{align}
which is the cocycle equation for the decoration data $n_1$.

\subsubsection{Symmetry condition}
\label{sec:1D:F:symm}

The $F$ move should also be consistent with the symmetry actions [see \eq{FUg}]. In 1D, we have the following commuting diagram
\begin{align}\label{1D:FUg}
\tikzfig{}{
\node (p1) at (0,0) {
\begin{tikzpicture}[]
\coordinate (0) at (0,0);
\coordinate (1) at (1,0);
\node[below] at (0) {$g_0$};
\node[below] at (1) {$g_2$};
\draw[line width=.9] (0,-.017)--(0,.1);
\draw[line width=.9] (1,-.017)--(1,.1);
\draw[->-=.8,line width=.9] (0)--(1);
\node[circle,draw=blue,fill=white,scale=.55] at ($(0)!.5!(1)$) {};
\pgfmathsetmacro{\dx}{.15};
\pgfmathsetmacro{\dy}{.2};
\end{tikzpicture}
};
\node (pp1) at (0,-2) {
\begin{tikzpicture}[]
\coordinate (0) at (0,0);
\coordinate (1) at (1,0);
\node[below] at (0) {$gg_0$};
\node[below] at (1) {$gg_2$};
\draw[line width=.9] (0,-.017)--(0,.1);
\draw[line width=.9] (1,-.017)--(1,.1);
\draw[->-=.8,line width=.9] (0)--(1);
\node[circle,draw=blue,fill=white,scale=.55] at (.5,0) {};
\pgfmathsetmacro{\dx}{.15};
\pgfmathsetmacro{\dy}{.2};
\end{tikzpicture}
};
\node (pp2) at (5.3,-2) {
\begin{tikzpicture}[]
\coordinate (0) at (0,0);
\coordinate (1) at (1,0);
\coordinate (2) at (2,0);
\node[below] at (0) {$gg_0$};
\node[below] at (1) {$gg_1$};
\node[below] at (2) {$gg_2$};
\draw[line width=.9] (0,-.017)--(0,.1);
\draw[line width=.9] (1,-.017)--(1,.1);
\draw[line width=.9] (2,-.017)--(2,.1);
\draw[->-=.8,line width=.9] (0)--(1);
\draw[->-=.8,line width=.9] (1)--(2);
\node[circle,draw=blue,fill=white,scale=.55] at ($(0)!.5!(1)$) {};
\node[circle,draw=blue,fill=white,scale=.55] at ($(1)!.5!(2)$) {};
\pgfmathsetmacro{\dx}{.15};
\pgfmathsetmacro{\dy}{.2};
\end{tikzpicture}
};
\node (p2) at (5.3,0) {
\begin{tikzpicture}[]
\coordinate (0) at (0,0);
\coordinate (1) at (1,0);
\coordinate (2) at (2,0);
\node[below] at (0) {$g_0$};
\node[below] at (1) {$g_1$};
\node[below] at (2) {$g_2$};
\draw[line width=.9] (0,-.017)--(0,.1);
\draw[line width=.9] (1,-.017)--(1,.1);
\draw[line width=.9] (2,-.017)--(2,.1);
\draw[->-=.8,line width=.9] (0)--(1);
\draw[->-=.8,line width=.9] (1)--(2);
\node[circle,draw=blue,fill=white,scale=.55] at ($(0)!.5!(1)$) {};
\node[circle,draw=blue,fill=white,scale=.55] at ($(1)!.5!(2)$) {};
\pgfmathsetmacro{\dx}{.15};
\pgfmathsetmacro{\dy}{.2};
\end{tikzpicture}
};
\draw[<-] (p1)--(p2) node [midway,above] {$F(g_0,g_1,g_2)$};
\draw[<-] (pp1)--(pp2) node [midway,above] {$F(gg_0,gg_1,gg_2)$};
\draw[->] (p1)--(pp1) node [midway,left] {$U(g)$};
\draw[->] (p2)--(pp2) node [midway,right] {$U(g)$};
}
\end{align}
or the symmetry condition for $F$ operators
\begin{align}\label{1D:Fg}
F(gg_0,gg_1,gg_2)=U(g)F(g_0,g_1,g_2)U(g)^\dagger.
\end{align}
As discussed in Section~\ref{sec:FSLU:symm}, the above equation can be viewed as a definition of the generic $F(g_0,g_1,g_2)$ in terms of the \emph{standard} $F$ move $F(e,g_0^{-1}g_1,g_0^{-1}g_2)$:
\begin{align}
F(g_0,g_1,g_2) = \emp^{g_0}F(e,g_0^{-1}g_1,g_0^{-1}g_2) := U(g_0) F(e,g_0^{-1}g_1,g_0^{-1}g_2) U(g_0)^\dagger.
\end{align}
Therefore, we only need to fix the expression of the standard $F$ move $F(e,g_0^{-1}g_1,g_0^{-1}g_2)$, and all other non-standard $F$ moves are obtained by a symmetry action on the standard one. The explicit expression of the standard $F$ move $F(e,g_0^{-1}g_1,g_0^{-1}g_2)$ is given by
\begin{align}\label{1D:Fe}
F(e,g_0^{-1}g_1,g_0^{-1}g_2)
=
|G_b|^{1/2}
\nu_2(g_0^{-1}g_1,g_1^{-1}g_2)
\big(c^{e}_{02}\big)^{\dagger n_1(e,g_0^{-1}g_2)}
\big(c^{e}_{01}\big)^{n_1(e,g_0^{-1}g_1)}
\left(c^{g_0^{-1}g_1}_{12}\right)^{n_1(g_0^{-1}g_1,g_0^{-1}g_2)}
.
\end{align}
Note that the $U(1)$ coefficient in the standard $F$ move is chosen to be the inhomogeneous cochain $\nu_2(g_0^{-1}g_1,g_1^{-1}g_2):=\nu_2(e,g_0^{-1}g_1,g_0^{-1}g_2)$. And we do not impose the condition ``$\nu_2(gg_0,gg_1,gg_2)=\nu_2(g_0,g_1,g_2)$'' in priori. In fact, as shown below, this condition does not hold in general.

We can apply a $U(g_0)$-action on the standard $F$ move \eq{1D:Fe}, and compare it with the generic expression \eq{1D:F}. The symmetry conditions for $n_1$ and $\nu_2$ are
\begin{align}
n_1(g_0,g_1) &= n_1(e,g_0^{-1}g_1) = n_1(g_0^{-1}g_1),\\\label{1D:symm:nu2}
\nu_2(g_0,g_1,g_2) &= \emp^{g_0}\nu_2(g_0^{-1}g_1,g_1^{-1}g_2)
= \nu_2(g_0^{-1}g_1,g_1^{-1}g_2)^{1-2s_1(g_0)} \cdot (-1)^{(\om_2\smile n_1 
)(g_0,g_0^{-1}g_1,g_1^{-1}g_2)},
\end{align}
where the term $(-1)^{\om_2\smile n_1}$ comes from the symmetry transformations of $c_{12}^{g_0^{-1}g_1}$. We also introduced new notations to relate the homogeneous cochain $\nu_2(g_0,g_1,g_2) = \emp^{g_0}\nu_2(g_0^{-1}g_1,g_1^{-1}g_2)$ and the inhomogeneous cochain $\nu_2(g_0^{-1}g_1,g_1^{-1}g_2)=\emp^e\nu_2(g_0^{-1}g_1,g_1^{-1}g_2)$. In the following, when we write the cochain $\nu_d$ without arguments, we will always mean the inhomogeneous one, i.e., $\nu_d=\emp^e\nu_d$.

\subsection{Associativity and twisted cocycle equations}
\label{sec:1D:dnu}

The $F$ move reduces three vertices on the lattice to two vertices. If one consider reducing four vertices to two vertices, there are two inequivalent ways to do that. The final results should be independent of the two ways. This gives us the consistency equation for Pachner moves:
\begin{align}
\tikzfig{}{
\node (L) at (0,0) {
\begin{tikzpicture}
\coordinate (0) at (0,0);
\coordinate (1) at (1,0);
\node[below] at (0) {$g_0$};
\node[below] at (1) {$g_3$};
\draw[line width=.9] (0,-.017)--(0,.1);
\draw[line width=.9] (1,-.017)--(1,.1);
\draw[->-=.6,line width=.9] (0)--(1);
\end{tikzpicture}
};
\node (U) at (3.5,1) {
\begin{tikzpicture}
\coordinate (0) at (0,0);
\coordinate (1) at (1,0);
\coordinate (2) at (2,0);
\node[below] at (0) {$g_0$};
\node[below] at (1) {$g_1$};
\node[below] at (2) {$g_3$};
\draw[line width=.9] (0,-.017)--(0,.1);
\draw[line width=.9] (1,-.017)--(1,.1);
\draw[line width=.9] (2,-.017)--(2,.1);
\draw[->-=.6,line width=.9] (0)--(1);
\draw[->-=.6,line width=.9] (1)--(2);
\end{tikzpicture}
};
\node (D) at (3.5,-1) {
\begin{tikzpicture}
\coordinate (0) at (0,0);
\coordinate (1) at (1,0);
\coordinate (2) at (2,0);
\node[below] at (0) {$g_0$};
\node[below] at (1) {$g_2$};
\node[below] at (2) {$g_3$};
\draw[line width=.9] (0,-.017)--(0,.1);
\draw[line width=.9] (1,-.017)--(1,.1);
\draw[line width=.9] (2,-.017)--(2,.1);
\draw[->-=.6,line width=.9] (0)--(1);
\draw[->-=.6,line width=.9] (1)--(2);
\end{tikzpicture}
};
\node (R) at (8,0) {
\begin{tikzpicture}
\coordinate (0) at (0,0);
\coordinate (1) at (1,0);
\coordinate (2) at (2,0);
\coordinate (3) at (3,0);
\node[below] at (0) {$g_0$};
\node[below] at (1) {$g_1$};
\node[below] at (2) {$g_2$};
\node[below] at (3) {$g_3$};
\draw[line width=.9] (0,-.017)--(0,.1);
\draw[line width=.9] (1,-.017)--(1,.1);
\draw[line width=.9] (2,-.017)--(2,.1);
\draw[line width=.9] (3,-.017)--(3,.1);
\draw[->-=.6,line width=.9] (0)--(1);
\draw[->-=.6,line width=.9] (1)--(2);
\draw[->-=.6,line width=.9] (2)--(3);
\end{tikzpicture}
};
\draw[<-] (L)--(U) node [midway,above,xshift=-3] {$F(013)$};
\draw[<-] (L)--(D) node [midway,below,xshift=-3] {$F(023)$};
\draw[<-] (U)--(R) node [midway,above,xshift=3] {$F(123)$};
\draw[<-] (D)--(R) node [midway,below,xshift=3] {$F(012)$};
}
\end{align}
In terms of $F$ operators, the above commuting diagram means
\begin{align}\label{1D:dF}
F(g_0,g_2,g_3) \cdot F(g_0,g_1,g_2) = F(g_0,g_1,g_3) \cdot F(g_1,g_2,g_3).
\end{align}
Similar to the standard $F$ symbol which can be used to derive all other non-standard ones by a symmetry action, we can also assume $g_0=e$ in the above equation. All other consistency equations with generic $g_0$ can be deduced from this standard equation by a $U(g_0)$-action. Therefore, we only need to consider the consistency equation
\begin{align}\nonumber\label{1D:dFe}
F(e,g_0^{-1}g_2,g_0^{-1}g_3) \cdot F(e,g_0^{-1}g_1,g_0^{-1}g_2)
&= F(e,g_0^{-1}g_1,g_0^{-1}g_3) \cdot F(g_0^{-1}g_1,g_0^{-1}g_2,g_0^{-1}g_3)\\
&= F(e,g_0^{-1}g_1,g_0^{-1}g_3) \cdot \emp^{(g_0^{-1}g_1)}F(e,g_1^{-1}g_2,g_1^{-1}g_3).
\end{align}
The above equation is simpler than the generic one \eq{1D:dF}, since only the last $F$ symbol is non-standard.

Substituting the standard $F$ move \eq{1D:Fe}, the consistency equation \eq{1D:dFe} becomes
\begin{align}
\nu_2(g_0^{-1}g_2,g_2^{-1}g_3) \nu_2(g_0^{-1}g_1,g_1^{-1}g_2)
=
\nu_2(g_0^{-1}g_1,g_1^{-1}g_3) \nu_2(g_1^{-1}g_2,g_2^{-1}g_3) (-1)^{(\om_2\smile n_1 
)(g_0^{-1}g_1,g_1^{-1}g_2,g_2^{-1}g_3)},
\end{align}
where the last term $(-1)^{\om_2\smile n_1 
}$ comes from the $U(g_0^{-1}g_1)$-action on $F(e,g_1^{-1}g_2,g_1^{-1}g_3)$ [see \eq{1D:symm:nu2}]. Note that the complex fermions do not contribute any fermion signs. So we have the twisted cocycle equation for inhomogeneous $\nu_2$:
\begin{align}\label{1D:dnu2}
\dd \nu_2 &= \mathcal O_3[n_1],
\end{align}
with obstruction function 
\begin{align}\label{1D:O3}
\mathcal O_3[n_1] = (-1)^{\omega_2\smile n_1}.
\end{align}

In summary, the associativity condition for the $F$ moves in 1D gives us the twisted cocycle equation \eqs{1D:dnu2}{1D:O3} for inhomogeneous cochain $\nu_2$.

\subsection{Classification of 1D FSPT phases}

The general classification of 1D FSPT phases is as follow. We first calculate the cohomology groups $H^1(G_b,\Z_2)$ and $H^2(G_b,U(1)_T)$. For each $n_1\in H^1(G_b,\Z_2)$, we solve the twisted cocycle equation \eq{1D:eq} for $\nu_2$. If $\nu_2$ is in the trivialization subgroup $\Gamma^2$ in \eq{1D:Gamma}, it is known to be trivialized by complex fermion decoration\cite{GuWen2014}, see Appendix \ref{sec:0D} for more details. So the obstruction-free $n_1$ and trivialization-free $\nu_2$ fully classify the 1D FSPT phases.

We note that we can use the FSLU transformations to construct the commuting projector parent Hamiltonians. Each term of the Hamiltonian is a sequence of fermionic $F$ moves that changes the label of a vertex from $g_\ast$ to $g_\ast'$. All the terms commute with each other for our FSPT wave function is at the fixed-point.

\section{Fixed-point wave function and classification of FSPT states in 2D}
\label{sec:2D}
\NewFigName{figures/Fig_2D_}

In this section, we construct and classify FSPT states in two spacial dimensions. 
The fixed-point wave function is again a superposition of all basis states $|\{g_i\}\rangle$ with fermion decorations:
\begin{align}\label{2D:wf}
|\Psi\rangle = \sum_{\{g_i\}} \Psi(\{g_i\}) \ |\{g_i\}\rangle.
\end{align}
The basis state $|\{g_i\}\rangle$ is a state (with vertex $i$ labelled by $g_i\in G_b$) decorated by complex fermions $c_{ij}^\s$ at link $\langle ij\rangle$ and Majorana fermions $\g_{i,A}^\s$ and $\g_{i,B}^\s$ near vertex $i$ according to several designed rules. So the fixed-point wave function would looks like
\begin{align}\label{2D:wf_}
|\Psi\rangle = \sum_{\text{all conf.}} \Psi\left(
\tikzfig{scale=1.5}{
\coordinate (0) at (0,0);
\coordinate (1) at (1.5,0);
\coordinate (2) at (1.5,1);
\coordinate (3) at (0,1);
\draw (0)--(1)--(2)--(3)--cycle;
\coordinate (4) at (.4,.3);
\coordinate (5) at (.6,.7);
\coordinate (6) at (1,.4);
\coordinate (7) at (0,.5);
\coordinate (8) at (.6,0);
\coordinate (9) at (1.1,0);
\coordinate (11) at (.55,1);
\coordinate (12) at (1.2,1);
\coordinate (13) at (1.5,.5);
\draw[] (0)--(4);
\draw[] (7)--(4);
\draw[] (7)--(5);
\draw[] (3)--(5);
\draw[] (4)--(5);
\draw[] (5)--(6);
\draw[] (8)--(4);
\draw[] (9)--(6);
\draw[] (11)--(5);
\draw[] (12)--(5);
\draw[] (12)--(6);
\draw[] (13)--(6);
\draw[] (1)--(6);
\draw[] (12)--(13);
\draw[] (4)--(6);
\draw[] (8)--(6);
\draw[green,line width=1.2] ($($(0)!.5!(4)$)!.333!(8)$)--($($(6)!.5!(4)$)!.333!(8)$)--($($(6)!.5!(4)$)!.333!(5)$)--($($(7)!.5!(4)$)!.333!(5)$)--($($(7)!.5!(4)$)!.333!(0)$)--cycle;
\draw[green,line width=1.2] ($(3)!.5!(7)$)--($($(3)!.5!(5)$)!.333!(7)$)--($($(3)!.5!(5)$)!.333!(11)$)--($(3)!.5!(11)$);
\draw[green,line width=1.2] ($(2)!.5!(12)$)--($($(2)!.5!(13)$)!.333!(12)$)--($($(13)!.5!(6)$)!.333!(12)$)--($($(13)!.5!(6)$)!.333!(1)$)--($($(9)!.5!(6)$)!.333!(1)$)--($(1)!.5!(9)$);
\node[circle,shading=ball,scale=.4] at ($($(0)!.5!(4)$)!.333!(8)$) {};
\node[circle,shading=ball,scale=.4] at ($($(3)!.5!(5)$)!.333!(7)$) {};
\node[circle,shading=ball,scale=.4] at ($($(4)!.5!(5)$)!.333!(6)$) {};
\node[circle,shading=ball,scale=.4] at ($($(6)!.5!(5)$)!.333!(12)$) {};
\node[circle,shading=ball,scale=.4] at ($($(6)!.5!(1)$)!.333!(13)$) {};
\node[circle,draw=blue,fill=white,scale=.35] at ($($(0)!.5!(4)$)!.333!(7)$) {};
\node[circle,draw=blue,fill=white,scale=.35] at ($($(5)!.5!(4)$)!.333!(7)$) {};
\node[circle,draw=blue,fill=white,scale=.35] at ($($(3)!.5!(5)$)!.333!(11)$) {};
\node[circle,draw=blue,fill=white,scale=.35] at ($($(5)!.5!(11)$)!.333!(12)$) {};
\node[circle,draw=blue,fill=white,scale=.35] at ($($(4)!.5!(6)$)!.333!(8)$) {};
\node[circle,draw=blue,fill=white,scale=.35] at ($($(9)!.5!(6)$)!.333!(8)$) {};
\node[circle,draw=blue,fill=white,scale=.35] at ($($(9)!.5!(1)$)!.333!(6)$) {};
\node[circle,draw=blue,fill=white,scale=.35] at ($($(12)!.5!(13)$)!.333!(6)$) {};
\node[circle,draw=blue,fill=white,scale=.35] at ($($(12)!.5!(13)$)!.333!(2)$) {};
}
\right) \stretchleftright{\Bigg|}{\ 
\vcenter{\hbox{\includegraphics[scale=1]{\FigName\arabic{Num}.pdf}}}
\ }{\Big\rangle}.
\end{align}

\subsection{Three layers of degrees of freedom}

In 2D, we decorate two layers of fermionic degrees of freedom to the BSPT states. Therefore, there are three layers of degrees of freedom, including the bosonic ones, in our 2D triangulation lattice model:
\begin{itemize}
\item
$|G_b|$ level bosonic (spin) state $|g_i\rangle$ ($g_i\in G_b$) on each vertex $i$.
\item
$|G_b|$ species of complex fermions $c_{ijk}^\s$ ($\s\in G_b$) at the center of each triangle $\langle ijk\rangle$.
\item
$|G_b|$ species of complex fermions (split to Majorana fermions) $a_{ij}^\s=(\g_{ij,A}^\s+i\g_{ij,B}^\s)/2$ ($\s\in G_b$) on the two sides of each link $\langle ij\rangle$.
\end{itemize}
These three layers of degrees of freedom are summarized in one triangle in the following figure:
\begin{align}\label{fig:2D:012}
\tikzfig{scale=2.3}{
\coordinate (g1) at (0,0);
\coordinate (g2) at (1,0);
\coordinate (g3) at (1/2,1.732/2);
\coordinate (center) at (1/2,1/2/1.732);
\pgfmathsetmacro{\s}{.11}
\coordinate (11) at (3/4-\s*1.732/2, 1.732/4-\s/2);
\coordinate (12) at (3/4+\s*1.732/2, 1.732/4+\s/2);
\coordinate (21) at (1/4+\s*1.732/2, 1.732/4-\s/2);
\coordinate (22) at (1/4-\s*1.732/2, 1.732/4+\s/2);
\coordinate (31) at (1/2, 0+\s);
\coordinate (32) at (1/2, 0-\s);
%
%
\draw [->-=.6,thick] (g1) node[scale=1,below left]{$g_0$} -- (g2);
\draw [->-=.6,thick] (g2) node[scale=1,below right]{$g_1$} -- (g3) node[scale=1,above]{$g_2$};
\draw [->-=.6,thick] (g1) -- (g3);
\draw[->-=3.2/4,thick,red,densely dotted] (11)--(12);
\draw[->-=3.2/4,thick,red,densely dotted] (22)--(21);
\draw[->-=3.2/4,thick,red,densely dotted] (31)--(32);
\draw[->-=3/4,thick,red,densely dotted] (21)--(11);
\draw[->-=3/4,thick,red,densely dotted] (21)--(31);
\draw[->-=3.2/4,thick,red,densely dotted] (11)--(31);
\draw[->-=3.2/4,thick,red,densely dotted] (22)--(21);
\fill [red] (11) circle (.6pt);
\fill [red] (12) circle (.6pt);
\fill [red] (21) circle (.6pt);
\fill [red] (22) circle (.6pt);
\fill [red] (31) circle (.6pt);
\fill [red] (32) circle (.6pt);
%
\node[red,scale=.7,below right,xshift=-6] at (11){$\g_{12A}^\sigma$};
\node[red,scale=.7,right] at (12){$\g_{12B}^\sigma$};
\node[red,scale=.7,below left,xshift=8] at (21){$\g_{02B}^\sigma$};
\node[red,scale=.7,left] at (22){$\g_{02A}^\sigma$};
\node[red,scale=.7,right] at (31){$\g_{01A}^\sigma$};
\node[red,scale=.7,below] at (32){$\g_{01B}^\sigma$};
\node[circle,shading=ball,scale=.6] at (.5,0.2886) {};
\node[blue,above,scale=.8,yshift=7] at (.5,0.2886) {$c_{012}^\sigma$};
}
\end{align}
Here, the three vertices of the triangle are labelled by $g_0,g_1,g_2\in G_b$. Blue ball is the complex fermion $c_{012}^\s$ ($\s\in G_b$) at the center of triangle $\langle 012\rangle$. Red dots represent Majorana fermions $\g_{ij,A}^\s$ and $\g_{ij,B}^\s$ ($\s\in G_b$) on the two sides of link $\langle ij\rangle$. 

The symmetry transformations of these degrees of freedom under $G_b$ are summarize them as follows ($g,g_i,\s\in G_b$):
\begin{align}
\label{2D:symm:gi}
U(g) |g_i\rangle &= |gg_i\rangle,\\
\label{2D:symm:c}
U(g) c_{ijk}^\s U(g)^\dagger &= (-1)^{\om_2(g,\s)} c_{ijk}^{g\s},\\
\label{2D:symm:A}
U(g) \g_{ij,A}^\s U(g)^\dagger &= (-1)^{\om_2(g,\s)} \g_{ij,A}^{g\s},\\
\label{2D:symm:B}
U(g) \g_{ij,B}^\s U(g)^\dagger &= (-1)^{\om_2(g,\s)+s_1(g)} \g_{ij,B}^{g\s}.
\end{align}
As in the 1D case, the bosonic degrees of freedom form a linear representation of $G_b$ (and $G_f$). On the other hand, the fermion modes support projective representations of $G_b$ with coefficient $(-1)^{\om_2}$, and linear representations of $G_f$ defined by \eq{UU}.

In the simpler case of $G_f=\Zf\times G_b$, all the flavors of fermions have the same transformation rule for different group element label $\s$. So we can suppress the species labels. This is again the previous group supercohomology models \cite{GuWen2014,WangGu2017}.

\subsection{Decorations of fermion layers}

In this section, we would give a systematic procedure to decorate Kitaev chains and complex fermions to the basis state $|\{g_i\}\rangle$ labelled by $g_i\in G_b$ for each vertex $i$. Similar to the 1D case, we only decorate (at most) one species of fermions to the state, although the Hilbert space is spanned by $|G_b|$ copies of fermions. Again, the decorations should be designed to respect the symmetry.

\subsubsection{Kitaev chain decoration}
\label{2D:decoration:Maj}

The Kitaev chain decoration in 2D is similar to the constructions in the pioneering works Refs.~\onlinecite{Tarantino2016,Ware2016}. However, we will adopt the more general procedures in Ref.~\onlinecite{WangGu2017}, which can deal with arbitrary triangulations of the 2D spacial manifold. The generalization in this paper for symmetry group $G_f=\Zf\timesw G_b$ is that we put (at most) one of the $G_b$ species Majorana fermions into nontrivial pairings and all others vacuum pairings. If we consider the symmetry group $G_b=\Z_2^T$ and nontrivial 2-cocycle $\om_2(T,T)=1$, our construction on a fixed triangular lattice will reproduce to the exactly solvable $T^2=-1$ topological superconductor model in Ref.~\onlinecite{WangNingChen2017}.

To simplify our notations and make it easier to generalize to higher dimensions, we present some notations for Majorana fermion pairings. For two Majorana fermions $\g_{i,C}^e$ and $\g_{j,D}^{g^{-1}h}$ at vertices $i$ and $j$ ($g, h \in G_b$ and $C,D=A,B$), we can choose the pairing such that
\begin{align}\label{Mpairing0}
-i \g_{i,C}^e \g_{j,D}^{g^{-1}h} = 1,
\end{align}
when acting on this state. We will call it \emph{standard} pairing, as the first Majorana fermion is labelled by the identity element $e\in G_b$. The standard pairing will be illustrated in figures by a red arrow pointing from vertex $iC$ to vertex $jD$. For the non-standard pairing between $\g_{i,C}^g$ and $\g_{j,D}^{h}$, we can use a $U(g)$-action on both sides of \eq{Mpairing0} and obtain
\begin{align}\label{Majpairing}
-i \g_{i,C}^g \g_{j,D}^{h} = (-1)^{\om_2(g,g^{-1}h) + s_1(g) (1 + \delta_{CB} + \delta_{DB})},
\end{align}
where $\delta_{CB}=1$ ($=0$) if the Majorana fermion $\g_{i,C}^\s$ is the $B$ type ($A$ type) one. This difference comes from the symmetry transformations of $A$ and $B$ type Majorana fermions [see \eqs{2D:symm:A}{2D:symm:B}]. 
For simplicity in describing the pairing, we introduce the projection operator of the Majorana fermion pairing as
\begin{align}\label{proj}
P_{iC,jD}^{g,h} := U(g) P_{iC,jD}^{e,g^{-1}h} U(g)^\dagger = U(g) \frac{1}{2}\left(1-i\g_{i,C}^e \g_{j,D}^{g^{-1}h}\right) U(g)^\dagger
= \frac{1}{2} \left[ 1-(-1)^{\om_2(g,g^{-1}h) + s_1(g) (1 + \delta_{CB} + \delta_{DB})}i\g_{i,C}^g \g_{j,D}^{h} \right].
\end{align}
This generic pairing projection operator $P_{iC,jD}^{g,h}$ is obtained from a $U(g)$-action on the standard pairing projection operator $P_{iC,jD}^{e,g^{-1}h} = \left(1-i\g_{i,C}^e \g_{j,D}^{g^{-1}h}\right)/2$. So the symmetric nature of the pairings can be easily seen from the symmetry transformations of the projection operators ($g,h,k\in G_b$):
\begin{align}\label{P_symm}
U(g) P_{iC,jD}^{h,k} U(g)^\dagger = P_{iC,jD}^{gh,gk}.
\end{align}
In 2D and higher dimensions, we will use the generic pairing rule \eq{Majpairing} and the projection operator \eq{proj} to construct $G_b$-symmetric states.

\paragraph{Decoration procedure.}

\begin{figure}[ht]
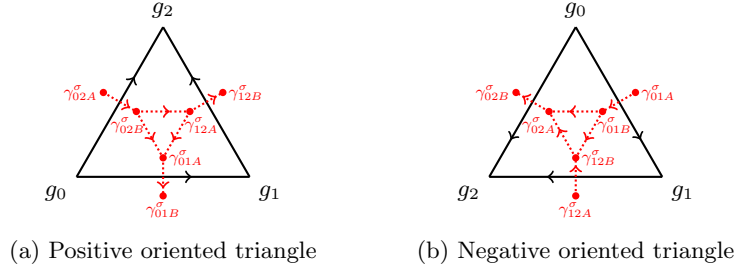

\centering
\begin{subfigure}[h]{.3\textwidth}
\centering
$\tikzfig{scale=2.3}{
\coordinate (g0) at (0,0);
\coordinate (g1) at (1,0);
\coordinate (g2) at (1/2,1.732/2);
\coordinate (center) at (1/2,1/2/1.732);
\draw [->-=2/3,thick] (g0) node[below left]{$g_0$} -- (g1);
\draw [->-=2/3,thick] (g1) node[below right]{$g_1$} -- (g2) node[above]{$g_2$};
\draw [->-=2/3,thick] (g0) -- (g2);
\pgfmathsetmacro{\s}{.11}
\coordinate (01) at (3/4-\s*1.732/2, 1.732/4-\s/2);
\coordinate (02) at (3/4+\s*1.732/2, 1.732/4+\s/2);
\coordinate (11) at (1/4+\s*1.732/2, 1.732/4-\s/2);
\coordinate (12) at (1/4-\s*1.732/2, 1.732/4+\s/2);
\coordinate (21) at (1/2, 0+\s);
\coordinate (22) at (1/2, 0-\s);
\fill [red] (01) circle (.6pt);
\fill [red] (02) circle (.6pt);
\fill [red] (11) circle (.6pt);
\fill [red] (12) circle (.6pt);
\fill [red] (21) circle (.6pt);
\fill [red] (22) circle (.6pt);
\draw[->-=3.2/4,thick,red,densely dotted] (01)--(02);
\draw[->-=3.2/4,thick,red,densely dotted] (12)
--(11);
\draw[->-=3.2/4,thick,red,densely dotted] (21)--(22);
\draw[->-=2.5/4,thick,red,densely dotted] (11)--(01);
\draw[->-=2.5/4,thick,red,densely dotted] (11)--(21);
\draw[->-=2.5/4,thick,red,densely dotted] (01)--(21);
\node[red,scale=.7,below right,xshift=-6] at (01){$\g_{12A}^\sigma$};
\node[red,scale=.7,right] at (02){$\g_{12B}^\sigma$};
\node[red,scale=.7,below left,xshift=8] at (11){$\g_{02B}^\sigma$};
\node[red,scale=.7,left] at (12){$\g_{02A}^\sigma$};
\node[red,scale=.7,right] at (21){$\g_{01A}^\sigma$};
\node[red,scale=.7,below] at (22){$\g_{01B}^\sigma$};
}$
\caption{Positive oriented triangle}
\end{subfigure}
\begin{subfigure}[h]{.3\textwidth}
\centering
$\tikzfig{scale=2.3}{
\coordinate (g2) at (0,0);
\coordinate (g1) at (1,0);
\coordinate (g0) at (1/2,1.732/2);
\coordinate (center) at (1/2,1/2/1.732);
\draw [->-=.75,thick] (g0) node[above]{$g_0$} -- (g1);
\draw [->-=2/3,thick] (g1) node[below right]{$g_1$} -- (g2) node[below left]{$g_2$};
\draw [->-=.75,thick] (g0) -- (g2);
\pgfmathsetmacro{\s}{.11}
\coordinate (01) at (3/4-\s*1.732/2, 1.732/4-\s/2);
\coordinate (02) at (3/4+\s*1.732/2, 1.732/4+\s/2);
\coordinate (11) at (1/4+\s*1.732/2, 1.732/4-\s/2);
\coordinate (12) at (1/4-\s*1.732/2, 1.732/4+\s/2);
\coordinate (21) at (1/2, 0+\s);
\coordinate (22) at (1/2, 0-\s);
\fill [red] (01) circle (.6pt);
\fill [red] (02) circle (.6pt);
\fill [red] (11) circle (.6pt);
\fill [red] (12) circle (.6pt);
\fill [red] (21) circle (.6pt);
\fill [red] (22) circle (.6pt);
\draw[->-=3.2/4,thick,red,densely dotted] (02)
--(01);
\draw[->-=3.2/4,thick,red,densely dotted] (11)--(12);
\draw[->-=3.2/4,thick,red,densely dotted] (22)
--(21);
\draw[->-=2.5/4,thick,red,densely dotted] (01)--(11);
\draw[->-=2.5/4,thick,red,densely dotted] (21)--(11);
\draw[->-=2.5/4,thick,red,densely dotted] (01)--(21);
\node[red,scale=.7,below right,xshift=-6] at (01){$\g_{01B}^\sigma$};
\node[red,scale=.7,right] at (02){$\g_{01A}^\sigma$};
\node[red,scale=.7,below left,xshift=7] at (11){$\g_{02A}^\sigma$};
\node[red,scale=.7,left] at (12){$\g_{02B}^\sigma$};
\node[red,scale=.7,right] at (21){$\g_{12B}^\sigma$};
\node[red,scale=.7,below] at (22){$\g_{12A}^\sigma$};
}$
\caption{Negative oriented triangle}
\end{subfigure}
\caption{Kasteleyn orientations of the resolved dual lattice. For a given triangulation of the 2D spacial spin manifold (shown by black links), we can construct a resolved dual lattice (shown by red links). The Majorana fermion pairings should respect the red link arrows in the figures.}
\label{fig:Kasteleyn}
\end{figure}

Our Majorana fermions $\g_{ij,A}^\s$ and $\g_{ij,B}^\s$ ($\s\in G_b$) are on the two sides of each link $\langle ij\rangle$. We use the convention that the Majorana fermion on left-hand-side (right-hand-side) of the oriented link $\langle ij\rangle$ is $\g_{ij,A}^\s$ ($\g_{ij,B}^\s$). The vacuum pairing between them is from $A$ to $B$: $-i\g_{ij,A}^\s \g_{ij,B}^\s = 1$. To decorate Kitaev chains on the lattice, we should also add arrows to the small red triangle inside each triangle $\langle 012\rangle$ (see \fig{fig:Kasteleyn}). These red arrows are constructed from the discrete spin structures (a choose of trivialization of Stiefel-Whitney homology class $w_0$ dual to cohomology class $w^2$) of the 2D spacial spin manifold triangulation. The Majorana fermions are designed to pair up with each other according to these red arrows. The red arrows constructed have the property that the number of counterclockwise arrows in a loop with even red links is always odd. This is crucial for the decorated Kitaev chain to have fixed fermion parity. For details of the Kasteleyn orientations for arbitrary triangulation, we refer the interested readers to Ref.~\onlinecite{WangGu2017}.

The Kitaev chain decoration is specified by $n_1(g_i,g_j)\in \Z_2$, which is a function of two group elements $g_i,g_j\in G_b$. If $n_1(g_i,g_j)=0$, the Majorana fermions $\g_{ij,A}^\s$ and $\g_{ij,B}^\s$ on the two sides of link $\langle ij\rangle$ are in vacuum pairings: $-i\g_{ij,A}^\s \g_{ij,B}^\s = 1$ (for all $\s\in G_b$). On the other hand, if $n_1(g_i,g_j)=1$, there is a domain wall along the direction dual to link $\langle ij\rangle$, where a Kitaev chain will be decorated [see the green belt shown in \eq{2D:n1}]. For all $|G_b|$ species of Majorana fermions, we only put $\g_{ij,A}^{g_i}$ and $\g_{ij,B}^{g_i}$ to be in the nontrivial pairing. And all other $|G_b|-1$ species of Majorana fermions $\g_{ij,A}^{\s}$ and $\g_{ij,B}^\s$ with $\s\neq g_i$ are still in vacuum pairings. Here is an example of the Kitaev chain decoration around the vertex $g_2$ inside a triangle (we omit the operator labels of Majorana fermions which are in vacuum pairings):
\begin{align}\label{2D:n1}
\tikzfig{scale=2.3}{
\coordinate (g0) at (0,0);
\coordinate (g1) at (1,0);
\coordinate (g2) at (1/2,1.732/2);
\coordinate (center) at (1/2,1/2/1.732);
\pgfmathsetmacro{\s}{.11}
\coordinate (01) at (3/4-\s*1.732/2, 1.732/4-\s/2);
\coordinate (02) at (3/4+\s*1.732/2, 1.732/4+\s/2);
\coordinate (11) at (1/4+\s*1.732/2, 1.732/4-\s/2);
\coordinate (12) at (1/4-\s*1.732/2, 1.732/4+\s/2);
\coordinate (21) at (1/2, 0+\s);
\coordinate (22) at (1/2, 0-\s);
\draw [green!30,line width=10pt] ($(12)!-.2!(11)$)--(12)--(11)--(01)--(02)--($(02)!-.2!(01)$);
\Ellipse{gray!50}{11}{01}{1.4}{.27};
\Ellipse{gray!50}{21}{22}{1.5}{.4};
\Ellipse{gray!50}{01}{02}{1.5}{.4};
\Ellipse{gray!50}{11}{12}{1.5}{.4};
\draw [->-=2/3,thick] (g0) node[below left]{$g_0$} -- (g1);
\draw [->-=2/3,thick] (g1) node[below right]{$g_1$} -- (g2) node[above]{$g_2$};
\draw [->-=2/3,thick] (g0) -- (g2);
\draw[->-=3.2/4,very thick,red,densely dashed] (01)--(02);
\draw[->-=3.2/4,very thick,red,densely dashed] (12)--(11);
\draw[->-=3.2/4,very thick,red,densely dashed] (21)--(22);
\draw[->-=2.5/4,thick,red,densely dotted] (11)--(21);
\draw[->-=2.5/4,thick,red,densely dotted] (01)--(21);
\draw[very thick,red] (11)--(01);
\draw[->-=2.5/4,very thick,blue] ($(11)!.625!(01)$)--($(11)!.626!(01)$);
\fill [red] (01) circle (.6pt);
\fill [red] (02) circle (.6pt);
\fill [red] (11) circle (.6pt);
\fill [red] (12) circle (.6pt);
\fill [red] (21) circle (.6pt);
\fill [red] (22) circle (.6pt);
\node[red,scale=.7,below right,xshift=-6] at (01){$\g_{12A}^{g_1}$};
\node[red,scale=.7,right] at (02){$\g_{12B}^{g_1}$};
\node[red,scale=.7,below left,xshift=8] at (11){$\g_{02B}^{g_0}$};
\node[red,scale=.7,left] at (12){$\g_{02A}^{g_0}$};
\node[red,scale=.7,right] at (21){$\g_{01A}^{\sigma}$};
\node[red,scale=.7,below] at (22){$\g_{01B}^{\sigma}$};
}
\end{align}
The domain wall decorated by a Kitaev chain is indicated by a green belt. Trivial (vacuum) pairings and nontrivial pairings are represented by dashed red lines and solid red lines, respectively. And the red (blue) arrows show that the trivial (nontrivial) pairing directions of Majorana fermions:
\begin{align}
&\tikzfig{scale=1}{
\coordinate (g0) at (0,0);
\coordinate (g1) at (1,0);
\Ellipse{gray!50}{g0}{g1}{1.4}{.27};
\draw[red,densely dashed,line width=1](0,0)--(1,0);
\node[circle,draw=red,fill=red,scale=.3] at (0,0){};
\node[circle,draw=red,fill=red,scale=.3] at (1,0){};
\node[red,xshift=-2,yshift=7,scale=.8] at (0,0){$\g_{ij,A}^{\s}$};
\node[red,xshift=-2,yshift=7,scale=.8] at (1,0){$\g_{ij,B}^{\s}$};
\draw[->-=.7,red,line width=1] ($(0,0)!.625!(1,0)$)--($(0,0)!.6251!(1,0)$);
\node[]at(0,0){};
}
\ \ \iff
\quad
-i \g_{ij,A}^{\s} \g_{ij,B}^{\s} = 1
\\
\label{2D:Mpair2}
&\tikzfig{scale=1}{
\coordinate (g0) at (0,0);
\coordinate (g1) at (1,0);
\Ellipse{gray!50}{g0}{g1}{1.4}{.27};
\draw[red,line width=1](0,0)--(1,0);
\node[circle,draw=red,fill=red,scale=.3] at (0,0){};
\node[circle,draw=red,fill=red,scale=.3] at (1,0){};
\node[red,xshift=-2,yshift=7,scale=.8] at (0,0){$\g_{02B}^{g_0}$};
\node[red,xshift=-2,yshift=7,scale=.8] at (1,0){$\g_{12A}^{g_1}$};
\draw[->-=.7,blue,line width=1] ($(0,0)!.625!(1,0)$)--($(0,0)!.6251!(1,0)$);
\node[]at(0,0){};
}
\ \ \iff
\quad
-i \g_{02B}^{g_0} \g_{12A}^{g_{1}} = (-1)^{\om_2(g_0,g_0^{-1}g_1)}.
\end{align}
We will discuss more about the pairing directions and why they are symmetric later.

\paragraph{Consistency condition.}

According to our decoration rule,  the number of decorated Kitaev chains going though the boundary of a given triangle $\langle 012\rangle$ is
\begin{align}
(\dd n_1)(g_0,g_1,g_2) = n_1(g_1,g_2) + n_1(g_0,g_2) + n_1(g_0,g_1).
\end{align}
Since we are constructing gapped state without intrinsic topological order, there should be no dangling free Majorana fermions inside any triangle. Therefore, we have the (mod 2) equation
%
\begin{align}\label{2D:dn1=0}
\dd n_1 = 0.
\end{align}
This is the consistency condition for the Kitaev chain decoration data $n_1$.

\paragraph{Symmetric pairing directions.}

Now let us turn back to the details of Majorana fermion pairings inside the triangle $\langle 012\rangle$. The strategy of constructing $G_b$-symmetric pairings is the same as in the 1D case: we first consider the \emph{standard} triangle of $g_0=e$, and then apply a $U(g_0)$-action to obtain all other non-standard triangles. The Majorana fermion pairings constructed in this way are automatically symmetric, due to the symmetry transformation rule of the pairing projection operators \eq{P_symm}. For the standard triangle, the Majorana fermions are paired (trivially or nontrivially) according to the Kasteleyn orientations indicated by red arrows. The pairings in the non-standard triangle is obtained by a $U(g_0)$-action as follows:
\begin{align}\label{2D:g_symm}
\tikzfig{scale=2.3}{
\coordinate (g0) at (0,0);
\coordinate (g1) at (1,0);
\coordinate (g2) at (1/2,1.732/2);
\coordinate (center) at (1/2,1/2/1.732);
\draw [->-=2/3,thick] (g0) node[below left]{$e$} -- (g1);
\draw [->-=2/3,thick] (g1) node[below right]{$g_0^{-1}g_1$} -- (g2) node[above]{$g_0^{-1}g_2$};
\draw [->-=2/3,thick] (g0) -- (g2);
\pgfmathsetmacro{\s}{.11}
\coordinate (01) at (3/4-\s*1.732/2, 1.732/4-\s/2);
\coordinate (02) at (3/4+\s*1.732/2, 1.732/4+\s/2);
\coordinate (11) at (1/4+\s*1.732/2, 1.732/4-\s/2);
\coordinate (12) at (1/4-\s*1.732/2, 1.732/4+\s/2);
\coordinate (21) at (1/2, 0+\s);
\coordinate (22) at (1/2, 0-\s);
\fill [red] (01) circle (.6pt);
\fill [red] (02) circle (.6pt);
\fill [red] (11) circle (.6pt);
\fill [red] (12) circle (.6pt);
\fill [red] (21) circle (.6pt);
\fill [red] (22) circle (.6pt);
\draw[->-=3.2/4,thick,red,densely dotted] (01)--(02);
\draw[->-=3.2/4,thick,red,densely dotted] (12)
--(11);
\draw[->-=3.2/4,thick,red,densely dotted] (21)--(22);
\draw[->-=2.5/4,thick,red,densely dotted] (11)--(01);
\draw[->-=2.5/4,thick,red,densely dotted] (11)--(21);
\draw[->-=2.5/4,thick,red,densely dotted] (01)--(21);
\node[red,scale=.7,below right,xshift=-5,yshift=4] at (01){$\g_{12A}^{g_0^{-1}g_1}$};
\node[red,scale=.7,right] at (02){$\g_{12B}^{g_0^{-1}g_1}$};
\node[red,scale=.7,below left,xshift=8] at (11){$\g_{02B}^{e}$};
\node[red,scale=.7,left] at (12){$\g_{02A}^{e}$};
\node[red,scale=.7,right] at (21){$\g_{01A}^{e}$};
\node[red,scale=.7,below] at (22){$\g_{01B}^{e}$};
}
\underrightarrow{\ \ U(g_0)\ \ }
\tikzfig{scale=2.3}{
\coordinate (g0) at (0,0);
\coordinate (g1) at (1,0);
\coordinate (g2) at (1/2,1.732/2);
\coordinate (center) at (1/2,1/2/1.732);
\draw [->-=2/3,thick] (g0) node[below left]{$g_0$} -- (g1);
\draw [->-=2/3,thick] (g1) node[below right]{$g_1$} -- (g2) node[above]{$g_2$};
\draw [->-=2/3,thick] (g0) -- (g2);
\pgfmathsetmacro{\s}{.11}
\coordinate (01) at (3/4-\s*1.732/2, 1.732/4-\s/2);
\coordinate (02) at (3/4+\s*1.732/2, 1.732/4+\s/2);
\coordinate (11) at (1/4+\s*1.732/2, 1.732/4-\s/2);
\coordinate (12) at (1/4-\s*1.732/2, 1.732/4+\s/2);
\coordinate (21) at (1/2, 0+\s);
\coordinate (22) at (1/2, 0-\s);
\fill [red] (01) circle (.6pt);
\fill [red] (02) circle (.6pt);
\fill [red] (11) circle (.6pt);
\fill [red] (12) circle (.6pt);
\fill [red] (21) circle (.6pt);
\fill [red] (22) circle (.6pt);
\draw[->-=3.2/4,thick,red,densely dotted] (01)--(02);
\draw[->-=3.2/4,thick,red,densely dotted] (12)
--(11);
\draw[->-=3.2/4,thick,red,densely dotted] (21)--(22);
\draw[thick,red,densely dotted] (11)--(01);
\draw[->-=2.5/4,thick,blue,densely dotted] ($(11)!.625!(01)$)--($(11)!.626!(01)$);
\draw[->-=2.5/4,thick,red,densely dotted] (11)--(21);
\draw[thick,red,densely dotted] (01)--(21);
\draw[->-=2.5/4,thick,blue,densely dotted] ($(01)!.625!(21)$)--($(01)!.626!(21)$);
\node[red,scale=.7,below right,xshift=-6] at (01){$\g_{12A}^{g_1}$};
\node[red,scale=.7,right] at (02){$\g_{12B}^{g_1}$};
\node[red,scale=.7,below left,xshift=8] at (11){$\g_{02B}^{g_0}$};
\node[red,scale=.7,left] at (12){$\g_{02A}^{g_0}$};
\node[red,scale=.7,right] at (21){$\g_{01A}^{g_0}$};
\node[red,scale=.7,below] at (22){$\g_{01B}^{g_0}$};
}
\end{align}
Note that Majorana fermions $\g_{ij,A}^\s$ and $\g_{ij,B}^\s$ ($\s\neq g_i$) of link $\langle ij\rangle$ are always in vacuum pairings $\left(-i\g_{ij,A}^\s\g_{ij,B}^\s=1\right)$, independent of the $n_1$ configurations. So their pairing directions always follow the red arrow Kasteleyn orientations in both figures of the above equation. For the two Majorana fermions $\g_{ij,A}^{g_i}$ and $\g_{ij,B}^{g_i}$ of the link $\langle ij\rangle$, there are two possibilities. If $n_1(g_i,g_j)=0$, these two Majorana fermions are also in vacuum pairing, with direction indicated by the red arrow and projection operator
\begin{align}
P_{ijA,ijB}^{g_i,g_i} = U(g_0) P_{ijA,ijB}^{g_0^{-1}g_i,g_0^{-1}g_i} U(g_0)^{-1} = \frac{1}{2} \left(1-i\g_{ij,A}^{g_i} \g_{ij,B}^{g_i}\right).
\end{align}
On the other hand, if $n_1(g_i,g_j)=1$, we will pair the Majorana fermion inside the triangle with another one belonging to another link with also $n_1=1$ [for example, $\g_{02B}^{g_0}$ and $\g_{12A}^{g_1}$ are paired in \eq{2D:n1}]. Note that there are always even number of Majorana fermions in nontrivial pairing among the three ones ($\g_{12A}^{g_1}$, $\g_{02B}^{g_0}$ and $\g_{01A}^{g_0}$) inside the triangle $\langle 012\rangle$, for we have $(\dd n_1)(g_0,g_1,g_2)=0$ (mod 2) from \eq{2D:dn1=0}. There are three possible nontrivial pairings inside the triangle $\langle 012\rangle$, with Majorana pairing projection operators
\begin{align}\label{2D:02B01A}
P_{02B,01A}^{g_0,g_0} &= U(g_0) P_{02B,01A}^{e,e} U(g_0)^\dagger
= \frac{1}{2}\,\big(1-i\g_{02B}^{g_0}\g_{01A}^{g_0}\big),\\
P_{02B,12A}^{g_0,g_1} &= U(g_0) P_{02B,12A}^{e,g_0^{-1}g_1} U(g_0)^\dagger
= \frac{1}{2} \left[1-(-1)^{\om_2(g_0,g_0^{-1}g_1)}i\g_{02B}^{g_0}\g_{12A}^{g_1}\right],\\\label{2D:12_01}
P_{12A,01A}^{g_1,g_0} &= U(g_0) P_{12A,01A}^{g_0^{-1}g_1,e} U(g_0)^\dagger
= \frac{1}{2} \left[1-(-1)^{\om_2(g_0,g_0^{-1}g_1)+s_1(g_0)}i\g_{12A}^{g_1}\g_{01A}^{g_0}\right].
\end{align}
Among the three possible nontrivial pairings, only the last two may change their directions in the non-standard triangle. They are indicated by blue arrows in the right-hand-side figure of \eq{2D:g_symm}. This can be understood from the following facts: The $(-1)^{\om_2}$ term appears in the projection operators when the pairing is between Majorana fermions with different group element labels; And the $(-1)^{s_1}$ term appears when the pairing is between the same $A/B$ type Majorana fermions. The pairing \eq{2D:02B01A} between $\g_{02B}^{g_0}$ and $\g_{01A}^{g_0}$ belongs to neither of the above two cases. So their pairing direction is the same as the red arrow even after $U(g_0)$-action.

\paragraph{Majorana fermion parity.}

Since the symmetry action may change the pairing directions inside a triangle, the Majorana fermion parity of this triangle may also be changed. The fermion parity difference between the standard and non-standard triangles can be calculated from the number of pairing arrows that are reversed by $U(g_0)$-action, which of course depends on the $n_1$ configurations. We can use, for example, $n_1(g_0,g_1) n_1(g_1,g_2) = 0,1$ to indicate whether $\g_{12A}^{g_1}$ and $\g_{01A}^{g_0}$ are paired or not. So the Majorana fermion parity change inside the triangle is in general given by
\begin{align}\nonumber\label{2D:Pf_M}
\Delta P_f^\g (012) &= (-1)^{\om_2(g_0,g_0^{-1}g_1) [n_1(g_0,g_2) n_1(g_1,g_2) + n_1(g_0,g_1) n_1(g_1,g_2)] + s_1(g_0) n_1(g_0,g_1) n_1(g_1,g_2)}\\\nonumber
&= (-1)^{\om_2(g_0,g_0^{-1}g_1) n_1(g_1,g_2) + s_1(g_0) n_1(g_0,g_1) n_1(g_1,g_2)}\\
&= (-1)^{(\om_2\smile n_1 + s_1\smile n_1\smile n_1)(g_0,g_0^{-1}g_1,g_1^{-1}g_2)},
\end{align}
where we have used $(\dd n_1)(g_0,g_1,g_2)=0$ from \eq{2D:dn1=0} in the second step. The above equation is a summary of phase factors from Eqs.~(\ref{2D:02B01A})-(\ref{2D:12_01}). We note that the above expression is also true for negative oriented triangles. All the fermion parity change cases involve the particular Majorana fermion $\g_{12A}^{g_1}$ ($\g_{12B}^{g_1}$ for negative oriented triangles). We will use it later in the definition of $F$ symbol to compensate the fermion parity changes of the Majorana fermion pairing projection operators.

Although the Majorana fermion parity of a given triangle may be changed, the fermion parity of the whole system is fixed under the global $U(g)$-action. Since the fermion parity of the vacuum pairings are not changed by symmetry actions, we only need to consider the $n_1$ domain walls decorated by Kitaev chains. For a particular (closed) Kitaev chain, the Majorana fermion parity is the same as the vacuum pairings if the pairings are constructed according to Kasteleyn orientations of the resolved dual lattice. It is also not hard to show that symmetry action will always change the arrow even times, following the pairing rules of \eq{2D:g_symm}. Therefore, we conclude all closed Kitaev chains will have even fermion parity, although the local fermion parity of a triangle may be changed compare to the Kasteleyn orientations.

To sum up, among the $|G_b|$ species of Majorana fermions, we decorate exactly one Kitaev chain to each symmetry domain wall specified by the $n_1$ configurations of the state. The decoration is symmetric under symmetry actions. The Majorana fermion parity of a triangle is changed according to \eq{2D:Pf_M} compared to the Kasteleyn oriented pairings.

\subsubsection{Complex fermion decoration}
\label{sec:2D:c}

The rules of  complex fermion decoration are much simpler than the pairings of Majorana fermions. The decoration is specified by a $\Z_2$-valued 2-cochain $n_2\in C^1(G_b,\Z_2)$. If $n_2(g_i,g_j,g_k)=0$, all the modes of complex fermions $c_{ijk}^\s$ ($\s\in G_b$) at the center of triangle $\langle ijk\rangle$ ($i<j<k$) are unoccupied. On the other hand, if $n_2(g_i,g_j,g_k)=1$, exactly one complex fermion mode $c_{ijk}^{g_i}$ will be decorated at the center of triangle $\langle ijk\rangle$. All other complex fermions $c_{ijk}^\s$ ($\s\neq g_i$) are still in vacuum states.

The complex fermion decoration rule is $G_b$-symmetric. Under a $U(g)$-action, the vertex labels $\{g_i\}$ becomes $\{gg_i\}$. According to our decoration rule, the decorated complex fermion [if $n_2(gg_i,gg_j,gg_k) = n_2(g_i,g_j,g_k) = 1$] should be $c_{ijk}^{gg_i}$, which is exactly the complex fermion $c_{ijk}^{g_i}$ by a $U(g)$-action.

\subsection{F moves}

To compare the states on different triangulations of the 2D spacial manifold, we should consider the 2D Pachner move, which is essentially the retriangulation of a rectangle. The Pachner move induces a FSLU transformation of the FSPT wave functions from the right-hand-side triangulation lattice $\cT'$ to the left-hand-side lattice $\cT$. We can first define the \emph{standard} $F$ move for rectangle with $g_0=e$, then other non-standard ones can be obtained by simply a $U(g_0)$-action. The standard $F$ move is given by:
\begin{align}
\label{2D:Fmove}
\Psi\left(
\!\!\!\!
\tikzfig{scale=2.3}{
\coordinate (g0) at (0,0);
\coordinate (g1) at (1,0);
\coordinate (g2) at (1,1);
\coordinate (g3) at (0,1);
\pgfmathsetmacro{\s}{.11}
\coordinate (02-1) at (1/2+\s/1.414, 1/2-\s/1.414);
\coordinate (02-2) at (1/2-\s/1.414, 1/2+\s/1.414);
\coordinate (03-1) at (0+\s, 1/2);
\coordinate (03-2) at (0-\s, 1/2);
\coordinate (12-1) at (1-\s, 1/2);
\coordinate (12-2) at (1+\s, 1/2);
\coordinate (01-1) at (1/2, 0+\s);
\coordinate (01-2) at (1/2, 0-\s);
\coordinate (23-1) at (1/2, 1-\s);
\coordinate (23-2) at (1/2, 1+\s);
\draw [green!30,line width=10pt] ($(03-2)!-.2!(03-1)$)--(03-2)--(03-1)--(02-2)--(02-1)--(01-1)--(01-2)--($(01-2)!-.2!(01-1)$);
\Ellipse{gray!50}{12-1}{12-2}{1.4}{.4};
\Ellipse{gray!50}{23-1}{23-2}{1.4}{.4};
\Ellipse{gray!50}{02-2}{03-1}{1.4}{.3};
\Ellipse{gray!50}{01-1}{02-1}{1.4}{.3};
\Ellipse{gray!50}{01-1}{01-2}{1.4}{.4};
\Ellipse{gray!50}{03-1}{03-2}{1.4}{.4};
\Ellipse{gray!50}{02-1}{02-2}{1.4}{.4};
\fill [red] (01-1) circle (.6pt);
\fill [red] (01-2) circle (.6pt);
\fill [red] (02-1) circle (.6pt);
\fill [red] (02-2) circle (.6pt);
\fill [red] (03-1) circle (.6pt);
\fill [red] (03-2) circle (.6pt);
\fill [red] (12-1) circle (.6pt);
\fill [red] (12-2) circle (.6pt);
\fill [red] (23-1) circle (.6pt);
\fill [red] (23-2) circle (.6pt);
\draw [->-=2/3,thick] (g0) -- (g1);
\draw [->-=2/3,thick] (g2)--(g1) node[scale=.85,below right,yshift=4]{$g_0^{-1}g_3$};
\draw [->-=2/3,thick] (g3)node[scale=.85,above left,yshift=-4]{$g_0^{-1}g_1$}--(g2) node[scale=.85,above right,yshift=-4]{$g_0^{-1}g_2$};
\draw [->-=2/3,thick] (g0)node[scale=.85,below left]{$e$} -- (g3);
\draw [->-=2/3,thick] (g0) -- (g2);
\draw[->-=1.1/2,thick,red,densely dotted] (23-1)--(02-2);
\draw[->-=1.1/2,very thick,red] (03-1)--(02-2);
\draw[->-=1.4/2,thick,red,densely dotted] (03-1)--(23-1);
\draw[->-=1.4/2,thick,red,densely dotted] (12-1)--(01-1);
\draw[->-=1.1/2,very thick,red] (02-1)--(01-1);
\draw[->-=1.1/2,thick,red,densely dotted] (02-1)--(12-1);
\draw[->-=3.2/4,very thick,red,densely dashed] (03-2)node[scale=.65,red,below]{$\g_{01A}^e$}--(03-1)node[scale=.65,red,below]{$\g_{01B}^e$};
\draw[->-=3.2/4,very thick,red,densely dotted] (01-1)node[scale=.65,red,left]{$\g_{03A}^e$}--(01-2)node[scale=.65,red,left]{$\g_{03B}^e$};
\draw[->-=3.2/4,very thick,red,densely dotted] (12-2)node[scale=.65,red,above,xshift=6]{$\g_{23A}^{g_0^{-1}g_2}$}--(12-1)node[scale=.65,red,above,xshift=-3]{$\g_{23B}^{g_0^{-1}g_2}$};
\draw[->-=3.2/4,very thick,red,densely dotted] (23-2)node[scale=.65,red,right]{$\g_{12A}^{g_0^{-1}g_1}$}--(23-1)node[scale=.65,red,right]{$\g_{12B}^{g_0^{-1}g_1}$};
\draw[->-=3.2/4,very thick,red,densely dotted] (02-2)node[scale=.65,red,xshift=12]{$\g_{02A}^e$}--(02-1)node[scale=.65,red,xshift=-12]{$\g_{02B}^e$};
\node[circle,shading=ball,scale=.6] at ($(.5,.5)!.3!(g1)$) {};
\node[below right,blue,scale=.8] at ($(.5,.5)!.3!(g1)$) {$c_{023}^{e}$};
\node[circle,shading=ball,scale=.6] at ($(.5,.5)!.3!(g3)$) {};
\node[above left,blue,scale=.8] at ($(.5,.5)!.3!(g3)$) {$c_{012}^{e}$};
}
\!\!\!
\right)
\quad = \quad
F(e,g_0^{-1}g_1,g_0^{-1}g_2,g_0^{-1}g_3)
\quad
\Psi\left(
\!\!\!\!
\tikzfig{scale=2.3}{
\coordinate (g0) at (0,0);
\coordinate (g1) at (1,0);
\coordinate (g2) at (1,1);
\coordinate (g3) at (0,1);
\pgfmathsetmacro{\s}{.11}
\coordinate (13-1) at (1/2-\s/1.414, 1/2-\s/1.414);
\coordinate (13-2) at (1/2+\s/1.414, 1/2+\s/1.414);
\coordinate (03-1) at (0+\s, 1/2);
\coordinate (03-2) at (0-\s, 1/2);
\coordinate (12-1) at (1-\s, 1/2);
\coordinate (12-2) at (1+\s, 1/2);
\coordinate (01-1) at (1/2, 0+\s);
\coordinate (01-2) at (1/2, 0-\s);
\coordinate (23-1) at (1/2, 1-\s);
\coordinate (23-2) at (1/2, 1+\s);
\draw [green!30,line width=10pt] ($(03-2)!-.2!(03-1)$)--(03-2)--(03-1)--(01-1)--(01-2)--($(01-2)!-.2!(01-1)$);
\Ellipse{gray!50}{13-1}{13-2}{1.4}{.4};
\Ellipse{gray!50}{12-1}{12-2}{1.4}{.4};
\Ellipse{gray!50}{23-2}{23-1}{1.4}{.4};
\Ellipse{gray!50}{01-1}{03-1}{1.3}{.2};
\Ellipse{gray!50}{01-1}{01-2}{1.4}{.4};
\Ellipse{gray!50}{03-1}{03-2}{1.4}{.4};
\draw [->-=2/3,thick] (g0) node[scale=.85,below left,yshift=4]{$e$} -- (g1);
\draw [->-=2/3,thick] (g2)--(g1) node[scale=.85,below right,yshift=4]{$g_0^{-1}g_3$};
\draw [->-=2/3,thick] (g3)--(g2) node[scale=.85,above right,yshift=-4]{$g_0^{-1}g_2$};
\draw [->-=2/3,thick] (g0) -- (g3)node[scale=.85,above left,yshift=-4]{$g_0^{-1}g_1$};
\draw [->-=2/3,thick] (g3)--(g1);
\fill [red] (01-1) circle (.6pt);
\fill [red] (01-2) circle (.6pt);
\fill [red] (13-1) circle (.6pt);
\fill [red] (13-2) circle (.6pt);
\fill [red] (03-1) circle (.6pt);
\fill [red] (03-2) circle (.6pt);
\fill [red] (12-1) circle (.6pt);
\fill [red] (12-2) circle (.6pt);
\fill [red] (23-1) circle (.6pt);
\fill [red] (23-2) circle (.6pt);
\draw[->-=1.1/2,thick,red,densely dotted] (03-1)--(13-1);
\draw[->-=1.4/2,very thick,red] (03-1)--(01-1);
\draw[->-=1.1/2,thick,red,densely dotted] (13-1)--(01-1);
\draw[thick,red,densely dotted] (23-1)--(12-1);
\draw[->-=.5,thick,blue] ($(23-1)!.7!(12-1)$) -- ($(23-1)!.701!(12-1)$);
\draw[->-=1.1/2,thick,red,densely dotted] (23-1)--(13-2);
\draw[thick,red,densely dotted] (12-1)--(13-2);
\draw[->-=.5,thick,blue] ($(12-1)!.55!(13-2)$) -- ($(12-1)!.56!(13-2)$);
\draw[->-=3.2/4,very thick,red,densely dotted] (03-2)node[scale=.65,red,above]{$\g_{01A}^e$}--(03-1)node[scale=.65,red,above]{$\g_{01B}^e$};
\draw[->-=3.2/4,very thick,red,densely dotted] (01-1)node[scale=.65,red,right]{$\g_{03A}^e$}--(01-2)node[scale=.65,red,right]{$\g_{03B}^e$};
\draw[->-=3.2/4,very thick,red,densely dotted] (12-2)node[scale=.65,red,below,xshift=6]{$\g_{23A}^{g_0^{-1}g_2}$}--(12-1)node[scale=.65,red,below,xshift=-3]{$\g_{23B}^{g_0^{-1}g_2}$};
\draw[->-=3.2/4,very thick,red,densely dotted] (23-2)node[scale=.65,red,left]{$\g_{12A}^{g_0^{-1}g_1}$}--(23-1)node[scale=.65,red,left]{$\g_{12B}^{g_0^{-1}g_1}$};
\draw[->-=3.2/4,very thick,red] (13-2)node[scale=.65,red,xshift=-14,yshift=5]{$\g_{13A}^{g_0^{-1}g_1}$}--(13-1)node[scale=.65,red,xshift=14,yshift=-4]{$\g_{13B}^{g_0^{-1}g_1}$};
\node[circle,shading=ball,scale=.6] at ($(.5,.5)!.3!(g0)$) {};
\node[below left,blue,scale=.8] at ($(.5,.5)!.3!(g0)$) {$c_{013}^{e}$};
\node[circle,shading=ball,scale=.6] at ($(.5,.5)!.3!(g2)$) {};
\node[above right,blue,scale=.8] at ($(.5,.5)!.3!(g2)$) {$c_{123}^{g_0^{-1}g_1}$};
}
\!\!\!
\right),
\end{align}
where the FSLU $F$ operator is defined as
\begin{align}\label{2D:F}
F(e,\bar 01,\bar 02,\bar 03)
= \nu_3(\bar 01,\bar 12,\bar 23) \big(c_{012}^{e\dagger}\big)^{ n_2(012)} \big(c_{023}^{e\dagger}\big)^{ n_2(023)} \big(c_{013}^{e}\big)^{n_2(013)} \left(c_{123}^{g_0^{-1}g_1}\right)^{n_2(123)} X_{0123}[n_1].
\end{align}
We used the abbreviation $\bar ij$ for $g_i^{-1}g_j$ in the arguments of $F$. And $n_2(ijk)$ represents $n_2(g_i,g_j,g_k)=n_2(g_i^{-1}g_j,g_j^{-1}g_k)$ for short.

The $U(1)$ phase factor $\nu_3(\bar 01,\bar 12,\bar 23) = \nu_3(g_0^{-1}g_1,g_1^{-1}g_2,g_2^{-1}g_3)$ in the front of $F$ symbol is a inhomogeneous 3-cochain depending on three group elements. By definition, it is related to the homogeneous cochain by
\begin{align}
\nu_3(g_0^{-1}g_1,g_1^{-1}g_2,g_2^{-1}g_3) = \nu_3(e,g_0^{-1}g_1,g_0^{-1}g_2,g_0^{-1}g_3),
\end{align}
with the first argument of homogeneous cochain to be the identity element $e\in G_b$. Later, we will use symmetry conditions to relate $\nu_3(e,g_0^{-1}g_1,g_0^{-1}g_2,g_0^{-1}g_3)$ and $\nu_3(g_0,g_1,g_2,g_3)$.

The complex fermion term of the form $c^\dagger c^\dagger c c$ annihilate two complex fermions at the two triangles of the right-hand-side figure and create two on the left-hand-side figure of \eq{2D:Fmove}. According to our decoration rules developed in section~\ref{sec:2D:c}, the triangle $\langle ijk\rangle$ is decorated by complex fermion $c_{ijk}^{g_i}$. So in the standard $F$ move, only the last fermion $c_{123}^{g_0^{-1}g_1}$ has group element label $g_0^{-1}g_1$, and the other three fermions all have group element label $e$. We note that, different from the $G_f=\Zf\times G_b$ case \cite{WangGu2017}, the complex fermion parity of the 2D $F$ move does not need to be conserved in general.

The term $X_{0123}[n_1]$ is related to the Kitaev chain decorations. In terms of Majorana fermion pairing projection operators \eq{proj}, the general expression of $X$ operator is
\begin{align}
\label{2D:X}
X_{0123}[n_1]
& = P_{0123}[n_1] \cdot \left(\gamma_{23B}^{g_0^{-1}g_2}\right)^{\dd n_2(0123)},\\
\label{2D:P}
P_{0123}[n_1] &= \left(\prod_{\mathrm{loop\;}i}2^{(L_i-1)/2}\right) \left( \prod_{\mathrm{Majorana\; pairs\;} \langle a,b\rangle \mathrm{\;in\;}\cT} P_{a,b}^{g_a,g_b} \right) \left( \prod_{\mathrm{link\;}\langle ij\rangle \notin \cT} \prod_{\s\in G_b} P_{ijA,ijB}^{\s,\s} \right).
\end{align}
The Majorana fermion $\gamma_{23B}^{g_0^{-1}g_2}$ is added for fermion parity considerations (which will be discussed in detail in the next section). If we do not add this term, $X_{0123}[n_1]$ would project the right-hand-side state to zero, whenever the Majorana fermion parity is changed under this $F$ move. We choose the Majorana fermion $\gamma_{23B}^{g_0^{-1}g_2}$ because all the Majorana fermion parity change cases involve it in the standard $F$ move [see the blue arrows in \eq{2D:Fmove}]. The pairing projection operator $P_{0123}[n_1]$ in \eq{2D:P} has three terms. The first term is a normalization factor, where $2L_i$ is the length of the $i$-th loop in the transition graph of Majorana pairing dimer configurations on the left triangulation lattice $\cT$ and right lattice $\cT'$. The second term projects the state to the Majorana pairing configuration state in the left figure. And the third term is the product of the vacuum projection operators for those Majorana fermions that do not appear explicitly in the left figure. As an example, the explicit $X$ operator for the $n_1$ configurations shown in \eq{2D:Fmove} is
\begin{align}\label{2D:X_eg}
X_{0123}[n_1] = 2^{1/2} \left(P_{01B,02A}^{e,e} P_{02B,03A}^{e,e} \prod_{\s\neq e}P_{02A,02B}^{\s,\s} \right) \left( \prod_{\s\in G_b}P_{13A,13B}^{\s,\s} \right).
\end{align}
Since there is no pairings for the two blue arrow links in \eq{2D:Fmove}, the Majorana fermion parity is always conserved for this $n_1$ configuration.

The $F$ symbol constructed above should be a FSLU operator. So it should be both fermion parity even and symmetric under $G_b$-action. Similar to the 1D case, we can use these conditions to obtain several consistency equations for the cochains $n_1$, $n_2$ and $\nu_3$.

\subsubsection{Fermion parity conservation}

It is proved that the Majorana fermion parity is conserved under 2D $F$ move if they are paired according to the Kasteleyn orientations in 2D \cite{WangGu2017}. Nevertheless, some of the links are not Kasteleyn oriented in the standard $F$ move \eq{2D:Fmove}. This is because the triangle $\langle 123\rangle$ is non-standard, i.e., the group element label of the first vertex is not $e\in G_b$. It should be obtained from the standard one by a $U(g_0^{-1}g_1)$-action. So the blue arrows inside this triangle may change their directions according to our symmetric pairing rules. The Majorana fermion parity change of this triangle can be calculated from \eq{2D:Pf_M}. Note that the three group element labels of the vertices are now $g_0^{-1}g_1$, $g_0^{-1}g_2$ and $g_0^{-1}g_3$. So the Majorana fermion parity change under the standard $F$ move is
\begin{align}\label{2D:Pf_Maj}
\Delta P_f^\g (F) = (-1)^{(\om_2\smile n_1 + s_1\smile n_1\smile n_1)(g_0^{-1}g_1,g_1^{-1}g_2,g_2^{-1}g_3)}.
\end{align}

On the other hand, the complex fermion parity change under the $F$ move can be simply calculated by counting the complex fermion numbers of the two sides:
\begin{align}
\Delta P_f^c (F) = (-1)^{n_2(g_0^{-1}g_1,g_1^{-1}g_2) + n_2(g_0^{-1}g_2,g_2^{-1}g_3) + n_2(g_0^{-1}g_1,g_1^{-1}g_3) + n_2(g_1^{-1}g_2,g_2^{-1}g_3)} = (-1)^{\dd n_2(g_0^{-1}g_1,g_1^{-1}g_2,g_2^{-1}g_3)}.
\end{align}
As a result, the conservation of total fermion parity $\Delta P_f=\Delta P_f^\g\cdot \Delta P_f^c = 1$ under the $F$ move enforces the condition
\begin{align}\label{2D:dn2}
\dd n_2 = \om_2 \smile n_1 + s_1 \smile n_1 \smile n_1.
\end{align}
It shows that the Majorana fermions and complex fermions are coupled to each other.

This is very different from the 2D FSPT states with unitary group $G_f=\Zf\times G_b$ (i.e., $\om_2=0$ and $s_1=0$) \cite{WangGu2017}, where the fermion parities of the Majorana fermions and complex fermions are conserved separately. So \eq{2D:dn2} is reduced to a simple cocycle equation $\dd n_2=0$. In the case of $T^2=-1$ topological superconductors \cite{WangNingChen2017}, although both $\om_2$ and $s_1$ are nontrivial, there combination $\om_2+ s_1 \smile n_1=0$ is also trivial. So we still have $\dd n_2=0$. That is the reason why it admits exactly solvable model with only Kitaev chain decorations.

\subsubsection{Symmetry condition}

In the previous discussions, we only considered the standard $F$ move with $g_0=e$. The non-standard $F$ moves are constructed by symmetry actions on the standard one. From \eq{FUg}, we have the following commuting diagram:
\begin{align}\label{2D:FUg}
\tikzfig{}{
\node (p1) at (0,0) {
\begin{tikzpicture}[scale=2.]
\coordinate (g0) at (0,0);
\coordinate (g1) at (1,0);
\coordinate (g2) at (1,1);
\coordinate (g3) at (0,1);
\pgfmathsetmacro{\s}{.08}
\coordinate (02-1) at (1/2+\s/1.414, 1/2-\s/1.414);
\coordinate (02-2) at (1/2-\s/1.414, 1/2+\s/1.414);
\coordinate (03-1) at (0+\s, 1/2);
\coordinate (03-2) at (0-\s, 1/2);
\coordinate (12-1) at (1-\s, 1/2);
\coordinate (12-2) at (1+\s, 1/2);
\coordinate (01-1) at (1/2, 0+\s);
\coordinate (01-2) at (1/2, 0-\s);
\coordinate (23-1) at (1/2, 1-\s);
\coordinate (23-2) at (1/2, 1+\s);
\fill [red] (01-1) circle (.8pt);
\fill [red] (01-2) circle (.8pt);
\fill [red] (02-1) circle (.8pt);
\fill [red] (02-2) circle (.8pt);
\fill [red] (03-1) circle (.8pt);
\fill [red] (03-2) circle (.8pt);
\fill [red] (12-1) circle (.8pt);
\fill [red] (12-2) circle (.8pt);
\fill [red] (23-1) circle (.8pt);
\fill [red] (23-2) circle (.8pt);
\draw [->-=2/3,thick] (g0) -- (g1);
\draw [->-=2/3,thick] (g2)--(g1) node[scale=.9,below right,yshift=4]{$g_0^{-1}g_3$};
\draw [->-=2/3,thick] (g3)node[scale=.9,above left,yshift=-4]{$g_0^{-1}g_1$}--(g2) node[scale=.9,above right,yshift=-4]{$g_0^{-1}g_2$};
\draw [->-=2/3,thick] (g0)node[scale=.9,below left]{$e$} -- (g3);
\draw [->-=2/3,thick] (g0) -- (g2);
\node[circle,shading=ball,scale=.6] at ($(.5,.5)!.4!(g1)$) {};
\node[circle,shading=ball,scale=.6] at ($(.5,.5)!.4!(g3)$) {};
\end{tikzpicture}
};
\node (pp1) at (0,-3.8) {
\begin{tikzpicture}[scale=2.]
\coordinate (g0) at (0,0);
\coordinate (g1) at (1,0);
\coordinate (g2) at (1,1);
\coordinate (g3) at (0,1);
\pgfmathsetmacro{\s}{.08}
\coordinate (02-1) at (1/2+\s/1.414, 1/2-\s/1.414);
\coordinate (02-2) at (1/2-\s/1.414, 1/2+\s/1.414);
\coordinate (03-1) at (0+\s, 1/2);
\coordinate (03-2) at (0-\s, 1/2);
\coordinate (12-1) at (1-\s, 1/2);
\coordinate (12-2) at (1+\s, 1/2);
\coordinate (01-1) at (1/2, 0+\s);
\coordinate (01-2) at (1/2, 0-\s);
\coordinate (23-1) at (1/2, 1-\s);
\coordinate (23-2) at (1/2, 1+\s);
\fill [red] (01-1) circle (.8pt);
\fill [red] (01-2) circle (.8pt);
\fill [red] (02-1) circle (.8pt);
\fill [red] (02-2) circle (.8pt);
\fill [red] (03-1) circle (.8pt);
\fill [red] (03-2) circle (.8pt);
\fill [red] (12-1) circle (.8pt);
\fill [red] (12-2) circle (.8pt);
\fill [red] (23-1) circle (.8pt);
\fill [red] (23-2) circle (.8pt);
\draw [->-=2/3,thick] (g0) -- (g1);
\draw [->-=2/3,thick] (g2)--(g1) node[scale=.9,below right,yshift=4]{$g_3$};
\draw [->-=2/3,thick] (g3)node[scale=.9,above left,yshift=-4]{$g_1$}--(g2) node[scale=.9,above right,yshift=-4]{$g_2$};
\draw [->-=2/3,thick] (g0)node[scale=.9,below left]{$g_0$} -- (g3);
\draw [->-=2/3,thick] (g0) -- (g2);
\node[circle,shading=ball,scale=.6] at ($(.5,.5)!.4!(g1)$) {};
\node[circle,shading=ball,scale=.6] at ($(.5,.5)!.4!(g3)$) {};
\end{tikzpicture}
};
\node (pp2) at (8.3,-3.8) {
\begin{tikzpicture}[scale=2.]
\coordinate (g0) at (0,0);
\coordinate (g1) at (1,0);
\coordinate (g2) at (1,1);
\coordinate (g3) at (0,1);
\pgfmathsetmacro{\s}{.08}
\coordinate (13-1) at (1/2-\s/1.414, 1/2-\s/1.414);
\coordinate (13-2) at (1/2+\s/1.414, 1/2+\s/1.414);
\coordinate (03-1) at (0+\s, 1/2);
\coordinate (03-2) at (0-\s, 1/2);
\coordinate (12-1) at (1-\s, 1/2);
\coordinate (12-2) at (1+\s, 1/2);
\coordinate (01-1) at (1/2, 0+\s);
\coordinate (01-2) at (1/2, 0-\s);
\coordinate (23-1) at (1/2, 1-\s);
\coordinate (23-2) at (1/2, 1+\s);
\draw [->-=2/3,thick] (g0) node[scale=.9,below left,yshift=4]{$g_0$} -- (g1);
\draw [->-=2/3,thick] (g2)--(g1) node[scale=.9,below right,yshift=4]{$g_3$};
\draw [->-=2/3,thick] (g3)--(g2) node[scale=.9,above right,yshift=-4]{$g_2$};
\draw [->-=2/3,thick] (g0) -- (g3)node[scale=.9,above left,yshift=-4]{$g_1$};
\draw [->-=2/3,thick] (g3)--(g1);
\fill [red] (01-1) circle (.8pt);
\fill [red] (01-2) circle (.8pt);
\fill [red] (13-1) circle (.8pt);
\fill [red] (13-2) circle (.8pt);
\fill [red] (03-1) circle (.8pt);
\fill [red] (03-2) circle (.8pt);
\fill [red] (12-1) circle (.8pt);
\fill [red] (12-2) circle (.8pt);
\fill [red] (23-1) circle (.8pt);
\fill [red] (23-2) circle (.8pt);
\node[circle,shading=ball,scale=.6] at ($(.5,.5)!.4!(g0)$) {};
\node[circle,shading=ball,scale=.6] at ($(.5,.5)!.4!(g2)$) {};
\end{tikzpicture}
};
\node (p2) at (8.3,0) {
\begin{tikzpicture}[scale=2.]
\coordinate (g0) at (0,0);
\coordinate (g1) at (1,0);
\coordinate (g2) at (1,1);
\coordinate (g3) at (0,1);
\pgfmathsetmacro{\s}{.08}
\coordinate (13-1) at (1/2-\s/1.414, 1/2-\s/1.414);
\coordinate (13-2) at (1/2+\s/1.414, 1/2+\s/1.414);
\coordinate (03-1) at (0+\s, 1/2);
\coordinate (03-2) at (0-\s, 1/2);
\coordinate (12-1) at (1-\s, 1/2);
\coordinate (12-2) at (1+\s, 1/2);
\coordinate (01-1) at (1/2, 0+\s);
\coordinate (01-2) at (1/2, 0-\s);
\coordinate (23-1) at (1/2, 1-\s);
\coordinate (23-2) at (1/2, 1+\s);
\draw [->-=2/3,thick] (g0) node[scale=.9,below left,yshift=4]{$e$} -- (g1);
\draw [->-=2/3,thick] (g2)--(g1) node[scale=.9,below right,yshift=4]{$g_0^{-1}g_3$};
\draw [->-=2/3,thick] (g3)--(g2) node[scale=.9,above right,yshift=-4]{$g_0^{-1}g_2$};
\draw [->-=2/3,thick] (g0) -- (g3)node[scale=.9,above left,yshift=-4]{$g_0^{-1}g_1$};
\draw [->-=2/3,thick] (g3)--(g1);
\fill [red] (01-1) circle (.8pt);
\fill [red] (01-2) circle (.8pt);
\fill [red] (13-1) circle (.8pt);
\fill [red] (13-2) circle (.8pt);
\fill [red] (03-1) circle (.8pt);
\fill [red] (03-2) circle (.8pt);
\fill [red] (12-1) circle (.8pt);
\fill [red] (12-2) circle (.8pt);
\fill [red] (23-1) circle (.8pt);
\fill [red] (23-2) circle (.8pt);
\node[circle,shading=ball,scale=.6] at ($(.5,.5)!.4!(g0)$) {};
\node[circle,shading=ball,scale=.6] at ($(.5,.5)!.4!(g2)$) {};
\end{tikzpicture}
};
\draw[<-] (p1)--(p2) node [midway,above] {$F(e,g_0^{-1}g_1,g_0^{-1}g_2,g_0^{-1}g_3)$};
\draw[<-] (pp1)--(pp2) node [midway,above] {$F(g_0,g_1,g_2,g_3)$};
\draw[->] (p1)--(pp1) node [midway,left] {$U(g_0)$};
\draw[->] (p2)--(pp2) node [midway,right] {$U(g_0)$};
}
\end{align}
So the non-standard $F$ operator is defined as
\begin{align}\label{2D:Fg}
F(g_0,g_1,g_2,g_3) = \emp^{g_0}F(e,g_0^{-1}g_1,g_0^{-1}g_2,g_0^{-1}g_3) := U(g_0) F(e,g_0^{-1}g_1,g_0^{-1}g_2,g_0^{-1}g_3) U(g_0)^\dagger.
\end{align}
The $F$ moves constructed in this way are automatically symmetric, because we can derive the transformation rule
\begin{align}
F(gg_0,gg_1,gg_2,gg_3) = U(g) F(g_0,g_1,g_2,g_3) U(g)^\dagger,
\end{align}
using \eq{U_proj} and the fact $F$ is fermion parity even.

After a $U(g_0)$-action on the standard $F$ operator \eq{2D:F}, we can obtain the generic $F$ symbol expression:
\begin{align}\label{2D:F_}
F(g_0,g_1,g_2,g_3)
= \nu_3(g_0,g_1,g_2,g_3) \big(c_{012}^{g_0\dagger}\big)^{ n_2(012)} \big(c_{023}^{g_0\dagger}\big)^{ n_2(023)} \big(c_{013}^{g_0}\big)^{n_2(013)} \big(c_{123}^{g_1}\big)^{n_2(123)} X_{0123}[n_1].
\end{align}
The complex fermions now have group element labels $g_0$ or $g_1$. And the $X$ operator is
\begin{align}\label{2D:X_}
X_{0123}[n_1] = P_{0123}[n_1] \cdot \big(\g_{23B}^{g_2}\big)^{\dd n_2(0123)},
\end{align}
with added Majorana fermion $\gamma_{23B}^{g_2}$ rather than $\gamma_{23B}^{g_0^{-1}g_2}$. $P_{0123}[n_1]$ has similar expression as \eq{2D:P} that project the Majorana fermions to the pairing state on the left-hand-side figure (the group element labels are changed appropriately).

From the decoration rules of Majorana fermions and complex fermions, $n_1$ and $n_2$ are invariant under symmetry actions. The generic homogeneous cochain $\nu_3$ in \eq{2D:F_} is a combination of the inhomogeneous $\nu_3$ in the standard $F$ move and the $\pm 1$ signs appearing in the symmetry action. So we have the following symmetry conditions for $n_1$, $n_2$ and $\nu_3$:
\begin{align}
n_1(g_0,g_1) &= n_1(e,g_0^{-1}g_1) = n_1(g_0^{-1}g_1),\\
n_2(g_0,g_1,g_2) &= n_2(e,g_0^{-1}g_1,g_0^{-1}g_2) = n_2(g_0^{-1}g_1,g_1^{-1}g_2),\\\label{2D:nu3_symm}
\nu_3(g_0,g_1,g_2,g_3) &= \emp^{g_0}\nu_3(g_0^{-1}g_1,g_1^{-1}g_2,g_2^{-1}g_3)
= \nu_3(g_0^{-1}g_1,g_1^{-1}g_2,g_2^{-1}g_3)^{1-2s_1(g_0)} \cdot \mathcal O_4^\mathrm{symm}(g_0,g_1,g_2,g_3).
\end{align}
The symmetry sign $\mathcal O_4^\mathrm{symm}$ in the last equation is given by
\begin{align}\label{2D:Osymm}\nonumber
\mathcal O_4^\mathrm{symm}(g_0,g_1,g_2,g_3)
&= (-1)^{\om_2(g_0,g_0^{-1}g_1)n_2(123) + [s_1(g_0)+\om_2(g_0,g_0^{-1}g_2)] \dd n_2(0123)}\\
&= (-1)^{(\om_2\smile n_2 + s_1\smile \dd n_2 )(g_0,g_0^{-1}g_1,g_1^{-1}g_2,g_2^{-1}g_3) +\om_2(g_0,g_0^{-1}g_2) \dd n_2(g_0^{-1}g_1,g_1^{-1}g_2,g_2^{-1}g_3)},
\end{align}
where the sign $(-1)^{\om_2(g_0,g_0^{-1}g_1)n_2(123)}$ comes from the symmetry transformation \eq{2D:symm:c} of $c_{123}^{g_0^{-1}g_1}$, and the sign $(-1)^{[s_1(g_0)+\om_2(g_0,g_0^{-1}g_2)] \dd n_2(0123)}$ comes from the symmetry transformation \eq{2D:symm:B} of $\g_{23B}^{g_0^{-1}g_2}$ in the $X$ operator. We note that the last term $\om_2\dd n_2$ in \eq{2D:Osymm} is not a cup product or cup-1 product form. This symmetry sign $\mathcal O_4^\mathrm{symm}$ will appear later in the twisted cocycle equation for $\nu_3$ as part of the obstruction function [see \eq{2D:O4_4}].

\subsection{Super pentagon and twisted cocycle equations}

The $F$ moves should satisfy a consistency condition known as pentagon equation for fusion categories. In fermionic setting, it is a super pentagon equation with some fermion sign twist for superfusion categories \cite{Gu2015,Brundan2017,Usher2016}. The 2D FSPT states correspond to a special kind of superfusion category in which all the simple objects are invertible. So the classification of 2D FSPT states can be understood mathematically as the classification of pointed superfusion categories corresponding to a given symmetry group.

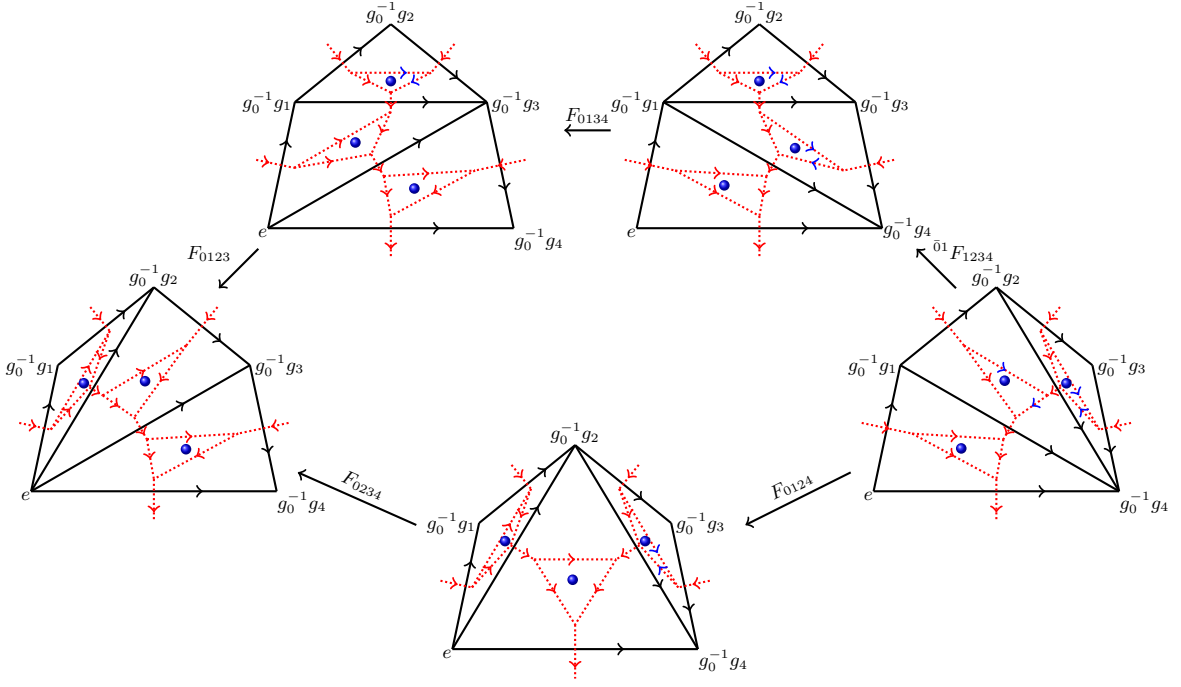
\begin{figure}[ht]
\centering
$
\tikzfig{scale=3.5}{
\pgfmathsetmacro{\th}{360/5}
\pgfmathsetmacro{\s}{.15}
\pgfmathsetmacro{\ss}{.12}
\pgfmathsetmacro{\out}{.4}
\pgfmathsetmacro{\ssmall}{.03}
\pgfmathsetmacro{\small}{.1}
\pgfmathsetmacro{\middle}{.2}
\pgfmathsetmacro{\large}{.4}
\node[inner sep=0pt] (p1) at (-1.6,0)
{
 \begin{tikzpicture}[scale=1.5]
\coordinate (0) at ($({-90-\th/2}:1)-(.5,0)$);
\coordinate (1) at ($({-90-\th/2-\th}:1)+(.1,0)$);
\coordinate (2) at ({-90-\th/2-2*\th}:1);
\coordinate (3) at ($({-90-\th/2-3*\th}:1)-(.1,0)$);
\coordinate (4) at ($({-90-\th/2-4*\th}:1)+(.5,0)$);
\draw[->-=.7,thick] (0)node[scale=.8,below left]{$e$}--(1)node[scale=.8,left,xshift=-1]{$g_0^{-1}g_1$};
\draw[->-=.7,thick] (1)--(2)node[scale=.8,above]{$g_0^{-1}g_2$};
\draw[->-=.7,thick] (2)--(3)node[scale=.8,right,xshift=2]{$g_0^{-1}g_3$};
\draw[->-=.7,thick] (3)--(4)node[scale=.8,below right]{$g_0^{-1}g_4$};
\draw[->-=.7,thick] (0)--(4);
\draw[->-=.7,thick] (0) -- (2);
\draw[->-=.7,thick] (0) -- (3);
\midpt{01a}{\out}{0}{1};
\midpt{01b}{\small}{1}{0};
\draw[->-=.55,thick,red,densely dotted] (01a)--(01b);
\midpt{12a}{\out}{1}{2};
\midpt{12b}{\small}{2}{1};
\draw[->-=.55,thick,red,densely dotted] (12a) -- (12b);
\midpt{23a}{\out}{2}{3};
\midpt{23b}{\large}{3}{2};
\draw[->-=.35,thick,red,densely dotted] (23a) -- (23b);
\midpt{34a}{\out}{3}{4};
\midpt{34b}{\large}{4}{3};
\draw[->-=.35,thick,red,densely dotted] (34a) -- (34b);
\midpt{04a}{.6*\out}{4}{0};
\midpt{04b}{\small}{0}{4};
\draw[->-=.8,thick,red,densely dotted] (04b)
-- (04a);
\midpt{02a}{\ssmall}{0}{2};
\midpt{02b}{\small}{2}{0};
\draw[->-=.8,thick,red,densely dotted] (02a)
-- (02b);
\midpt{03a}{\small}{0}{3};
\midpt{03b}{\small}{3}{0};
\draw[->-=.8,thick,red,densely dotted] (03a)
-- (03b);
\draw[->-=.5,thick,red,densely dotted] (01b) -- (02a);
\draw[->-=.65,thick,red,densely dotted] (01b) -- (12b);
\draw[->-=.5,thick,red,densely dotted] (12b) -- (02a);
\draw[->-=.5,thick,red,densely dotted] (03b) -- (04b);
\draw[->-=.5,thick,red,densely dotted] (03b) -- (34b);
\draw[->-=.5,thick,red,densely dotted] (34b) -- (04b);
\draw[->-=.5,thick,red,densely dotted] (02b) -- (03a);
\draw[->-=.5,thick,red,densely dotted] (02b) -- (23b);
\draw[->-=.5,thick,red,densely dotted] (23b) -- (03a);
\node[circle,shading=ball,scale=4] at ($($(0)!.5!(2)$)!.25!(1)$) {};
\node[circle,shading=ball,scale=4] at ($($(2)!.5!(3)$)!.333!(0)$) {};
\node[circle,shading=ball,scale=4] at ($($(4)!.5!(3)$)!.333!(0)$) {};
 \end{tikzpicture}
};
\node[inner sep=0pt] (p2) at (-.7,1)
{
 \begin{tikzpicture}[scale=1.5]
\draw[->-=.7,thick] (0)node[scale=.8,below left]{$e$}--(1)node[scale=.8,left,xshift=-1]{$g_0^{-1}g_1$};
\draw[->-=.7,thick] (1)--(2)node[scale=.8,above]{$g_0^{-1}g_2$};
\draw[->-=.7,thick] (2)--(3)node[scale=.8,right,xshift=2]{$g_0^{-1}g_3$};
\draw[->-=.7,thick] (3)--(4)node[scale=.8,below right]{$g_0^{-1}g_4$};
\draw[->-=.7,thick] (0)--(4);
\draw[->-=.7,thick] (1) -- (3);
\draw[->-=.7,thick] (0) -- (3);
\midpt{01b}{\middle}{1}{0};
\draw[->-=.35,thick,red,densely dotted] (01a) -- (01b);
\midpt{12b}{\middle}{2}{1};
\draw[->-=.55,thick,red,densely dotted] (12a) -- (12b);
\midpt{23b}{\middle}{3}{2};
\draw[->-=.55,thick,red,densely dotted] (23a) -- (23b);
\midpt{34b}{\large}{4}{3};
\draw[->-=.55,thick,red,densely dotted] (34a) -- (34b);
\midpt{04b}{\small}{0}{4};
\draw[->-=.8,thick,red,densely dotted] (04b)
-- (04a);
\midpt{13a}{\small}{1}{3};
\midpt{13b}{\small}{3}{1};
\draw[->-=.8,thick,red,densely dotted] (13a)
-- (13b);
\midpt{03a}{\small}{0}{3};
\midpt{03b}{\small}{3}{0};
\draw[->-=.8,thick,red,densely dotted] (03a)
-- (03b);
\draw[->-=.5,thick,red,densely dotted] (12b) -- (13a);
\draw[thick,red,densely dotted] (12b) -- (23b);
\draw[->-=.5,blue,thick] ($(12b)!.7!(23b)$) -- ($(12b)!.701!(23b)$);
\draw[thick,red,densely dotted] (23b) -- (13a);
\draw[->-=.5,blue,thick] ($(23b)!.5!(13a)$) -- ($(23b)!.501!(13a)$);
\draw[->-=.5,thick,red,densely dotted] (03b) -- (04b);
\draw[->-=.5,thick,red,densely dotted] (03b) -- (34b);
\draw[->-=.5,thick,red,densely dotted] (34b) -- (04b);
\draw[->-=.5,thick,red,densely dotted] (01b) -- (03a);
\draw[->-=.5,thick,red,densely dotted] (01b) -- (13b);
\draw[->-=.5,thick,red,densely dotted] (13b) -- (03a);
\node[circle,shading=ball,scale=4] at ($($(1)!.5!(3)$)!.27!(2)$) {};
\node[circle,shading=ball,scale=4] at ($($(1)!.5!(0)$)!.36!(3)$) {};
\node[circle,shading=ball,scale=4] at ($($(3)!.5!(4)$)!.37!(0)$) {};
 \end{tikzpicture}
};
\node[inner sep=0pt] (p3) at (.7,1)
{
 \begin{tikzpicture}[scale=1.5]
\draw[->-=.7,thick] (0)node[scale=.8,below left]{$e$}--(1)node[scale=.8,left,xshift=-1]{$g_0^{-1}g_1$};
\draw[->-=.7,thick] (1)--(2)node[scale=.8,above]{$g_0^{-1}g_2$};
\draw[->-=.7,thick] (2)--(3)node[scale=.8,right,xshift=2]{$g_0^{-1}g_3$};
\draw[->-=.7,thick] (3)--(4)node[scale=.8,right]{$g_0^{-1}g_4$};
\draw[->-=.7,thick] (0)--(4);
\draw[->-=.7,thick] (1) -- (3);
\draw[->-=.7,thick] (1) -- (4);
\midpt{01b}{\large}{1}{0};
\draw[->-=.55,thick,red,densely dotted] (01a) -- (01b);
\midpt{12b}{\middle}{2}{1};
\draw[->-=.55,thick,red,densely dotted] (12a) -- (12b);
\midpt{23b}{\middle}{3}{2};
\draw[->-=.55,thick,red,densely dotted] (23a) -- (23b);
\midpt{34b}{\large}{4}{3};
\draw[->-=.35,thick,red,densely dotted] (34a) -- (34b);
\midpt{04b}{\small}{0}{4};
\draw[->-=.8,thick,red,densely dotted] (04b)
-- (04a);
\midpt{13a}{\small}{1}{3};
\midpt{13b}{\small}{3}{1};
\draw[->-=.8,thick,red,densely dotted] (13a)
-- (13b);
\midpt{14a}{\small}{1}{4};
\midpt{14b}{\small}{4}{1};
\draw[->-=.8,thick,red,densely dotted] (14a)
-- (14b);
\draw[->-=.5,thick,red,densely dotted] (12b) -- (13a);
\draw[thick,red,densely dotted] (12b) -- (23b);
\draw[->-=.5,blue,thick] ($(12b)!.7!(23b)$) -- ($(12b)!.701!(23b)$);
\draw[thick,red,densely dotted] (23b) -- (13a);
\draw[->-=.5,blue,thick] ($(23b)!.5!(13a)$) -- ($(23b)!.501!(13a)$);
\draw[->-=.5,thick,red,densely dotted] (01b) -- (04b);
\draw[->-=.5,thick,red,densely dotted] (01b) -- (14b);
\draw[->-=.5,thick,red,densely dotted] (14b) -- (04b);
\draw[->-=.5,thick,red,densely dotted] (13b) -- (14a);
\draw[thick,red,densely dotted] (13b) -- (34b);
\draw[->-=.5,blue,thick] ($(13b)!.65!(34b)$) -- ($(13b)!.651!(34b)$);
\draw[thick,red,densely dotted] (34b) -- (14a);
\draw[->-=.5,blue,thick] ($(34b)!.5!(14a)$) -- ($(34b)!.501!(14a)$);
\node[circle,shading=ball,scale=4] at ($($(1)!.5!(3)$)!.27!(2)$) {};
\node[circle,shading=ball,scale=4] at ($($(1)!.5!(4)$)!.27!(3)$) {};
\node[circle,shading=ball,scale=4] at ($($(1)!.5!(0)$)!.32!(4)$) {};
 \end{tikzpicture}
};
\node[inner sep=0pt] (p4) at (0,-.6)
{
 \begin{tikzpicture}[scale=1.5]
\draw[->-=.7,thick] (0)node[scale=.8,below left]{$e$}--(1)node[scale=.8,left,xshift=-1]{$g_0^{-1}g_1$};
\draw[->-=.7,thick] (1)--(2)node[scale=.8,above]{$g_0^{-1}g_2$};
\draw[->-=.7,thick] (2)--(3)node[scale=.8,right,xshift=2]{$g_0^{-1}g_3$};
\draw[->-=.7,thick] (3)--(4)node[scale=.8,below right]{$g_0^{-1}g_4$};
\draw[->-=.7,thick] (0)--(4);
\draw[->-=.7,thick] (0) -- (2);
\draw[->-=.7,thick] (2) -- (4);
\midpt{01b}{\small}{1}{0};
\draw[->-=.55,thick,red,densely dotted] (01a) -- (01b);
\midpt{12b}{\small}{2}{1};
\draw[->-=.55,thick,red,densely dotted] (12a) -- (12b);
\midpt{23b}{\small}{3}{2};
\draw[->-=.55,thick,red,densely dotted] (23a) -- (23b);
\midpt{34b}{\small}{4}{3};
\draw[->-=.55,thick,red,densely dotted] (34a) -- (34b);
\midpt{04b}{\middle}{0}{4};
\draw[->-=.8,thick,red,densely dotted] (04b)
-- (04a);
\midpt{02a}{\ssmall}{0}{2};
\midpt{02b}{\middle}{2}{0};
\draw[->-=.8,thick,red,densely dotted] (02a)
-- (02b);
\midpt{24a}{\ssmall}{2}{4};
\midpt{24b}{\middle}{4}{2};
\draw[->-=.8,thick,red,densely dotted] (24a)
-- (24b);
\draw[->-=.5,thick,red,densely dotted] (01b) -- (02a);
\draw[->-=.65,thick,red,densely dotted] (01b) -- (12b);
\draw[->-=.5,thick,red,densely dotted] (12b) -- (02a);
\draw[->-=.5,thick,red,densely dotted] (23b) -- (24a);
\draw[thick,red,densely dotted] (23b) -- (34b);
\draw[->-=.5,blue,thick] ($(23b)!.65!(34b)$) -- ($(23b)!.651!(34b)$);
\draw[thick,red,densely dotted] (34b) -- (24a);
\draw[->-=.5,blue,thick] ($(34b)!.5!(24a)$) -- ($(34b)!.501!(24a)$);
\draw[->-=.5,thick,red,densely dotted] (02b) -- (04b);
\draw[->-=.5,thick,red,densely dotted] (02b) -- (24b);
\draw[->-=.5,thick,red,densely dotted] (24b) -- (04b);
\node[circle,shading=ball,scale=4] at ($($(2)!.5!(0)$)!.32!(4)$) {};
\node[circle,shading=ball,scale=4] at ($($(0)!.5!(2)$)!.25!(1)$) {};
\node[circle,shading=ball,scale=4] at ($($(4)!.5!(2)$)!.25!(3)$) {};
 \end{tikzpicture}
};
\node[inner sep=0pt] (p5) at (1.6,0)
{
 \begin{tikzpicture}[scale=1.5]
\draw[->-=.7,thick] (0)node[scale=.8,below left]{$e$}--(1)node[scale=.8,left,xshift=-1]{$g_0^{-1}g_1$};
\draw[->-=.7,thick] (1)--(2)node[scale=.8,above]{$g_0^{-1}g_2$};
\draw[->-=.7,thick] (2)--(3)node[scale=.8,right,xshift=2]{$g_0^{-1}g_3$};
\draw[->-=.7,thick] (3)--(4)node[scale=.8,below right]{$g_0^{-1}g_4$};
\draw[->-=.7,thick] (0)--(4);
\draw[->-=.7,thick] (2) -- (4);
\draw[->-=.7,thick] (1) -- (4);
\midpt{01b}{\large}{1}{0};
\draw[->-=.55,thick,red,densely dotted] (01a) -- (01b);
\midpt{12b}{\large}{2}{1};
\draw[->-=.35,thick,red,densely dotted] (12a) -- (12b);
\midpt{23b}{\small}{3}{2};
\draw[->-=.55,thick,red,densely dotted] (23a) -- (23b);
\midpt{34b}{\small}{4}{3};
\draw[->-=.55,thick,red,densely dotted] (34a) -- (34b);
\midpt{04b}{\small}{0}{4};
\draw[->-=.8,thick,red,densely dotted] (04b)
-- (04a);
\midpt{14a}{\small}{1}{4};
\midpt{14b}{\small}{4}{1};
\draw[->-=.8,thick,red,densely dotted] (14a)
-- (14b);
\midpt{24a}{\ssmall}{2}{4};
\midpt{24b}{\small}{4}{2};
\draw[->-=.8,thick,red,densely dotted] (24a)
-- (24b);
\draw[->-=.5,thick,red,densely dotted] (01b) -- (04b);
\draw[->-=.5,thick,red,densely dotted] (01b) -- (14b);
\draw[->-=.5,thick,red,densely dotted] (14b) -- (04b);
\draw[->-=.5,thick,red,densely dotted] (23b) -- (24a);
\draw[thick,red,densely dotted] (23b) -- (34b);
\draw[->-=.5,blue,thick] ($(23b)!.65!(34b)$) -- ($(23b)!.651!(34b)$);
\draw[thick,red,densely dotted] (34b) -- (24a);
\draw[->-=.5,blue,thick] ($(34b)!.5!(24a)$) -- ($(34b)!.501!(24a)$);
\draw[->-=.5,thick,red,densely dotted] (12b) -- (14a);
\draw[thick,red,densely dotted] (12b) -- (24b);
\draw[->-=.5,blue,thick] ($(12b)!.5!(24b)$) -- ($(12b)!.501!(24b)$);
\draw[thick,red,densely dotted] (24b) -- (14a);
\draw[->-=.5,blue,thick] ($(24b)!.5!(14a)$) -- ($(24b)!.501!(14a)$);
\node[circle,shading=ball,scale=4] at ($($(1)!.5!(0)$)!.32!(4)$) {};
\node[circle,shading=ball,scale=4] at ($($(4)!.5!(2)$)!.25!(3)$) {};
\node[circle,shading=ball,scale=4] at ($($(1)!.5!(2)$)!.33!(4)$) {};
 \end{tikzpicture}
};
\draw[<-,thick] (p2.east) -- (p3.west) node[scale=.8,midway,above] {$F_{0134}$};
\draw[<-,thick] (-1.1,-.3) -- (-.65,-.5) node[scale=.8,midway,above,sloped] {$F_{0234}$};
\draw[<-,thick] (.6,-.5) -- (1.,-.3) node[scale=.8,midway,above,sloped] {$F_{0124}$};
\draw[<-,thick] (-1.4,.4) -- (-1.25,.55) node[scale=.8,midway,above left] {$F_{0123}$};
\draw[<-,thick] (1.25,.55) -- (1.4,.4) node[scale=.8,midway,above right,xshift=-5] {$\emp^{\bar 01}F_{1234}$};
}
$
\caption{Standard super pentagon equation. The dual trivalent graph of the triangulation is the usual string diagram pentagon equation for tensor category. Algebraically, this standard super pentagon condition corresponds to \eq{2D:dF}. Since the group element label of the first vertex is $e\in G_b$, all the fermionic $F$ moves are standard except $\emp^{\bar 01}F_{1234}$. Note that we used a simpler notation $F_{ijkl}=F(e,\bar ij,\bar jk,\bar kl)$ in the figure. Blue arrows indicate that the Majorana fermion pairing directions may be changed compared to the red arrow Kasteleyn orientations.}
\label{fig:pentagon}
\end{figure}

Similar to previous discussions, we only need to consider the \emph{standard} super pentagon equation with $g_0=e$. All other super pentagon equations can be obtained from it by a $U(g_0)$ symmetry action. So it is enough to merely consider the standard super pentagon as coherence conditions. This standard super pentagon equation is shown in \fig{fig:pentagon}. Algebraically, we have the following equation
\begin{align}\nonumber\label{2D:dF}
F(e,\bar 02,\bar 03,\bar 04) \cdot F(e,\bar 01,\bar 02,\bar 04)
&= F(e,\bar 01,\bar 02,\bar 03) \cdot F(e,\bar 01,\bar 03,\bar 04) \cdot F(\bar 01,\bar 02,\bar 03,\bar 04)\\
&= F(e,\bar 01,\bar 02,\bar 03) \cdot F(e,\bar 01,\bar 03,\bar 04) \cdot \emp^{\bar 01}F(e,\bar 12,\bar 13,\bar 14),
\end{align}
where we used $\bar ij$ to denote $g_i^{-1}g_j$. Note that only the last $F$ symbol is non-standard in the above equation.

Now we can substitute the explicit expression of the standard $F$ move \eq{2D:F} into the standard super pentagon equation \eq{2D:dF}. After eliminating all complex fermions and Majorana fermions, we would obtain a twisted cocycle equation for the inhomogeneous 3-cochain $\nu_3$. In general, the twisted cocycle equation reads
\begin{align}
\dd \nu_3 = \mathcal O_4[n_2],
\end{align}
where $\mathcal O_4[n_2]$ is a functional of $n_2$ only (as well as $\om_2$ and $s_1$ parametrizing the symmetry group, of course). The $n_1$ dependence of $\mathcal O_4$ is though $\dd n_2$ by \eq{2D:dn2}.
Since the fermion parities of Majorana fermions and complex fermions are coupled to each other, $\mathcal O_4[n_2]$ is much more complicated than the special result $\mathcal O_4[n_2] = (-1)^{n_2\smile n_2}$ for unitary $G_f = \Zf \times G_b$ \cite{GuWen2014,Gu2015,WangGu2017}.

From general considerations, the obstruction function $\mathcal O_4[n_2]$ consists of four terms:
\begin{align}\label{2D:O4_4}
\mathcal O_4[n_2] = \mathcal O_4^\mathrm{symm}[n_2] \cdot \mathcal O_4^{c}[n_2] \cdot \mathcal O_4^{c\gamma}[\dd n_2] \cdot \mathcal O_4^{\gamma}[\dd n_2].
\end{align}
These four terms have different physical meanings and are summarized as
\begin{align}\label{2D:O4_s}
\mathcal O_4^\mathrm{symm}[n_2](01234) &= (-1)^{(\om_2\smile n_2+s_1\smile\dd n_2)(01234)+\om_2(013)\dd n_2(1234)},\\\label{2D:O4_c}
\mathcal O_4^{c}[n_2] &= (-1)^{n_2\smile n_2+ \dd n_2 \smile_1 n_2},\\\label{2D:O4_cg}
\mathcal O_4^{c\gamma}[\dd n_2] &= (-1)^{\dd n_2\smile_2\dd n_2},\\\label{2D:O4_g}
\mathcal O_4^{\gamma}[\dd n_2](01234) &= (-1)^{\dd n_2(0124)\dd n_2(0234)}(-i)^{\dd n_2(0123)[1-\dd n_2(0124)]\text{ (mod 2)}}.
\end{align}
Note that the $\dd n_2$ terms in the exponent of $(-i)$ in the last equation should be understood as taking mod 2 values (can only be 0 or 1). And the notation $\smile_i$ is the higher cup product by Steenrod \cite{Steenrod1947}. By adding a coboundary $(-1)^{\dd(s_1\smile n_2+n_2\smile_2\dd n_2)}$ to the obstruction function and shifting $\nu_3\rightarrow \nu_3(-1)^{s_1\smile n_2+n_2\smile_2\dd n_2}$, we can simplify the above obstruction function to
\begin{align}\label{2D:O4}
\mathcal O_4[n_2](01234) = (-1)^{(\om_2\smile n_2 + n_2\smile n_2 + n_2 \smile_1 \dd n_2)(01234) + \om_2(013)\dd n_2(1234) + \dd n_2(0124)\dd n_2(0234)} (-i)^{\dd n_2(0123)[1-\dd n_2(0124)]\text{ (mod 2)}}.
\end{align}
We note that only the first three terms $\om_2\smile n_2 + n_2\smile n_2 + n_2 \smile_1 \dd n_2$ in the exponent are expressed as (higher) cup product form, while other terms are not. If we consider the special case of $\om_2=0$ and $s_1=0$, then we have $\dd n_2=0$ from \eq{2D:dn2}. So the above equation will reduce to the known sign twist $\mathcal O_4[n_2] = (-1)^{n_2\smile n_2}$ in the super pentagon or super-cohomology equation \cite{GuWen2014,Gu2015}.

Before calculating the obstruction function in detail, we would note that we have checked numerically that the claimed expression \eq{2D:O4} of $\mathcal O_4[n_2]$ is a cocycle, i.e. $\dd \mathcal O_4 = 1$, for arbitrary choices of $s_1$, $\om_2$, $n_1$ and $n_2$ satisfying the corresponding consistency equations. This is a consistency check, because the super pentagon equation \eq{2D:dF} always implies a one higher dimensional equation involving one more vertex.

\subsubsection{Calculations of obstruction function $\mathcal O_4[n_2]$}

In this subsection, we would give explicit calculations of the four terms of the obstruction function $\mathcal O_4[n_2]$ in \eq{2D:O4_4}.

The first term $\mathcal O_4^\mathrm{symm}[n_2]$ comes from the $U(\bar 01)$ symmetry action on $F(e,\bar 12,\bar 13,\bar 14)$ in the last term of \eq{2D:dF}. The homogeneous $\nu_3$ in the non-standard $F$ move is related to the inhomogeneous $\nu_3$ of the standard $F$ move by a symmetry action [see \eq{2D:nu3_symm}]. So using the explicit expression \eq{2D:Osymm}, we have
\begin{align}\nonumber
\mathcal O_4^\mathrm{symm}[n_2](\bar 01,\bar 02,\bar 03,\bar 04)
= (-1)^{(\om_2\smile n_2 + s_1\smile \dd n_2 )(\bar 01,\bar 12,\bar 23,\bar 34) +\om_2(\bar 01,\bar 13) \dd n_2(\bar 12,\bar 23,\bar 34)},
\end{align}
which is exactly \eq{2D:O4_s} claimed above.

The second term $\mathcal O_4^{c}[n_2]$ is the fermion sign from reordering the complex fermion operators $\left(c_{ijk}^{g_0^{-1}g_i}\right)^{n_2(g_i,g_j,g_k)}$ in \eq{2D:dF}. To compare the complex fermion operators on the two sides of the super pentagon equation, one have to rearrange these operators and finally obtain the fermion sign
\begin{align}\nonumber
\mathcal O_4^{c}[n_2](01234) &= (-1)^{(n_2\smile n_2+ \dd n_2 \smile_1 n_2)(01234)}\\
&= (-1)^{n_2(012) n_2(234) + \left[\dd n_2(0234) n_2(012) + \dd n_2(0134) n_2(123) + \dd n_2(0124) n_2(234)\right]}.
\end{align}
This is a generalization of the usual sign twist $(-1)^{n_2\smile n_2}$ for 2-cocycle $n_2$. If $\dd n_2\neq 0$, we would have another term $(-1)^{\dd n_2 \smile_1 n_2}$.

The third term $\mathcal O_4^{c\gamma}[\dd n_2]$ of the obstruction function originates from reordering the complex fermion and the Majorana fermions. For instance, to put all complex fermion operators to the front of Majorana fermion operators on left-hand-side of \eq{2D:dF}, we have to switch the $X$ operator of $F(0234)$ and the complex fermions of $F(0124)$. So there is a fermion sign $(-1)^{\dd n_2(0234)\dd n_2(0124)}$. Combining it with the fermion signs from the right-hand-side, we have the total sign
\begin{align}\nonumber
\mathcal O_4^{c\gamma}[\dd n_2](01234) &= (-1)^{(\dd n_2\smile_2\dd n_2)(01234)}\\
&= (-1)^{\dd n_2(0123)\dd n_2(0134) + \dd n_2(0234)\dd n_2(0124) + \dd n_2(0123)\dd n_2(1234) + \dd n_2(0134)\dd n_2(1234)}.
\end{align}
Since the fermion parities of the $X$ operator and the complex fermion operator $c^\dagger c^\dagger c c$ are only related to $\dd n_2$, this obstruction function $\mathcal O_4^{c\gamma}[\dd n_2]$ is a functional of $\dd n_2$ (rather than $n_2$ directly).

In the rest of this subsection, we would calculate the most complicated part $\mathcal O_4^\g[\dd n_2]$ of the obstruction function. In addition to $\pm 1$, this Majorana fermion term can also take values in $\pm i$. Whenever the Majorana fermion parity of the $F$ move is changed, i.e. $\dd n_2(0123)=1$, there will be a dangling Majorana fermion $\g_{23B}^{g_0^{-1}g_2}$ in the $X$ operator \eq{2D:X}. The presence of Majorana fermions depends only on $\dd n_2$. So similar to $\mathcal O_4^{c\gamma}[\dd n_2]$, we expect $\mathcal O_4^\g[\dd n_2]$ to be a functional of $\dd n_2$ only.

We can denote the five $X$ operators in the standard super pentagon equation \eq{2D:dF} by $X_{1234}=P_2 \big(\g_{34B}^{\bar 03}\big)^{\dd n_2(1234)}$, $X_{0234}=P_4 \big(\g_{34B}^{\bar 03}\big)^{\dd n_2(0234)}$, $X_{0134}=P_3 \big(\g_{34B}^{\bar 03}\big)^{\dd n_2(0134)}$, $X_{0124}=P_5 \big(\g_{24B}^{\bar 02}\big)^{\dd n_2(0124)}$ and $X_{0123}=P_4 \big(\g_{23B}^{\bar 02}\big)^{\dd n_2(0123)}$. Here, $P_i$ is the Majorana pairing projection operator of the corresponding $i$-th figure ($1\leq i\leq 5$) in the super pentagon equation \fig{fig:pentagon}. We use the convention that the rightmost figure is the first one with projection operator $P_1$, and the other four figures are counted counterclockwise. We also use the simpler notation
\begin{align}
\g_{ijB}^{\bar 0i} := \g_{ijB}^{g_0^{-1}g_i}.
\end{align}
Using these $X$ operators, the obstruction function coming from Majorana fermions can be calculated by
\begin{align}\nonumber\label{2D:O4g}
\mathcal O_4^\g[\dd n_2] &= \langle X_{1234}^\dagger X_{0134}^\dagger X_{0123}^\dagger X_{0234} X_{0124}\rangle\\
&= \left\langle P_1
\big(\g_{34B}^{\bar 03}\big)^{\dd n_2(1234)} P_2
\big(\g_{34B}^{\bar 03}\big)^{\dd n_2(0134)} P_3
\big(\g_{23B}^{\bar 02}\big)^{\dd n_2(0123)} P_4
P_4 \big(\g_{34B}^{\bar 03}\big)^{\dd n_2(0234)}
P_5 \big(\g_{24B}^{\bar 02}\big)^{\dd n_2(0124)}
P_1 \right\rangle.
\end{align}
The average is taken over the Majorana fermion state of the rightmost figure in \fig{fig:pentagon}. We also inserted $P_1$, which is 1 acting on the rightmost state, at the first and the last places of the operator string.

\eq{2D:O4g} should be calculated separately for different Majorana fermion configurations. Among the five dangling Majorana fermions of the five $X$ operators, only three of them are different:
\begin{align}
\left(\g_{34B}^{\bar 03}, \g_{24B}^{\bar 02}, \g_{23B}^{\bar 02}\right).
\end{align}
So we can use the triple of their number
\begin{align}
[\dd n_2(\hat 0)+\dd n_2(\hat 1)+\dd n_2(\hat 2),\ \ \dd n_2(\hat 3),\ \ \dd n_2(\hat 4)]\ \ \ \text{(mod 2)}
\end{align}
to indicate the presence or absence of the three Majorana fermions in \eq{2D:O4g}. For simplicity, we have used the notation
\begin{align}
\dd n_2(\hat i) := \dd n_2(01...\hat i...34),
\end{align}
where $\hat i$ means that the number $i$ is removed from the argument. Each element of the triple corresponds to the Majorana fermion parity change of one or several $F$ moves. There are in total $2^3/2=4$ different possibilities for the Majorana fermion parity changes (see the first column of \tab{tab:O4g}), since the total Majorana parity of the five $F$ moves should be even. We can now calculate $\mathcal O_4^\g[\dd n_2]$ for these four cases separately.

\begin{table}[ht]
\centering
\begin{tabular}{|c|c|c|c|c|}
\hline
$P_f^\g$ changes & expression of $\mathcal O_4^\g[\dd n_2]$ & $P_1$ & $P_4$ & $\mathcal O_4^\g[\dd n_2]$ \\
\hline\hline
$(0,0,0)$ & $\left\langle \big(\g_{34B}^{\bar 03}\big)^{\dd n_2(\hat 0)+\dd n_2(\hat 2)+\dd n_2(\hat 1)} \right\rangle$ & $1$ & $1$ & $1$ \\
\hline
$(1,0,1)$ & $\left\langle \big(\g_{34B}^{\bar 03}\big)^{\dd n_2(\hat 0)+\dd n_2(\hat 2)} \g_{23B}^{\bar 02} \big(\g_{34B}^{\bar 03}\big)^{\dd n_2(\hat 1)} P_1 \right\rangle$ & $(-1)^{s_1(\bar 02)+\om_2(\bar 02,\bar 23)} \big(-i\g_{23B}^{\bar 02} \g_{34B}^{\bar 03}\big)$ & $1$ & $-i$ \\
\hline
$(1,1,0)$ & $\left\langle \big(\g_{34B}^{\bar 03}\big)^{\dd n_2(\hat 0)+\dd n_2(\hat 2)} P_4 \big(\g_{34B}^{\bar 03}\big)^{\dd n_2(\hat 1)} \g_{24B}^{\bar 02} P_1 \right\rangle$ & $(-1)^{\om_2(\bar 02,\bar 23)} \big(-i\g_{34B}^{\bar 03} \g_{24A}^{\bar 02}\big)$ & $-i\g_{24A}^{\bar 02} \g_{24B}^{\bar 02}$ & $(-1)^{\dd n_2(\hat 1)}$ \\
\hline
$(0,1,1)$ & $\left\langle \big(\g_{34B}^{\bar 03}\big)^{\dd n_2(\hat 0)+\dd n_2(\hat 2)} \g_{23B}^{\bar 02} P_4 \big(\g_{34B}^{\bar 03}\big)^{\dd n_2(\hat 1)} \g_{24B}^{\bar 02} P_1 \right\rangle$ & $-i\g_{23B}^{\bar 02} \g_{24A}^{\bar 02}$ & $-i\g_{24A}^{\bar 02} \g_{24B}^{\bar 02}$ & $(-1)^{\dd n_2(\hat 1)}$ \\
\hline
\end{tabular}
\caption{Calculations of $\mathcal O_4^\gamma[\dd n_2]$ for all possible Kitaev chain configurations in the super pentagon equation \fig{fig:pentagon}. The first column is the Majorana fermion parity change triple $[\dd n_2(\hat 0)+\dd n_2(\hat 1)+\dd n_2(\hat 2), \dd n_2(\hat 3), \dd n_2(\hat 4)]$ (mod 2). There are in total 4 different cases. The second column is a simplified version of \eq{2D:O4g} for each case. The third and forth columns are the Majorana pairing projection operators we used in calculation. And the last column is the final result of $\mathcal O_4^\gamma[\dd n_2]$, which can be summarized by \eq{2D:O4g_}.}
\label{tab:O4g}
\end{table}

As an example, let us calculate $\mathcal O_4^\g[\dd n_2]$ for the second case with Majorana fermion parity changes $(1,0,1)$ (the third raw of \tab{tab:O4g}). The dangling Majorana fermions present in \eq{2D:O4g} are $\g_{34B}^{\bar 03}$ and $\g_{23B}^{\bar 02}$. We can expand the projection operators $P_i$ to Majorana fermion operators. Since $\g_{34B}^{\bar 03}$ and $\g_{23B}^{\bar 02}$ are paired in the triangle $\langle 234\rangle$ of the rightmost figure in \fig{fig:pentagon}, we can consider only their pairing term [recall \eq{proj}]
\begin{align}\label{2D:O4g:P}
P_{23B,34B}^{\bar 02,\bar 03} = \frac{1}{2} \left[ 1 - (-1)^{\om_2(\bar 02,\bar 23)+s_1(\bar 02)} i \g_{23B}^{\bar 02} \g_{34B}^{\bar 03} \right]
\end{align}
 in $P_1$. And all other projection operators $P_i$ with $i\neq 1$ can be chosen to be 1. So the obstruction function \eq{2D:O4g} can be calculated as
\begin{align}
\mathcal O_4^\g[\dd n_2](01234) \big|_{(1,0,1)}
&= \left\langle \big(\g_{34B}^{\bar 03}\big)^{\dd n_2(\hat 0)+\dd n_2(\hat 2)} \g_{23B}^{\bar 02} \big(\g_{34B}^{\bar 03}\big)^{\dd n_2(\hat 1)} P_1 \right\rangle\\\label{2D:O4g:ex2}
&= \left\langle \big(\g_{34B}^{\bar 03}\big)^{\dd n_2(\hat 0)+\dd n_2(\hat 2)} \g_{23B}^{\bar 02} \big(\g_{34B}^{\bar 03}\big)^{\dd n_2(\hat 1)}
(-1)^{\om_2(\bar 02,\bar 23)+s_1(\bar 02)} \big(-i\g_{23B}^{\bar 02} \g_{34B}^{\bar 03}\big) \right\rangle\\\label{2D:O4g:ex3}
&= (-i) (-1)^{\om_2(\bar 02,\bar 23) + s_1(\bar 02) + \dd n_2(\hat 1)}\\
&= -i.
\end{align}
Note that we replaced $P_1$ by the second term of \eq{2D:O4g:P} to obtain \eq{2D:O4g:ex2} (see the third column of \tab{tab:O4g}). In this way, the Majorana fermions all appear in \eq{2D:O4g:ex2} even times. After switching the Majorana fermions $\g_{23B}^{\bar 02}$ and $\big(\g_{34B}^{\bar 03}\big)^{\dd n_2(\hat 1)}$, we obtain a fermion sign $(-1)^{\dd n_2(\hat 1)}$. And then we can eliminate all Majorana fermion operators since their square is one. To simplify the phase factor \eq{2D:O4g:ex3}, we observe that the conditions
\begin{align}
\dd n_2(\hat 3) &= [\om_2(\bar 01,\bar 12)+s_1(\bar 01)n_1(\bar 12)] n_1(\bar 24) = 0,\\
\dd n_2(\hat 4) &= [\om_2(\bar 01,\bar 12)+s_1(\bar 01)n_1(\bar 12)] n_1(\bar 23) = 1,
\end{align}
imply $n_1(\bar 23)=1$ and $n_1(\bar 24)=0$. We also have $n_1(\bar 34)=1$ from $\dd n_1(234)=0$. Using the relation
\begin{align}\nonumber
\dd n_2(\hat 1)
&= \om_2(\bar 02,\bar 23)n_1(\bar 34) + s_1(\bar 02)n_1(\bar 23)n_1(\bar 34)\\
&= \om_2(\bar 02,\bar 23) + s_1(\bar 02),
\end{align}
the exponent of $(-1)$ in \eq{2D:O4g:ex3} is in fact $0$. We therefore have the final result $\mathcal O_4^\g[\dd n_2](01234) |_{(1,0,1)} = -i$.

Similarly, we can calculate $\mathcal O_4^\g[\dd n_2]$ for all other cases of Majorana fermion parity changes. The information we need in the calculation is shown in \tab{tab:O4g}. The final results shown in the last column of \tab{tab:O4g} can be summarized into a simple expression (which is a functional of $\dd n_2$ only):
\begin{align}\nonumber\label{2D:O4g_}
\mathcal O_4^{\gamma}[\dd n_2](01234) &= (-1)^{\dd n_2(\hat 3)\dd n_2(\hat 1)} (-i)^{\dd n_2(\hat 4)\left[1-\dd n_2(\hat 3)\right]\text{ (mod 2)}}\\
&= (-1)^{\dd n_2(0124)\dd n_2(0234)} (-i)^{\dd n_2(0123)[1-\dd n_2(0124)]\text{ (mod 2)}}.
\end{align}
The $\dd n_2$ terms in the exponent of $(-i)$ should be understood as taking mod 2 values (can only be 0 or 1). This is exactly the result claimed previously in \eq{2D:O4_g}.

\subsection{Boundary ASPT states in $\Gamma^2$}
\label{sec:ASPT:1D}

We have constructed 2D FSPT states in the above discussions. However, not all of them correspond to distinct FSPT phases. In the following subsection, we will construct explicitly a FSLU transformation path to connect an FSPT state with $n_2=\om_2$ and a state without complex fermion decorations. The physical understanding is that there is a gapped symmetric boundary for the 2D FSPT state. So we conclude the state with $n_2$, which is in the new coboundary subgroup
\begin{align}\label{A:G2}
\Gamma^2 = \{\omega_2 \in H^2(G_b,\Z_2) \},
\end{align}
should be considered as in the trivial FSPT phase.

\subsubsection{FSLU to trivialize the 2D bulk}
\label{sec:ASPT:1D_}

Let us fix the symmetry group $G_f$ with given $G_b$, $\om_2$ and $s_1$. We consider the special group supercohomology 2D FSPT state constructed from $(n_2,\nu_3)$ data satisfying $\dd n_2=0$ and $\dd \nu_3=\mathcal O_4=(-1)^{n_2\smile n_2}$ \cite{GuWen2014}. We will show that the 2D FSPT state with $n_2=\om_2$ can be connected to a product state by FSLU transformations.

\begin{figure}[ht]
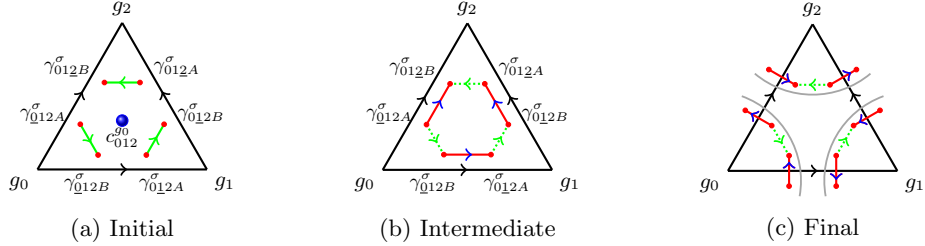

\centering
\begin{subfigure}[h]{.25\textwidth}
\centering
$\tikzfig{scale=1.3}{
\coordinate (0) at ({90+120}:1);
\coordinate (1) at ({-30}:1);
\coordinate (2) at ({90}:1);
\draw[->-=.53,black,thick](0)--(1);
\draw[->-=.53,black,thick](1)--(2);
\draw[->-=.53,black,thick](0)--(2);
\node[below left,scale=.9]at(0){$g_0$};
\node[below right,scale=.9]at(1){$g_1$};
\node[above,scale=.9]at(2){$g_2$};
\pgfmathsetmacro{\r}{.43};\pgfmathsetmacro{\theta}{35};
\coordinate (12_1) at ({30-\theta}:\r);\coordinate (12_2) at ({30+\theta}:\r);
\coordinate (02_1) at ({150+\theta}:\r);\coordinate (02_2) at ({150-\theta}:\r);
\coordinate (01_1) at ({-90-\theta}:\r);\coordinate (01_2) at ({-90+\theta}:\r);
\draw[->-=.6,green,thick](01_2)--(12_1);
\draw[->-=.6,green,thick](12_2)--(02_2);
\draw[->-=.6,green,thick](02_1)--(01_1);
\node[xshift=-4,yshift=-11,scale=.8] at (01_1) {$\g_{\underline 012B}^\s$};
\node[xshift=6,yshift=-11,scale=.8] at (01_2) {$\g_{0\underline 12A}^\s$};
\node[xshift=14,yshift=4,scale=.8] at (12_1) {$\g_{0\underline 12B}^\s$};
\node[xshift=14,yshift=6,scale=.8] at (12_2) {$\g_{01\underline 2A}^\s$};
\node[xshift=-14,yshift=4,scale=.8] at (02_1) {$\g_{\underline 012A}^\s$};
\node[xshift=-14,yshift=6,scale=.8] at (02_2) {$\g_{01\underline 2B}^\s$};
\node[circle,draw=red,fill=red,scale=.2] at (01_1) {};
\node[circle,draw=red,fill=red,scale=.2] at (01_2) {};
\node[circle,draw=red,fill=red,scale=.2] at (02_1) {};
\node[circle,draw=red,fill=red,scale=.2] at (02_2) {};
\node[circle,draw=red,fill=red,scale=.2] at (12_1) {};
\node[circle,draw=red,fill=red,scale=.2] at (12_2) {};
\node[circle,draw=blue,shading=ball,scale=.5] at (0,0) {};
\node[below,scale=.8] at (0,0) {$c_{012}^{g_0}$};
}$
\caption{Initial}
\label{fig:2DFSLUa}
\end{subfigure}
\begin{subfigure}[h]{.25\textwidth}
\centering
$\tikzfig{scale=1.3}{
\coordinate (0) at ({90+120}:1);
\coordinate (1) at ({-30}:1);
\coordinate (2) at ({90}:1);
\draw[->-=.5,black,thick](0)--(1);
\draw[->-=.5,black,thick](1)--(2);
\draw[->-=.5,black,thick](0)--(2);
\node[below left,scale=.9]at(0){$g_0$};
\node[below right,scale=.9]at(1){$g_1$};
\node[above,scale=.9]at(2){$g_2$};
\pgfmathsetmacro{\r}{.42};\pgfmathsetmacro{\theta}{35};
\coordinate (12_1) at ({30-\theta}:\r);\coordinate (12_2) at ({30+\theta}:\r);
\coordinate (02_1) at ({150+\theta}:\r);\coordinate (02_2) at ({150-\theta}:\r);
\coordinate (01_1) at ({-90-\theta}:\r);\coordinate (01_2) at ({-90+\theta}:\r);
\draw[red,thick](12_1)--(12_2);
\draw[red,thick](02_1)--(02_2);
\draw[red,thick](01_1)--(01_2);
\draw[->-=.6,blue,thick]($(12_1)!.6!(12_2)$)--($(12_1)!.601!(12_2)$);
\draw[->-=.6,blue,thick]($(02_1)!.6!(02_2)$)--($(02_1)!.601!(02_2)$);
\draw[->-=.6,blue,thick]($(01_1)!.6!(01_2)$)--($(01_1)!.601!(01_2)$);
\draw[->-=.6,green,densely dotted,thick](01_2)--(12_1);
\draw[->-=.6,green,densely dotted,thick](12_2)--(02_2);
\draw[->-=.6,green,densely dotted,thick](02_1)--(01_1);
\node[xshift=-4,yshift=-11,scale=.8] at (01_1) {$\g_{\underline 012B}^\s$};
\node[xshift=6,yshift=-11,scale=.8] at (01_2) {$\g_{0\underline 12A}^\s$};
\node[xshift=14,yshift=4,scale=.8] at (12_1) {$\g_{0\underline 12B}^\s$};
\node[xshift=14,yshift=6,scale=.8] at (12_2) {$\g_{01\underline 2A}^\s$};
\node[xshift=-14,yshift=4,scale=.8] at (02_1) {$\g_{\underline 012A}^\s$};
\node[xshift=-14,yshift=6,scale=.8] at (02_2) {$\g_{01\underline 2B}^\s$};
\node[circle,draw=red,fill=red,scale=.2] at (01_1) {};
\node[circle,draw=red,fill=red,scale=.2] at (01_2) {};
\node[circle,draw=red,fill=red,scale=.2] at (02_1) {};
\node[circle,draw=red,fill=red,scale=.2] at (02_2) {};
\node[circle,draw=red,fill=red,scale=.2] at (12_1) {};
\node[circle,draw=red,fill=red,scale=.2] at (12_2) {};
}$
\caption{Intermediate}
\label{fig:2DFSLUb}
\end{subfigure}
\begin{subfigure}[h]{.25\textwidth}
\centering
$\tikzfig{scale=1.3}{
\coordinate (0) at ({90+120}:1);
\coordinate (1) at ({-30}:1);
\coordinate (2) at ({90}:1);
\draw[->-=.53,black,thick](0)--(1);
\draw[->-=.53,black,thick](1)--(2);
\draw[->-=.53,black,thick](0)--(2);
\node[below left,scale=.9]at(0){$g_0$};
\node[below right,scale=.9]at(1){$g_1$};
\node[above,scale=.9]at(2){$g_2$};
\pgfmathsetmacro{\r}{.42};\pgfmathsetmacro{\theta}{35};
\coordinate (12_1) at ({30-\theta}:\r);\coordinate (12_2) at ({30+\theta}:\r);
\coordinate (02_1) at ({150+\theta}:\r);\coordinate (02_2) at ({150-\theta}:\r);
\coordinate (01_1) at ({-90-\theta}:\r);\coordinate (01_2) at ({-90+\theta}:\r);
\coordinate (12_1o) at ($ (30:.5)!-1!(12_2) $);
\coordinate (12_2o) at ($ (30:.5)!-1!(12_1) $);
\coordinate (02_1o) at ($ (150:.5)!-1!(02_2) $);
\coordinate (02_2o) at ($ (150:.5)!-1!(02_1) $);
\coordinate (01_1o) at ($ (-90:.5)!-1!(01_2) $);
\coordinate (01_2o) at ($ (-90:.5)!-1!(01_1) $);
\draw[->-=.8,red,thick](01_1o)--(01_1);
\draw[->-=.8,red,thick](01_2)--(01_2o);
\draw[->-=.8,red,thick](02_1)--(02_1o);
\draw[->-=.8,red,thick](02_2o)--(02_2);
\draw[->-=.8,red,thick](12_1o)--(12_1);
\draw[->-=.8,red,thick](12_2)--(12_2o);
\draw[->-=.8,blue,thick]($(01_1o)!.8!(01_1)$)--($(01_1o)!.801!(01_1)$);
\draw[->-=.8,blue,thick]($(01_2)!.8!(01_2o)$)--($(01_2)!.801!(01_2o)$);
\draw[->-=.8,blue,thick]($(02_1)!.8!(02_1o)$)--($(02_1)!.801!(02_1o)$);
\draw[->-=.8,blue,thick]($(02_2o)!.8!(02_2)$)--($(02_2o)!.801!(02_2)$);
\draw[->-=.8,blue,thick]($(12_1o)!.8!(12_1)$)--($(12_1o)!.801!(12_1)$);
\draw[->-=.8,blue,thick]($(12_2)!.8!(12_2o)$)--($(12_2)!.801!(12_2o)$);
\draw[->-=.6,green,densely dotted,thick](01_2)--(12_1);
\draw[->-=.6,green,densely dotted,thick](12_2)--(02_2);
\draw[->-=.6,green,densely dotted,thick](02_1)--(01_1);
\node[circle,draw=red,fill=red,scale=.2] at (01_1) {};
\node[circle,draw=red,fill=red,scale=.2] at (01_2) {};
\node[circle,draw=red,fill=red,scale=.2] at (02_1) {};
\node[circle,draw=red,fill=red,scale=.2] at (02_2) {};
\node[circle,draw=red,fill=red,scale=.2] at (12_1) {};
\node[circle,draw=red,fill=red,scale=.2] at (12_2) {};
\node[circle,draw=red,fill=red,scale=.2] at (01_1o) {};
\node[circle,draw=red,fill=red,scale=.2] at (01_2o) {};
\node[circle,draw=red,fill=red,scale=.2] at (02_1o) {};
\node[circle,draw=red,fill=red,scale=.2] at (02_2o) {};
\node[circle,draw=red,fill=red,scale=.2] at (12_1o) {};
\node[circle,draw=red,fill=red,scale=.2] at (12_2o) {};
\pgfmathsetmacro{\theta}{10};
\draw[thick,gray!70!white,line width=.7] ([shift=(-\theta:.92)]{90+120}:1.2) arc (-\theta:{60+\theta}:.92);
\draw[thick,gray!70!white,line width=.7] ([shift=({120-\theta}:.92)]{-30}:1.2) arc ({120-\theta}:{180+\theta}:.92);
\draw[thick,gray!70!white,line width=.7] ([shift=(-120-\theta:.92)]{90}:1.2) arc ({-120-\theta}:{-60+\theta}:.92);
}$
\caption{Final}
\label{fig:2DFSLUc}
\end{subfigure}
\caption{FSLU transforms a 2D ``FSPT'' state to a fermion product state. There are two FSLU transformations $(a)\xrightarrow{U_1}(b)\xrightarrow{U_2}(c)$ to trivialize the initial 2D FSPT state. (a) Majorana fermions are in vacuum pairs (green arrows). (b) There is exactly one Kitaev's Majorana chain inside the triangle (red links). (c) There is one Kitaev's Majorana chain around each vertex (inside the gray arc). And we can shrink it to the vertex and redefine the basis state $|g_i\rangle$ for the vertex.
}
\label{fig:2DFSLU}
\end{figure}

Consider a 2D triangulation lattice of a closed oriented spacial manifold. The FSPT wave function is a superposition of the basis states with coefficients related to $\nu_3$ [see \eq{2D:wf}]. The $n_2$ data specifies the decorations of complex fermion $c_{ijk}^{g_i}$ at the center of each triangle $\langle ijk\rangle$ of the bosonic basis state $|\{g_i\}\rangle$. So the fermionic basis state can be expressed as
\begin{align}\label{A:a}
|(a)\rangle &= \prod_{\langle ijk\rangle} \big(c_{ijk}^{g_i\dagger}\big)^{n_2(ijk)} |\{g_i\}\rangle.
\end{align}
We will show in the following that the above state can be transformed by two FSLU as
\begin{align}
|(a)\rangle \xrightarrow{\ U_1\ } |(b)\rangle \xrightarrow{\ U_2\ } |(c)\rangle,
\end{align}
where the schematic figures of these three states are shown in \fig{fig:2DFSLU}. The final state $|(c)\rangle$ is obtained from the bosonic state $|\{g_i\}\rangle$ by decorating a small Kitaev chain around each vertex (see the Kitaev chain along the gray arc in \fig{fig:2DFSLUc}). After shrinking the small Kitaev chain to each vertex $i$, we can view the state $|g_i\rangle'$ with a fermion mode as a new basis state. So the final state has the expression
\begin{align}\label{A:c}
|(c)\rangle &= |\{g_i\}\rangle' = \otimes_{i} |g_i\rangle',
\end{align}
which is a fermionic product state without complex fermion decorations on the triangles. Therefore, using the two FSLU $U_1$ and $U_2$, we have connect an FSPT state with complex fermion decorations specified by $n_2=\om_2$ to another FSPT state without complex fermion decorations. So the decoration data $n_2=\om_2$ for complex fermion layer is trivialized.

The following is the detailed constructions for the two FSLU transformations.

(1) The first FSLU transformation $U_1$ from $|(a)\rangle$ to $|(b)\rangle$.

Apart from the complex fermion $c_{012}^\s$ (blue dots) at the center of the triangle in \fig{fig:2DFSLUa}, we also add $3|G_b|$ fermion modes ($a_{\underline 012}^\s$, $a_{0\underline 12}^\s$ and $a_{01\underline 2}^\s$) near the three vertices ($0$, $1$ and $2$) of the triangle and split them into $6|G_b|$ Majorana fermions (red dots in \fig{fig:2DFSLUa}). The Majorana fermions are paired from $\g_A^\s$ to $\g_B^\s$ (vacuum pair) near each vertex, respecting the right-hand rule of the triangle orientation (green arrows). Since all the Majorana fermions are in vacuum pairings, we does not change the initial state \eq{A:a} with only complex fermions $c_{ijk}^{g_i}$.

After the transformation by $U_1$, the initial state is changed to the intermediate state $|(b)\rangle$ shown in \fig{fig:2DFSLUb}. In the state $|(b)\rangle$, there is exactly one nontrivial Majorana chain (red lines) and $|G_b|-1$ trivial Majorana chains (dotted green arrow) along the boundary of the triangle. The Majorana fermions forming nontrivial Kitaev chain are labelled by group elements of the nearby vertices. And the pairing directions are chosen to respect the $G_b$ symmetry. So the projection operators \eq{proj} for the nontrivial pairings (red lines) inside the triangle $\langle 012\rangle$ in \fig{fig:2DFSLUb} are
\begin{align}
P_{\underline 012B,0\underline 12A}^{g_0,g_1}
= U(g_0) P_{\underline 012B,0\underline 12A}^{e,g_0^{-1}g_1} U(g_0)^\dagger
= \frac{1}{2} \left[ 1-(-1)^{\om_2(g_0,g_0^{-1}g_1)}i\g_{\underline 012B}^{e} \g_{0\underline 12A}^{g_0^{-1}g_1} \right],\\
P_{0\underline 12B,01\underline 2A}^{g_1,g_2}
= U(g_0) P_{0\underline 12B,01\underline 2A}^{e,g_1^{-1}g_2} U(g_0)^\dagger
= \frac{1}{2} \left[ 1-(-1)^{\om_2(g_1,g_1^{-1}g_2)}i\g_{0\underline 12B}^{e} \g_{01\underline 2A}^{g_1^{-1}g_2} \right],\\
P_{\underline 012A,01\underline 2B}^{g_0,g_2}
= U(g_0) P_{\underline 012A,01\underline 2B}^{e,g_0^{-1}g_2} U(g_0)^\dagger
= \frac{1}{2} \left[ 1-(-1)^{\om_2(g_0,g_0^{-1}g_2)}i\g_{\underline 012A}^{e} \g_{01\underline 2B}^{g_0^{-1}g_2} \right].
\end{align}
The blue arrow on the red link in \fig{fig:2DFSLUb} means that the actual arrow direction is obtained by a $G_b$-action and depends on $\om_2$. The above Kitaev chain decoration procedure along the boundary of a triangle is very similar to the 1D FSPT construction with only three vertices. The Majorana pairings have the following properties: (i) Both the vacuum and the nontrivial pairings are invariant under $G_b$-action. (ii) The fermion parity change of Majorana chains from \fig{fig:2DFSLUa} to \fig{fig:2DFSLUb} inside the triangle is exactly $(-1)^{\om_2(g_0^{-1}g_1,g_1^{-1}g_2)}$. Since we have chosen $n_2=\om_2$, we conclude the complex fermion mode $c_{ijk}^{g_i}$ at the center of the triangle should be totally annihilated, to make sure that $U_1$ is fermion parity even.

In such a way, the FSLU transformation $U_1$ annihilate the fermions at the center of each triangle and create a small Kitaev's Majorana chain along the boundary of the triangle. The explicit expression of the standard FSLU transformation $U_1$ with $g_0=e$ is
\begin{align}\label{eq:U1}
&U_1 =
\prod_{\langle ijk\rangle}
\big(c_{ijk}^{g_i}\big)^{n_2(ijk)}
\left(
\prod_{\s\ne g_i}P_{\underline ijkA,\underline ijkB}^{\s,\s}
\prod_{\s\ne g_j}P_{i\underline jkA,i\underline jkB}^{\s,\s}
\prod_{\s\ne g_k}P_{ij\underline kA,ij\underline kB}^{\s,\s}
\right)
\left(
P_{\underline ijkB,i\underline jkA}^{g_i,g_j}
P_{i\underline jkB,ij\underline kA}^{g_j,g_k}
P_{\underline ijkA,ij\underline kB}^{g_i,g_k}
\right)
\big(\g_{\underline ijkA}^{g_i} \big)^{n_2(ijk)}.
\end{align}
The first term $c_{ijk}^{g_k}$ is used to remove the complex fermions at the center of each triangle. And other terms are Majorana fermion pairing projection operators \eq{proj} to decorate one Kitaev chain along the boundary of the triangle. The last dangling Majorana operator is inserted for fermion parity considerations. Other non-standard $U_1$ can be obtained from the standard one by a $U(g_0)$-action. Therefore, the operator $U_1$ from $|(a)\rangle$ to $|(b)\rangle$ is both fermion parity even and symmetric under $G_b$-action.

(2) The second FSLU transformation $U_2$ from $|(b)\rangle$ to $|(c)\rangle$.

Since the state is on a closed oriented surface, there are four Majorana fermions on the two sides of the oriented link $\langle ij\rangle$. For convenience, we now relabel them by $\g_{\underline ij1}^{g_i},\g_{i\underline j1}^{g_j}$ on the right-hand side, and $\g_{\underline ij2}^{g_i},\g_{i\underline j2}^{g_j}$ on the left-hand side of the oriented link $\langle ij\rangle$ (see the four red dots near each link). We can use an FSLU $U_2$ to change the Majorana fermion pairings from $P_{\underline ij1,i\underline j1}^{g_i,g_j}=P_{\underline ij2,i\underline j2}^{g_i,g_j}=1$ (see red links in \fig{fig:2DFSLUb}) to $P_{\underline ij1,\underline ij2}^{g_i,g_i}=P_{i\underline j2,i\underline j1}^{g_j,g_j}=1$ (see red links in \fig{fig:2DFSLUc}). These four Majorana fermions near link $\langle ij\rangle$ form a loop with Kasteleyn orientations. So the Majorana fermion parity is unchanged under this FSLU.

The expression for the FSLU $U_2$ is simply
\begin{align}\label{eq:U2}
U_2 &= \prod_{\langle ij\rangle 
} P_{\underline ij1,\underline ij2}^{g_i,g_i} P_{i\underline j2,i\underline j1}^{g_j,g_j}.
\end{align}
Note that the actual direction of the blue arrow in \fig{fig:2DFSLUc} near vertex $i$ depends on $P_{\underline ij1,\underline ij2}^{g_i,g_i}=U(g_i)\frac{1}{2}(1-i\g_{\underline ij1}^e\g_{\underline ij2}^e)U(g_i)^\dagger$. So the arrow direction will be reversed if $s_1(g_i)=1$, and $\g_{\underline ij1}^{g_i}$ and $\g_{\underline ij2}^{g_i}$ are Majorana fermions belonging to the same $A/B$ type.

After the above two FSLU transformations $U_1$ and $U_2$, we have a state where each vertex is surrounded by one nontrivial Majorana chain (red arrows) and $|G_b|-1$ trivial Majorana chains (green arrows). The fermion parity of this vertex depends on the number and the orientations of the triangles sharing this vertex. We can define a new state $|g_i\rangle'$ around the vertex $i$ as the combination of the original bosonic state $|g_i\rangle$ and the neighboring Majorana fermions (the degrees of freedom inside the gray circles in \fig{fig:2DFSLUc}). It is easy to check that $|g_i\rangle'$ has the same $G_b$-transformation property as $|g_i\rangle$, i.e., $|g_i\rangle'\rightarrow|gg_i\rangle'$. So the final state $|(c)\rangle$ is a fermionic product state \eq{A:c}.

In summary, using the two FSLU transformations Eqs.~(\ref{eq:U1}) and (\ref{eq:U2}), we can remove the complex fermions at the triangles of an FSPT state with $n_2=\om_2$, and obtain an FSPT state with $n_2=0$. Therefore, the complex fermion decoration layer with $n_2=\om_2$ is trivialized by these FSLU transformations.

\subsubsection{Boundary ASPT of the 2D bulk}

We have shown the 2D FSPT state with $n_2=\om_2$ on a closed surface can be connected to an FSPT state with $n_2=0$. However, for a system with boundary, there is something unusual left.

\begin{figure}[ht]
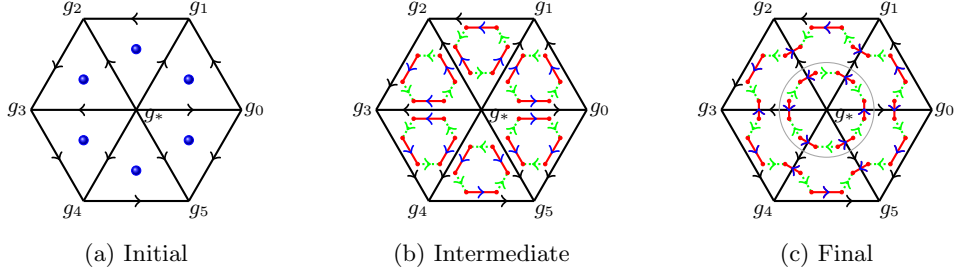

\centering
\begin{subfigure}[h]{.25\textwidth}
\centering
$\tikzfig{scale=1.4}{
\foreach \i in {0,...,5}{
\coordinate (\i) at ({\i*60}:1);
\draw[->-=.55,black,thick](0,0)--({\i*60}:1);
\node[circle,draw=blue,shading=ball,scale=.4] at (\i*60+30:0.577) {};
}
\foreach \i in {0,...,4}{\draw[->-=.55,black,thick]({\i*60}:1)--({\i*60+60}:1);}
\draw[->-=.55,black,thick](0)--(5);
\node[right,yshift=-2,scale=.9] at (0,0) {$g_\ast$};
\node[xshift=5,scale=.9] at (0) {$g_0$};
\node[xshift=4,yshift=4,scale=.9] at (1) {$g_1$};
\node[xshift=-4,yshift=4,scale=.9] at (2) {$g_2$};
\node[xshift=-5,scale=.9] at (3) {$g_3$};
\node[xshift=-4,yshift=-4,scale=.9] at (4) {$g_4$};
\node[xshift=4,yshift=-4,scale=.9] at (5) {$g_5$};
}$
\caption{Initial}
\label{fig:2DFSLU_boundarya}
\end{subfigure}
\begin{subfigure}[h]{.25\textwidth}
\centering
$\tikzfig{scale=1.4}{
\pgfmathsetmacro{\arrow}{.75};
\foreach \i in {0,...,5}{
\coordinate (\i) at ({\i*60}:1);
\draw[->-=\arrow,black,thick](0,0)--({\i*60}:1);
}
\foreach \i in {0,...,4}{\draw[->-=\arrow,black,thick]({\i*60}:1)--({\i*60+60}:1);}
\draw[->-=\arrow,black,thick](0)--(5);
\pgfmathsetmacro{\r}{.25};\pgfmathsetmacro{\theta}{35};
\begin{scope}[shift={(.5,0.2886)}]
\coordinate (12_1) at ({30-\theta}:\r);\coordinate (12_2) at ({30+\theta}:\r);
\coordinate (02_1) at ({150+\theta}:\r);\coordinate (02_2) at ({150-\theta}:\r);
\coordinate (01_1) at ({-90-\theta}:\r);\coordinate (01_2) at ({-90+\theta}:\r);
\draw[red,thick](12_1)--(12_2);
\draw[red,thick](02_1)--(02_2);
\draw[red,thick](01_1)--(01_2);
\draw[->-=.6,blue,thick]($(12_1)!.6!(12_2)$)--($(12_1)!.601!(12_2)$);
\draw[->-=.6,blue,thick]($(02_1)!.6!(02_2)$)--($(02_1)!.601!(02_2)$);
\draw[->-=.6,blue,thick]($(01_1)!.6!(01_2)$)--($(01_1)!.601!(01_2)$);
\draw[->-=.6,green,densely dotted,thick](01_2)--(12_1);
\draw[->-=.6,green,densely dotted,thick](12_2)--(02_2);
\draw[->-=.6,green,densely dotted,thick](02_1)--(01_1);
\node[circle,draw=red,fill=red,scale=.15] at (01_1) {};
\node[circle,draw=red,fill=red,scale=.15] at (01_2) {};
\node[circle,draw=red,fill=red,scale=.15] at (02_1) {};
\node[circle,draw=red,fill=red,scale=.15] at (02_2) {};
\node[circle,draw=red,fill=red,scale=.15] at (12_1) {};
\node[circle,draw=red,fill=red,scale=.15] at (12_2) {};
\end{scope}
\begin{scope}[shift={(-.5,0.2886)}]
\coordinate (12_1) at ({30-\theta}:\r);\coordinate (12_2) at ({30+\theta}:\r);
\coordinate (02_1) at ({150+\theta}:\r);\coordinate (02_2) at ({150-\theta}:\r);
\coordinate (01_1) at ({-90-\theta}:\r);\coordinate (01_2) at ({-90+\theta}:\r);
\draw[red,thick](12_1)--(12_2);
\draw[red,thick](02_1)--(02_2);
\draw[red,thick](01_1)--(01_2);
\draw[->-=.6,blue,thick]($(12_1)!.6!(12_2)$)--($(12_1)!.601!(12_2)$);
\draw[->-=.6,blue,thick]($(02_2)!.6!(02_1)$)--($(02_2)!.601!(02_1)$);
\draw[->-=.6,blue,thick]($(01_2)!.6!(01_1)$)--($(01_2)!.601!(01_1)$);
\draw[->-=.6,green,densely dotted,thick](01_2)--(12_1);
\draw[->-=.6,green,densely dotted,thick](12_2)--(02_2);
\draw[->-=.6,green,densely dotted,thick](02_1)--(01_1);
\node[circle,draw=red,fill=red,scale=.15] at (01_1) {};
\node[circle,draw=red,fill=red,scale=.15] at (01_2) {};
\node[circle,draw=red,fill=red,scale=.15] at (02_1) {};
\node[circle,draw=red,fill=red,scale=.15] at (02_2) {};
\node[circle,draw=red,fill=red,scale=.15] at (12_1) {};
\node[circle,draw=red,fill=red,scale=.15] at (12_2) {};
\end{scope}
\begin{scope}[shift={(0,-0.5773)}]
\coordinate (12_1) at ({30-\theta}:\r);\coordinate (12_2) at ({30+\theta}:\r);
\coordinate (02_1) at ({150+\theta}:\r);\coordinate (02_2) at ({150-\theta}:\r);
\coordinate (01_1) at ({-90-\theta}:\r);\coordinate (01_2) at ({-90+\theta}:\r);
\draw[red,thick](12_1)--(12_2);
\draw[red,thick](02_1)--(02_2);
\draw[red,thick](01_1)--(01_2);
\draw[->-=.6,blue,thick]($(12_2)!.6!(12_1)$)--($(12_2)!.601!(12_1)$);
\draw[->-=.6,blue,thick]($(02_2)!.6!(02_1)$)--($(02_2)!.601!(02_1)$);
\draw[->-=.6,blue,thick]($(01_1)!.6!(01_2)$)--($(01_1)!.601!(01_2)$);
\draw[->-=.6,green,densely dotted,thick](01_2)--(12_1);
\draw[->-=.6,green,densely dotted,thick](12_2)--(02_2);
\draw[->-=.6,green,densely dotted,thick](02_1)--(01_1);
\node[circle,draw=red,fill=red,scale=.15] at (01_1) {};
\node[circle,draw=red,fill=red,scale=.15] at (01_2) {};
\node[circle,draw=red,fill=red,scale=.15] at (02_1) {};
\node[circle,draw=red,fill=red,scale=.15] at (02_2) {};
\node[circle,draw=red,fill=red,scale=.15] at (12_1) {};
\node[circle,draw=red,fill=red,scale=.15] at (12_2) {};
\end{scope}
\begin{scope}[shift={(0,0.5773)}]
\coordinate (12_1) at ({30-\theta+60}:\r);\coordinate (12_2) at ({30+\theta+60}:\r);
\coordinate (02_1) at ({150+\theta+60}:\r);\coordinate (02_2) at ({150-\theta+60}:\r);
\coordinate (01_1) at ({-90-\theta+60}:\r);\coordinate (01_2) at ({-90+\theta+60}:\r);
\draw[red,thick](12_1)--(12_2);
\draw[red,thick](02_1)--(02_2);
\draw[red,thick](01_1)--(01_2);
\draw[->-=.6,blue,thick]($(12_1)!.6!(12_2)$)--($(12_1)!.601!(12_2)$);
\draw[->-=.6,blue,thick]($(02_1)!.6!(02_2)$)--($(02_1)!.601!(02_2)$);
\draw[->-=.6,blue,thick]($(01_1)!.6!(01_2)$)--($(01_1)!.601!(01_2)$);
\draw[->-=.6,green,densely dotted,thick](01_2)--(12_1);
\draw[->-=.6,green,densely dotted,thick](12_2)--(02_2);
\draw[->-=.6,green,densely dotted,thick](02_1)--(01_1);
\node[circle,draw=red,fill=red,scale=.15] at (01_1) {};
\node[circle,draw=red,fill=red,scale=.15] at (01_2) {};
\node[circle,draw=red,fill=red,scale=.15] at (02_1) {};
\node[circle,draw=red,fill=red,scale=.15] at (02_2) {};
\node[circle,draw=red,fill=red,scale=.15] at (12_1) {};
\node[circle,draw=red,fill=red,scale=.15] at (12_2) {};
\end{scope}
\begin{scope}[shift={(-.5,-0.2886)}]
\coordinate (12_1) at ({30-\theta+60}:\r);\coordinate (12_2) at ({30+\theta+60}:\r);
\coordinate (02_1) at ({150+\theta+60}:\r);\coordinate (02_2) at ({150-\theta+60}:\r);
\coordinate (01_1) at ({-90-\theta+60}:\r);\coordinate (01_2) at ({-90+\theta+60}:\r);
\draw[red,thick](12_1)--(12_2);
\draw[red,thick](02_1)--(02_2);
\draw[red,thick](01_1)--(01_2);
\draw[->-=.6,blue,thick]($(12_1)!.6!(12_2)$)--($(12_1)!.601!(12_2)$);
\draw[->-=.6,blue,thick]($(02_2)!.6!(02_1)$)--($(02_2)!.601!(02_1)$);
\draw[->-=.6,blue,thick]($(01_2)!.6!(01_1)$)--($(01_2)!.601!(01_1)$);
\draw[->-=.6,green,densely dotted,thick](01_2)--(12_1);
\draw[->-=.6,green,densely dotted,thick](12_2)--(02_2);
\draw[->-=.6,green,densely dotted,thick](02_1)--(01_1);
\node[circle,draw=red,fill=red,scale=.15] at (01_1) {};
\node[circle,draw=red,fill=red,scale=.15] at (01_2) {};
\node[circle,draw=red,fill=red,scale=.15] at (02_1) {};
\node[circle,draw=red,fill=red,scale=.15] at (02_2) {};
\node[circle,draw=red,fill=red,scale=.15] at (12_1) {};
\node[circle,draw=red,fill=red,scale=.15] at (12_2) {};
\end{scope}
\begin{scope}[shift={(.5,-0.2886)}]
\coordinate (12_1) at ({30-\theta+60}:\r);\coordinate (12_2) at ({30+\theta+60}:\r);
\coordinate (02_1) at ({150+\theta+60}:\r);\coordinate (02_2) at ({150-\theta+60}:\r);
\coordinate (01_1) at ({-90-\theta+60}:\r);\coordinate (01_2) at ({-90+\theta+60}:\r);
\draw[red,thick](12_1)--(12_2);
\draw[red,thick](02_1)--(02_2);
\draw[red,thick](01_1)--(01_2);
\draw[->-=.6,blue,thick]($(12_2)!.6!(12_1)$)--($(12_2)!.601!(12_1)$);
\draw[->-=.6,blue,thick]($(02_2)!.6!(02_1)$)--($(02_2)!.601!(02_1)$);
\draw[->-=.6,blue,thick]($(01_2)!.6!(01_1)$)--($(01_2)!.601!(01_1)$);
\draw[->-=.6,green,densely dotted,thick](12_1)--(01_2);
\draw[->-=.6,green,densely dotted,thick](02_2)--(12_2);
\draw[->-=.6,green,densely dotted,thick](01_1)--(02_1);
\node[circle,draw=red,fill=red,scale=.15] at (01_1) {};
\node[circle,draw=red,fill=red,scale=.15] at (01_2) {};
\node[circle,draw=red,fill=red,scale=.15] at (02_1) {};
\node[circle,draw=red,fill=red,scale=.15] at (02_2) {};
\node[circle,draw=red,fill=red,scale=.15] at (12_1) {};
\node[circle,draw=red,fill=red,scale=.15] at (12_2) {};
\end{scope}
\node[right,yshift=-2,scale=.9] at (0,0) {$g_\ast$};
\node[xshift=5,scale=.9] at (0) {$g_0$};
\node[xshift=4,yshift=4,scale=.9] at (1) {$g_1$};
\node[xshift=-4,yshift=4,scale=.9] at (2) {$g_2$};
\node[xshift=-5,scale=.9] at (3) {$g_3$};
\node[xshift=-4,yshift=-4,scale=.9] at (4) {$g_4$};
\node[xshift=4,yshift=-4,scale=.9] at (5) {$g_5$};
}$
\caption{Intermediate}
\label{fig:2DFSLU_boundaryb}
\end{subfigure}
\begin{subfigure}[h]{.25\textwidth}
\centering
$\tikzfig{scale=1.4}{
\pgfmathsetmacro{\arrow}{.75};
\foreach \i in {0,...,5}{
\coordinate (\i) at ({\i*60}:1);
\draw[->-=.55,black,thick](0,0)--({\i*60}:1);
}
\foreach \i in {0,...,4}{\draw[->-=\arrow,black,thick]({\i*60}:1)--({\i*60+60}:1);}
\draw[->-=\arrow,black,thick](0)--(5);
\pgfmathsetmacro{\r}{.25};\pgfmathsetmacro{\theta}{35};
\foreach \i in {0,...,5}{
\draw[->-=.6,red,thick] ([shift=({30+60*\i}:0.5773)] {150+\theta+60*\i}:\r)--([shift=({90+60*\i}:0.5773)] {-30-\theta+60*\i}:\r);
\draw[->-=.6,blue,thick] ($([shift=({30+60*\i}:0.5773)] {150+\theta+60*\i}:\r)!.6!([shift=({90+60*\i}:0.5773)] {-30-\theta+60*\i}:\r)$) -- ($([shift=({30+60*\i}:0.5773)] {150+\theta+60*\i}:\r)!.601!([shift=({90+60*\i}:0.5773)] {-30-\theta+60*\i}:\r)$);
\draw[->-=.6,red,thick] ([shift=({90+60*\i}:0.5773)] {-30+\theta+60*\i}:\r)--([shift=({30+60*\i}:0.5773)] {150-\theta+60*\i}:\r);
\draw[->-=.6,blue,thick] ($([shift=({90+60*\i}:0.5773)] {-30+\theta+60*\i}:\r)!.6!([shift=({30+60*\i}:0.5773)] {150-\theta+60*\i}:\r)$) -- ($([shift=({90+60*\i}:0.5773)] {-30+\theta+60*\i}:\r)!.601!([shift=({30+60*\i}:0.5773)] {150-\theta+60*\i}:\r)$);
}
\foreach \i in {0,...,4}{
\draw[->-=.55,green,densely dotted,thick] ([shift=({30+60*\i}:0.5773)] {150+\theta+60*\i}:\r)--([shift=({30+60*\i}:0.5773)] {-90-\theta+60*\i}:\r);
\draw[->-=.6,red,thick] ([shift=({30+60*\i}:0.5773)] {30-\theta+60*\i}:\r)--([shift=({30+60*\i}:0.5773)] {30+\theta+60*\i}:\r);
\draw[->-=.6,blue,thick] ($([shift=({30+60*\i}:0.5773)] {30-\theta+60*\i}:\r)!.6!([shift=({30+60*\i}:0.5773)] {30+\theta+60*\i}:\r)$)--($([shift=({30+60*\i}:0.5773)] {30-\theta+60*\i}:\r)!.601!([shift=({30+60*\i}:0.5773)] {30+\theta+60*\i}:\r)$);
\draw[->-=.6,green,densely dotted,thick] ([shift=({30+60*\i}:0.5773)] {30+\theta+60*\i}:\r)--([shift=({30+60*\i}:0.5773)] {150-\theta+60*\i}:\r);
\draw[->-=.6,green,densely dotted,thick] ([shift=({30+60*\i}:0.5773)] {-90+\theta+60*\i}:\r)--([shift=({30+60*\i}:0.5773)] {30-\theta+60*\i}:\r);
}
\foreach \i in {5}{
\draw[->-=.55,green,densely dotted,thick] ([shift=({30+60*\i}:0.5773)] {-90-\theta+60*\i}:\r)--([shift=({30+60*\i}:0.5773)] {150+\theta+60*\i}:\r);
\draw[->-=.6,red,thick] ([shift=({30+60*\i}:0.5773)] {30+\theta+60*\i}:\r)--([shift=({30+60*\i}:0.5773)] {30-\theta+60*\i}:\r);
\draw[->-=.6,blue,thick] ($([shift=({30+60*\i}:0.5773)] {30+\theta+60*\i}:\r)!.6!([shift=({30+60*\i}:0.5773)] {30-\theta+60*\i}:\r)$)--($([shift=({30+60*\i}:0.5773)] {30+\theta+60*\i}:\r)!.601!([shift=({30+60*\i}:0.5773)] {30-\theta+60*\i}:\r)$);
\draw[->-=.6,green,densely dotted,thick] ([shift=({30+60*\i}:0.5773)] {150-\theta+60*\i}:\r)--([shift=({30+60*\i}:0.5773)] {30+\theta+60*\i}:\r);
\draw[->-=.6,green,densely dotted,thick] ([shift=({30+60*\i}:0.5773)] {30-\theta+60*\i}:\r)--([shift=({30+60*\i}:0.5773)] {-90+\theta+60*\i}:\r);
}
\foreach \i in {0,...,5}{
\foreach \j in {0,120,240}{
\node[circle,draw=red,fill=red,scale=.15] at ([shift=({30+60*\i}:0.5773)] {30-\theta+60*\i+\j}:\r) {};
\node[circle,draw=red,fill=red,scale=.15] at ([shift=({30+60*\i}:0.5773)] {30+\theta+60*\i+\j}:\r) {};
}}
\node[right,yshift=-2,scale=.9] at (0,0) {$g_\ast$};
\node[xshift=5,scale=.9] at (0) {$g_0$};
\node[xshift=4,yshift=4,scale=.9] at (1) {$g_1$};
\node[xshift=-4,yshift=4,scale=.9] at (2) {$g_2$};
\node[xshift=-5,scale=.9] at (3) {$g_3$};
\node[xshift=-4,yshift=-4,scale=.9] at (4) {$g_4$};
\node[xshift=4,yshift=-4,scale=.9] at (5) {$g_5$};
\draw[gray!70!white] (0,0) circle (.45);
}$
\caption{Final}
\label{fig:2DFSLU_boundaryc}
\end{subfigure}
\caption{FSLU transformations for ``FSPT'' state on 2D surface with boundary. The degrees of freedom inside the gray circle in (c) are combined to be the new basis state $|g_\ast\rangle'$. There is a 1D ASPT state along the boundary of the 2D bulk shown in (c).}
\label{fig:2DFSLU_boundary}
\end{figure}

\begin{figure}[ht]
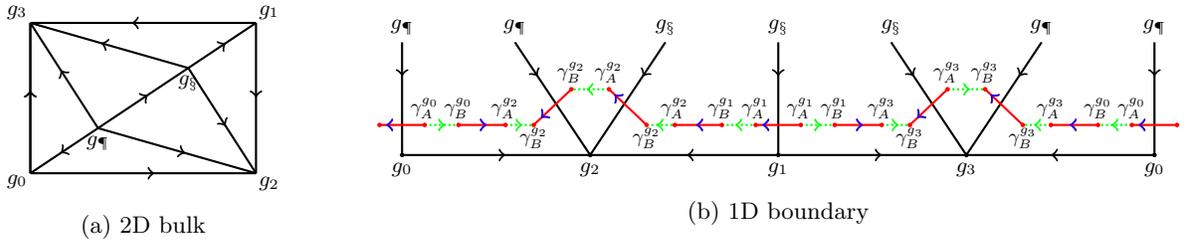

\centering
\begin{subfigure}[h]{.28\textwidth}
\centering
$\tikzfig{scale=2}{
\coordinate (0) at (0,0);
\coordinate (1) at (1.5,0);
\coordinate (2) at (1.5,1);
\coordinate (3) at (0,1);
\coordinate (a) at ($(0)!.3!(2)$);
\coordinate (b) at ($(0)!.7!(2)$);
\draw[->-=.55,black,thick](0)--(1);
\draw[-<-=.55,black,thick](1)--(2);
\draw[->-=.55,black,thick](2)--(3);
\draw[->-=.55,black,thick](0)--(3);
\draw[->-=.55,black,thick](0)--(3);
\draw[->-=.55,black,thick](a)--(0);
\draw[->-=.55,black,thick](a)--(1);
\draw[->-=.55,black,thick](a)--(3);
\draw[->-=.55,black,thick](a)--(b);
\draw[->-=.55,black,thick](b)--(1);
\draw[->-=.55,black,thick](b)--(2);
\draw[->-=.55,black,thick](b)--(3);
\node [xshift=-5,yshift=-3,scale=.9] at (0) {$g_0$};
\node [xshift=5,yshift=-4,scale=.9] at (1) {$g_2$};
\node [xshift=5,yshift=4,scale=.9] at (2) {$g_1$};
\node [xshift=-5,yshift=4,scale=.9] at (3) {$g_3$};
\node [below,scale=.9] at (a) {$g_{\P}$};
\node [below,scale=.9] at (b) {$g_{\S}$};
}$
\caption{2D bulk}
\label{fig:1Dboundary_a}
\end{subfigure}
\begin{subfigure}[h]{.65\textwidth}
\centering
$\tikzfig{scale=2.5}{
\coordinate (0) at (0,0);
\coordinate (1) at (1,0);
\coordinate (2) at (2,0);
\coordinate (3) at (3,0);
\coordinate (4) at (4,0);
\coordinate (a) at ($(0)!.3!(2)$);
\coordinate (b) at ($(0)!.7!(2)$);
\draw[->-=.55,black,thick](0)--(1);
\draw[-<-=.55,black,thick](1)--(2);
\draw[->-=.55,black,thick](2)--(3);
\draw[->-=.55,black,thick](4)--(3);
\pgfmathsetmacro{\yA}{.6};
\coordinate (0A) at (0,\yA);
\coordinate (1A) at (.6,\yA);
\coordinate (1B) at (1.4,\yA);
\coordinate (2A) at (2,\yA);
\coordinate (3A) at (2.6,\yA);
\coordinate (3B) at (3.4,\yA);
\coordinate (4A) at (4,\yA);
\draw[->-=.3,black,thick](0A)--(0);
\draw[->-=.3,black,thick](1A)--(1);
\draw[->-=.3,black,thick](1B)--(1);
\draw[->-=.3,black,thick](2A)--(2);
\draw[->-=.3,black,thick](3A)--(3);
\draw[->-=.3,black,thick](3B)--(3);
\draw[->-=.3,black,thick](4A)--(4);
\node [above,scale=.9] at (0A) {$g_{\P}$};
\node [above,scale=.9] at (1A) {$g_{\P}$};
\node [above,scale=.9] at (1B) {$g_{\S}$};
\node [above,scale=.9] at (2A) {$g_{\S}$};
\node [above,scale=.9] at (3A) {$g_{\S}$};
\node [above,scale=.9] at (3B) {$g_{\P}$};
\node [above,scale=.9] at (4A) {$g_{\P}$};
\pgfmathsetmacro{\ya}{.16};
\pgfmathsetmacro{\yb}{.35};
\coordinate (0a) at (-.12,\ya);
\coordinate (0b) at (.12,\ya);
\draw[->-=.85,red,thick](0b)--(0a);
\draw[->-=.85,blue,thick]($(0b)!.85!(0a)$)--($(0b)!.851!(0a)$);
\coordinate (2a) at (2-.12,\ya);
\coordinate (2b) at (2+.12,\ya);
\draw[->-=.85,red,thick](2b)--(2a);
\draw[->-=.85,blue,thick]($(2b)!.85!(2a)$)--($(2b)!.851!(2a)$);
\coordinate (4a) at (4-.12,\ya);
\coordinate (4b) at (4+.12,\ya);
\draw[->-=.85,red,thick](4b)--(4a);
\draw[->-=.85,blue,thick]($(4b)!.85!(4a)$)--($(4b)!.851!(4a)$);
\coordinate (1Aa) at (1-.3,\ya);
\coordinate (1Ab) at (1-.1,\yb);
\draw[->-=.8,red,thick](1Ab)--(1Aa);
\draw[->-=.8,blue,thick]($(1Ab)!.8!(1Aa)$)--($(1Ab)!.801!(1Aa)$);
\coordinate (1Ba) at (1+.1,\yb);
\coordinate (1Bb) at (1+.3,\ya);
\draw[->-=.85,red,thick](1Bb)--(1Ba);
\draw[->-=.85,blue,thick]($(1Bb)!.85!(1Ba)$)--($(1Bb)!.851!(1Ba)$);
\coordinate (3Aa) at (3-.3,\ya);
\coordinate (3Ab) at (3-.1,\yb);
\draw[->-=.8,red,thick](3Ab)--(3Aa);
\draw[->-=.8,blue,thick]($(3Ab)!.8!(3Aa)$)--($(3Ab)!.801!(3Aa)$);
\coordinate (3Ba) at (3+.1,\yb);
\coordinate (3Bb) at (3+.3,\ya);
\draw[->-=.85,red,thick](3Bb)--(3Ba);
\draw[->-=.85,blue,thick]($(3Bb)!.85!(3Ba)$)--($(3Bb)!.851!(3Ba)$);
\coordinate (01-0) at (0+.3,\ya);
\coordinate (01-1) at (0+.55,\ya);
\draw[->-=.6,red,thick](01-0)--(01-1);
\draw[->-=.6,blue,thick]($(01-0)!.6!(01-1)$)--($(01-0)!.601!(01-1)$);
\coordinate (23-0) at (2+.3,\ya);
\coordinate (23-1) at (2+.55,\ya);
\draw[->-=.6,red,thick](23-0)--(23-1);
\draw[->-=.6,blue,thick]($(23-0)!.6!(23-1)$)--($(23-0)!.601!(23-1)$);
\coordinate (12-0) at (1+1-.55,\ya);
\coordinate (12-1) at (1+1-.3,\ya);
\draw[-<-=.6,red,thick](12-0)--(12-1);
\draw[-<-=.6,blue,thick]($(12-0)!.6!(12-1)$)--($(12-0)!.601!(12-1)$);
\coordinate (34-0) at (3+1-.55,\ya);
\coordinate (34-1) at (3+1-.3,\ya);
\draw[->-=.6,red,thick](34-1)--(34-0);
\draw[->-=.6,blue,thick]($(34-1)!.6!(34-0)$)--($(34-1)!.601!(34-0)$);
\draw[->-=.6,densely dotted,green,thick](0b)--(01-0);
\draw[->-=.6,densely dotted,green,thick](01-1)--(1Aa);
\draw[->-=.6,densely dotted,green,thick](1Ba)--(1Ab);
\draw[-<-=.6,densely dotted,green,thick](1Bb)--(12-0);
\draw[-<-=.6,densely dotted,green,thick](12-1)--(2a);
\draw[->-=.6,densely dotted,green,thick](2b)--(23-0);
\draw[->-=.6,densely dotted,green,thick](23-1)--(3Aa);
\draw[->-=.6,densely dotted,green,thick](3Ab)--(3Ba);
\draw[->-=.6,densely dotted,green,thick](34-0)--(3Bb);
\draw[->-=.6,densely dotted,green,thick](4a)--(34-1);
\node [above,scale=.8] at (0b) {$\g_A^{g_0}$};
\node [above,scale=.8] at (01-0) {$\g_B^{g_0}$};
\node [above,scale=.8] at (01-1) {$\g_A^{g_2}$};
\node [yshift=-5,scale=.8] at (1Aa) {$\g_B^{g_2}$};
\node [above,scale=.8] at (1Ba) {$\g_A^{g_2}$};
\node [above,scale=.8] at (1Ab) {$\g_B^{g_2}$};
\node [yshift=-5,scale=.8] at (1Bb) {$\g_B^{g_2}$};
\node [above,scale=.8] at (12-0) {$\g_A^{g_2}$};
\node [above,scale=.8] at (12-1) {$\g_B^{g_1}$};
\node [above,scale=.8] at (2a) {$\g_A^{g_1}$};
\node [above,scale=.8] at (2b) {$\g_A^{g_1}$};
\node [above,scale=.8] at (23-0) {$\g_B^{g_1}$};
\node [above,scale=.8] at (23-1) {$\g_A^{g_3}$};
\node [yshift=-5,scale=.8] at (3Aa) {$\g_B^{g_3}$};
\node [above,scale=.8] at (3Ab) {$\g_A^{g_3}$};
\node [above,scale=.8] at (3Ba) {$\g_B^{g_3}$};
\node [above,scale=.8] at (34-0) {$\g_A^{g_3}$};
\node [yshift=-5,scale=.8] at (3Bb) {$\g_B^{g_3}$};
\node [above,scale=.8] at (4a) {$\g_A^{g_0}$};
\node [above,scale=.8] at (34-1) {$\g_B^{g_0}$};
\node[circle,draw=red,fill=red,scale=.15] at (0a) {};
\node[circle,draw=red,fill=red,scale=.15] at (0b) {};
\node[circle,draw=red,fill=red,scale=.15] at (2a) {};
\node[circle,draw=red,fill=red,scale=.15] at (2b) {};
\node[circle,draw=red,fill=red,scale=.15] at (4a) {};
\node[circle,draw=red,fill=red,scale=.15] at (4b) {};
\node[circle,draw=red,fill=red,scale=.15] at (1Aa) {};
\node[circle,draw=red,fill=red,scale=.15] at (1Ab) {};
\node[circle,draw=red,fill=red,scale=.15] at (1Ba) {};
\node[circle,draw=red,fill=red,scale=.15] at (1Bb) {};
\node[circle,draw=red,fill=red,scale=.15] at (3Aa) {};
\node[circle,draw=red,fill=red,scale=.15] at (3Ab) {};
\node[circle,draw=red,fill=red,scale=.15] at (3Ba) {};
\node[circle,draw=red,fill=red,scale=.15] at (3Bb) {};
\node[circle,draw=red,fill=red,scale=.15] at (01-0) {};
\node[circle,draw=red,fill=red,scale=.15] at (01-1) {};
\node[circle,draw=red,fill=red,scale=.15] at (12-0) {};
\node[circle,draw=red,fill=red,scale=.15] at (12-1) {};
\node[circle,draw=red,fill=red,scale=.15] at (23-0) {};
\node[circle,draw=red,fill=red,scale=.15] at (23-1) {};
\node[circle,draw=red,fill=red,scale=.15] at (34-0) {};
\node[circle,draw=red,fill=red,scale=.15] at (34-1) {};
\node[circle,draw=black,fill=black,scale=.15] at (0) {};
\node[circle,draw=black,fill=black,scale=.15] at (1) {};
\node[circle,draw=black,fill=black,scale=.15] at (2) {};
\node[circle,draw=black,fill=black,scale=.15] at (3) {};
\node[circle,draw=black,fill=black,scale=.15] at (4) {};
\node [below,scale=.9] at (0) {$g_0$};
\node [below,scale=.9] at (1) {$g_2$};
\node [below,scale=.9] at (2) {$g_1$};
\node [below,scale=.9] at (3) {$g_3$};
\node [below,scale=.9] at (4) {$g_0$};
%
}$
\caption{1D boundary}
\label{fig:1Dboundary_b}
\end{subfigure}
\caption{ASPT on the 1D boundary of 2D bulk. On the boundary, the Majorana fermions forming nontrivial Kitaev chain (red dots) are labelled by $A/B$ and $g\in G_b$. Blue arrow indicates that the pairing directions may be changed under symmetry action.}
\label{fig:1Dboundary}
\end{figure}

Consider a state defined on a 2D triangulation lattice with boundary. We can perform FSLU transformations similar to Eqs.~(\ref{eq:U1}) and (\ref{eq:U2}). The only difference is that the link $\langle ij\rangle$ in the product in \eq{eq:U2} is not on the boundary of the space manifold, since there are only two Majorana fermions near the each boundary link (see \fig{fig:2DFSLU_boundary} for an example with only one interior vertex labelled by $g_\ast$). After the transformations, the bulk state becomes a tensor product of interior vertex state $|g_i\rangle'$ (see \fig{fig:1Dboundary_a} for an example). But the boundary is only transformed under $U_1$ and is a so-called 1D ASPT state (see \fig{fig:1Dboundary_b} for an example). This boundary state is again a combination of one nontrivial Majorana chain (see red links in \fig{fig:1Dboundary_b}) and $|G_b|-1$ trivial ones (see green links in \fig{fig:1Dboundary_b}).

The anomalous feature of the boundary can be seen from the symmetry action on the boundary. Under a $U(g)$ symmetry action, the Majorana fermions are transformed as $\g_A^{g_i}\rightarrow (-1)^{\om_2(g,g_i)}\g_A^{gg_i}$, $\g_B^{g_i}\rightarrow (-1)^{\om_2(g,g_i)+s_1(g)}\g_B^{gg_i}$. Since there are three types of Majorana pairings on the boundary (see \fig{fig:1Dboundary_b}), we should analyze their symmetry transformations separately:
\begin{enumerate}[(i)]
\item
Vacuum pairings (green lines) $-i\g_A^{g_i} \g_B^{g_i}=1$ are transformed trivially under $G_b$-action.
\item
The nontrivial Majorana pairings (red lines) parallel to link $\langle ij\rangle$ are always $A$-$B$ type pairing with different group element labels $g_i$ and $g_j$. Under the $g\in G_b$ action, the pairing arrow is changed according to $(-1)^{\om_2(g,g_i)+\om_2(g,g_j)}$.
\item
The nontrivial Majorana pairings (red lines) crossing a black lattice link are always labelled by the same group element and can be of different $A/B$ types. So the pairing arrow is changed as $(-1)^{s_1(g)}$ if the pairing is $A$-$A$ or $B$-$B$ type.
\end{enumerate}
Therefore, depending on $\om_2(g,g_i)$, $\om_2(g,g_j)$ and $s_1(g)$, the local Majorana fermion parity (pairing direction) for the second and third types pairings may be changed.

For a closed 1D array of Majorana fermions as the boundary of a 2D bulk (see \fig{fig:1Dboundary_b} for example), it is not hard to show that the total Majorana fermion parity is always fixed under $G_b$-action \footnote{For any vertex $i$, there are always two Majorana fermions having a second type pairing with another site. So the fermion parity changes such as $(-1)^{\om_2(g,g_i)}$ appear in pairs. For the third type pairing, the number of $A$-$A$ and $B$-$B$ pairing is always even for a closed chain. So the total fermion parity is fixed.}.
However, if we want to define a similar state on an open chain (such as \fig{fig:1Dboundary_b} with open boundary condition), the total fermion parity may be violated under $U(g)$-action. This is simply because the direction of the link crossing the boundary may be changed. So the symmetry action is incompatible with the fermion parity of the open 1D ASPT chain. It implies that the 1D ASPT state can exist only on the boundary of a 2D bulk state.

\subsubsection{Boundary $F$ move and fermion parity violation}

There is another way to understand the anomalous feature of the 1D boundary. We can try to construct such 1D state without 2D bulk directly, and find out the inconsistency of the state.

We consider the 1D state with a Kitaev chain, and without complex fermion decoration ($n_1=0$). We put $|G_b|$ species of Majorana fermions near each vertex similar to the 2D construction. But only one of them is in nontrivial pairings between different vertices. The $F$ move for this state is given by
\begin{align}
\label{A:Fmove}
\Psi\left(
\!\!\!\!
\tikzfig{scale=1}{
\coordinate (0) at (0,0);
\coordinate (1) at (1,0);
\node[below] at (0) {$g_0$};
\node[below] at (1) {$g_2$};
\draw[line width=1] (0,-.017)--(0,.1);
\draw[line width=1] (1,-.017)--(1,.1);
\draw[->-=.8,line width=1] (0)--(1);
\pgfmathsetmacro{\dx}{.15};
\pgfmathsetmacro{\dy}{.2};
\node[circle,draw=red,fill=red,scale=.3] at (0-\dx,\dy){};
\node[circle,draw=red,fill=red,scale=.3] at (0+\dx,\dy){};
\node[circle,draw=red,fill=red,scale=.3] at (1-\dx,\dy){};
\node[circle,draw=red,fill=red,scale=.3] at (1+\dx,\dy){};
\node[red,xshift=-2,yshift=7,scale=.8] at (0-\dx,\dy){$\gamma_{0A}^{g_0}$};
\node[red,xshift=3,yshift=7,scale=.8] at (0+\dx,\dy){$\gamma_{0B}^{g_0}$};
\node[red,xshift=-2,yshift=7,scale=.8] at (1-\dx,\dy){$\gamma_{2A}^{g_2}$};
\node[red,xshift=3,yshift=7,scale=.8] at (1+\dx,\dy){$\gamma_{2B}^{g_2}$};
\draw[red,densely dashed,line width=1] (0-\dx,\dy)--(0+\dx,\dy);
\draw[red,densely dashed,line width=1] (1-\dx,\dy)--(1+\dx,\dy);
\coordinate (a) at (0+\dx,\dy);
\coordinate (b) at (0-\dx+1,\dy);
\draw[red,line width=1](a)--(b);
\draw[->-=.7,blue,line width=1] ($(a)!.65!(b)$)--($(a)!.651!(b)$);
\node[]at(0,\dy){};
}
\!\!\!
\right)
\quad = \quad
F(g_0,g_1,g_2)
\quad
\Psi\left(
\!\!\!\!
\tikzfig{scale=1}{
\coordinate (0) at (0,0);
\coordinate (1) at (1,0);
\coordinate (2) at (2,0);
\node[below] at (0) {$g_0$};
\node[below] at (1) {$g_1$};
\node[below] at (2) {$g_2$};
\draw[line width=1] (0,-.017)--(0,.1);
\draw[line width=1] (1,-.017)--(1,.1);
\draw[line width=1] (2,-.017)--(2,.1);
\draw[->-=.8,line width=1] (0)--(1);
\draw[->-=.8,line width=1] (1)--(2);
\pgfmathsetmacro{\dx}{.15};
\pgfmathsetmacro{\dy}{.2};
\node[circle,draw=red,fill=red,scale=.3] at (0-\dx,\dy){};
\node[circle,draw=red,fill=red,scale=.3] at (0+\dx,\dy){};
\node[circle,draw=red,fill=red,scale=.3] at (1-\dx,\dy){};
\node[circle,draw=red,fill=red,scale=.3] at (1+\dx,\dy){};
\node[circle,draw=red,fill=red,scale=.3] at (2-\dx,\dy){};
\node[circle,draw=red,fill=red,scale=.3] at (2+\dx,\dy){};
\node[red,xshift=-2,yshift=7,scale=.8] at (0-\dx,\dy){$\gamma_{0A}^{g_0}$};
\node[red,xshift=3,yshift=7,scale=.8] at (0+\dx,\dy){$\gamma_{0B}^{g_0}$};
\node[red,xshift=-2,yshift=7,scale=.8] at (1-\dx,\dy){$\gamma_{1A}^{g_1}$};
\node[red,xshift=3,yshift=7,scale=.8] at (1+\dx,\dy){$\gamma_{1B}^{g_1}$};
\node[red,xshift=-2,yshift=7,scale=.8] at (2-\dx,\dy){$\gamma_{2A}^{g_2}$};
\node[red,xshift=3,yshift=7,scale=.8] at (2+\dx,\dy){$\gamma_{2B}^{g_2}$};
\draw[red,densely dashed,line width=1] (0-\dx,\dy)--(0+\dx,\dy);
\draw[red,densely dashed,line width=1] (1-\dx,\dy)--(1+\dx,\dy);
\draw[red,densely dashed,line width=1] (2-\dx,\dy)--(2+\dx,\dy);
\coordinate (a) at (0+\dx,\dy);
\coordinate (b) at (0-\dx+1,\dy);
\draw[red,line width=1](a)--(b);
\draw[->-=.7,blue,line width=1] ($(a)!.65!(b)$)--($(a)!.651!(b)$);
\coordinate (a) at (1+\dx,\dy);
\coordinate (b) at (1-\dx+1,\dy);
\draw[red,line width=1](a)--(b);
\draw[->-=.7,blue,line width=1] ($(a)!.65!(b)$)--($(a)!.651!(b)$);
\node[]at(0,\dy){};
}
\!\!\!
\right),
\end{align}
where the FSLU $F$ operator is defined as
\begin{align}
\label{A:F}
F(g_0,g_1,g_2) &= 
|G_b|^{1/2}
\nu_2(g_0,g_1,g_2)
X
.\quad\text{(violate fermion parity)}
\end{align}
$X$ is some Majorana fermion projection operators to impose the Majorana pairings. Since the Majorana fermion parity change of the $F$ move is $(-1)^{\om_2(g_0^{-1}g_1,g_1^{-1}g_2)}$ (one can check directly using Kasteleyn orientations), the above $F$ symbol may violate the Majorana fermion parity. So we the state is obstructed if $[\om_2]$ is nontrivial. 

However, we can introduce a 2D bulk to the above 1D state we are constructing. The difference is that we can use a complex fermion from the 2D bulk to compensate the Majorana fermion parity of the 1D $F$ move \eq{A:F}. So the new $F$ move reads
\begin{align}
\label{A:F_}
F(g_0,g_1,g_2) &= 
|G_b|^{1/2}
\nu_2(g_0,g_1,g_2)
\big(c_{012}^{g_0}\big)^{n_2(g_0,g_1,g_2)} X
.
\end{align}
We have to impose the condition
\begin{align}\label{A:n2n0}
n_2 = \om_2.
\end{align}
to make the new $F$ move total fermion parity even. So there is no longer fermion parity inconsistency for the 1D ASPT state on the boundary of a 2D bulk.

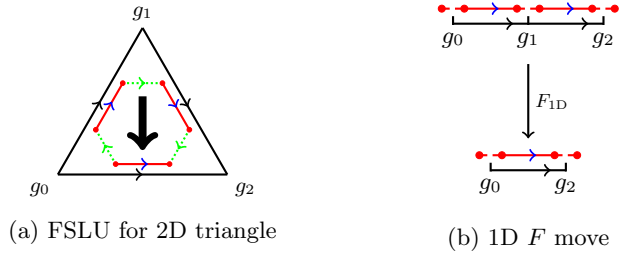
\begin{figure}[ht]
\centering
\begin{subfigure}[h]{.28\textwidth}
\centering
$\tikzfig{scale=1.3}{
\coordinate (0) at ({90+120}:1);
\coordinate (1) at ({-30}:1);
\coordinate (2) at ({90}:1);
\draw[->-=.5,black,thick](0)--(1);
\draw[-<-=.5,black,thick](1)--(2);
\draw[->-=.5,black,thick](0)--(2);
\node[below left,scale=.9]at(0){$g_0$};
\node[below right,scale=.9]at(1){$g_2$};
\node[above,scale=.9]at(2){$g_1$};
\pgfmathsetmacro{\r}{.48};\pgfmathsetmacro{\theta}{35};
\coordinate (12_1) at ({30-\theta}:\r);\coordinate (12_2) at ({30+\theta}:\r);
\coordinate (02_1) at ({150+\theta}:\r);\coordinate (02_2) at ({150-\theta}:\r);
\coordinate (01_1) at ({-90-\theta}:\r);\coordinate (01_2) at ({-90+\theta}:\r);
\draw[red,thick](12_1)--(12_2);
\draw[red,thick](02_1)--(02_2);
\draw[red,thick](01_1)--(01_2);
\draw[-<-=.4,blue,thick]($(12_1)!.6!(12_2)$)--($(12_1)!.601!(12_2)$);
\draw[->-=.6,blue,thick]($(02_1)!.6!(02_2)$)--($(02_1)!.601!(02_2)$);
\draw[->-=.6,blue,thick]($(01_1)!.6!(01_2)$)--($(01_1)!.601!(01_2)$);
\draw[-<-=.6,green,densely dotted,thick](01_2)--(12_1);
\draw[-<-=.6,green,densely dotted,thick](12_2)--(02_2);
\draw[-<-=.6,green,densely dotted,thick](02_1)--(01_1);
\node[circle,draw=red,fill=red,scale=.2] at (01_1) {};
\node[circle,draw=red,fill=red,scale=.2] at (01_2) {};
\node[circle,draw=red,fill=red,scale=.2] at (02_1) {};
\node[circle,draw=red,fill=red,scale=.2] at (02_2) {};
\node[circle,draw=red,fill=red,scale=.2] at (12_1) {};
\node[circle,draw=red,fill=red,scale=.2] at (12_2) {};
\coordinate (x) at (0,.3);
\coordinate (y) at (0,-.25);
\draw[->,line width=3pt] (x) -- (y);
}$
\caption{FSLU for 2D triangle}
\label{fig:2D1D_a}
\end{subfigure}
\begin{subfigure}[h]{.28\textwidth}
\centering
$\tikzfig{scale=1.3}{
\node (up) at (0,0){
\begin{tikzpicture}[scale=1]
\coordinate (0) at (0,0);
\coordinate (1) at (1,0);
\coordinate (2) at (2,0);
\node[below] at (0) {$g_0$};
\node[below] at (1) {$g_1$};
\node[below] at (2) {$g_2$};
\draw[thick] (0,-.017)--(0,.1);
\draw[thick] (1,-.017)--(1,.1);
\draw[thick] (2,-.017)--(2,.1);
\draw[->-=.8,thick] (0)--(1);
\draw[->-=.8,thick] (1)--(2);
\pgfmathsetmacro{\dx}{.15};
\pgfmathsetmacro{\dy}{.2};
\node[circle,draw=red,fill=red,scale=.3] at (0-\dx,\dy){};
\node[circle,draw=red,fill=red,scale=.3] at (0+\dx,\dy){};
\node[circle,draw=red,fill=red,scale=.3] at (1-\dx,\dy){};
\node[circle,draw=red,fill=red,scale=.3] at (1+\dx,\dy){};
\node[circle,draw=red,fill=red,scale=.3] at (2-\dx,\dy){};
\node[circle,draw=red,fill=red,scale=.3] at (2+\dx,\dy){};
\draw[red,densely dashed,thick] (0-\dx,\dy)--(0+\dx,\dy);
\draw[red,densely dashed,thick] (1-\dx,\dy)--(1+\dx,\dy);
\draw[red,densely dashed,thick] (2-\dx,\dy)--(2+\dx,\dy);
\coordinate (a) at (0+\dx,\dy);
\coordinate (b) at (0-\dx+1,\dy);
\draw[red,thick](a)--(b);
\draw[->-=.7,blue,thick] ($(a)!.65!(b)$)--($(a)!.651!(b)$);
\coordinate (a) at (1+\dx,\dy);
\coordinate (b) at (1-\dx+1,\dy);
\draw[red,thick](a)--(b);
\draw[->-=.7,blue,thick] ($(a)!.65!(b)$)--($(a)!.651!(b)$);
\node[]at(0,\dy){};
\end{tikzpicture}
};
\node (down) at (0,-1.5){
\begin{tikzpicture}[scale=1]
\coordinate (0) at (0,0);
\coordinate (1) at (1,0);
\node[below] at (0) {$g_0$};
\node[below] at (1) {$g_2$};
\draw[thick] (0,-.017)--(0,.1);
\draw[thick] (1,-.017)--(1,.1);
\draw[->-=.8,thick] (0)--(1);
\pgfmathsetmacro{\dx}{.15};
\pgfmathsetmacro{\dy}{.2};
\node[circle,draw=red,fill=red,scale=.3] at (0-\dx,\dy){};
\node[circle,draw=red,fill=red,scale=.3] at (0+\dx,\dy){};
\node[circle,draw=red,fill=red,scale=.3] at (1-\dx,\dy){};
\node[circle,draw=red,fill=red,scale=.3] at (1+\dx,\dy){};
\draw[red,densely dashed,thick] (0-\dx,\dy)--(0+\dx,\dy);
\draw[red,densely dashed,thick] (1-\dx,\dy)--(1+\dx,\dy);
\coordinate (a) at (0+\dx,\dy);
\coordinate (b) at (0-\dx+1,\dy);
\draw[red,thick](a)--(b);
\draw[->-=.7,blue,thick] ($(a)!.65!(b)$)--($(a)!.651!(b)$);
\node[]at(0,\dy){};
\end{tikzpicture}
};
\draw[->,thick] (up) -- (down) node[scale=.8,midway,right] {$F_\mathrm{1D}$};
}$
\caption{1D $F$ move}
\label{fig:2D1D_b}
\end{subfigure}
\caption{Relation between FSLU $U_1$ for a 2D triangle and 1D $F$ move. The 2D triangle can be viewed as a 1D $F$ move from the upper two links to the lower link. It relates the trivialization of 2D $n_2$ data to the obstruction of 1D Kitaev chain. (a) The FSLU transformation $U_1$ in \eq{eq:U1} for a single triangle. This FSLU changes the fermion parities of the Majorana fermions and the complex fermions by $(-1)^{\om_2(\bar 01,\bar 23)}$ and $(-1)^{n_2(012)}$ respectively. (b) The 1D $F$ move \eq{A:F_} for the boundary ASPT state. This $F$ move changes the fermion parities of the Majorana fermions and the complex fermions by $(-1)^{\om_2(\bar 01,\bar 23)}$ and $(-1)^{n_2(012)}$ respectively.}
\label{fig:2D1D}
\end{figure}

In fact, the 1D $F$ move \eqs{A:Fmove}{A:F_} can be understood as the FSLU transformation $U_1$ \eq{eq:U1} for a single triangle $\langle 012\rangle$ (see \fig{fig:2D1D}). The upper two links $\langle 01\rangle$ and $\langle 12\rangle$ correspond to the right-hand-side of 1D $F$ move. And the lower link $\langle 02\rangle$ corresponds to the left-hand-side of 1D $F$ move. The additional complex fermion $c_{012}^{g_0}$ in 1D $F$ move \eq{A:F_} is merely the decorate complex fermion at the center of the 2D triangle. This picture relates the 1D FSPT obstruction and the 2D FSPT trivialization, and explains the trivialization relation \eq{A:n2n0} in a simple way.

\subsection{Classification of 2D FSPT phases}

The general classification of 2D FSPT phases is as follow. We first calculate the cohomology groups $H^1(G_b,\Z_2)$, $H^2(G_b,\Z_2)$ and $H^3(G_b,U(1)_T)$. For each $n_1\in H^1(G_b,\Z_2)$, we solve the twisted cocycle equation \eq{2D:eq} for $n_2$. For each solution $n_2$, we solve the twisted cocycle equation \eq{2D:eq} for $\nu_3$. If $n_2$ and $\nu_3$ are in the trivialization subgroup $\Gamma^2$ and $\Gamma^3$ in \eq{2D:Gamma}, then they are trivialized by boundary ASPT states. So the obstruction-free and trivialization-free $(n_1,n_2,\nu_3)$ fully classify the 2D FSPT phases. 

Similar to the bosonic case, we can also use the 2D FSLU transformations to construct the commuting projector parent Hamiltonians. The procedure is tedious but straightforward. For the case of complex fermion decorations only, it is given explictly in Refs.~\cite{Gu2015}. The terms of the Hamiltonian are sequences of fermionic $F$ moves that changes the group element label of a vertex from one to another. All the terms commute with each other, because our FSPT wave function is at the fixed-point.

\section{Fixed-point wave function and classification of FSPT states in 3D}
\label{sec:3D}
\NewFigName{figures/Fig_3D_}

The fixed-point wave function for FSPT state in 3D have four layers of degrees of freedom. It is a superposition of all possible basis states as (we omit the 2D $p+ip$ chiral superconductor layer in the figure for simplicity):
\begin{align}
|\Psi\rangle = \sum_{\text{all conf.}} \Psi\left( 
\tikzfig{scale=2}{
\pgfmathsetmacro{\x}{.2};
\pgfmathsetmacro{\y}{.3};
\draw[dashed] (\x,\y)--(1+\x,\y)--(1+\x,1+\y)--(\x,1+\y)--cycle;
\draw[dashed] (0,0)--(\x,\y);
\draw[] (1,0)--(1+\x,\y);
\draw[] (0,1)--(\x,1+\y);
\draw[] (1,1)--(1+\x,1+\y);
\coordinate (C0) at (.65,.75);
\coordinate (C1) at (.3,.9);
\coordinate (C2) at (.4,.5);
\coordinate (C3) at (.85,.4);
\coordinate (C4) at (1,1.1);
\draw[green,densely dotted,line width=1.5] (C0)--(C1);
\draw[green,densely dotted,line width=1.5] (C0)--(C3);
\draw[green,densely dotted,line width=1.5] (C1)--($(C1)+(0,.5)$);
\draw[green,densely dotted,line width=1.5] (C3)--($(C3)+(0,-.3)$);
\draw[green,densely dotted,line width=1.5] (C4)--($(C4)+(.3,0)$);
\draw[green,densely dotted,line width=1.5] (C4)--($(C4)+(0,.3)$);
%
\draw[dashed] (0,0)--(\x,1+\y)--(1+\x,\y)--cycle;
\node[circle,shading=ball,scale=.4] at (C0) {};
\node[circle,shading=ball,scale=.4] at (C1) {};
\node[circle,shading=ball,scale=.4] at (C2) {};
\node[circle,draw=blue,fill=white,scale=.35] at (C3) {};
\node[circle,draw=blue,fill=white,scale=.35] at (C4) {};
\filldraw[white,opacity=0.2] (0,1)--(1,1)--(1+\x,1+\y)--(\x,1+\y)--cycle;
\draw[] (0,1)--(1,1)--(1+\x,1+\y)--(\x,1+\y)--cycle;
\filldraw[white,opacity=0.2] (1,0)--(1,1)--(1+\x,1+\y)--(1+\x,\y)--cycle;
\draw[] (1,0)--(1,1)--(1+\x,1+\y)--(1+\x,\y)--cycle;
\filldraw[white,opacity=0.2] (0,0)--(1,0)--(1,1)--(0,1)--cycle;
\draw[] (0,0)--(1,0)--(1,1)--(0,1)--cycle;
\draw[] (1,1)--(0,0); \draw[] (1,1)--(1+\x,\y); \draw[] (1,1)--(\x,1+\y);
\draw[green,line width=1.5] ($(C1)+(0,.2)$)--($(C1)+(0,.5)$);
\draw[green,line width=1.5] ($(C4)+(.13,0)$)--($(C4)+(.3,0)$);
\draw[green,line width=1.5] ($(C4)+(0,.12)$)--($(C4)+(0,.3)$);
}
\right) \stretchleftright{\Bigg|\ }{
\vcenter{\hbox{\includegraphics[scale=1]{\FigName\arabic{Num}.pdf}}}
}{\Big\rangle}.
\end{align}
The basis state is a labelled by group elements of $G_b$ on each vertex. On the plane dual to each link, we put 2D $p+ip$ chiral superconductor. Different from other layers, the $p+ip$ superconductor layer does not have fixed-point wave function on discrete lattice \cite{KapustinFidkowski2018}. So we will not discuss this layer decoration until at the end of this section. Along the dual link of each triangle, we decorate Kitaev chains (see green lines in the above equation). And at the center of each tetrahedron, we decorate some complex fermions (see blue dots in the above equation).

\subsection{Four layers of degrees of freedom}
\label{sec:3D:dof}

\begin{figure}[ht]
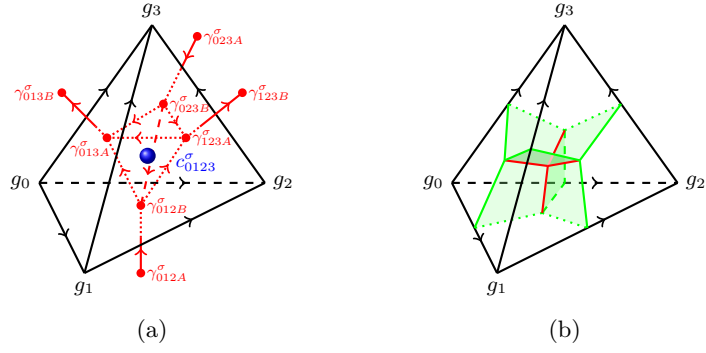

\centering
\begin{subfigure}[h]{.3\textwidth}
\centering
$
\tikzfig{scale=3}{
\coordinate (0) at (0,0);
\coordinate (1) at (.2,-.4);
\coordinate (2) at (1,0);
\coordinate (3) at (.5,.7);
\coordinate (center) at (.48,.15);
\coordinate (012-1) at (.45,-.4);\coordinate (012-2) at (.45,-.1);
\coordinate (012-c) at ($(012-1)!0.52!(012-2)$);
\coordinate (013-1) at (.3,.2);\coordinate (013-2) at (.1,.4);
\coordinate (013-c) at ($(013-1)!0.25!(013-2)$);
\coordinate (123-1) at (.65,.2);\coordinate (123-2) at (.9,.4);
\coordinate (123-c) at ($(123-1)!0.25!(123-2)$);
\coordinate (023-1) at (0.55,0.35);\coordinate (023-2) at (.7,.65);
\coordinate (023-c) at ($(023-1)!0.52!(023-2)$);
%
%
\draw[->-=.6,thick](0)node[scale=.9,left]{$g_0$}--(1)node[scale=.9,below]{$g_1$};
\draw[->-=.65,thick,dashed](0)--(2)node[scale=.9,right]{$g_2$};
\draw[->-=.6,thick](0)--(3)node[scale=.9,above]{$g_3$};
\draw[->-=.7,thick,red](023-2)node[red,scale=.7,right]{$\g_{023A}^\s$}--(023-c);
\draw[thick,densely dotted,red](023-c)--(023-1)node[red,scale=.7,right]{$\g_{023B}^\s$};
\draw[->-=.6,thick](2)--(3);
\draw[->-=.7,thick,red](012-1)node[red,right,scale=.7]{$\g_{012A}^\s$}--(012-c);
\draw[thick,densely dotted,red](012-c)--(012-2)node[red,right,scale=.7]{$\g_{012B}^\s$};
\draw[->-=.6,thick](1)--(2);
\draw[thick,densely dotted,red](013-1)node[red,scale=.7,xshift=-9,yshift=-6]{$\g_{013A}^\s$}--(013-c);
\draw[->-=.7,thick,red](013-c)--(013-2)node[red,scale=.7,left]{$\g_{013B}^\s$};
\draw[thick,densely dotted,red](123-1)node[red,scale=.7,right]{$\g_{123A}^\s$}--(123-c);
\draw[->-=.7,thick,red](123-c)--(123-2)node[red,scale=.7,right]{$\g_{123B}^\s$};
\draw[->-=.7,thick,dashed,red](023-1)--(012-2);
\node[circle,shading=ball,scale=.7] (ball) at (.48,.12) {};
\node[blue,scale=.8,xshift=23,yshift=-3] at (.48,.12) {$c_{0123}^\s$};
\draw[->-=.6,thick,densely dotted,red](023-1)--(123-1);
\draw[->-=.65,thick,densely dotted,red](123-1)--(013-1);
\draw[->-=.6,thick,densely dotted,red](012-2)--(123-1);
\draw[->-=.55,thick,densely dotted,red](023-1)--(013-1);
\draw[->-=.55,thick,densely dotted,red](012-2)--(013-1);
\fill [red] (012-1) circle (.55pt);
\fill [red] (012-2) circle (.55pt);
\fill [red] (013-1) circle (.55pt);
\fill [red] (013-2) circle (.55pt);
\fill [red] (123-1) circle (.55pt);
\fill [red] (123-2) circle (.55pt);
\fill [red] (023-1) circle (.55pt);
\fill [red] (023-2) circle (.55pt);
\draw[->-=.7,thick](1)--(3);
}
$
\caption{}
\label{fig:3D:dof_a}
\end{subfigure}
\begin{subfigure}[h]{.3\textwidth}
\centering
$
\tikzfig{scale=3}{
\coordinate (0) at (0,0);
\coordinate (1) at (.2,-.4);
\coordinate (2) at (1,0);
\coordinate (3) at (.5,.7);
\coordinate (center) at (.48,.15);
\draw[->-=.65,thick](0)node[scale=.9,left]{$g_0$}--(1)node[scale=.9,below]{$g_1$};
\draw[->-=.65,thick,dashed](0)--(2)node[scale=.9,right]{$g_2$};
\draw[->-=.6,thick](0)--(3)node[scale=.9,above]{$g_3$};
\draw[->-=.6,thick](1)--(2);
\draw[->-=.6,thick](2)--(3);
\coordinate (01) at ($(0)!.5!(1)$);
\coordinate (02) at ($(0)!.5!(2)$);
\coordinate (03) at ($(0)!.5!(3)$);
\coordinate (12) at ($(1)!.5!(2)$);
\coordinate (13) at ($(1)!.5!(3)$);
\coordinate (23) at ($(2)!.5!(3)$);
\coordinate (012) at ($(01)!.3333!(2)$);
\coordinate (013) at ($(01)!.3333!(3)$);
\coordinate (023) at ($(02)!.3333!(3)$);
\coordinate (123) at ($(12)!.3333!(3)$);
\coordinate (0123) at ($(012)!.25!(3)$);
\fill[green!80,nearly transparent] (0123) -- (012) -- (02) -- (023) -- cycle;
\fill[green!40,nearly transparent] (0123) -- (012) -- (01) -- (013) -- cycle;
\fill[green!40,nearly transparent] (0123) -- (012) -- (12) -- (123) -- cycle;
\fill[green!40,nearly transparent] (0123) -- (013) -- (03) -- (023) -- cycle;
\fill[green!40,nearly transparent] (0123) -- (123) -- (23) -- (023) -- cycle;
\fill[green!60!black,nearly transparent] (0123) -- (123) -- (13) -- (013) -- cycle;
\draw[thick,green,dashed](012)--(02);
\draw[thick,green,dotted](012)--(01);
\draw[thick,green,dotted](012)--(12);
\draw[thick,green,dashed](023)--(02);
\draw[thick,green,dotted](023)--(03);
\draw[thick,green,dotted](023)--(23);
\draw[thick,red!30] ($(0123)!.35!(023)$) -- (0123);
\draw[thick,red] ($(0123)!.35!(023)$) -- (023);
\draw[thick,red](0123)--(123);
\draw[thick,red](0123)--(012);
\draw[thick,red](0123)--(013);
\draw[thick,green](013)--(01);
\draw[thick,green](013)--(03);
\draw[thick,green](013)--(13);
\draw[thick,green](123)--(12);
\draw[thick,green](123)--(13);
\draw[thick,green](123)--(23);
\draw[->-=.7,thick](1)--(3);
}
$
\caption{}
\label{fig:3D:dof_b}
\end{subfigure}
\caption{Four layers of degrees of freedom in a tetrahedron of 3D triangulation lattice. (a) Layers of bosonic state $|g_i\rangle$ on each vertex $i$, complex fermions $c_{ijkl}^\s$ at the center of each tetrahedron $\langle ijkl\rangle$ and Majorana fermions $\g_{ijk,A/B}^\s$ on the two sides of each triangle $\langle ijk\rangle$. (b) Layer of 2D $p+ip$ chiral superconductor on the (green) plane dual to each link $\langle ij\rangle$. The boundary chiral Majorana modes of the $p+ip$ superconductors are along the (red) intersecting lines of the (green) planes.}
\label{fig:3D:dof}
\end{figure}

Similar to 2D, we construct FSPT states on 3D lattice by decorating complex fermions, Kitaev chains and $p+ip$ superconductors to the BSPT states. Therefore, there are four layers of degrees of freedom including the bosonic ones on 3D triangulation lattice:
\begin{itemize}
\item
$|G_b|$ level bosonic (spin) state $|g_i\rangle$ ($g_i\in G_b$) on each vertex $i$.
\item
$|G_b|$ species of complex fermions $c_{ijkl}^\s$ ($\s\in G_b$) at the center of each tetrahedron $\langle ijkl\rangle$.
\item
$|G_b|$ species of complex fermions (splitted to Majorana fermions) $a_{ijk}^\s=(\g_{ijk,A}^\s+i\g_{ijk,B}^\s)/2$ ($\s\in G_b$) on the two sides of each triangle $\langle ijk\rangle$.
\item
$|G_b|$ species of 2D $p+ip$ chiral superconductors [may have several copies indicated by $n_1(g_i,g_j)\in\Z_T$] on the plane dual to link $\langle ij\rangle$. The boundary chiral Majorana modes (along the link dual to some triangle) are $\psi_{ij,L;\al}^\s$ or $\psi_{ij,R;\al}^\s$ ($\s\in G_b$) depending on the chirality (left/right-hand-rule with respect to the oriented link $\langle ij\rangle$). Here $\al$ labels the number of the chiral Majorana modes ($\al=1,2,...,|n_1(g_i,g_j)|$).
\end{itemize}
The four layers of degrees of freedom are summarized in one tetrahedron of the 3D triangulation lattice in \fig{fig:3D:dof}. The four vertices of the tetrahedron are labelled by $g_0,g_1,g_2,g_3\in G_b$. In \fig{fig:3D:dof_a}, blue ball is the complex fermion $c_{0123}^\s$ ($\s\in G_b$) at the center of the tetrahedron. Red dots represent Majorana fermions $\g_{ijk,A}^\s$ and $\g_{ijk,B}^\s$ ($\s\in G_b$) on the two sides of the triangle $\langle ijk\rangle$. We use the convention that the direction pointing from $\g_{ijk,A}^\s$ to $\g_{ijk,B}^\s$ is the same as the right-hand rule orientation of the triangle $\langle ijk\rangle$. In \fig{fig:3D:dof_b}, each green area dual to link $\langle ij\rangle$ represents the decorated 2D $p+ip$ superconductors. There are $|n_1(g_i,g_j)|$ right-moving or left-moving chiral Majorana modes $\psi_{i,R/L;\al}^{g_i}$ ($\al=1,2,...,|n_1|$) along the boundary (red line) of the green area dual to each link $\langle ij\rangle$.

As discussed in section~\ref{sec:FSLU:dof}, the symmetry transformation rules of these degrees of freedom under $G_b$ are summarize as follows ($g,g_i,\s\in G_b$):
\begin{align}
\label{3D:symm:gi}
U(g) |g_i\rangle &= |gg_i\rangle,\\
\label{3D:symm:c}
U(g) c_{ijkl}^\s U(g)^\dagger &= (-1)^{\om_2(g,\s)} c_{ijkl}^{g\s},\\
\label{3D:symm:A}
U(g) \g_{ijk,A}^\s U(g)^\dagger &= (-1)^{\om_2(g,\s)} \g_{ijk,A}^{g\s},\\
\label{3D:symm:B}
U(g) \g_{ijk,B}^\s U(g)^\dagger &= (-1)^{\om_2(g,\s)+s_1(g)} \g_{ijk,B}^{g\s},\\
\label{3D:symm:R}
U(g) \psi_{ij,R;\al}^\s U(g)^\dagger &= (-1)^{\om_2(g,\s)} \psi_{ij,g(R);\al}^{g\s},\\
\label{3D:symm:L}
U(g) \psi_{ij,L;\al}^\s U(g)^\dagger &= (-1)^{\om_2(g,\s)+s_1(g)} \psi_{ij,g(L);\al}^{g\s}.
\end{align}
The bosonic degrees of freedom always form a linear representation of $G_b$ (and $G_f$). On the other hand, the fermion modes support projective representations of $G_b$ with coefficient $(-1)^{\om_2}$, and hence linear representations of $G_f$ by \eq{UU}.

\subsection{Decorations of fermion layers}

In this section, we would construct systematic procedures of decorating Kitaev chains and complex fermions to the bosonic basis state $|\{g_i\}\rangle$. Each layer of the degrees of freedom will twist the consistent equations for the next layer. The decoration rules should respect the symmetry in all layers.

We will focus on the Kitaev chain and complex fermion decoration here, and the decorations of 2D $p+ip$ superconductors will be discussed at the end of this section.

\subsubsection{Kitaev chain decoration}
\label{sec:3D:Maj}

The Kitaev chain decoration in 3D is similar to the constructions in the Ref.~\onlinecite{WangGu2017}. The difference is that we would put $|G_b|$ species of Majorana fermions rather than one. However, we still put only one species of Majorana fermions into nontrivial pairings along the decorated Kitaev chain.

\begin{figure}[ht]
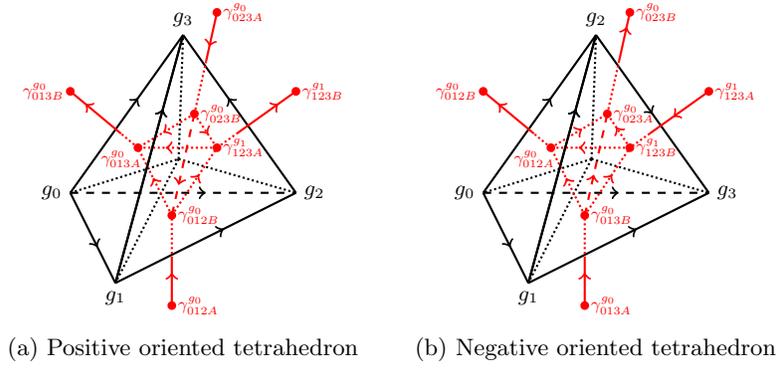

\centering
\begin{subfigure}[h]{.3\textwidth}
\centering
$\tikzfig{scale=3}{
\coordinate (0) at (0,0);
\coordinate (1) at (.2,-.4);
\coordinate (2) at (1,0);
\coordinate (3) at (.5,.7);
\coordinate (center) at (.48,.15);
\draw[->-=.6,thick](0)node[scale=.9,left]{$g_0$}--(1)node[scale=.9,below]{$g_1$};
\draw[->-=.6,thick,dashed](0)--(2)node[scale=.9,right]{$g_2$};
\draw[->-=.6,thick](0)--(3)node[scale=.9,above]{$g_3$};
\draw[->-=.6,thick](1)--(2);
\draw[->-=.7,thick](1)--(3);
\draw[->-=.6,thick](2)--(3);
\draw[thick,densely dotted](0)--(center);
\draw[thick,densely dotted](center)--(1);
\draw[thick,densely dotted](center)--(2);
\draw[thick,densely dotted](center)--(3);
\coordinate (012-1) at (.45,-.5);\coordinate (012-2) at (.45,-.1);
\coordinate (012-c) at ($(012-1)!0.52!(012-2)$);
\coordinate (013-1) at (.3,.2);\coordinate (013-2) at (0,.45);
\coordinate (013-c) at ($(013-1)!0.25!(013-2)$);
\coordinate (123-1) at (.65,.2);\coordinate (123-2) at (1,.45);
\coordinate (123-c) at ($(123-1)!0.25!(123-2)$);
\coordinate (023-1) at (0.55,0.35);\coordinate (023-2) at (0.65,.8);
\coordinate (023-c) at ($(023-1)!0.52!(023-2)$);
\fill [red] (012-1) circle (.55pt);
\fill [red] (012-2) circle (.55pt);
\fill [red] (013-1) circle (.55pt);
\fill [red] (013-2) circle (.55pt);
\fill [red] (123-1) circle (.55pt);
\fill [red] (123-2) circle (.55pt);
\fill [red] (023-1) circle (.55pt);
\fill [red] (023-2) circle (.55pt);
\draw[->-=.7,thick,red](012-1)node[right,scale=.7]{$\g_{012A}^{g_0}$}--(012-c);
\draw[thick,densely dotted,red](012-c)--(012-2)node[right,scale=.7]{$\g_{012B}^{g_0}$};
\draw[thick,densely dotted,red](013-1)node[scale=.7,xshift=-9,yshift=-6]{$\g_{013A}^{g_0}$}--(013-c);
\draw[->-=.7,thick,red](013-c)--(013-2)node[scale=.7,left]{$\g_{013B}^{g_0}$};
\draw[thick,densely dotted,red](123-1)node[scale=.7,right]{$\g_{123A}^{g_1}$}--(123-c);
\draw[->-=.7,thick,red](123-c)--(123-2)node[scale=.7,right]{$\g_{123B}^{g_1}$};
\draw[->-=.7,thick,red](023-2)node[scale=.7,right]{$\g_{023A}^{g_0}$}--(023-c);
\draw[thick,densely dotted,red](023-c)--(023-1)node[scale=.7,right]{$\g_{023B}^{g_0}$};
\draw[->-=.6,thick,densely dotted,red](023-1)--(123-1);
\draw[->-=.65,thick,densely dotted,red](123-1)--(013-1);
\draw[->-=.6,thick,densely dotted,red](012-2)--(123-1);
\draw[->-=.55,thick,densely dotted,red](023-1)--(013-1);
\draw[->-=.7,thick,dashed,red](023-1)--(012-2);
\draw[->-=.55,thick,densely dotted,red](012-2)--(013-1);
\draw[->-=.7,thick](1)--(3);
}$
\caption{Positive oriented tetrahedron}
\label{fig:3DKasteleyn1}
\end{subfigure}
\begin{subfigure}[h]{.3\textwidth}
\centering
$\tikzfig{scale=3}{
\coordinate (0) at (0,0);
\coordinate (1) at (.2,-.4);
\coordinate (2) at (1,0);
\coordinate (3) at (.5,.7);
\coordinate (center) at (.48,.15);
\draw[->-=.6,thick](0)node[scale=.9,left]{$g_0$}--(1)node[scale=.9,below]{$g_1$};
\draw[->-=.6,thick,dashed](0)--(2)node[scale=.9,right]{$g_3$};
\draw[->-=.6,thick](0)--(3)node[scale=.9,above]{$g_2$};
\draw[->-=.6,thick](1)--(2);
\draw[->-=.7,thick](1)--(3);
\draw[->-=.5,thick](3)--(2);
\draw[thick,densely dotted](center)--(0);
\draw[thick,densely dotted](center)--(1);
\draw[thick,densely dotted](center)--(2);
\draw[thick,densely dotted](center)--(3);
\coordinate (012-1) at (.45,-.5);\coordinate (012-2) at (.45,-.1);
\coordinate (012-c) at ($(012-1)!0.52!(012-2)$);
\coordinate (013-1) at (.3,.2);\coordinate (013-2) at (0,.45);
\coordinate (013-c) at ($(013-1)!0.25!(013-2)$);
\coordinate (123-1) at (.65,.2);\coordinate (123-2) at (1,.45);
\coordinate (123-c) at ($(123-1)!0.25!(123-2)$);
\coordinate (023-1) at (0.55,0.35);\coordinate (023-2) at (0.65,.8);
\coordinate (023-c) at ($(023-1)!0.52!(023-2)$);
\fill [red] (012-1) circle (.55pt);
\fill [red] (012-2) circle (.55pt);
\fill [red] (013-1) circle (.55pt);
\fill [red] (013-2) circle (.55pt);
\fill [red] (123-1) circle (.55pt);
\fill [red] (123-2) circle (.55pt);
\fill [red] (023-1) circle (.55pt);
\fill [red] (023-2) circle (.55pt);
\draw[->-=.7,thick,red](012-1)node[scale=.7,right]{$\g_{013A}^{g_0}$}--(012-c);
\draw[thick,densely dotted,red](012-c)--(012-2)node[scale=.7,right]{$\g_{013B}^{g_0}$};
\draw[thick,densely dotted,red](013-1)node[scale=.7,xshift=-9,yshift=-6]{$\g_{012A}^{g_0}$}--(013-c);
\draw[->-=.7,thick,red](013-c)--(013-2)node[scale=.7,left]{$\g_{012B}^{g_0}$};
\draw[->-=.6,thick,red](123-2)node[scale=.7,right]{$\g_{123A}^{g_1}$}--(123-c);
\draw[thick,densely dotted,red](123-c)--(123-1)node[scale=.7,right]{$\g_{123B}^{g_1}$};
\draw[thick,densely dotted,red](023-1)node[scale=.7,right]{$\g_{023A}^{g_0}$}--(023-c);
\draw[->-=.7,thick,red](023-c)--(023-2)node[scale=.7,right]{$\g_{023B}^{g_0}$};
\draw[->-=.6,thick,densely dotted,red](123-1)--(023-1);
\draw[->-=.65,thick,densely dotted,red](123-1)--(013-1);
\draw[->-=.6,thick,densely dotted,red](012-2)--(123-1);
\draw[->-=.55,thick,densely dotted,red](013-1)--(023-1);
\draw[->-=.4,thick,dashed,red](012-2)--(023-1);
\draw[->-=.55,thick,densely dotted,red](012-2)--(013-1);
\draw[->-=.7,thick](1)--(3);
}$
\caption{Negative oriented tetrahedron}
\label{fig:3DKasteleyn2}
\end{subfigure}
\caption{Local Kasteleyn orientations of the resolved dual lattice. For a given triangulation of the 3D spacial spin manifold (shown by black links), we can construct a resolved dual lattice (shown by red links). The Majorana fermion pairings (in the standard tetrahedron) should respect the red link arrows in the figures.}
\label{fig:3DKasteleyn}
\end{figure}

\paragraph{Decoration procedure.}

For a given 3D triangulation lattice, we first construct the resolved dual lattice (red lattice shown in \fig{fig:3DKasteleyn}). Our Majorana fermions $\g_{ijk,A}^\s$ and $\g_{ijk,B}^\s$ ($\s\in G_b$) are at the (red) vertices on the two sides of each (black) triangle $\langle ijk\rangle$. The red arrow is follow the convention that the direction from $\g_{ijk,A}^\s$ to $\g_{ijk,B}^\s$ is the same as the right-hand rule orientation of the triangle $\langle ijk\rangle$. The direction of vacuum pairing between them is from $A$ to $B$: $-i\g_{ijk,A}^\s \g_{ijk,B}^\s = 1$ when acting on the state. To decorate Kitaev chains on the red lattice, we also should add arrows to the small red tetrahedron inside each black tetrahedron (see \fig{fig:3DKasteleyn}). These red arrows are constructed from the discrete spin structures (a choose of trivialization of Stiefel-Whitney homology class $w_1$ dual to cohomology class $w^2$) of the 3D spacial spin manifold triangulation. The Majorana fermions are always paired according to these red arrows on the red lattice. The red arrows have the property that the number of counterclockwise arrows in the smallest red loop around each black link is always odd. For details of the local Kasteleyn orientations for arbitrary triangulation, we refer the interested readers to Ref.~\onlinecite{WangGu2017}.

The Kitaev chain decoration on the red lattice is specified by $n_2(g_i,g_j,g_k)\in \Z_2$, which is a function of three group elements $g_i,g_j,g_k\in G_b$. If $n_2(g_i,g_j,g_k)=0$, the Majorana fermions $\g_{ijk,A}^\s$ and $\g_{ijk,B}^\s$ on the two sides of the triangle $\langle ijk\rangle$ are in vacuum pairings: $-i\g_{ijk,A}^\s \g_{ijk,B}^\s = 1$ (for all $\s\in G_b$). On the other hand, if $n_2(g_i,g_j,g_k)=1$, we will decorate a Kitaev chain going though the triangle $\langle ijk\rangle$. For all $|G_b|$ species of Majorana fermions, we only put $\g_{ijk,A}^{g_i}$ and $\g_{ijk,B}^{g_i}$ to be in the nontrivial pairing. And all other $|G_b|-1$ species of Majorana fermions $\g_{ijk,A}^{\s}$ and $\g_{ijk,B}^\s$ with $\s\neq g_i$ are still in vacuum pairings. Here is an example of the decoration of Kitaev chain going though triangles $\langle 013\rangle$ and $\langle 023\rangle$ of the tetrahedron $\langle 0123\rangle$ (we omit the operator labels of Majorana fermions which are in vacuum pairings along the Kitaev chain):
\begin{align}\label{3D:n2}
\tikzfig{scale=3}{
\coordinate (0) at (0,0);
\coordinate (1) at (.2,-.4);
\coordinate (2) at (1,0);
\coordinate (3) at (.5,.7);
\coordinate (center) at (.48,.15);
\coordinate (012-1) at (.45,-.4);\coordinate (012-2) at (.45,-.1);
\coordinate (012-c) at ($(012-1)!0.52!(012-2)$);
\coordinate (013-1) at (.3,.2);\coordinate (013-2) at (.1,.4);
\coordinate (013-c) at ($(013-1)!0.25!(013-2)$);
\coordinate (123-1) at (.65,.2);\coordinate (123-2) at (.9,.4);
\coordinate (123-c) at ($(123-1)!0.25!(123-2)$);
\coordinate (023-1) at (0.55,0.35);\coordinate (023-2) at (.7,.65);
\coordinate (023-c) at ($(023-1)!0.52!(023-2)$);
\Ellipse{gray!50}{012-2}{012-1}{1.4}{.25};
\Ellipse{gray!50}{023-2}{023-1}{1.4}{.25};
\Ellipse{gray!50}{123-1}{013-1}{1.4}{.2};
\draw[->-=.6,thick](0)node[scale=.9,left]{$g_0$}--(1)node[scale=.9,below]{$g_1$};
\draw[->-=.6,thick,dashed](0)--(2)node[scale=.9,right]{$g_2$};
\draw[->-=.6,thick](0)--(3)node[scale=.9,above]{$g_3$};
\draw[->-=.7,thick](1)--(3);
\Ellipse{gray!50}{013-2}{013-1}{1.4}{.25};
\draw[->-=.7,thick,red](023-2)node[red,scale=.7,right]{$\g_{023A}^{\s}$}--(023-c);
\draw[thick,densely dotted,red](023-c)--(023-1);
\draw[->-=.6,thick](2)--(3);
\Ellipse{gray!50}{123-2}{123-1}{1.4}{.25};
\node[red,scale=.7,right,xshift=-1] at (023-1) {$\g_{023B}^{\s}$};
\draw[->-=.7,thick,red](012-1)node[red,right,scale=.7]{$\g_{012A}^{\s}$}--(012-c);
\draw[thick,densely dotted,red](012-c)--(012-2)node[red,right,scale=.7]{$\g_{012B}^{\s}$};
\draw[->-=.6,thick](1)--(2);
\draw[very thick,densely dotted,green](013-1)node[red,scale=.7,below,xshift=-10]{$\g_{013A}^{g_0}$}--(013-c);
\draw[->-=.7,very thick,green](013-c)--(013-2)node[red,scale=.7,left]{$\g_{013B}^{g_0}$};
\draw[very thick,densely dotted,green](123-1)node[red,scale=.7,right,yshift=-3]{$\g_{123A}^{g_1}$}--(123-c);
\draw[->-=.7,very thick,green](123-c)--(123-2)node[red,scale=.7,right]{$\g_{123B}^{g_1}$};
\draw[thick,dashed,red](023-1)--(012-2);
\draw[->-=2.5/4,thick,blue] ($(023-1)!.7!(012-2)$)--($(023-1)!.701!(012-2)$);
\draw[thick,densely dotted,red](023-1)--(123-1);
\draw[->-=2.5/4,thick,blue] ($(023-1)!.6!(123-1)$)--($(023-1)!.601!(123-1)$);
\draw[very thick,densely dotted,green](123-1)--(013-1);
\draw[->-=2.5/4,very thick,blue] ($(123-1)!.65!(013-1)$)--($(123-1)!.651!(013-1)$);
\draw[thick,densely dotted,red](012-2)--(123-1);
\draw[->-=2.5/4,thick,blue] ($(012-2)!.6!(123-1)$)--($(012-2)!.601!(123-1)$);
\draw[->-=.55,thick,densely dotted,red](023-1)--(013-1);
\draw[->-=.55,thick,densely dotted,red](012-2)--(013-1);
\fill [red] (012-1) circle (.55pt);
\fill [red] (012-2) circle (.55pt);
\fill [red] (013-1) circle (.55pt);
\fill [red] (013-2) circle (.55pt);
\fill [red] (123-1) circle (.55pt);
\fill [red] (123-2) circle (.55pt);
\fill [red] (023-1) circle (.55pt);
\fill [red] (023-2) circle (.55pt);
\draw[->-=.7,thick](1)--(3);
}
\end{align}
The decorated Kitaev chain is indicated by a green line ($\g_{013B}^{g_0}$-$\g_{013A}^{g_0}$-$\g_{123A}^{g_1}$-$\g_{123B}^{g_1}$). Along the Kitaev chain, there is a nontrivial pairing between $\g_{123A}^{g_1}$ and $\g_{013A}^{g_0}$. And the Majorana fermions $\g_{013A/B}^\s$ ($\s\neq g_0$) and $\g_{123A/B}^\s$ ($\s\neq g_1$) are all in vacuum pairings. For the triangles without Kitaev chain going though (triangles $\langle 012\rangle$ and $\langle 023\rangle$), the Majorana fermions on their two sides are all in vacuum pairings ($\g_{012A/B}^{\s}$ and $\g_{023A/B}^{\s}$ for all $\s\in G_b$). In summary, we have the following Majorana fermion pairings:
\begin{align}
&\tikzfig{scale=1}{
\coordinate (g0) at (0,0);
\coordinate (g1) at (1,0);
\Ellipse{gray!50}{g0}{g1}{1.4}{.27};
\draw[red,densely dashed,line width=1](0,0)--(1,0);
\node[circle,draw=red,fill=red,scale=.3] at (0,0){};
\node[circle,draw=red,fill=red,scale=.3] at (1,0){};
\node[red,xshift=-2,yshift=7,scale=.8] at (0,0){$\g_{ijk,A}^{\s}$};
\node[red,xshift=-2,yshift=7,scale=.8] at (1,0){$\g_{ijk,B}^{\s}$};
\draw[->-=.7,red,line width=1] ($(0,0)!.625!(1,0)$)--($(0,0)!.6251!(1,0)$);
\node[]at(0,0){};
}
\ \ \iff
\quad
-i \g_{ijk,A}^{\s} \g_{ijk,B}^{\s} = 1
\\
\label{3D:Mpair2}
&\tikzfig{scale=1}{
\coordinate (g0) at (0,0);
\coordinate (g1) at (1,0);
\Ellipse{gray!50}{g0}{g1}{1.4}{.27};
\draw[red,line width=1](0,0)--(1,0);
\node[circle,draw=red,fill=red,scale=.3] at (0,0){};
\node[circle,draw=red,fill=red,scale=.3] at (1,0){};
\node[red,xshift=-2,yshift=7,scale=.8] at (0,0){$\g_{123A}^{g_1}$};
\node[red,xshift=-2,yshift=7,scale=.8] at (1,0){$\g_{013A}^{g_0}$};
\draw[->-=.7,blue,line width=1] ($(0,0)!.625!(1,0)$)--($(0,0)!.6251!(1,0)$);
\node[]at(0,0){};
}
\ \ \iff
\quad
-i \g_{123A}^{g_1} \g_{013A}^{g_0} = (-1)^{\om_2(g_0,g_0^{-1}g_1)+s_1(g_0)}.
\end{align}
Both the trivial and nontrivial Majorana fermion pairings are indicated by gray ellipses. The blue arrow means that the pairing direction of Majorana fermions may be changed compared to the local Kasteleyn orientation indicated by red arrow. We will discuss more about the detailed pairing directions and why they are symmetric later.

\paragraph{Consistency condition.}

According to our decoration rule,  the total number of decorated Kitaev chains going though the four boundary triangles of a given tetrahedron $\langle 0123\rangle$ is
\begin{align}
(\dd n_2)(g_0,g_1,g_2,g_3) = n_2(g_1,g_2,g_3) + n_2(g_0,g_2,g_3) + n_2(g_0,g_1,g_3) + n_2(g_0,g_1,g_2).
\end{align}
Since we are constructing gapped state without intrinsic topological order, there should be no dangling free Majorana fermions inside any tetrahedron. So the number of total Kitaev chains 
going though the boundary of a tetrahedron should be even. We therefore have the (mod 2) equation
\begin{align}\label{3D:dn2}
\dd n_2 = 0,
\end{align}
as the consistency condition for Kitaev chain decorations.


\paragraph{Symmetric pairing directions.}
\label{sec:3D:Maj:dir}

Now let us turn back to the details of symmetric Majorana fermion pairings inside each tetrahedron of the 3D triangulation lattice. Our strategy of constructing $G_b$-symmetric pairings is the same as in the 2D case: we first consider the \emph{standard} tetrahedron with $g_0=e$, and then apply a $U(g_0)$-action to obtain all other non-standard tetrahedra with generic group element labels. In this way, the Majorana fermion pairings are automatically symmetric, because of the symmetry transformation rule of the pairing projection operators \eq{P_symm}. For the standard tetrahedron, the Majorana fermions are paired according to the local Kasteleyn orientations indicated by red arrows. And the pairings in the non-standard tetrahedron is obtained by a $U(g_0)$-action as follows:
\begin{align}\label{3D:g_symm}
\tikzfig{scale=3}{
\coordinate (0) at (0,0);
\coordinate (1) at (.2,-.4);
\coordinate (2) at (1,0);
\coordinate (3) at (.5,.7);
\coordinate (center) at (.48,.15);
\coordinate (012-1) at (.45,-.4);\coordinate (012-2) at (.45,-.1);
\coordinate (012-c) at ($(012-1)!0.52!(012-2)$);
\coordinate (013-1) at (.3,.2);\coordinate (013-2) at (.1,.4);
\coordinate (013-c) at ($(013-1)!0.25!(013-2)$);
\coordinate (123-1) at (.65,.2);\coordinate (123-2) at (.9,.4);
\coordinate (123-c) at ($(123-1)!0.25!(123-2)$);
\coordinate (023-1) at (0.55,0.35);\coordinate (023-2) at (.7,.65);
\coordinate (023-c) at ($(023-1)!0.52!(023-2)$);
%
%
\draw[->-=.6,thick](0)node[scale=.9,left]{$e$}--(1)node[scale=.9,below]{$g_0^{-1}g_1$};
\draw[->-=.6,thick,dashed](0)--(2)node[scale=.9,right]{$g_0^{-1}g_2$};
\draw[->-=.7,thick,red](023-2)node[red,scale=.7,right]{$\g_{023A}^{e}$}--(023-c);
\draw[thick,densely dotted,red](023-c)--(023-1);
\draw[->-=.6,thick](2)--(3);
\draw[->-=.7,thick,red](012-1)node[red,right,scale=.7]{$\g_{012A}^{e}$}--(012-c);
\draw[thick,densely dotted,red](012-c)--(012-2)node[red,right,scale=.7]{$\g_{012B}^{e}$};
\draw[->-=.6,thick](0)--(3)node[scale=.9,above]{$g_0^{-1}g_3$};
\draw[thick,densely dotted,red](013-1)node[red,scale=.7,below,xshift=-10]{$\g_{013A}^{e}$}--(013-c);
\draw[->-=.7,thick,red](013-c)--(013-2)node[red,scale=.7,left]{$\g_{013B}^{e}$};
\draw[thick,densely dotted,red](123-1)node[red,scale=.7,right,xshift=-5,yshift=-5]{$\g_{123A}^{g_0^{-1}g_1}$}--(123-c);
\draw[->-=.7,thick,red](123-c)--(123-2)node[red,scale=.7,right]{$\g_{123B}^{g_0^{-1}g_1}$};
\node[red,scale=.7,right,xshift=-1]at (023-1) {$\g_{023B}^{e}$};
\draw[thick,dashed,red](023-1)--(012-2);
\draw[->-=2.5/4,thick,red] ($(023-1)!.7!(012-2)$)--($(023-1)!.701!(012-2)$);
\draw[thick,densely dotted,red](023-1)--(123-1);
\draw[->-=2.5/4,thick,red] ($(023-1)!.6!(123-1)$)--($(023-1)!.601!(123-1)$);
\draw[thick,densely dotted,red](123-1)--(013-1);
\draw[->-=2.5/4,thick,red] ($(123-1)!.65!(013-1)$)--($(123-1)!.651!(013-1)$);
\draw[thick,densely dotted,red](012-2)--(123-1);
\draw[->-=2.5/4,thick,red] ($(012-2)!.6!(123-1)$)--($(012-2)!.601!(123-1)$);
\draw[->-=.55,thick,densely dotted,red](023-1)--(013-1);
\draw[->-=.55,thick,densely dotted,red](012-2)--(013-1);
\fill [red] (012-1) circle (.55pt);
\fill [red] (012-2) circle (.55pt);
\fill [red] (013-1) circle (.55pt);
\fill [red] (013-2) circle (.55pt);
\fill [red] (123-1) circle (.55pt);
\fill [red] (123-2) circle (.55pt);
\fill [red] (023-1) circle (.55pt);
\fill [red] (023-2) circle (.55pt);
\draw[->-=.6,thick](1)--(2);
\draw[->-=.7,thick](1)--(3);
}
\underrightarrow{\ \ U(g_0)\ \ }
\tikzfig{scale=3}{
\coordinate (0) at (0,0);
\coordinate (1) at (.2,-.4);
\coordinate (2) at (1,0);
\coordinate (3) at (.5,.7);
\coordinate (center) at (.48,.15);
\coordinate (012-1) at (.45,-.4);\coordinate (012-2) at (.45,-.1);
\coordinate (012-c) at ($(012-1)!0.52!(012-2)$);
\coordinate (013-1) at (.3,.2);\coordinate (013-2) at (.1,.4);
\coordinate (013-c) at ($(013-1)!0.25!(013-2)$);
\coordinate (123-1) at (.65,.2);\coordinate (123-2) at (.9,.4);
\coordinate (123-c) at ($(123-1)!0.25!(123-2)$);
\coordinate (023-1) at (0.55,0.35);\coordinate (023-2) at (.7,.65);
\coordinate (023-c) at ($(023-1)!0.52!(023-2)$);
%
 %
\draw[->-=.6,thick](0)node[scale=.9,left]{$g_0$}--(1)node[scale=.9,below]{$g_1$};
\draw[->-=.6,thick,dashed](0)--(2)node[scale=.9,right]{$g_2$};
\draw[->-=.7,thick,red](023-2)node[red,scale=.7,right]{$\g_{023A}^{g_0}$}--(023-c);
\draw[thick,densely dotted,red](023-c)--(023-1);
\draw[->-=.6,thick](2)--(3);
\draw[->-=.7,thick,red](012-1)node[red,right,scale=.7]{$\g_{012A}^{g_0}$}--(012-c);
\draw[thick,densely dotted,red](012-c)--(012-2)node[red,right,scale=.7]{$\g_{012B}^{g_0}$};
\draw[->-=.6,thick](0)--(3)node[scale=.9,above]{$g_3$};
\draw[thick,densely dotted,red](013-1)node[red,scale=.7,below,xshift=-10]{$\g_{013A}^{g_0}$}--(013-c);
\draw[->-=.7,thick,red](013-c)--(013-2)node[red,scale=.7,left]{$\g_{013B}^{g_0}$};
\draw[thick,densely dotted,red](123-1)node[red,scale=.7,right,yshift=-3]{$\g_{123A}^{g_1}$}--(123-c);
\draw[->-=.7,thick,red](123-c)--(123-2)node[red,scale=.7,right]{$\g_{123B}^{g_1}$};
\node[red,scale=.7,right,xshift=-1] at (023-1) {$\g_{023B}^{g_0}$};
\draw[thick,dashed,red](023-1)--(012-2);
\draw[->-=2.5/4,thick,blue] ($(023-1)!.7!(012-2)$)--($(023-1)!.701!(012-2)$);
\draw[thick,densely dotted,red](023-1)--(123-1);
\draw[->-=2.5/4,thick,blue] ($(023-1)!.6!(123-1)$)--($(023-1)!.601!(123-1)$);
\draw[thick,densely dotted,red](123-1)--(013-1);
\draw[->-=2.5/4,thick,blue] ($(123-1)!.65!(013-1)$)--($(123-1)!.651!(013-1)$);
\draw[thick,densely dotted,red](012-2)--(123-1);
\draw[->-=2.5/4,thick,blue] ($(012-2)!.6!(123-1)$)--($(012-2)!.601!(123-1)$);
\draw[->-=.55,thick,densely dotted,red](023-1)--(013-1);
\draw[->-=.55,thick,densely dotted,red](012-2)--(013-1);
\fill [red] (012-1) circle (.55pt);
\fill [red] (012-2) circle (.55pt);
\fill [red] (013-1) circle (.55pt);
\fill [red] (013-2) circle (.55pt);
\fill [red] (123-1) circle (.55pt);
\fill [red] (123-2) circle (.55pt);
\fill [red] (023-1) circle (.55pt);
\fill [red] (023-2) circle (.55pt);
\draw[->-=.6,thick](1)--(2);
\draw[->-=.7,thick](1)--(3);
}
\end{align}
Note that the Majorana fermions $\g_{ijk,A}^\s$ and $\g_{ijk,B}^\s$ ($\s\neq g_i$) on the two sides of triangle $\langle ijk\rangle$ are always in vacuum pairings $\big(-i\g_{ijk,A}^\s\g_{ijk,B}^\s=1\big)$, independent of the $n_2$ configurations. So their pairing directions always follow the red arrow local Kasteleyn orientations in both figures of the above equation. For the two Majorana fermions $\g_{ijk,A}^{g_i}$ and $\g_{ijk,B}^{g_i}$ of the triangle $\langle ijk\rangle$, there are two possibilities. If $n_2(g_i,g_j,g_k)=0$ (there is no Kitaev chain going though this triangle), these two Majorana fermions are also in vacuum pairing, with direction indicated by the red arrow and projection operator
\begin{align}
P_{ijkA,ijkB}^{g_i,g_i} = U(g_0) P_{ijkA,ijkB}^{g_0^{-1}g_i,g_0^{-1}g_i} U(g_0)^{-1} = \frac{1}{2} \left(1-i\g_{ijk,A}^{g_i} \g_{ijk,B}^{g_i}\right).
\end{align}
On the other hand, if $n_2(g_i,g_j,g_k)=1$ (there is a Kitaev chain going though this triangle), we will pair the Majorana fermion inside the triangle with another one belonging to another triangle with also $n_2=1$ [for example, $\g_{123A}^{g_1}$ and $\g_{013A}^{g_0}$ are paired in \eq{3D:n2}]. Note that there are always even number of Majorana fermions in nontrivial pairing among the four Majorana fermions ($\g_{123A}^{g_1}$, $\g_{023B}^{g_0}$, $\g_{013A}^{g_0}$ and $\g_{012B}^{g_0}$) inside the tetrahedron $\langle 0123\rangle$, for we have $(\dd n_2)(g_0,g_1,g_2,g_3)=0$ (mod 2) from \eq{3D:dn2}. There are in total $(4\times 3)/2=6$ possible nontrivial pairings inside the tetrahedron $\langle 0123\rangle$ (i.e., the six links of the small red tetrahedron inside the big tetrahedron $\langle0123 \rangle$). The Majorana pairing projection operators of them are as follows:
\begin{align}\label{3D:t1}
P_{012B,013A}^{g_0,g_0} &= U(g_0) P_{012B,013A}^{e,e} U(g_0)^\dagger
= \frac{1}{2}\,\big(1-i\g_{012B}^{g_0}\g_{013A}^{g_0}\big),\\\label{3D:t2}
P_{023B,013A}^{g_0,g_0} &= U(g_0) P_{023B,013A}^{e,e} U(g_0)^\dagger
= \frac{1}{2}\,\big(1-i\g_{023B}^{g_0}\g_{013A}^{g_0}\big),\\\label{3D:t3}
P_{023B,012B}^{g_0,g_0} &= U(g_0) P_{023B,012B}^{e,e} U(g_0)^\dagger
= \frac{1}{2} \left[1-(-1)^{s_1(g_0)}i\g_{023B}^{g_0}\g_{012B}^{g_0}\right],\\\label{3D:t4}
P_{023B,123A}^{g_0,g_1} &= U(g_0) P_{023B,123A}^{e,g_0^{-1}g_1} U(g_0)^\dagger
= \frac{1}{2} \left[1-(-1)^{\om_2(g_0,g_0^{-1}g_1)}i\g_{023B}^{g_0}\g_{123A}^{g_1}\right],\\\label{3D:t5}
P_{012B,123A}^{g_0,g_1} &= U(g_0) P_{012B,123A}^{e,g_0^{-1}g_1} U(g_0)^\dagger
= \frac{1}{2} \left[1-(-1)^{\om_2(g_0,g_0^{-1}g_1)}i\g_{012B}^{g_0}\g_{123A}^{g_1}\right],\\\label{3D:t6}
P_{123A,013A}^{g_1,g_0} &= U(g_0) P_{123A,013A}^{g_0^{-1}g_1,e} U(g_0)^\dagger
= \frac{1}{2} \left[1-(-1)^{\om_2(g_0,g_0^{-1}g_1)+s_1(g_0)}i\g_{123A}^{g_1}\g_{013A}^{g_0}\right].
\end{align}
Among the six possible nontrivial pairings, only the last four may change their directions in the non-standard triangle. They are indicated by blue arrows in the right-hand-side figure of \eq{3D:g_symm}. This can be understood from the following facts from the symmetry transformation on projection operators \eq{proj}: The $(-1)^{\om_2}$ term appears in the projection operators when the pairing is between Majorana fermions with different group element labels [see Eqs.~(\ref{3D:t4}), (\ref{3D:t5}) and (\ref{3D:t6})]; And the $(-1)^{s_1}$ term appears when the pairing is between the same $A/B$ type Majorana fermions [see \eqs{3D:t3}{3D:t6}]. The first two pairings \eqs{3D:t1}{3D:t2} belongs to neither of the above two cases. So their pairing direction is the same as the red arrow Kasteleyn orientations even after $U(g_0)$-action.

\begin{figure}[ht]
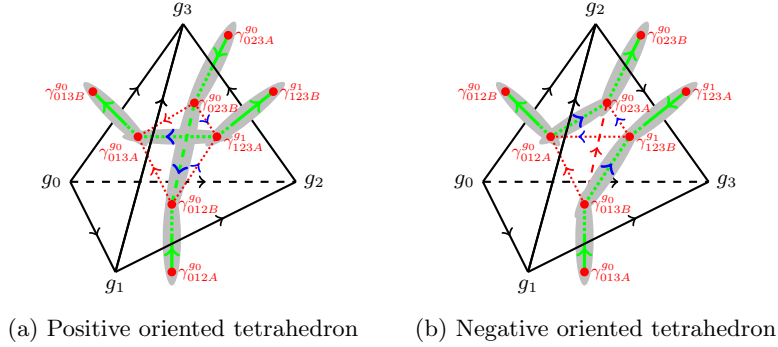

\centering
\begin{subfigure}[h]{.3\textwidth}
\centering
$
\tikzfig{scale=3}{
\coordinate (0) at (0,0);
\coordinate (1) at (.2,-.4);
\coordinate (2) at (1,0);
\coordinate (3) at (.5,.7);
\coordinate (center) at (.48,.15);
\coordinate (012-1) at (.45,-.4);\coordinate (012-2) at (.45,-.1);
\coordinate (012-c) at ($(012-1)!0.52!(012-2)$);
\coordinate (013-1) at (.3,.2);\coordinate (013-2) at (.1,.4);
\coordinate (013-c) at ($(013-1)!0.25!(013-2)$);
\coordinate (123-1) at (.65,.2);\coordinate (123-2) at (.9,.4);
\coordinate (123-c) at ($(123-1)!0.25!(123-2)$);
\coordinate (023-1) at (0.55,0.35);\coordinate (023-2) at (.7,.65);
\coordinate (023-c) at ($(023-1)!0.52!(023-2)$);
\Ellipse{gray!50}{012-2}{012-1}{1.4}{.27};
\Ellipse{gray!50}{023-2}{023-1}{1.4}{.25};
\Ellipse{gray!50}{123-1}{013-1}{1.4}{.2};
\draw[->-=.6,thick,dashed](0)--(2)node[scale=.9,right]{$g_2$};
\Ellipse{gray!50}{023-1}{012-2}{1.4}{.18};
\draw[->-=.6,thick](0)node[scale=.9,left]{$g_0$}--(1)node[scale=.9,below]{$g_1$};
\draw[->-=.6,thick](0)--(3)node[scale=.9,above]{$g_3$};
\draw[->-=.7,thick](1)--(3);
\Ellipse{gray!50}{013-2}{013-1}{1.4}{.27};
\draw[->-=.7,very thick,green](023-2)node[red,scale=.7,right]{$\g_{023A}^{g_0}$}--(023-c);
\draw[very thick,densely dotted,green](023-c)--(023-1);
\draw[->-=.6,thick](2)--(3);
\Ellipse{gray!50}{123-2}{123-1}{1.4}{.25};
\node[red,scale=.7,right,xshift=-1] at (023-1) {$\g_{023B}^{g_0}$};
\draw[->-=.7,very thick,green](012-1)node[red,right,scale=.7]{$\g_{012A}^{g_0}$}--(012-c);
\draw[very thick,densely dotted,green](012-c)--(012-2)node[red,right,scale=.7]{$\g_{012B}^{g_0}$};
\draw[->-=.6,thick](1)--(2);
\draw[very thick,densely dotted,green](013-1)node[red,scale=.7,below,xshift=-10]{$\g_{013A}^{g_0}$}--(013-c);
\draw[->-=.7,very thick,green](013-c)--(013-2)node[red,scale=.7,left]{$\g_{013B}^{g_0}$};
\draw[very thick,densely dotted,green](123-1)node[red,scale=.7,right,yshift=-3]{$\g_{123A}^{g_1}$}--(123-c);
\draw[->-=.7,very thick,green](123-c)--(123-2)node[red,scale=.7,right]{$\g_{123B}^{g_1}$};
\draw[very thick,dashed,green](023-1)--(012-2);
\draw[->-=2.5/4,very thick,blue] ($(023-1)!.7!(012-2)$)--($(023-1)!.701!(012-2)$);
\draw[thick,densely dotted,red](023-1)--(123-1);
\draw[->-=2.5/4,thick,blue] ($(023-1)!.6!(123-1)$)--($(023-1)!.601!(123-1)$);
\draw[very thick,densely dotted,green](123-1)--(013-1);
\draw[->-=2.5/4,very thick,blue] ($(123-1)!.65!(013-1)$)--($(123-1)!.651!(013-1)$);
\draw[thick,densely dotted,red](012-2)--(123-1);
\draw[->-=2.5/4,thick,blue] ($(012-2)!.6!(123-1)$)--($(012-2)!.601!(123-1)$);
\draw[->-=.55,thick,densely dotted,red](023-1)--(013-1);
\draw[->-=.55,thick,densely dotted,red](012-2)--(013-1);
\fill [red] (012-1) circle (.55pt);
\fill [red] (012-2) circle (.55pt);
\fill [red] (013-1) circle (.55pt);
\fill [red] (013-2) circle (.55pt);
\fill [red] (123-1) circle (.55pt);
\fill [red] (123-2) circle (.55pt);
\fill [red] (023-1) circle (.55pt);
\fill [red] (023-2) circle (.55pt);
\draw[->-=.7,thick](1)--(3);
}
$
\caption{Positive oriented tetrahedron}
\label{fig:3Dresolve_1}
\end{subfigure}
\begin{subfigure}[h]{.3\textwidth}
\centering
$
\tikzfig{scale=3}{
\coordinate (0) at (0,0);
\coordinate (1) at (.2,-.4);
\coordinate (2) at (1,0);
\coordinate (3) at (.5,.7);
\coordinate (center) at (.48,.15);
\coordinate (012-1) at (.45,-.4);\coordinate (012-2) at (.45,-.1);
\coordinate (012-c) at ($(012-1)!0.52!(012-2)$);
\coordinate (013-1) at (.3,.2);\coordinate (013-2) at (.1,.4);
\coordinate (013-c) at ($(013-1)!0.25!(013-2)$);
\coordinate (123-1) at (.65,.2);\coordinate (123-2) at (.9,.4);
\coordinate (123-c) at ($(123-1)!0.25!(123-2)$);
\coordinate (023-1) at (0.55,0.35);\coordinate (023-2) at (.7,.65);
\coordinate (023-c) at ($(023-1)!0.52!(023-2)$);
\Ellipse{gray!50}{012-2}{012-1}{1.4}{.27};
\Ellipse{gray!50}{023-2}{023-1}{1.4}{.25};
\Ellipse{gray!50}{023-1}{013-1}{1.4}{.25};
\draw[->-=.6,thick,dashed](0)--(2)node[scale=.9,right]{$g_3$};
\draw[thick,dashed,red](023-1)--(012-2);
\draw[-<-=2.5/4,thick,red] ($(023-1)!.5!(012-2)$)--($(023-1)!.501!(012-2)$);
\Ellipse{gray!50}{123-1}{012-2}{1.4}{.23};
\draw[->-=.6,thick](0)node[scale=.9,left]{$g_0$}--(1)node[scale=.9,below]{$g_1$};
\draw[->-=.6,thick](0)--(3)node[scale=.9,above]{$g_2$};
\draw[->-=.7,thick](1)--(3);
\Ellipse{gray!50}{013-2}{013-1}{1.4}{.27};
\draw[-<-=.7,very thick,green](023-2)node[red,scale=.7,right]{$\g_{023B}^{g_0}$}--(023-c);
\draw[very thick,densely dotted,green](023-c)--(023-1);
\draw[-<-=.6,thick](2)--(3);
\Ellipse{gray!50}{123-2}{123-1}{1.4}{.25};
\node[red,scale=.7,right,xshift=-1] at (023-1) {$\g_{023A}^{g_0}$};
\draw[->-=.7,very thick,green](012-1)node[red,right,scale=.7]{$\g_{013A}^{g_0}$}--(012-c);
\draw[very thick,densely dotted,green](012-c)--(012-2)node[red,right,scale=.7]{$\g_{013B}^{g_0}$};
\draw[->-=.6,thick](1)--(2);
\draw[very thick,densely dotted,green](013-1)node[red,scale=.7,below,xshift=-10]{$\g_{012A}^{g_0}$}--(013-c);
\draw[->-=.7,very thick,green](013-c)--(013-2)node[red,scale=.7,left]{$\g_{012B}^{g_0}$};
\draw[very thick,densely dotted,green](123-1)node[red,scale=.7,right,yshift=-3]{$\g_{123B}^{g_1}$}--(123-c);
\draw[-<-=.7,very thick,green](123-c)--(123-2)node[red,scale=.7,right]{$\g_{123A}^{g_1}$};
%
\draw[thick,densely dotted,red](023-1)--(123-1);
\draw[-<-=2.5/4,thick,blue] ($(023-1)!.6!(123-1)$)--($(023-1)!.601!(123-1)$);
\draw[thick,densely dotted,red](123-1)--(013-1);
\draw[->-=2.5/4,thick,blue] ($(123-1)!.65!(013-1)$)--($(123-1)!.651!(013-1)$);
\draw[very thick,densely dotted,green](012-2)--(123-1);
\draw[->-=2.5/4,very thick,blue] ($(012-2)!.6!(123-1)$)--($(012-2)!.601!(123-1)$);
\draw[very thick,densely dotted,green](023-1)--(013-1);
\draw[-<-=2.5/4,very thick,blue] ($(023-1)!.55!(013-1)$)--($(023-1)!.551!(013-1)$);
\draw[->-=.55,thick,densely dotted,red](012-2)--(013-1);
\fill [red] (012-1) circle (.55pt);
\fill [red] (012-2) circle (.55pt);
\fill [red] (013-1) circle (.55pt);
\fill [red] (013-2) circle (.55pt);
\fill [red] (123-1) circle (.55pt);
\fill [red] (123-2) circle (.55pt);
\fill [red] (023-1) circle (.55pt);
\fill [red] (023-2) circle (.55pt);
\draw[->-=.7,thick](1)--(3);
}
$
\caption{Negative oriented tetrahedron}
\label{fig:3Dresolve_2}
\end{subfigure}
\caption{Resolvation of four strings of Kitaev chains meeting at one tetrahedron. If four strings (green lines) meet at one tetrahedron, we should pair the Majorana fermions $\g_{123A/B}^{g_1}$ to $\g_{013A/B}^{g_0}$, and $\g_{023B/A}^{g_0}$ to $\g_{012B/A}^{g_0}$ (gray ellipses) for positive/negative oriented tetrahedron. We omit the labels of Majorana fermions ($\g_{ijk,A/B}^{\s}$ with $\s\neq g_i$) that are in vacuum pairings in the figure.}
\label{fig:3Dresolve}
\end{figure}

There is another subtlety when $n_2=1$ for all the four triangles of a tetrahedron. There are four strings of Kitaev chains meeting at the tetrahedron $\langle 0123\rangle$. In this case, we should resolve the crossing point of the four strings. We use the convention that the Majorana fermions $\g_{123A}^{g_1}$ and $\g_{013A}^{g_0}$ are paired, and $\g_{023B}^{g_0}$ and $\g_{012B}^{g_0}$ are paired (see \fig{fig:3Dresolve_1}). Of course, all other Majorana fermions $\g_{ijk,A/B}^\s$ with $\s\neq g_i$ are still in vacuum pairings. This resolvation convention is the same as \Ref{WangGu2017}.

\paragraph{Majorana fermion parity.}
\label{sec:3D:Maj:Pf}

Since the symmetry action may change the pairing directions inside a tetrahedron, the Majorana fermion parity of this tetrahedron may also be changed. We can calculate the fermion parity difference between the standard and non-standard tetrahedra by counting the number of pairing arrows that are reversed by $U(g_0)$-action. It of course depends on the $n_2$ configurations. We can use, for example, $n_2(g_0,g_1,g_2) n_2(g_1,g_2,g_3) = 0,1$ to indicate whether $\g_{012B}^{g_1}$ and $\g_{123A}^{g_1}$ are paired or not. So the Majorana fermion parity change inside the triangle is in general given by
\begin{align}\nonumber\label{3D:Pf_M}
\Delta P_f^\g (0123) &= (-1)^{\om_2(g_0,g_0^{-1}g_1) n_2(g_1,g_2,g_3) + s_1(g_0) [ n_2(g_0,g_2,g_3) n_2(g_0,g_1,g_2) + n_2(g_0,g_1,g_3) n_2(g_1,g_2,g_3) ]}\\
&= (-1)^{\left[\om_2\smile n_2 + s_1\smile (n_2\smile_1 n_2)\right](g_0,g_0^{-1}g_1,g_1^{-1}g_2,g_2^{-1}g_3)},
\end{align}
where we have used the higher cup product definition $(n_2\Smile_1 n_2)(0123)=n_2(023)n_2(012)+n_2(013)n_2(123)$. The above equation is a summary of the phase factors appearing in Eqs.~(\ref{3D:t1})-(\ref{3D:t6}). The first term $(-1)^{\om_2}$ appears iff $n_2(g_1,g_2,g_3)=1$. And the second term $(-1)^{s_1(g_0)}$ in the first line of \eq{3D:Pf_M} appears when $n_2(g_0,g_2,g_3) n_2(g_0,g_1,g_2)=1$ or $n_2(g_0,g_1,g_3) n_2(g_1,g_2,g_3)=1$. Note that different from the 2D case, we can not find a single Majorana fermion involving all the Majorana fermion parity configurations.


The above discussions of Majorana fermion pairings also apply to the negative orientated tetrahedra. The red arrow local Kasteleyn orientations inside a tetrahedron is given in \fig{fig:3DKasteleyn2}. The symmetric Majorana fermion pairing in a non-standard tetrahedron is also obtained by a $U(g_0)$-action from the standard tetrahedron. We also have Majorana pairing projection operators similar to Eqs.~(\ref{3D:t1})-(\ref{3D:t6}) for negative orientated tetrahedron. For $n_2$ configurations with four Kitaev chains meeting at one negative oriented tetrahedron, we would use the resolvation convention shown in \fig{fig:3Dresolve_2}. And the expression of Majorana fermion parity changes \eq{3D:Pf_M} is valid for both positive and negative oriented tetrahedra.

To sum up, although there are $|G_b|$ species of Majorana fermions, we decorate exactly one Kitaev chain to each intersection line of symmetry domain walls specified by $n_2$ configurations. The decoration is compatible with symmetry actions. The Majorana fermion parity of a tetrahedron is changed according to \eq{3D:Pf_M} compared to the local Kasteleyn oriented pairings.

\subsubsection{Complex fermion decoration}
\label{sec:3D:c}

The rules of  complex fermion decoration are much simpler than that of the Majorana fermions. The decoration is specified by a $\Z_2$-valued 3-cochain $n_3\in C^1(G_b,\Z_2)$. If $n_3(g_i,g_j,g_k,g_l)=0$, all the modes of complex fermions $c_{ijkl}^\s$ ($\s\in G_b$) at the center of tetrahedron $\langle ijkl\rangle$ ($i<j<k<l$) are unoccupied. On the other hand, if $n_3(g_i,g_j,g_k,g_l)=1$, exactly one complex fermion mode $c_{ijkl}^{g_i}$ will be decorated at the center of tetrahedron $\langle ijkl\rangle$ (see \fig{fig:3D:dof_a}). All other complex fermions $c_{ijkl}^\s$ ($\s\neq g_i$) are still in vacuum states.

It is simple to check that the complex fermion decoration is $G_b$-symmetric. Under a $U(g)$-action, the bosonic vertex labels $\{g_i\}$ becomes $\{gg_i\}$. And we should decorate $c_{ijkl}^{gg_i}$ to the tetrahedron $\langle ijkl\rangle$ if $n_3(gg_i,gg_j,gg_k,gg_l) = n_3(g_i,g_j,g_k,g_l) = 1$. This is exactly the complex fermion $c_{ijkl}^{g_i}$ by a $U(g)$-action.

\subsection{F moves}

The 3D Pachner move for different triangulations of 3D spacial manifold induces FSLU transformation of the FSPT wave functions on the lattices. Since FSPT state is invertible, we only need to consider one of the many Pachner moves. Other Pachner moves can be derived from this one using the invertibility and unitarity of the move. We can first define the \emph{standard} $F$ move with $g_0=e$, then other non-standard ones can be obtained by simply a $U(g_0)$-action. The standard $F$ move is given by
\begin{align}
\label{3D:Fmove}
\Psi\left(
\tikzfig{scale=2}{
\coordinate (0) at (.55,-.7);
\coordinate (1) at (0,0);
\coordinate (2) at (.3,-.15);
\coordinate (3) at (1,0);
\coordinate (4) at (.55,.7);
\draw[-<-=.7,thick](0)node[scale=.8,below,xshift=4]{$g_0^{-1}g_1$}--(1)node[scale=.8,left]{$e$};
\draw[->-=.65,thick](0)--(2)node[scale=.8,left,below,xshift=-25,yshift=5]{$g_0^{-1}g_2$};
\draw[->-=.6,thick](0)--(3)node[scale=.8,right]{$g_0^{-1}g_4$};
\draw[->-=.6,thick](1)--(2);
\draw[->-=.65,thick,dashed](1)--(3);
\draw[->-=.4,thick](1)--(4)node[scale=.8,above,xshift=5]{$g_0^{-1}g_3$};
\draw[->-=.5,thick](2)--(3);
\draw[->-=.35,thick](2)--(4);
\draw[-<-=.4,thick](3)--(4);
\coordinate (center1234) at (.48,.15);
\coordinate (124) at (.1,.6);
\coordinate (234) at (.9,.6);
\coordinate (134) at (.4,.8);
\coordinate (center0123) at (.48,-.15);
\coordinate (012) at (0,-.5);
\coordinate (023) at (.9,-.6);
\coordinate (013) at (.4,-.8);
\node [scale=.5,above] at (124) {$\langle 023\rangle$};
\node [scale=.5,above] at (234) {$\langle 234\rangle$};
\node [scale=.5,above,xshift=-8] at (134) {$\langle 034\rangle$};
\node [scale=.5,below] at (012) {$\langle 012\rangle$};
\node [scale=.5,below] at (023) {$\langle 124\rangle$};
\node [scale=.5,left] at (013) {$\langle 014\rangle$};
\node [scale=.5,right] at ($(center1234)!.2!(center0123)$) {$\langle 024\rangle$};
\draw[thick,red,densely dotted](center1234)--(124);
\draw[->-=.7,thick,red]($(center1234)!.35!(124)$)--(124);
\draw[very thick,green,densely dotted](center1234)--(234);
\draw[-<-=.75,very thick,green]($(center1234)!.35!(234)$)--(234);
\draw[thick,red,densely dotted](center1234)--(134);
\draw[-<-=.7,thick,red](134)--($(134)!.4!(center1234)$);
\draw[very thick,green,densely dotted](012)
--(center0123);
\draw[-<-=.6,very thick,green](012)--($(center0123)!.4!(012)$);
\draw[thick,red,densely dotted](023)
--(center0123);
\draw[->-=.4,thick,red](023)--($(center0123)!.35!(023)$);
\draw[thick,red,densely dotted](center0123)--(013);
\draw[-<-=.75,thick,red]($(center0123)!.62!(013)$)--(013);
\draw[->-=.8,very thick,green,densely dotted](center0123)
--(center1234);
}
\right)\quad = \quad
F(e,g_0^{-1}g_1,g_0^{-1}g_2,g_0^{-1}g_3,g_0^{-1}g_4)
\quad
\Psi\left(
\tikzfig{scale=2}{
\coordinate (0) at (.55,-.7);
\coordinate (1) at (0,0);
\coordinate (2) at (.3,-.15);
\coordinate (3) at (1,0);
\coordinate (4) at (.55,.7);
\draw[-<-=.6,thick](0)node[scale=.8,below,xshift=4]{$g_0^{-1}g_1$}--(1)node[scale=.8,left]{$e$};
\draw[->-=.7,thick](0)--(2)node[scale=.8,below,xshift=-28,yshift=8]{$g_0^{-1}g_2$};
\draw[->-=.7,thick](0)--(3);
\draw[->-=.6,thick](1)--(2);
\draw[->-=.75,thick,dashed](1)--(3);
\draw[->-=.5,thick](1)--(4);
\draw[->-=.7,thick](2)--(3)node[scale=.8,right]{$g_0^{-1}g_4$};
\draw[->-=.5,thick](2)--(4)node[scale=.8,above,xshift=4]{$g_0^{-1}g_3$};
\draw[-<-=.4,thick](3)--(4);
\draw[->-=.3,thick,dashed](0)--(4);
\coordinate (center0124) at (.25,-.06);
\coordinate (124) at (.1,.6);
\coordinate (234) at (.8,.65);
\coordinate (134) at (.4,.8);
\coordinate (center0234) at (.65,-.05);
\coordinate (012) at (0,-.5);
\coordinate (023) at (.8,-.65);
\coordinate (013) at (.4,-.8);
\coordinate (center0134) at (.5,.1);
\node [scale=.5,above] at (124) {$\langle 023\rangle$};
\node [scale=.5,above,xshift=8] at (234) {$\langle 234\rangle$};
\node [scale=.5,above,xshift=-8] at (134) {$\langle 034\rangle$};
\node [scale=.5,below] at (012) {$\langle 012\rangle$};
\node [scale=.5,below] at (023) {$\langle 124\rangle$};
\node [scale=.5,left] at (013) {$\langle 014\rangle$};
\node [scale=.5,below] at ($(center0124)!.5!(center0234)$) {$\langle 123\rangle$};
\draw[thick,red,densely dotted](center0124)--(124);
\draw[->-=.7,thick,red]($(center0124)!.25!(124)$)--(124);
\draw[very thick,green,densely dotted](center0234)--(234);
\draw[-<-=.75,very thick,green]($(center0234)!.35!(234)$)--(234);
\draw[->-=.8,thick,red,densely dotted](center0134)--(134);
\draw[thick,red]($(134)!.35!(center0134)$)--(134);
\draw[very thick,green,densely dotted](012)
--(center0124);
\draw[-<-=.6,very thick,green](012)--($(center0124)!.24!(012)$);
\draw[thick,red,densely dotted](023)
--(center0234);
\draw[->-=.4,thick,red](023)--($(center0234)!.35!(023)$);
\draw[thick,red,densely dotted](center0134)--(013);
\draw[-<-=.75,thick,red]($(center0134)!.73!(013)$)--(013);
\draw[->-=.7,very thick,green,densely dotted](center0234)--(center0124);
\draw[-<-=.7,thick,red,densely dotted](center0124)
--(center0134);
\draw[->-=.6,thick,red,densely dotted](center0234)
--(center0134);
%
\node [scale=.5,above,xshift=-.1cm] at ($(center0124)!.5!(center0134)$) {$\langle 013\rangle$};
\node [scale=.5,above,xshift=.2cm] at ($(center0234)!.5!(center0134)$) {$\langle 134\rangle$};
}
\right)
\end{align}
on the (black) triangulation lattice, and
\begin{align}
\label{3D:Fmove2}
\Psi\left(
\tikzfig{scale=2}{
\coordinate (124) at (-.4,.7);
\coordinate (234) at (.4,.7);
\coordinate (134) at (0,.8);
\coordinate (012) at (-.4,-.7);
\coordinate (023) at (.4,-.7);
\coordinate (013) at (0,-.8);
\coordinate (A1) at (.2,.3);
\coordinate (A2) at (0,.45);
\coordinate (A3) at (-.2,.3);
\coordinate (A4) at (0,.15);
\coordinate (B0) at (0,-.15);
\coordinate (B1) at (.2,-.3);
\coordinate (B2) at (0,-.45);
\coordinate (B3) at (-.2,-.3);
\Ellipse{gray!50}{124}{A3}{1.2}{.2};
\Ellipse{gray!50}{134}{A2}{1.2*1.1}{.2*1.1};
\Ellipse{gray!50}{013}{B2}{1.2*1.1}{.2*1.1};
\Ellipse{gray!50}{023}{B1}{1.2}{.2}; 
\Ellipse{gray!75}{A4}{A1}{1.2*1.2}{.2*1.4};
\Ellipse{gray!75}{B0}{B3}{1.2*1.2}{.2*1.4};
\Ellipse{gray!75}{234}{A1}{1.2}{.2};
\Ellipse{gray!75}{B0}{A4}{1.2*1.2}{.2*1.4};
\Ellipse{gray!75}{B3}{012}{1.2}{.2};
\draw[-<-=.55,thick,red](B1)--(B2);
\draw[->-=.7,thick,red](A3)--(124);
\draw[-<-=.7,very thick,green](A1)--(234);
\draw[-<-=.7,thick,red](134)
--(A2);
\draw[-<-=.7,very thick,green](012)
--(B3);
\draw[->-=.7,thick,red](023)
--(B1);
\draw[-<-=.7,thick,red](B2)--(013);
\draw[->-=.7,very thick,green](B0)
--(A4);
\draw[-<-=.55,thick,red](A2)--(A1);
\draw[-<-=.55,thick,red](A2)--(A3);
\draw[->-=.7,thick,red](A1)--(A3);
\draw[->-=.6,very thick,green](A4)--(A1);
\draw[->-=.6,thick,red](A4)--(A3);
\draw[-<-=.8,thick,red,densely dotted](A2)--(A4);
\draw[->-=.6,thick,red](B1)--(B0);
\draw[->-=.6,very thick,green](B3)--(B0);
\draw[->-=.7,thick,red](B1)--(B3);
\draw[-<-=.55,thick,red](B3)--(B2);
\draw[-<-=.8,thick,red,densely dotted](B0)--(B2);
\fill [red] (124) circle (.7pt);
\fill [red] (234) circle (.7pt);
\fill [red] (134) circle (.7pt);
\fill [red] (012) circle (.7pt);
\fill [red] (023) circle (.7pt);
\fill [red] (013) circle (.7pt);
\fill [red] (A1) circle (.7pt);
\fill [red] (A2) circle (.7pt);
\fill [red] (A3) circle (.7pt);
\fill [red] (A4) circle (.7pt);
\fill [red] (B0) circle (.7pt);
\fill [red] (B1) circle (.7pt);
\fill [red] (B2) circle (.7pt);
\fill [red] (B3) circle (.7pt);
\node [scale=.5,above] at (124) {$023B$};
\node [scale=.5,above] at (234) {$234A$};
\node [scale=.5,above] at (134) {$034B$};
\node [scale=.5,below] at (012) {$012B$};
\node [scale=.5,below] at (023) {$124A$};
\node [scale=.5,below] at (013) {$014A$};
\node [scale=.5,right] at (A1){$234B$};
\node [scale=.5,xshift=.45cm,yshift=.15cm] at (A2){$034A$};
\node [scale=.5,left] at (A3){$023A$};
\node [scale=.5,below left] at (A4){$024B$};
\node [scale=.5,xshift=.45cm,yshift=.15cm] at (B0){$024A$};
\node [scale=.5,right] at (B1){$124B$};
\node [scale=.5,below left] at (B2){$014B$};
\node [scale=.5,left] at (B3){$012A$};
}
\right)\quad = \quad
F(e,g_0^{-1}g_1,g_0^{-1}g_2,g_0^{-1}g_3,g_0^{-1}g_4)
\quad
\Psi\left(
\tikzfig{scale=2}{
\coordinate (124) at (-.6,.7);
\coordinate (234) at (.6,.7);
\coordinate (134) at (0,.8);
\coordinate (012) at (-.6,-.7);
\coordinate (023) at (.6,-.7);
\coordinate (013) at (0,-.8);
\coordinate (A0) at (-.5,0);
\coordinate (A1) at (-.18,-.2);
\coordinate (A2) at (-.255,.05);
\coordinate (A4) at (-.42,-.23);
\coordinate (B0) at (.5,0);
\coordinate (B2) at (.255,.05);
\coordinate (B3) at (.18,-.2);
\coordinate (B4) at (.42,-.23);
\coordinate (C0) at (0,.45);
\coordinate (C1) at (.18,.3);
\coordinate (C3) at (-.18,.3);
\coordinate (C4) at (0,.15);
\Ellipse{gray!50}{124}{A0}{1.1}{.15};
\Ellipse{gray!50}{134}{C0}{1.2*1.1}{.2*1.1};
\Ellipse{gray!50}{013}{C4}{1.1}{.1};
\Ellipse{gray!50}{023}{B4}{1.2}{.2};
\Ellipse{gray!50}{A2}{C3}{1.2*1.2}{.2*1.4};
\Ellipse{gray!50}{B2}{C1}{1.2*1.2}{.2*1.5};
\Ellipse{gray!75}{A1}{A4}{1.2*1.2}{.2*1.9};
\Ellipse{gray!75}{B0}{B3}{1.2*1.}{.2*1.3};
\Ellipse{gray!75}{234}{B0}{1.1}{.15};
\Ellipse{gray!75}{012}{A4}{1.2}{.2};
\Ellipse{gray!75}{A1}{B3}{1.2*1.}{.2*1.3};
\draw[-<-=.8,thick,red,densely dotted](B2)--(B4);
\draw[->-=.7,thick,red](A0)--(124);
\draw[-<-=.7,very thick,green](B0)--(234);
\draw[thick,red](C0)--(134);
\draw[->-=.7,thick,red]($(C0)!.7!(134)$)--($(C0)!.701!(134)$);
\draw[-<-=.7,very thick,green](012)
--(A4);
\draw[->-=.7,thick,red](023)
--(B4);
\draw[thick,red](C4)--($(C4)!.28!(013)$);
\draw[-<-=.7,thick,red]($(C4)!.47!(013)$)--(013);
\draw[-<-=.7,thick,red](A2)--(C3);
\draw[->-=.7,very thick,green](B3)--(A1);
\draw[thick,red](B2)--(C1);
\draw[->-=.7,thick,red]($(B2)!.7!(C1)$)--($(B2)!.701!(C1)$);
\draw[-<-=.55,thick,red](A0)--(A2);
\draw[->-=.6,thick,red](A4)--(A0);
\draw[->-=.7,thick,red](A1)
--(A0);
\draw[-<-=.55,thick,red](A1)--(A2);
\draw[->-=.6,very thick,green](A1)--(A4);
\draw[-<-=.8,thick,red,densely dotted](A4)--(A2);
\draw[thick,red](B2)--(B0);
\draw[-<-=.6,thick,blue]($(B2)!.6!(B0)$)--($(B2)!.601!(B0)$);
\draw[very thick,green](B0)--(B3);
\draw[->-=.7,very thick,blue]($(B0)!.7!(B3)$)--($(B0)!.701!(B3)$);
\draw[->-=.6,thick,red](B4)
--(B0);
\draw[->-=.6,thick,blue]($(B4)!.6!(B0)$)--($(B4)!.601!(B0)$);
\draw[-<-=.55,thick,red](B2)
--(B3);
\draw[-<-=.55,thick,blue]($(B2)!.55!(B3)$)--($(B2)!.551!(B3)$);
\draw[->-=.6,thick,red](B4)--(B3);
\draw[-<-=.55,thick,red](C0)--(C1);
\draw[->-=.6,thick,red](C3)--(C0);
\draw[-<-=.7,thick,red](C3)
--(C1);
\draw[-<-=.55,thick,red](C3)--(C4);
\draw[->-=.6,thick,red](C4)--(C1);
\draw[-<-=.8,thick,red,densely dotted](C0)--(C4);
\fill [red] (124) circle (.7pt);
\fill [red] (234) circle (.7pt);
\fill [red] (134) circle (.7pt);
\fill [red] (012) circle (.7pt);
\fill [red] (023) circle (.7pt);
\fill [red] (013) circle (.7pt);
\fill [red] (A0) circle (.7pt);
\fill [red] (A1) circle (.7pt);
\fill [red] (A2) circle (.7pt);
\fill [red] (A4) circle (.7pt);
\fill [red] (B0) circle (.7pt);
\fill [red] (B4) circle (.7pt);
\fill [red] (B2) circle (.7pt);
\fill [red] (B3) circle (.7pt);
\fill [red] (C0) circle (.7pt);
\fill [red] (C1) circle (.7pt);
\fill [red] (C3) circle (.7pt);
\fill [red] (C4) circle (.7pt);
\node [scale=.5,above] at (124) {$023B$};
\node [scale=.5,above] at (234) {$234A$};
\node [scale=.5,above] at (134) {$034B$};
\node [scale=.5,below] at (012) {$012B$};
\node [scale=.5,below] at (023) {$124A$};
\node [scale=.5,below] at (013) {$014A$};
\node [scale=.5,left] at (A0){$023A$};
\node [scale=.5,below] at (A1){$123B$};
\node [scale=.5,xshift=-.4cm,yshift=.2cm] at (A2){$013B$};
\node [scale=.5,left] at (A4){$012A$};
\node [scale=.5,right] at (B0){$234B$};
\node [scale=.5,xshift=.4cm,yshift=.2cm] at (B2){$134A$};
\node [scale=.5,below] at (B3){$123A$};
\node [scale=.5,right] at (B4){$124B$};
\node [scale=.5,xshift=.45cm,yshift=.15cm] at (C0){$034A$};
\node [scale=.5,right] at (C1){$134B$};
\node [scale=.5,left] at (C3){$013A$};
\node [scale=.5,below] at (C4){$014B$};
}
\right)
\end{align}
on the (red) resolved dual lattice. This Pachner move involves five vertices from $\langle 0\rangle$ to $\langle 4\rangle$, with group element labels $e,g_0^{-1}g_1,g_0^{-1}g_2,g_0^{-1}g_3$ and $g_0^{-1}g_4$ in $G_b$. There are three tetrahedra ($\langle 0123\rangle$, $\langle 0134\rangle$ and $\langle 1234\rangle$) in the right-hand-side figure, and two tetrahedra ($\langle 0124\rangle$ and $\langle 0234\rangle$) on the left-hand-side figure of \eq{3D:Fmove}. The Pachner move removes the link $\langle 13\rangle$ on the right. Note that we omit the $p+ip$ layer in the above figures. And for each tetrahedron, the meeting point of the four red/green strings in \eq{3D:Fmove} should be resolved to a small tetrahedron with local Kasteleyn orientations in \eq{3D:Fmove2} (using the convention \fig{fig:3DKasteleyn2}). The blue arrows of the red links inside the only non-standard tetrahedron $\langle 1234\rangle$ [see the right-hand-side figure of \eq{3D:Fmove2}] mean that their directions may be changed under $U(g_0^{-1}g_1)$-action compared to the red arrow local Kasteleyn orientations [see \eq{3D:g_symm}]. Green line represents the decorated Kitaev chain specified by the $n_2$ data. We also omit the blue ball symbols for the decorated complex fermions at the center of the tetrahedra in the figures.

The explicit expression of the standard FSLU $F$ operator for the above Pachner move is
\begin{align}\label{3D:F}
&\quad F(e,\bar 01,\bar 02,\bar 03,\bar 04) \\\nonumber
&= \nu_4(\bar 01,\bar 12,\bar 23,\bar 34)
\big(c^{e\dagger}_{0124}\big)^{ n_3(0124)} \big(c^{e\dagger}_{0234}\big)^{ n_3(0234)} \big(c^{e}_{0123}\big)^{n_3(0123)} \big(c^{e}_{0134}\big)^{n_3(0134)} \left(c^{g_0^{-1}g_1}_{1234}\right)^{n_3(1234)} X_{01234}[n_2].
\end{align}
We used the abbreviation $\bar ij$ for $g_i^{-1}g_j$ in the arguments of $F$ and $\nu_4$. Different from $\nu_4$, the 3-cochain $n_3$ is symmetric under $G_b$-action. So there is no difference between homogeneous and inhomogeneous $n_3$. We use $n_3(ijkl)$ to represents $n_3(g_i,g_j,g_k,g_l)=n_3(g_i^{-1}g_j,g_j^{-1}g_k,g_k^{-1}g_l)$ for short. In the following, we would explain the terms in $F$ operator \eq{3D:F} one by one. There is some subtlety about the Majorana fermion parity changes for the $X$ operator. And we will give the explicit expression for $X$ in the next subsection.

The $U(1)$ phase factor $\nu_4(\bar 01,\bar 12,\bar 23,\bar 34) = \nu_4(g_0^{-1}g_1,g_1^{-1}g_2,g_2^{-1}g_3,g_3^{-1}g_4)$ in the front of $F$ symbol is a inhomogeneous 4-cochain depending on four group elements of $G_b$. By definition, the inhomogeneous cochain is related to the homogeneous one by
\begin{align}
\nu_4(g_0^{-1}g_1,g_1^{-1}g_2,g_2^{-1}g_3,g_3^{-1}g_4) = \nu_4(e,g_0^{-1}g_1,g_0^{-1}g_2,g_0^{-1}g_3,g_0^{-1}g_4),
\end{align}
with the first argument of the homogeneous $\nu_4$ to be the identity element $e\in G_b$. As will be discussed below, we can relate $\nu_4(e,g_0^{-1}g_1,g_0^{-1}g_2,g_0^{-1}g_3,g_0^{-1}g_4)$ and $\nu_4(g_0,g_1,g_2,g_3,g_4)$ from the symmetry conditions of $F$. Different from the special case $G_f=\Z_f\times G_b$, they do not equal to each other in general.

The complex fermion term of the form $c^\dagger c^\dagger c c c$ annihilate three complex fermions at the three tetrahedra of the right-hand-side figure and create two on the left-hand-side figure of \eq{3D:Fmove}. Following section~\ref{sec:3D:c}, the tetrahedron $\langle ijkl\rangle$ is decorated by complex fermion $c_{ijkl}^{g_i}$. So in the standard $F$ move, only the last fermion $c_{1234}^{g_0^{-1}g_1}$ has group element label $g_0^{-1}g_1$, and all other four fermions have group element label $e$. We note that, different from the special case $G_f=\Zf\times G_b$ where we have $\dd n_3=n_2\smile n_2$ \cite{WangGu2017}, the complex fermion number $n_3$ has more complicated relation with the $n_2$ data in general.


\subsubsection{Majorana fermion parity and $X$ operator}

In this subsection, we will give explicit expression for the $X$ operator in the standard $F$ move \eq{3D:F}, which is related to the Kitaev chain decorations.

As a consequence of the local Kasteleyn orientation of the red lattice, the Majorana fermion parities for the two Kitaev chain decorated states on the two sides of \eq{3D:Fmove} may be different. It is shown in \Ref{WangGu2017} that, if the Majorana fermions are paired according to the local Kasteleyn orientations, the Majorana fermion parity difference under the $F$ move is
\begin{align}\label{3D:n22}
\Delta P_f^\g (F)\big |_{s_1=\om_2=0} = (-1)^{(n_2\smile n_2)(01234)}.
\end{align}
So among all the Kitaev chain decoration configurations of the $F$ move, only the ones with $n_2(012)=n_2(234)=1$ would change the Majorana fermion parity [see the green lines in \eq{3D:Fmove} for an example]. However, for generic symmetry group with nonzero $s_1$ and $\om_2$, the Majorana fermions are paired according to the rules designed in section~\ref{sec:3D:Maj:dir}. As shown by blue arrows in \eq{3D:g_symm}, the pairing directions inside the non-standard tetrahedron may be changed by the symmetry action compared to the local Kasteleyn orientations.

In the standard $F$ move \eq{3D:Fmove}, there are in total five relevant tetrahedra. Four of these tetrahedra are standard, with group element label of the first vertex $e\in G_b$. However, the tetrahedron $\langle 1234\rangle$ on the right-hand-side of \eq{3D:Fmove} is non-standard and has first vertex label $g_0^{-1}g_1$. Therefore, only inside the tetrahedron $\langle 1234\rangle$ of the standard $F$ move, the pairing directions of the Majorana fermions may be changed. The pairings are given by the projection operators Eqs.~(\ref{3D:t1})-(\ref{3D:t6}), with the replacement $(g_0,g_1,g_2,g_3)\rightarrow (g_0^{-1}g_1,g_0^{-1}g_2,g_0^{-1}g_3,g_0^{-1}g_4)$. Furthermore, as discussed also in section~\ref{sec:3D:Maj:Pf}, the Majorana fermion parity change for a non-standard tetrahedron $\langle 0123\rangle$ compared to the local Kasteleyn orientations is given by \eq{3D:Pf_M}. For the tetrahedron $\langle 1234\rangle$ of the standard $F$ move, the Majorana fermion parity change compared to the local Kasteleyn orientations is then
\begin{align}\label{3D:Pf_symm}
\Delta P_f^\g (1234) &= (-1)^{\left[\om_2\smile n_2 + s_1\smile (n_2\smile_1 n_2)\right](g_0^{-1}g_1,g_1^{-1}g_2,g_2^{-1}g_3,g_3^{-1}g_4)}.
\end{align}
This is obtained from \eq{3D:Pf_M} by simply the replacement $(g_0,g_1,g_2,g_3)\rightarrow (g_0^{-1}g_1,g_0^{-1}g_2,g_0^{-1}g_3,g_0^{-1}g_4)$. Combined it with \eq{3D:n22}, the total Majorana fermion parity change under the standard $F$ move \eq{3D:Fmove} is
\begin{align}\label{3D:Pf_F}
\Delta P_f^\g (F) = (-1)^{\left[n_2\smile n_2 + \om_2\smile n_2 + s_1\smile (n_2\smile_1 n_2)\right](01234)}.
\end{align}
We note that there is no difference between the homogeneous and inhomogeneous notations for $\Z_2$-valued cocycles $s_1,\om_2$ and cochain $n_2$, for they are symmetric under $G_b$-action. So we can just use $(01234)$ to represent $(g_0,g_1,g_2,g_3,g_4)$ or $(g_0^{-1}g_1,g_1^{-1}g_2,g_2^{-1}g_3,g_3^{-1}g_4)$ in the above equation. This is very different from the $U(1)$ phase factor $\nu_4$.

It is convenient to split the exponent of $(-1)$ in \eq{3D:Pf_F} into two parts:
\begin{align}\label{3D:al}
\al_4(01234) &:= (n_2\smile n_2+\om_2\smile n_2)(01234) + s_1(01)n_2(124)n_2(234),\quad\text{(mod 2)}\\\label{3D:be}
\be_4(01234) &:= s_1(01)n_2(134)n_2(123), \quad\text{(mod 2)}.
\end{align}
Their summation gives the total Majorana fermion parity change number
\begin{align}
\al_4 + \be_4 = n_2\smile n_2 + \om_2\smile n_2 + s_1\smile (n_2\smile_1 n_2),\quad\text{(mod 2)}.
\end{align}
The philosophy of this splitting is as follows. The first part $\alpha_4$ has the property that $\al_4(01234)=1$ implies $n_2(234)=1$, because all of the terms in $\alpha_4(01234)$ contain the factor $n_2(234)$. However, the second part $\be_4$ does not contain the factor $n_2(234)$. For a given Kitaev chain configuration $n_2$, if $\al_4(01234)=1$ and $\be_4(01234)=0$, we can conclude that the Majorana fermion parity of the Kitaev chain going through the link $\langle 234\rangle$ is changed. We can add a Majorana fermion operator $\g_{234B}^{g_0^{-1}g_2}$ to the $X$ operator of \eq{3D:F} to represent the correct fermion parity change. On the other hand, if $\al_4(01234)=0$ and $\be_4(01234)=1$, the Majorana fermion parity of the Kitaev chain going through the link $\langle 123\rangle$ is changed. We can add another Majorana fermion operator $\g_{123A}^{g_0^{-1}g_1}$ to the $X$ operator (we can also use $\g_{134A}^{g_0^{-1}g_1}$ as another convention). In this case, adding $\g_{234B}^{g_0^{-1}g_2}$ does not make sense, because the Kitaev chain may even do not go through this Majorana fermion.

With the above understanding, the explicit form of $X_{01234}[n_2]$ in the standard $F$ move \eq{3D:F} can be expressed by Majorana fermion pairing projection operators as:
\begin{align}
\label{3D:X}
X_{01234}[n_2]
& = P_{01234}[n_2] \cdot \left(\gamma_{234B}^{g_0^{-1}g_2}\right)^{\al_4(01234)} \left(\gamma_{123A}^{g_0^{-1}g_1}\right)^{\be_4(01234)},\\
\label{3D:P}
P_{01234}[n_1] &= \left(\prod_{\mathrm{loop\;}i}2^{(L_i-1)/2}\right) \left( \prod_{\mathrm{Majorana\; pairs\;} \langle a,b\rangle \mathrm{\;in\;}\cT} P_{a,b}^{g_a,g_b} \right) \left( \prod_{\mathrm{triangle\;}\langle ijk\rangle \notin \cT} \prod_{\s\in G_b} P_{ijkA,ijkB}^{\s,\s} \right).
\end{align}
The two Majorana fermion operators $\gamma_{234B}^{g_0^{-1}g_2}$ and $\gamma_{123A}^{g_0^{-1}g_1}$ are inserted when $\al_4(01234)=1$ and $\be_4(01234)=1$, respectively. The first part $P_{01234}[n_1]$ in the $X$ operator \eq{3D:X} is the Majorana fermion pairing projection operator which enforces the symmetric pairing rules for the left-hand-side Majorana state in \eq{3D:Fmove}. Similar to the 2D case, the general expression of $P_{01234}[n_1]$ \eq{3D:P} has three terms. The first term is a normalization factor, where $2L_i$ is the length of the $i$-th loop in the transition graph of Majorana pairing dimer configurations on the left triangulation lattice $\cT$ and right lattice $\cT'$. The second term projects the right-hand-side state to the left-hand-side state using the pairing projection operators \eq{proj}. And the third term is the vacuum projection operators for the Majorana fermions that do not appear explicitly in the left figure. For example, the explicit $X$ operator for the $n_2$ configurations shown in \eq{3D:Fmove} [only $n_2(012)=n_2(024)=n_2(123)=n_2(234)=1$] is
\begin{align}\label{3D:X_eg}
X_{01234}[\tilde n_2] = 2 \left( P_{024B,234B}^{e,g_0^{-1}g_2} P_{012A,024A}^{e,e} \prod_{\s\neq e} P_{024A,024B}^{\s,\s} \right) \left( \prod_{\s\in G_b} P_{013A,013B}^{\s,\s} P_{123A,123B}^{\s,\s} P_{134A,134B}^{\s,\s} \right) \left(\gamma_{234B}^{g_0^{-1}g_2}\right)^{\al_4(01234)},
\end{align}
where we do not need $\gamma_{123A}^{g_0^{-1}g_1}$, because the configuration $n_2(134)=0$ implies $\be_4(01234)=s_1(01)n_2(134)n_2(123)=0$.

The $F$ symbol \eq{3D:F} should be a FSLU operator. It should be both fermion parity even and symmetric under $G_b$-action. We can use these constraints to obtain several consistency equations for the cochains $n_2$, $n_3$ and $\nu_4$.

\subsubsection{Fermion parity conservation}

As discussed in previous subsection, the Majorana fermion parity change for the standard $F$ move is given by \eq{3D:n22}, if the pairings are according to the local Kasteleyn orientations. From the symmetry action on the only non-standard tetrahedron $\langle 1234\rangle$ of the standard $F$ move, there is an additional Majorana fermion parity change \eq{3D:Pf_symm}. By combine them, we obtain the total Majorana fermion parity change \eq{3D:Pf_F} for the standard $F$ move.

On the other hand, the complex fermion parity change under the standard $F$ move can be simply calculated by counting the complex fermions decorated at the five tetrahedra on the two sides:
\begin{align}\label{3D:Pf_c}
\Delta P_f^c (F) = (-1)^{n_3(1234)+n_3(0234)+n_3(0134)+n_3(0124)+n_3(0123)} = (-1)^{\dd n_3(01234)}.
\end{align}
As a FSLU transformation, the standard $F$ move should preserve the total fermion parity. So we have the constraint $\Delta P_f(F)=\Delta P_f^\g(F) \cdot \Delta P_f^c(F) = 1$. Using the explicit expressions \eqs{3D:Pf_F}{3D:Pf_c}, we have the following (mod 2) equation
\begin{align}\label{3D:dn3}
\dd n_3 = n_2\smile n_2 + \om_2\smile n_2 + s_1\smile (n_2\smile_1 n_2).
\end{align}
The decorations of Majorana fermions and complex fermions are not independent, and should satisfy the above constraint.

We note that, if we consider the special case of unitary symmetry group $G_f=\Zf\times G_b$ (i.e., $\om_2=0$ and $s_1=0$), the above equation is reduced to the previous known result $\dd n_3 = n_2\smile n_2$ \cite{Kapustin2017,WangGu2017,Morgan2018}.

\subsubsection{Symmetry condition}

In the previous constructions, we only considered the standard $F$ move \eq{3D:Fmove} with the first vertex label $e\in G_b$. The non-standard $F$ move is defined to be obtained from the standard one by a $U(g_0)$ symmetry action. In such way, the $F$ moves are symmetric under $G_b$-actions. In this subsection, we will derive the symmetry transformation rules for the $F$ move, and the $U(1)$ phase factor $\nu_4$ in the front of $F$ move.

According to \eq{FUg}, we have the following commuting diagram for the standard and non-standard $F$ moves (we omit the decorated fermion layers in the figures):
\begin{align}\label{3D:FUg}
\tikzfig{}{
\node (p1) at (0,0) {
\begin{tikzpicture}[scale=2.]
\coordinate (0) at (.55,-.7);
\coordinate (1) at (0,0);
\coordinate (2) at (.3,-.15);
\coordinate (3) at (1,0);
\coordinate (4) at (.55,.7);
\draw[-<-=.7,thick](0)node[scale=.8,below,xshift=4]{$g_0^{-1}g_1$}--(1)node[scale=.8,left,transparent]{$g_0$};
\node[scale=.8,left]at(1){$e$};
\draw[->-=.65,thick](0)--(2)node[scale=.8,right,xshift=2,yshift=-4]{$g_0^{-1}g_2$};
\draw[->-=.6,thick](0)--(3)node[scale=.8,right]{$g_0^{-1}g_4$};
\draw[->-=.6,thick](1)--(2);
\draw[->-=.6,thick,dashed](1)--(3);
\draw[->-=.4,thick](1)--(4)node[scale=.8,above,xshift=5]{$g_0^{-1}g_3$};
\draw[->-=.5,thick](2)--(3);
\draw[->-=.35,thick](2)--(4);
\draw[-<-=.4,thick](3)--(4);
\end{tikzpicture}
};
\node (pp1) at (0,-4.8) {
\begin{tikzpicture}[scale=2.]
\coordinate (0) at (.55,-.7);
\coordinate (1) at (0,0);
\coordinate (2) at (.3,-.15);
\coordinate (3) at (1,0);
\coordinate (4) at (.55,.7);
\draw[-<-=.7,thick](0)node[scale=.8,below,xshift=4]{$g_1$}--(1)node[scale=.8,left]{$g_0$};
\draw[->-=.65,thick](0)--(2)node[scale=.8,right,xshift=2,yshift=-4]{$g_2$};
\draw[->-=.6,thick](0)--(3)node[scale=.8,right,transparent]{$g_0^{-1}g_4$};
\node[scale=.8,right]at (3){$g_4$};
\draw[->-=.6,thick](1)--(2);
\draw[->-=.6,thick,dashed](1)--(3);
\draw[->-=.4,thick](1)--(4)node[scale=.8,above,xshift=5]{$g_3$};
\draw[->-=.5,thick](2)--(3);
\draw[->-=.35,thick](2)--(4);
\draw[-<-=.4,thick](3)--(4);
\end{tikzpicture}
};
\node (pp2) at (9.3,-4.8) {
\begin{tikzpicture}[scale=2.]
\coordinate (0) at (.55,-.7);
\coordinate (1) at (0,0);
\coordinate (2) at (.3,-.15);
\coordinate (3) at (1,0);
\coordinate (4) at (.55,.7);
\draw[-<-=.6,thick](0)node[scale=.8,below,xshift=4]{$g_1$}--(1)node[scale=.8,left]{$g_0$};
\draw[->-=.7,thick](0)--(2)node[scale=.8,below right,xshift=2,yshift=2]{$g_2$};
\draw[->-=.7,thick](0)--(3);
\draw[->-=.6,thick](1)--(2);
\draw[->-=.7,thick,dashed](1)--(3);
\draw[->-=.5,thick](1)--(4);
\draw[->-=.5,thick](2)--(3)node[scale=.8,right]{$g_4$};
\node[scale=.8,right,transparent] at (3) {$g_0^{-1}g_4$};
\draw[->-=.5,thick](2)--(4)node[scale=.8,above,xshift=4]{$g_3$};
\draw[-<-=.4,thick](3)--(4);
\draw[->-=.65,thick,dashed](0)--(4);
\end{tikzpicture}
};
\node (p2) at (9.3,0) {
\begin{tikzpicture}[scale=2.]
\coordinate (0) at (.55,-.7);
\coordinate (1) at (0,0);
\coordinate (2) at (.3,-.15);
\coordinate (3) at (1,0);
\coordinate (4) at (.55,.7);
\draw[-<-=.6,thick](0)node[scale=.8,below,xshift=4]{$g_0^{-1}g_1$}--(1)node[scale=.8,left,transparent]{$g_0$};
\node[scale=.8,left]at(1){$e$};
\draw[->-=.7,thick](0)--(2)node[scale=.8,below right,xshift=3,yshift=5]{$g_0^{-1}g_2$};
\draw[->-=.7,thick](0)--(3);
\draw[->-=.6,thick](1)--(2);
\draw[->-=.7,thick,dashed](1)--(3);
\draw[->-=.5,thick](1)--(4);
\draw[->-=.5,thick](2)--(3)node[scale=.8,right]{$g_0^{-1}g_4$};
\draw[->-=.5,thick](2)--(4)node[scale=.8,above,xshift=4]{$g_0^{-1}g_3$};
\draw[-<-=.4,thick](3)--(4);
\draw[->-=.65,thick,dashed](0)--(4);
\end{tikzpicture}
};
\draw[<-] (p1)--(p2) node [midway,above] {$F(e,g_0^{-1}g_1,g_0^{-1}g_2,g_0^{-1}g_3,g_0^{-1}g_4)$};
\draw[<-] (pp1)--(pp2) node [midway,above] {$F(g_0,g_1,g_2,g_3,g_4)$};
\draw[->] (p1)--(pp1) node [midway,left] {$U(g_0)$};
\draw[->] (p2)--(pp2) node [midway,right] {$U(g_0)$};
}
\end{align}
So the non-standard $F$ operator is defined as
\begin{align}\label{3D:Fg}
F(g_0,g_1,g_2,g_3,g_4) = \emp^{g_0}F(e,g_0^{-1}g_1,g_0^{-1}g_2,g_0^{-1}g_3,g_0^{-1}g_4) := U(g_0) F(e,g_0^{-1}g_1,g_0^{-1}g_2,g_0^{-1}g_3,g_0^{-1}g_4) U(g_0)^\dagger.
\end{align}
The non-standard $F$ moves constructed in this way are automatically symmetric, because one can show the transformation rule for the non-standard $F$ moves
\begin{align}
F(gg_0,gg_1,gg_2,gg_3,gg_4) = U(g) F(g_0,g_1,g_2,g_3,g_4) U(g)^\dagger,
\end{align}
using \eq{U_proj} and the fact that $F$ operator is fermion parity even.

Using a $U(g_0)$-action on the standard $F$ operator \eq{3D:F}, we can obtain the non-standard $F$ symbol expression as
\begin{align}\label{3D:F_}
&\quad F(g_0,g_1,g_2,g_3,g_4) \\\nonumber
&= \nu_4(g_0,g_1,g_2,g_3,g_4)
\big(c^{g_0\dagger}_{0124}\big)^{n_3(0124)} \big(c^{g_0\dagger}_{0234}\big)^{n_3(0234)} \big(c^{g_0}_{0123}\big)^{n_3(0123)} \big(c^{g_0}_{0134}\big)^{n_3(0134)} \big(c^{g_1}_{1234}\big)^{n_3(1234)} X_{01234}[n_2] Y_{01234}[n_1].
\end{align}
The decorated complex fermions now have group element labels $g_0$ (for the first four complex fermions) or $g_1$ (for the last complex fermion). And the $X$ operator is
\begin{align}\label{3D:X_}
X_{01234}[n_2]
& = P_{01234}[n_2] \cdot \big(\gamma_{234B}^{g_2}\big)^{\al_4(01234)} \big(\gamma_{123A}^{g_1}\big)^{\be_4(01234)},
\end{align}
with added Majorana fermion $\gamma_{234B}^{g_2}$ and $\gamma_{123A}^{g_1}$, rather than $\gamma_{234B}^{g_0^{-1}g_2}$ and $\gamma_{123A}^{g_0^{-1}g_1}$ in \eq{3D:X}. The operator $P_{01234}[n_2]$ projects the Majorana fermions to the pairing state on the left-hand-side figure. It has similar expression as \eq{3D:P}, and is a product of many Majorana pairing projection operators \eq{proj} with appropriate group element labels.

From the decoration rules of Majorana fermions and complex fermions constructed in previous sections, the data $n_1$, $n_2$ and $n_3$ are invariant under $G_b$ symmetry actions. The homogeneous cochain $\nu_4$ in the non-standard $F$ move \eq{3D:F_} is a combination of the inhomogeneous $\nu_4$ in the standard $F$ move \eq{3D:F} and the $\pm 1$ signs which appear from the symmetry action. Therefore, we have the following symmetry conditions for the data $n_1$, $n_2$, $n_3$ and $\nu_4$:
\begin{align}
n_1(g_0,g_1) &= n_1(e,g_0^{-1}g_1) = n_1(g_0^{-1}g_1),\\
n_2(g_0,g_1,g_2) &= n_2(e,g_0^{-1}g_1,g_0^{-1}g_2) = n_2(g_0^{-1}g_1,g_1^{-1}g_2),\\
n_3(g_0,g_1,g_2,g_3) &= n_3(e,g_0^{-1}g_1,g_0^{-1}g_2,g_0^{-1}g_3) = n_3(g_0^{-1}g_1,g_1^{-1}g_2,g_2^{-1}g_3),\\\nonumber
\nu_4(g_0,g_1,g_2,g_3,g_4) &= \emp^{g_0}\nu_4(e,g_0^{-1}g_1,g_0^{-1}g_2,g_0^{-1}g_3,g_0^{-1}g_4)
=\emp^{g_0}\nu_4(g_0^{-1}g_1,g_1^{-1}g_2,g_2^{-1}g_3,g_3^{-1}g_4)\\\label{3D:nu4_symm}
&= \nu_4(g_0^{-1}g_1,g_1^{-1}g_2,g_2^{-1}g_3,g_3^{-1}g_4)^{1-2s_1(g_0)} \cdot \mathcal O_5^\mathrm{symm}(g_0,g_1,g_2,g_3,g_4).
\end{align}
The last equation can be viewed as the definition of homogeneous $\nu_4$ in the non-standard $F$ move \eq{3D:F_} in terms of the inhomogeneous $\nu_4$ in the standard $F$ move \eq{3D:F}. The symmetry sign difference $\mathcal O_5^\mathrm{symm}$ is given by
\begin{align}\nonumber
\mathcal O_5^\mathrm{symm}(g_0,g_1,g_2,g_3,g_4)
&= (-1)^{\om_2(0,\bar 01)n_3(1234) + [\om_2(0,\bar 02)+s_1(0)]\al_4(01234) + \om_2(0,\bar 01)\be_4(01234)}\\\label{3D:Osymm}
&= (-1)^{(\om_2\smile n_3+s_1\smile \al_4)(0,\bar 01,\bar 12,\bar 23,\bar 34) + \om_2(0,\bar 02)\al_4(01234) + \om_2(0,\bar 01)\be_4(01234)}.
\end{align}
We note that some of the terms above can not be expressed as a cup product form. The above equation can be obtained by straightforward calculation using \eq{3D:Fg}. In the first line of \eq{3D:Osymm}, the first term of the form $(-1)^{\om_2 n_3}$ comes from the $U(g_0)$ symmetry transformation \eq{3D:symm:c} of the last complex fermion $c_{1234}^{g_0^{-1}g_1}$ in the standard $F$ move \eq{3D:F}. The second term of the form $(-1)^{(\om_2+s_1) \al_4}$ comes from the symmetry transformation \eq{3D:symm:B} of $\g_{234B}^{g_0^{-1}g_2}$ in the $X$ operator \eq{3D:X} of the standard $F$ move \eq{3D:F}. And the last sign of the form $(-1)^{\om_2 \be_4}$ comes from the symmetry transformation \eq{3D:symm:A} of $\g_{123A}^{g_0^{-1}g_1}$ in the $X$ operator \eq{3D:X} of the standard $F$ move. The 4-cochains $\al_4$ and $\be_4$ are defined in \eqs{3D:al}{3D:be}.

If we have $\be_4=0$ (for example, $s_1=0$ or $n_1=0$), we only need to insert the dangling Majorana fermion operator $\g_{234B}^{g_0^{-1}g_2}$ in the standard $X$ operator. The expressions of the standard $F$ move \eq{3D:F} and $X$ operator \eq{3D:X} are similar to the special case of $s_1=\om_2=0$ \cite{WangGu2017}. Using $\dd n_3=\al_4+\be_4=\al_4$, the symmetry sign \eq{3D:Osymm} is reduced to
\begin{align}\nonumber
\mathcal O_5^\mathrm{symm}(g_0,g_1,g_2,g_3,g_4)\big |_{\be_4=0}
&= (-1)^{\om_2(0,\bar 01)n_3(1234) + [\om_2(0,\bar 02)+s_1(0)]\dd n_3(01234)}\\\label{3D:Osymm_}
&= (-1)^{(\om_2\smile n_3 + s_1\smile \dd n_3)(0,\bar 01,\bar 12,\bar 23,\bar 34) + \om_2(0,\bar 02)\dd n_3(01234)}.
\end{align}
The symmetry sign $\mathcal O_5^\mathrm{symm}$ will appear later in the twisted cocycle equation for $\nu_4$ as part of the obstruction function [see \eq{3D:O5_4}]. In the special case of $\be_4=0$, the calculation of $\mathcal O_5$ is much simpler than the generic case.

\subsection{Super fusion hexagon and twisted cocycle equations}

Just as the 2D $F$ move should satisfy the super pentagon equation of superfusion category, the 3D $F$ move should satisfy a super hexagon equation of superfusion 2-category. One should distinguish it from the hexagon equation of braided tensor category, for the former is in the fusion level rather than the braided level. Our 3D FSPT states constructed correspond to some kind of pointed superfusion 2-categories with to a given symmetry group.

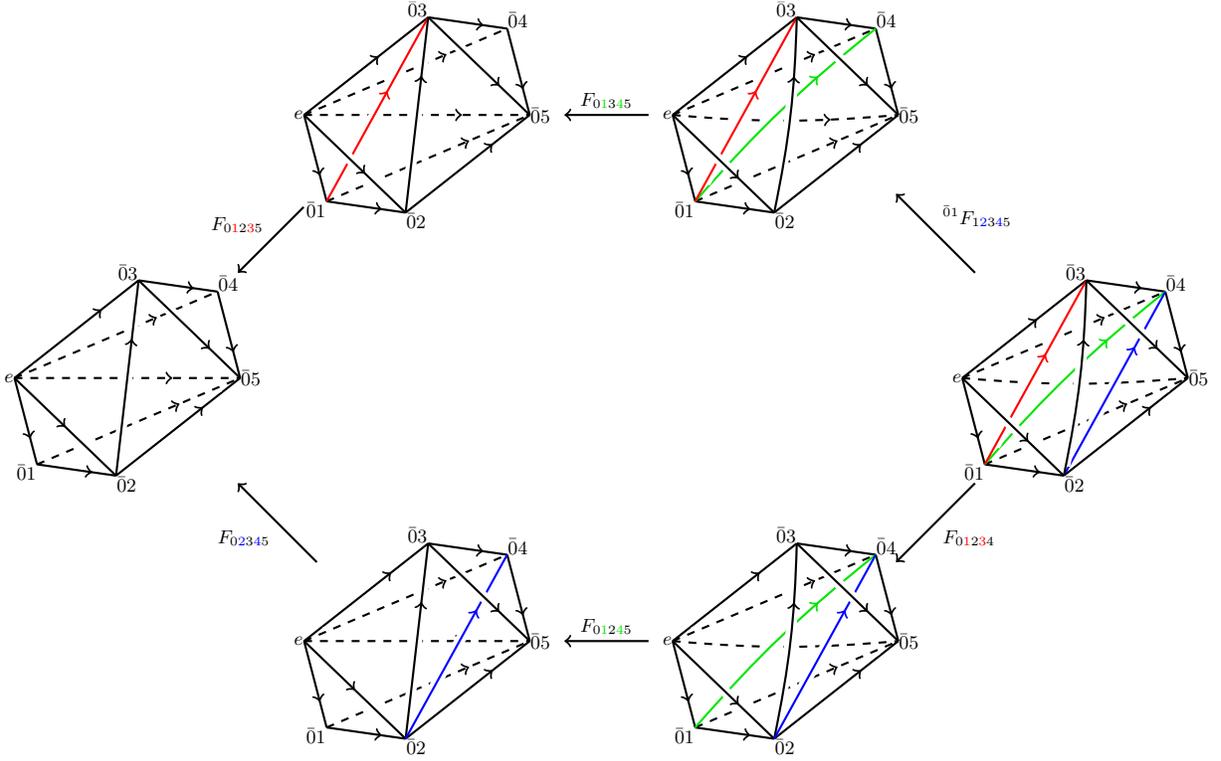
\begin{figure}[ht]
\centering
$
\tikzfig{scale=3.5}{
\pgfmathsetmacro{\th}{360/6}
\pgfmathsetmacro{\s}{.15}
\pgfmathsetmacro{\ss}{.12}
\node[inner sep=0pt] (p1) at (-1.8,0)
{
 \begin{tikzpicture}[scale=1.5]
\coordinate (1) at ($({-90-\th/2}:1)-(.3,-.1)$);
\coordinate (0) at ({-90-\th/2-\th}:1);
\coordinate (3) at ($({-90-\th/2-2*\th}:1)+(.6,0)$);
\coordinate (4) at ($({-90-\th/2-3*\th}:1)+(.3,-.1)$);
\coordinate (5) at ({-90-\th/2-4*\th}:1);
\coordinate (2) at ($({-90-\th/2-5*\th}:1)-(.6,0)$);
\draw[->-=.7,thick] (0)node[scale=.8,left]{$e$}--(1)node[scale=.8,below left]{$\bar 01$};
\draw[->-=.7,thick] (1)--(2)node[scale=.8,below right]{$\bar 02$};
\draw[->-=.7,thick] (2)--(5);
\draw[->-=.7,thick] (3)--(4)node[scale=.8,above right]{$\bar 04$};
\draw[->-=.7,thick] (4)--(5)node[scale=.8,right]{$\bar 05$};
\draw[->-=.7,thick] (0)--(3)node[scale=.8,above left]{$\bar 03$};
\Overline{0}{4}{.7}{dashed};
\Overline{1}{5}{.7}{dashed};
\Overline{0}{5}{.7}{dashed};
\Overline{0}{2}{.5}{};
\Overline{2}{3}{.7}{};
\Overline{3}{5}{.7}{};
 \end{tikzpicture}
};
\node[inner sep=0pt] (p2) at (-.7,1)
{
 \begin{tikzpicture}[scale=1.5]
\draw[->-=.7,thick] (0)node[scale=.8,left]{$e$}--(1)node[scale=.8,below left]{$\bar 01$};
\draw[->-=.7,thick] (1)--(2)node[scale=.8,below right]{$\bar 02$};
\draw[->-=.7,thick] (2)--(5);
\draw[->-=.7,thick] (3)--(4)node[scale=.8,above right]{$\bar 04$};
\draw[->-=.7,thick] (4)--(5)node[scale=.8,right]{$\bar 05$};
\draw[->-=.6,thick] (0)--(3)node[scale=.8,above left]{$\bar 03$};
\Overline{0}{4}{.7}{dashed};
\Overline{1}{5}{.7}{dashed};
\Overline{0}{5}{.7}{dashed};
\Overline{1}{3}{.6}{red};
\Overline{0}{2}{.6}{};
\Overline{2}{3}{.7}{};
\Overline{3}{5}{.7}{};
 \end{tikzpicture}
};
\node[inner sep=0pt] (p3) at (.7,1)
{
 \begin{tikzpicture}[scale=1.5]
\draw[->-=.7,thick] (0)node[scale=.8,left]{$e$}--(1)node[scale=.8,below left]{$\bar 01$};
\draw[->-=.7,thick] (1)--(2)node[scale=.8,below right]{$\bar 02$};
\draw[->-=.7,thick] (2)--(5);
\draw[->-=.7,thick] (3)--(4)node[scale=.8,above right]{$\bar 04$};
\draw[->-=.7,thick] (4)--(5)node[scale=.8,right]{$\bar 05$};
\draw[->-=.6,thick] (0)--(3)node[scale=.8,above left]{$\bar 03$};
\Overline{0}{4}{.7}{dashed};
\Overline{1}{5}{.7}{dashed};
\OverlineBend{0}{5}{.7}{right}{dashed,black};
\OverlineBend{1}{4}{.7}{left}{black!10!green};
\Overline{1}{3}{.6}{red};
\Overline{3}{5}{.7}{};
\Overline{0}{2}{.65}{};
\OverlineBend{2}{3}{.7}{right}{black};
 \end{tikzpicture}
};
\node[inner sep=0pt] (p2) at (-.7,-1)
{
 \begin{tikzpicture}[scale=1.5]
\draw[->-=.7,thick] (0)node[scale=.8,left]{$e$}--(1)node[scale=.8,below left]{$\bar 01$};
\draw[->-=.7,thick] (1)--(2)node[scale=.8,below right]{$\bar 02$};
\draw[->-=.7,thick] (2)--(5);
\draw[->-=.7,thick] (3)--(4)node[scale=.8,above right]{$\bar 04$};
\draw[->-=.7,thick] (4)--(5)node[scale=.8,right]{$\bar 05$};
\draw[->-=.7,thick] (0)--(3)node[scale=.8,above left]{$\bar 03$};
\Overline{0}{4}{.7}{dashed};
\Overline{1}{5}{.7}{dashed};
\Overline{0}{5}{.7}{dashed};
\Overline{0}{2}{.5}{};
\Overline{2}{3}{.7}{};
\Overline{2}{4}{.7}{blue};
\Overline{3}{5}{.7}{};
 \end{tikzpicture}
};
\node[inner sep=0pt] (p3) at (.7,-1)
{
 \begin{tikzpicture}[scale=1.5]
\draw[->-=.7,thick] (0)node[scale=.8,left]{$e$}--(1)node[scale=.8,below left]{$\bar 01$};
\draw[->-=.7,thick] (1)--(2)node[scale=.8,below right]{$\bar 02$};
\draw[->-=.7,thick] (2)--(5);
\draw[->-=.7,thick] (3)--(4)node[scale=.8,above right]{$\bar 04$};
\draw[->-=.7,thick] (4)--(5)node[scale=.8,right]{$\bar 05$};
\draw[->-=.7,thick] (0)--(3)node[scale=.8,above left]{$\bar 03$};
\Overline{0}{4}{.7}{dashed};
\Overline{1}{5}{.7}{dashed};
\OverlineBend{0}{5}{.7}{right}{dashed,black};
\OverlineBend{1}{4}{.7}{left}{black!10!green};
\Overline{0}{2}{.65}{};
\OverlineBend{2}{3}{.7}{right}{black};
\Overline{2}{4}{.7}{blue};
\Overline{3}{5}{.7}{};
 \end{tikzpicture}
};
\node[inner sep=0pt] (p5) at (1.8,0)
{
 \begin{tikzpicture}[scale=1.5]
\draw[->-=.7,thick] (0)node[scale=.8,left]{$e$}--(1)node[scale=.8,below left]{$\bar 01$};
\draw[->-=.7,thick] (1)--(2)node[scale=.8,below right]{$\bar 02$};
\draw[->-=.7,thick] (2)--(5);
\draw[->-=.7,thick] (3)--(4)node[scale=.8,above right]{$\bar 04$};
\draw[->-=.7,thick] (4)--(5)node[scale=.8,right]{$\bar 05$};
\draw[->-=.6,thick] (0)--(3)node[scale=.8,above left]{$\bar 03$};
\Overline{0}{4}{.7}{dashed};
\Overline{1}{5}{.7}{dashed};
\OverlineBend{0}{5}{.7}{right}{dashed,black};
\OverlineBend{1}{4}{.7}{left}{black!10!green};
\Overline{1}{3}{.6}{red};
\Overline{2}{4}{.7}{blue};
\Overline{3}{5}{.7}{};
\Overline{0}{2}{.65}{};
\OverlineBend{2}{3}{.7}{right}{black};
 \end{tikzpicture}
};
\draw[<-,thick] (-1.4,.4) -- (-1.15,.65) node[scale=.8,midway,above left] {$F_{0{\color{red}1}2{\color{red}3}5}$};
\draw[<-,thick] (-.16,1) -- (.16,1) node[scale=.8,midway,above] {$F_{0{\color{black!10!green}1}3{\color{black!10!green}4}5}$};
\draw[<-,thick] (1.1,.7) -- (1.4,.4) node[scale=.8,midway,above right] {$\emp^{\bar 01}F_{1{\color{blue}2}3{\color{blue}4}5}$};
\draw[<-,thick] (-1.4,-.4) -- (-1.1,-.7) node[scale=.8,midway,below left] {$F_{0{\color{blue}2}3{\color{blue}4}5}$};
\draw[<-,thick] (-.16,-1) -- (.16,-1) node[scale=.8,midway,above] {$F_{0{\color{black!10!green}1}2{\color{black!10!green}4}5}$};
\draw[<-,thick] (1.1,-.7) -- (1.4,-.4) node[scale=.8,midway,below right] {$F_{0{\color{red}1}2{\color{red}3}4}$};
}
$
\caption{Standard super fusion hexagon equation. Algebraically, this standard super pentagon condition corresponds to \eq{3D:dF}. The colored numbers $i$ and $j$ in the subscript of $F$ indicate that the link $\langle ij\rangle$ with the same color is added after this $F$ move. Since the group element label of the first vertex is $e\in G_b$, all the fermionic $F$ moves are standard except $\emp^{\bar 01}F_{12345}$. We used a simpler notation $F_{ijkl}=F(e,\bar ij,\bar jk,\bar kl)$ for the standard $F$ move. We also omit all the fermion layers in the figure.}
\label{fig:hexagon}
\end{figure}

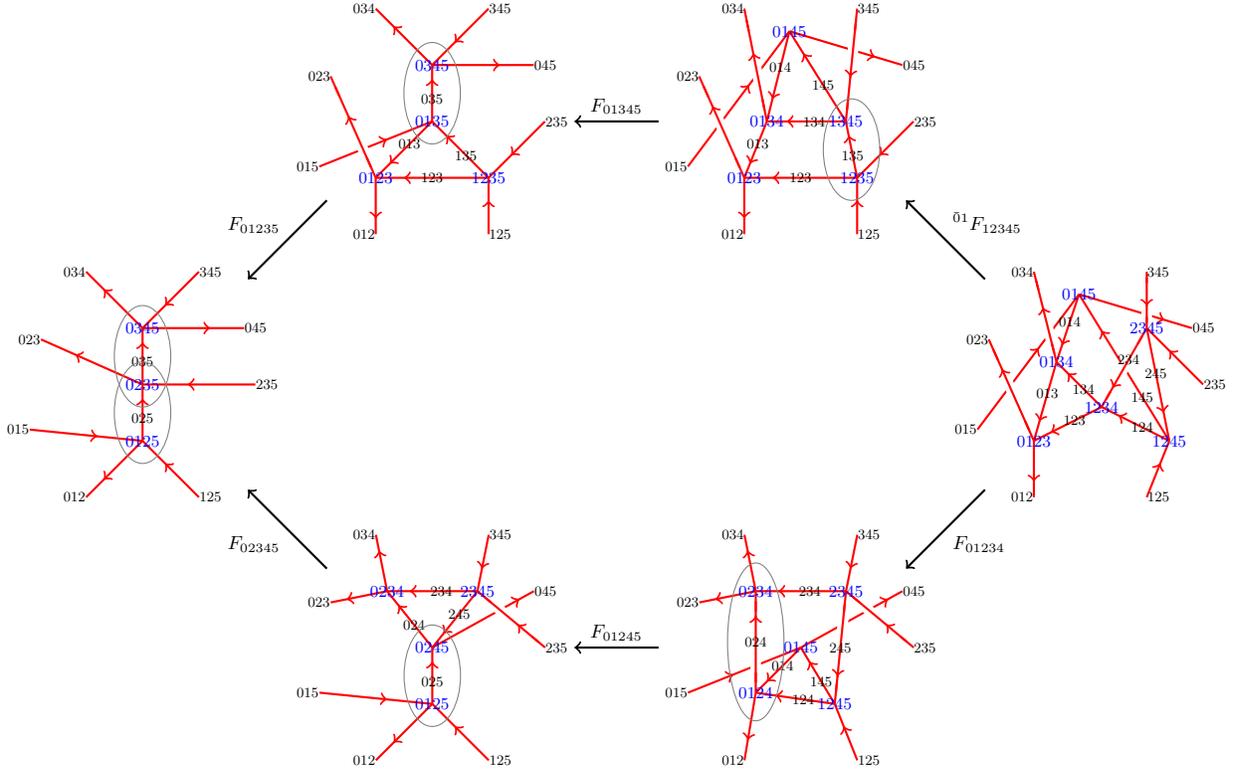
\begin{figure}[ht]
\centering
$
\tikzfig{scale=3.5}{
\pgfmathsetmacro{\th}{360/6}
\pgfmathsetmacro{\s}{.15}
\pgfmathsetmacro{\ss}{.12}
\node[inner sep=0pt] (p1) at (-1.8,0)
{
 \begin{tikzpicture}[scale=1.5]
\coordinate (012) at (-.5,-1);
\coordinate (015) at (-1,-.4);
\coordinate (023) at (-.9,.4);
\coordinate (034) at (-.5,1);
\coordinate (345) at (.5,1);
\coordinate (045) at (.9,.5);
\coordinate (235) at (1,0);
\coordinate (125) at (.5,-1);
\coordinate (0345) at (0,.5);
\coordinate (0235) at (0,0);
\coordinate (0125) at (0,-.5);
\draw[-<-=.4,thick,red] (012)node[scale=.6,left,black]{012}--(0125);
\draw[->-=.6,thick,red] (015)node[scale=.6,left,black]{015}--(0125);
\draw[->-=.6,thick,red] (125)node[scale=.6,right,black]{125}--(0125);
\draw[-<-=.4,thick,red] (023)node[scale=.6,left,black]{023}--(0235);
\draw[->-=.6,thick,red] (235)node[scale=.6,right,black]{235}--(0235);
\draw[-<-=.4,thick,red] (034)node[scale=.6,left,black]{034}--(0345);
\draw[->-=.6,thick,red] (345)node[scale=.6,right,black]{345}--(0345);
\draw[-<-=.4,thick,red] (045)node[scale=.6,right,black]{045}--(0345);
\draw[->-=.75,thick,red] (0235)--(0345) node[scale=.6,black,pos=0.4]{035};
\draw[->-=.75,thick,red] (0125)--(0235) node[scale=.6,black,pos=0.4]{025};
\node[scale=.7,blue] at (0345) {0345};
\node[scale=.7,blue] at (0235) {0235};
\node[scale=.7,blue] at (0125) {0125};
\draw [gray] ($(0125)!.5!(0235)$) ellipse [x radius=.45, y radius=.25, rotate=90];
\draw [gray] ($(0345)!.5!(0235)$) ellipse [x radius=.45, y radius=.25, rotate=90];
 \end{tikzpicture}
};
\node[inner sep=0pt] (p2) at (-.7,1)
{
 \begin{tikzpicture}[scale=1.5]
\coordinate (0123) at (-.5,-.5);
\coordinate (1235) at (.5,-.5);
\coordinate (0135) at (0,0);
\draw[-<-=.4,thick,red] (012)node[scale=.6,left,black]{012}--(0123);
\draw[->-=.6,thick,red] (015)node[scale=.6,left,black]{015}--(0135);
\draw[->-=.6,thick,red] (125)node[scale=.6,right,black]{125}--(1235);
\OverlineHex{0123}{023}{red}{.6};\node[scale=.6,left,black] at (023) {023};
\draw[->-=.6,thick,red] (235)node[scale=.6,right,black]{235}--(1235);
\draw[-<-=.4,thick,red] (034)node[scale=.6,left,black]{034}--(0345);
\draw[->-=.6,thick,red] (345)node[scale=.6,right,black]{345}--(0345);
\draw[-<-=.4,thick,red] (045)node[scale=.6,right,black]{045}--(0345);
\draw[->-=.75,thick,red] (0235)--(0345) node[scale=.6,black,pos=0.4]{035};
\draw[->-=.75,thick,red] (0135)--(0123) node[scale=.6,black,pos=0.4]{013};
\draw[->-=.75,thick,red] (1235)--(0135) node[scale=.6,black,pos=0.4]{135};
\draw[->-=.75,thick,red] (1235)--(0123) node[scale=.6,black,pos=0.5]{123};
\node[scale=.7,blue] at (0345) {0345};
\node[scale=.7,blue] at (0135) {0135};
\node[scale=.7,blue] at (0123) {0123};
\node[scale=.7,blue] at (1235) {1235};
\draw [gray] ($(0135)!.5!(0345)$) ellipse [x radius=.45, y radius=.25, rotate=90];
 \end{tikzpicture}
};
\node[inner sep=0pt] (p3) at (.7,1)
{
 \begin{tikzpicture}[scale=1.5]
\coordinate (0145) at (-.1,.8);
\coordinate (0134) at (-.3,0);
\coordinate (1345) at (.4,0);
\draw[-<-=.4,thick,red] (012)node[scale=.6,left,black]{012}--(0123);
\draw[->-=.6,thick,red] (015)node[scale=.6,left,black]{015}--(0145);
\draw[->-=.6,thick,red] (125)node[scale=.6,right,black]{125}--(1235);
\OverlineHex{0123}{023}{red}{.7};\node[scale=.6,left,black] at (023) {023};
\draw[->-=.6,thick,red] (235)node[scale=.6,right,black]{235}--(1235);
\OverlineHex{0134}{034}{red}{.6};\node[scale=.6,left,black] at (034) {034};
\draw[-<-=.3,thick,red] (045)node[scale=.6,right,black]{045}--(0145);
\OverlineHex{345}{1345}{red}{.6};\node[scale=.6,right,black] at (345) {345};
\draw[->-=.75,thick,red] (1235)--(0123) node[scale=.6,black,pos=0.5]{123};
\draw[->-=.75,thick,red] (0145)--(0134) node[scale=.6,black,pos=0.4]{014};
\draw[->-=.75,thick,red] (1345)--(0145) node[scale=.6,black,pos=0.4]{145};
\draw[->-=.75,thick,red] (1345)--(0134) node[scale=.6,black,pos=0.4]{134};
\draw[->-=.75,thick,red] (0134)--(0123) node[scale=.6,black,pos=0.4]{013};
\draw[->-=.75,thick,red] (1235)--(1345) node[scale=.6,black,pos=0.4]{135};
\node[scale=.7,blue] at (0123) {0123};
\node[scale=.7,blue] at (1235) {1235};
\node[scale=.7,blue] at (0145) {0145};
\node[scale=.7,blue] at (0134) {0134};
\node[scale=.7,blue] at (1345) {1345};
\draw [gray] ($(1235)!.5!(1345)$) ellipse [x radius=.45, y radius=.25, rotate=90];
 \end{tikzpicture}
};
\node[inner sep=0pt] (p2) at (-.7,-1)
{
 \begin{tikzpicture}[scale=1.5]
\coordinate (0234) at (-.4,.5);
\coordinate (2345) at (.4,.5);
\coordinate (0245) at (0,0);
\draw[-<-=.4,thick,red] (012)node[scale=.6,left,black]{012}--(0125);
\draw[->-=.6,thick,red] (015)node[scale=.6,left,black]{015}--(0125);
\draw[->-=.6,thick,red] (125)node[scale=.6,right,black]{125}--(0125);
\draw[-<-=.4,thick,red] (023)node[scale=.6,left,black]{023}--(0234);
\draw[-<-=.2,thick,red] (045)node[scale=.6,right,black]{045}--(0245);
\OverlineHex{235}{2345}{red}{.4};\node[scale=.6,right,black] at (235) {235};
\draw[-<-=.4,thick,red] (034)node[scale=.6,left,black]{034}--(0234);
\draw[->-=.6,thick,red] (345)node[scale=.6,right,black]{345}--(2345);
\draw[->-=.75,thick,red] (0125)--(0245) node[scale=.6,black,pos=0.4]{025};
\draw[->-=.75,thick,red] (2345)--(0234) node[scale=.6,black,pos=0.4]{234};
\draw[->-=.75,thick,red] (0245)--(0234) node[scale=.6,black,pos=0.4]{024};
\draw[->-=.75,thick,red] (2345)--(0245) node[scale=.6,black,pos=0.4]{245};
\node[scale=.7,blue] at (0125) {0125};
\node[scale=.7,blue] at (0245) {0245};
\node[scale=.7,blue] at (0234) {0234};
\node[scale=.7,blue] at (2345) {2345};
\draw [gray] ($(0125)!.5!(0245)$) ellipse [x radius=.45, y radius=.25, rotate=90];
 \end{tikzpicture}
};
\node[inner sep=0pt] (p3) at (.7,-1)
{
 \begin{tikzpicture}[scale=1.5]
\coordinate (0124) at (-.4,-.4);
\coordinate (0145) at (0,0);
\coordinate (1245) at (.3,-.5);
\draw[-<-=.4,thick,red] (012)node[scale=.6,left,black]{012}--(0124);
\draw[->-=.4,thick,red] (015)node[scale=.6,left,black]{015}--(0145);
\draw[->-=.6,thick,red] (125)node[scale=.6,right,black]{125}--(1245);
\draw[-<-=.4,thick,red] (023)node[scale=.6,left,black]{023}--(0234);
\draw[-<-=.2,thick,red] (045)node[scale=.6,right,black]{045}--(0145);
\OverlineHex{235}{2345}{red}{.4};\node[scale=.6,right,black] at (235) {235};
\draw[-<-=.4,thick,red] (034)node[scale=.6,left,black]{034}--(0234);
\draw[->-=.6,thick,red] (345)node[scale=.6,right,black]{345}--(2345);
\draw[->-=.75,thick,red] (2345)--(0234) node[scale=.6,black,pos=0.4]{234};
\OverlineHex{0124}{0234}{red}{.75};\node[scale=.6,black] at ($(0124)!.5!(0234)$) {024};
\draw[->-=.75,thick,red] (0145)--(0124) node[scale=.6,black,pos=0.4]{014};
\draw[->-=.75,thick,red] (1245)--(0124) node[scale=.6,black,pos=0.4]{124};
\OverlineHex{2345}{1245}{red}{.75};\node[scale=.6,black] at ($(2345)!.5!(1245)$) {245};
\draw[->-=.75,thick,red] (1245)--(0145) node[scale=.6,black,pos=0.4]{145};
\node[scale=.7,blue] at (0234) {0234};
\node[scale=.7,blue] at (2345) {2345};
\node[scale=.7,blue] at (0124) {0124};
\node[scale=.7,blue] at (0145) {0145};
\node[scale=.7,blue] at (1245) {1245};
\draw [gray] ($(0124)!.5!(0234)$) ellipse [x radius=.7, y radius=.25, rotate=90];
 \end{tikzpicture}
};
\node[inner sep=0pt] (p5) at (1.8,0)
{
 \begin{tikzpicture}[scale=1.5]
\coordinate (0134) at (-.3,.2);
\coordinate (1234) at (.1,-.2);
\coordinate (0145) at (-.1,.8);
\coordinate (1245) at (.7,-.5);
\coordinate (2345) at (.5,.5);
\draw[-<-=.4,thick,red] (012)node[scale=.6,left,black]{012}--(0123);
\draw[->-=.6,thick,red] (015)node[scale=.6,left,black]{015}--(0145);
\draw[->-=.6,thick,red] (125)node[scale=.6,right,black]{125}--(1245);
\OverlineHex{0123}{023}{red}{.7};\node[scale=.6,left,black] at (023) {023};
\draw[->-=.6,thick,red] (235)node[scale=.6,right,black]{235}--(2345);
\OverlineHex{0134}{034}{red}{.6};\node[scale=.6,left,black] at (034) {034};
\draw[-<-=.3,thick,red] (045)node[scale=.6,right,black]{045}--(0145);
\OverlineHex{345}{2345}{red}{.6};\node[scale=.6,right,black] at (345) {345};
\draw[->-=.75,thick,red] (1245)--(0145) node[scale=.6,black,pos=0.3]{145};
\draw[->-=.75,thick,red] (2345)--(1245) node[scale=.6,black,pos=0.4]{245};
\draw[->-=.75,thick,red] (0145)--(0134) node[scale=.6,black,pos=0.4]{014};
\draw[->-=.75,thick,red] (0134)--(0123) node[scale=.6,black,pos=0.4]{013};
\draw[->-=.75,thick,red] (1234)--(0123) node[scale=.6,black,pos=0.4]{123};
\draw[->-=.75,thick,red] (1245)--(1234) node[scale=.6,black,pos=0.4]{124};
\OverlineHex{2345}{1234}{red}{.75};\node[scale=.6,black] at ($(2345)!.4!(1234)$) {234};
\draw[->-=.75,thick,red] (1234)--(0134) node[scale=.6,black,pos=0.4]{134};
\node[scale=.7,blue] at (2345) {2345};
\node[scale=.7,blue] at (0145) {0145};
\node[scale=.7,blue] at (1245) {1245};
\node[scale=.7,blue] at (0123) {0123};
\node[scale=.7,blue] at (0134) {0134};
\node[scale=.7,blue] at (1234) {1234};
 \end{tikzpicture}
};
\draw[<-,thick] (-1.4,.4) -- (-1.1,.7) node[scale=.8,midway,above left] {$F_{01235}$};
\draw[<-,thick] (-.16,1) -- (.16,1) node[scale=.8,midway,above] {$F_{01345}$};
\draw[<-,thick] (1.1,.7) -- (1.4,.4) node[scale=.8,midway,above right] {$\emp^{\bar 01}F_{12345}$};
\draw[<-,thick] (-1.4,-.4) -- (-1.1,-.7) node[scale=.8,midway,below left] {$F_{02345}$};
\draw[<-,thick] (-.16,-1) -- (.16,-1) node[scale=.8,midway,above] {$F_{01245}$};
\draw[<-,thick] (1.1,-.7) -- (1.4,-.4) node[scale=.8,midway,below right] {$F_{01234}$};
}
$
\caption{Super fusion hexagon equation on the lattice dual to triangulation in \fig{fig:hexagon}. Each meeting point of four red links should be resolved to a small tetrahedron as in \fig{fig:3DKasteleyn2}. }
\label{fig:hexagon_dual}
\end{figure}

It is enough to merely consider the \emph{standard} super hexagon equations with first vertex label $e\in G_b$ as coherence conditions. This is because all other non-standard ones can be obtained from it by a $U(g_0)$ symmetry action. The standard super hexagon equation is shown in \fig{fig:hexagon} for the triangulation lattice and \fig{fig:hexagon_dual} for the dual lattice. Algebraically, we have the following equation
\begin{align}\nonumber\label{3D:dF}
F(e,\bar 02,\bar 03,\bar 04,\bar 05) \cdot F(e,\bar 01,\bar 02,\bar 04,\bar 05) \cdot F(e,\bar 01,\bar 02,\bar 03,\bar 04)
&= F(e,\bar 01,\bar 02,\bar 03,\bar 05) \cdot F(e,\bar 01,\bar 03,\bar 04,\bar 05) \cdot F(\bar 01,\bar 02,\bar 03,\bar 04,\bar 05)\\
&= F(e,\bar 01,\bar 02,\bar 03,\bar 05) \cdot F(e,\bar 01,\bar 03,\bar 04,\bar 05) \cdot \emp^{\bar 01}F(e,\bar 12,\bar 13,\bar 14,\bar 15),
\end{align}
where we again used $\bar ij$ to denote $g_i^{-1}g_j$. In the above equation, only the last $F$ symbol is non-standard. It can be obtained from the standard one by a $U(g_0^{-1}g_1)$ symmetry action [see \eq{3D:Fg}].

Standard super fusion hexagon equation. Algebraically, this standard super pentagon condition corresponds to \eq{3D:dF}. The colored numbers $i$ and $j$ in the subscript of $F$ indicate that the link $\langle ij\rangle$ with the same color is added after this $F$ move. Since the group element label of the first vertex is $e\in G_b$, all the fermionic $F$ moves are standard except $\emp^{\bar 01}F_{12345}$. We used a simpler notation $F_{ijkl}=F(e,\bar ij,\bar jk,\bar kl)$ for the standard $F$ move. We also omit all the fermion layers in the figure.

Using the explicit expression of the standard $F$ move \eq{3D:F}, we can unfold the standard super hexagon equation \eq{3D:dF}. 
By eliminating all complex fermions and Majorana fermions, we can obtain a twisted cocycle equation for the inhomogeneous 3-cochain $\nu_3$ in the standard $F$ move. In general, the twisted cocycle equation reads
\begin{align}
\dd \nu_4 = \mathcal O_5[n_3],
\end{align}
where $\mathcal O_5[n_3]$ is a functional of $n_3$ only (as well as $\om_2$ and $s_1$ parametrizing the given symmetry group). The $n_2$ dependence of $\mathcal O_5$ is though $\dd n_3$ by \eq{3D:dn3}. With nonzero $\om_2$ and $s_1$, the obstruction function $\mathcal O_5[n_3]$ is more complicated than the special result for unitary $G_f = \Zf \times G_b$ \cite{WangGu2017,Kapustin2017,Morgan2018}.

Similar to the 2D case, the obstruction function $\mathcal O_5[n_3]$ consists of four terms from general considerations:
\begin{align}\label{3D:O5_4}
\mathcal O_5[n_3] = \mathcal O_5^\mathrm{symm}[n_3] \cdot \mathcal O_5^{c}[n_3] \cdot \mathcal O_5^{c\gamma}[\dd n_3] \cdot \mathcal O_5^{\gamma}[\dd n_3].
\end{align}
The explicit expressions of these four terms are summarized as
\begin{align}\label{3D:O5_s}
\mathcal O_5^\mathrm{symm}[n_3](012345) &= (-1)^{(\om_2\smile n_3+s_1\smile \al_4)(012345) + \om_2(013)\al_4(12345) + \om_2(012)\be_4(12345)},\\\label{3D:O5_c}
\mathcal O_5^{c}[n_3] &= (-1)^{n_3\smile_1 n_3 + \dd n_3\smile_2 n_3},\\\label{3D:O5_cg}
\mathcal O_5^{c\gamma}[\dd n_3] &= (-1)^{\dd n_3\smile_3 \dd n_3},\\\label{3D:O5_g}
\mathcal O_5^{\gamma}[\dd n_3](012345) \big |_{\be_4=0}
&= (-1)^{\dd n_3(02345)\dd n_3(01235) + \om_2(023) \left[\dd n_3(01245) + \dd n_3(01235) + \dd n_3(01234)\right]} \\\nonumber
&\quad\times i^{\dd n_3(01245) \dd n_3(01234) \text{ (mod 2)}}
\times (-i)^{\left[\dd n_3(12345)+\dd n_3(02345)+\dd n_3(01345)\right] \dd n_3(01235) \text{ (mod 2)}}.
\end{align}
We note that the expression of the last term $\mathcal O_5^{\gamma}[\dd n_3]$ is obtained under the assumption of $\be_4=0$ [see \eq{3D:be}]. This is true for $s_1=0$ (unitary $G_b$) or $n_1=0$. The calculation of generic $\mathcal O_5^{\gamma}$ with nonzero $\be_4$ is much more complicated (but the procedures are the same), so leave it for the future. By adding a coboundary $(-1)^{\dd(s_1\smile n_3+n_3\smile_3\dd n_3)}$ to the obstruction function and shifting $\nu_4\rightarrow \nu_4(-1)^{s_1\smile n_3+n_3\smile_3\dd n_3}$, we can simplify the above obstruction function to
\begin{align}\label{3D:O5}\nonumber
\mathcal O_5[n_3](012345)\big |_{\be_4=0} &= (-1)^{(\om_2\smile n_3 + n_3\smile_1 n_3 + n_3\smile_2 \dd n_3)(012345) + \om_2(013)\dd n_3(12345)}\\\nonumber
&\quad\times (-1)^{\dd n_3(02345)\dd n_3(01235) + \om_2(023) \left[\dd n_3(01245) + \dd n_3(01235) + \dd n_3(01234)\right]} \\
&\quad\times i^{\dd n_3(01245) \dd n_3(01234) \text{ (mod 2)}}
\times (-i)^{\left[\dd n_3(12345)+\dd n_3(02345)+\dd n_3(01345)\right] \dd n_3(01235) \text{ (mod 2)}}.
\end{align}
If we consider the special case of $\om_2=s_1=0$, then we have $\dd n_3=n_2\smile n_2$ from \eq{3D:dn3}. And the above obstruction reduces to the known result for $G_f=\Zf\times G_b$ \cite{WangGu2017,Kapustin2017,Morgan2018}.

Before calculating the obstruction function $\mathcal O_5[n_3]$ in detail, we note that we have checked numerically that $\mathcal O_5[n_3]$ \eq{3D:O5} satisfies $\dd \mathcal O_5 = 1$. It should be true, because the super hexagon equation \eq{3D:dF} implies a one higher dimensional equation involving one more vertex.

\subsubsection{Calculations of obstruction function $\mathcal O_5[n_3]$}

In this subsection, we would give explicit calculations of the four terms of the obstruction function $\mathcal O_5[n_3]$ in \eq{3D:O5_4}, with the assumption of $\be_4=0$.

The first term $\mathcal O_5^\mathrm{symm}[n_3]$ comes from the $U(\bar 01)$ symmetry action on $F(e,\bar 12,\bar 13,\bar 14,\bar 15)$ in the last term of \eq{3D:dF}. Since the homogeneous $\nu_4$ of the non-standard $F$ move is obtained from the standard one by a symmetry action \eq{3D:nu4_symm}], there is a sign difference $\mathcal O_5^\mathrm{symm}$ between them. Using the replacement $(g_0,g_1,g_2,g_3,g_4)\rightarrow(g_0^{-1}g_1,g_0^{-1}g_2,g_0^{-1}g_3,g_0^{-1}g_4,g_0^{-1}g_5)$ in the explicit expression \eq{3D:Osymm}, we have
\begin{align}\nonumber
\mathcal O_5^\mathrm{symm}[n_3](\bar 01,\bar 02,\bar 03,\bar 04,\bar 05)
&= (-1)^{(\om_2\smile n_3+s_1\smile \al_4)(\bar 01,\bar 12,\bar 23,\bar 34,\bar 45) + \om_2(\bar 01,\bar 13)\al_4(12345) + \om_2(\bar 01,\bar 12)\be_4(12345)},
\end{align}
where $\al_4$ and $\be_4$ are defined in \eqs{3D:al}{3D:be}. This is exactly the result \eq{3D:O5_s} claimed above. If we consider the special case of $\be_4=0$, the result is reduced to
\begin{align}\nonumber
\mathcal O_5^\mathrm{symm}[n_3](\bar 01,\bar 02,\bar 03,\bar 04,\bar 05)\big |_{\be_4=0}
&= (-1)^{(\om_2\smile n_3+s_1\smile \dd n_3)(\bar 01,\bar 12,\bar 23,\bar 34,\bar 45) + \om_2(\bar 01,\bar 13)\dd n_3(12345)}.
\end{align}

The second term $\mathcal O_5^{c}[n_3]$ is the complex fermion sign from reordering the fermion operators $\left(c_{ijkl}^{g_0^{-1}g_i}\right)^{n_3(ijkl)}$ in \eq{3D:dF}. To compare the complex fermion operators on the two sides of the super hexagon equation, we have to rearrange these operators. The reordering give us the complex fermion sign
\begin{align}
&\quad\mathcal O_5^{c}[n_3](012345) = (-1)^{(n_3\smile_1 n_3+ \dd n_3 \smile_2 n_3)(01234)}\\\nonumber
&= (-1)^{n_3(0345)n_3(0123) + n_3(0145)n_3(1234) + n_3(0125)n_3(2345)}\\\nonumber
&\quad\cdot (-1)^{\dd n_3(01234)n_3(0145) +\dd n_3(02345)n_3(0125) +\dd n_3(01234)n_3(1245) +\dd n_3(01345)n_3(1235) +\dd n_3(01234)n_3(2345) +\dd n_3(01245)n_3(2345)}.
\end{align}
This is the result claimed in \eq{3D:O5_c}. It is a generalization of the sign $\mathcal O_5[n_3]\big |_{s_1=\om_2=n_1=n_2=0}=(-1)^{Sq^2(n_3)}=(-1)^{n_3\smile_1 n_3}$ in the special group supercohomology theory in \Ref{GuWen2014}. The general group supercohomology theory for unitary $G_f=\Zf\times G_b$ also has this complex fermion sign \cite{WangGu2017}.

The third obstruction function term $\mathcal O_4^{c\gamma}[\dd n_2]$ originates from reordering the complex fermion and the Majorana fermions. It is also present in \Ref{WangGu2017} and has the form
\begin{align}
&\quad\mathcal O_5^{c\gamma}[\dd n_3] = (-1)^{\dd n_3\smile_3 \dd n_3}\\\nonumber
&= (-1)^{\dd n_3(01245)\dd n_3(01234)+\dd n_3(01235)\dd n_3(01345)+\dd n_3(02345)\dd n_3(01234) + \dd n_3(02345)\dd n_3(01245)+\dd n_3(01235)\dd n_3(12345)+\dd n_3(01345)\dd n_3(12345)}.
\end{align}
This is exactly the result claimed previously in \eq{3D:O5_cg}. This obstruction function is a functional of $\dd n_3$ (rather than $n_3$ directly), because the fermion parities of the $X$ operator and the complex fermion operator $c^\dagger c^\dagger c c$ only depend on $\dd n_3$.

In the rest of this subsection, we will deal with the the most complicated part $\mathcal O_5^\g[\dd n_3]$, with the assumption $\be_4=0$. Similar to the Majorana fermion phase factor in \Ref{WangGu2017}, this obstruction function takes value in $\{\pm1,\pm i\}$. Since the presence of the dangling Majorana fermions in $X$ depends on $\dd n_3$, we expect that $\mathcal O_5^\g[\dd n_3]$ is a functional of $\dd n_3$ only.

Let us denote the six $X$ operators in the standard super hexagon equation \eq{3D:dF} by
$X_{12345}=P_2 \big(\g_{345B}^{\bar 03}\big)^{\dd n_3(12345)}$,
$X_{02345}=P_4 \big(\g_{345B}^{\bar 03}\big)^{\dd n_3(02345)}$, 
$X_{01345}=P_3 \big(\g_{345B}^{\bar 03}\big)^{\dd n_3(01345)}$, 
$X_{01245}=P_5 \big(\g_{245B}^{\bar 02}\big)^{\dd n_3(01245)}$, 
$X_{01235}=P_4 \big(\g_{235B}^{\bar 02}\big)^{\dd n_3(01235)}$ and 
$X_{01234}=P_6 \big(\g_{234B}^{\bar 02}\big)^{\dd n_3(01234)}$. Here, the operator $P_i$ is the Majorana pairing projection operator of the corresponding $i$-th figure ($1\leq i\leq 6$) in the super hexagon equation shown in Figs.~\ref{fig:hexagon} and \ref{fig:hexagon_dual}. We will use the convention that the rightmost figure is the first one with projection operator $P_1$, and the other five figures are counted counterclockwise. (Note that this labelling convention is different from \Ref{WangGu2017}.) We also used the simpler notations for the Majorana fermion operators
\begin{align}
\g_{ijkB}^{\bar 0i} := \g_{ijkB}^{g_0^{-1}g_i}.
\end{align}
Since the super hexagon equation only involves vertices from $\langle 0\rangle$ to $\langle 5\rangle$, we can simplify the argument of $\dd n_3$ as
\begin{align}
\dd n_3(\hat i) &:= \dd n_3(01...\hat i...45),
\end{align}
where the number $i$ is removed in the argument. The obstruction function $\mathcal O_5^\g$ coming from Majorana fermions can be calculated from the $X$ operators as
\begin{align}\label{3D:O5g}
\mathcal O_5^\g[\dd n_3] &= \langle X_{12345}^\dagger X_{01345}^\dagger X_{01235}^\dagger X_{02345} X_{01245} X_{01234}\rangle\\\nonumber
&= \left\langle P_1
\big(\g_{345B}^{\bar 03}\big)^{\dd n_3(\hat 0)} P_2
\big(\g_{345B}^{\bar 03}\big)^{\dd n_3(\hat 2)} P_3
\big(\g_{235B}^{\bar 02}\big)^{\dd n_3(\hat 4)} P_4
P_4 \big(\g_{345B}^{\bar 03}\big)^{\dd n_3(\hat 1)}
P_5 \big(\g_{245B}^{\bar 02}\big)^{\dd n_3(\hat 3)}
P_6 \big(\g_{234B}^{\bar 02}\big)^{\dd n_3(\hat 5)}
P_1 \right\rangle.
\end{align}
The average is taken over the Majorana fermion pairing state of the rightmost figure (with projection operator $P_1$) in Figs.~\ref{fig:hexagon} and \ref{fig:hexagon_dual}. We also inserted $P_1$, which is 1 acting on the rightmost state, at the first and the last places of the operator string in \eq{3D:O5g}.

We can calculate \eq{3D:O5g} separately for different Majorana fermion configurations specified by $n_2$. Among the six dangling Majorana fermions of the six $X$ operators in \eq{3D:O5g}, only four of them are different:
\begin{align}
\left(\g_{345B}^{\bar 03}, \g_{245B}^{\bar 02}, \g_{235B}^{\bar 02}, \g_{234B}^{\bar 02}\right).
\end{align}
We can use the quadruple of their number
\begin{align}
[\dd n_3(\hat 0)+\dd n_3(\hat 1)+\dd n_3(\hat 2),\ \ \dd n_3(\hat 3),\ \ \dd n_3(\hat 4),\ \ \dd n_3(\hat 5)]\ \ \ \text{(mod 2)}
\end{align}
to indicate the presence or absence of the four Majorana fermions in \eq{3D:O5g} respectively. Each element of the quadruple corresponds to the Majorana fermion parity change of one or several $F$ moves. In total, there are $2^4/2=8$ different possibilities for the Majorana fermion parity changes (see the first column of \tab{tab:O5g}), for the total Majorana fermion parity of the six $F$ moves should be even. For the 8 different cases, we can calculate $\mathcal O_5^\g[\dd n_3]$ separately.

\begin{table}[ht]
\centering
\begin{tabular}{|c|c|c|c|c|}
\hline
$P_f^\g$ changes 
& expression of $\mathcal O_5^\g[\dd n_3]$ 
& $P_4$ 
& $P_5$ 
& $\mathcal O_5^\g[\dd n_3]$ \\
\hline
\hline
$(0,0,0,0)$
& $\left\langle \big(\g_{345B}^{\bar 03}\big)^{\dd n_3(\hat 0)+\dd n_3(\hat 2)+\dd n_3(\hat 1)} \right\rangle$ 
& $1$ 
& $1$ 
& $1$ \\
\hline
$(1,0,0,1)$ 
& $\left\langle \big(\g_{345B}^{\bar 03}\big)^{\dd n_3(\hat 0)+\dd n_3(\hat 2)} P_4 \big(\g_{345B}^{\bar 03}\big)^{\dd n_3(\hat 1)} P_5 \g_{234B}^{\bar 02} \right\rangle$ 
& $P_{234A,234B}^{\bar 02,\bar 02}$ 
& $P_{345B,234A}^{\bar 03,\bar 02}$ 
& $(-1)^{\om_2(\bar 02,\bar 23)}$ \\
\hline
$(1,0,1,0)$ 
& $\left\langle \big(\g_{345B}^{\bar 03}\big)^{\dd n_3(\hat 0)+\dd n_3(\hat 2)} \g_{235B}^{\bar 02} \big(\g_{345B}^{\bar 03}\big)^{\dd n_3(\hat 1)} P_5 \right\rangle$ 
& $1$ 
& $P_{235B,345B}^{\bar 02,\bar 03}$ 
& $(-i)(-1)^{\dd n_3(\hat 1)+\om_2(\bar 02,\bar 23)}$ \\
\hline
$(0,0,1,1)$ 
& $\left\langle \big(\g_{345B}^{\bar 03}\big)^{\dd n_3(\hat 0)+\dd n_3(\hat 2)} \g_{235B}^{\bar 02} P_4 \big(\g_{345B}^{\bar 03}\big)^{\dd n_3(\hat 1)} P_5 \g_{234B}^{\bar 02} \right\rangle$ 
& $P_{234A,234B}^{\bar 02,\bar 02}$ 
& $P_{235B,234A}^{\bar 02,\bar 02}$ 
& $(-1)^{\dd n_3(\hat 1)}$ \\
\hline
$(1,1,0,0)$ 
& $\left\langle \big(\g_{345B}^{\bar 03}\big)^{\dd n_3(\hat 0)+\dd n_3(\hat 2)} P_4 \big(\g_{345B}^{\bar 03}\big)^{\dd n_3(\hat 1)} P_5 \g_{245B}^{\bar 02} \right\rangle$ 
& $P_{245A,245B}^{\bar 02,\bar 02}$ 
& $P_{345B,245A}^{\bar 03,\bar 02}$ 
& $(-1)^{\om_2(\bar 02,\bar 23)}$ \\
\hline
$(0,1,0,1)$ 
& $\left\langle \big(\g_{345B}^{\bar 03}\big)^{\dd n_3(\hat 0)+\dd n_3(\hat 2)} P_4 \big(\g_{345B}^{\bar 03}\big)^{\dd n_3(\hat 1)} P_5 \g_{245B}^{\bar 02} \g_{234B}^{\bar 02} \right\rangle$ 
& $P_{234A,234B}^{\bar 02,\bar 02} \atop P_{245A,245B}^{\bar 02,\bar 02}$ 
& $P_{234A,245A}^{\bar 02,\bar 02}$ 
& $i$ \\
\hline
$(0,1,1,0)$ 
& $\left\langle \big(\g_{345B}^{\bar 03}\big)^{\dd n_3(\hat 0)+\dd n_3(\hat 2)} \g_{235B}^{\bar 02} P_4 \big(\g_{345B}^{\bar 03}\big)^{\dd n_3(\hat 1)} P_5 \g_{245B}^{\bar 02} \right\rangle$ 
& $P_{245A,245B}^{\bar 02,\bar 02}$ 
& $P_{235B,245A}^{\bar 02,\bar 02}$ 
& $(-1)^{\dd n_3(\hat 1)}$ \\
\hline
$(1,1,1,1)$ 
& $\left\langle \big(\g_{345B}^{\bar 03}\big)^{\dd n_3(\hat 0)+\dd n_3(\hat 2)} \g_{235B}^{\bar 02} P_4 \big(\g_{345B}^{\bar 03}\big)^{\dd n_3(\hat 1)} P_5 \g_{245B}^{\bar 02} \g_{234B}^{\bar 02} \right\rangle$ 
& $P_{234A,234B}^{\bar 02,\bar 02} \atop P_{245A,245B}^{\bar 02,\bar 02}$ 
& $P_{235B,345B}^{\bar 02,\bar 03} \atop P_{234A,245A}^{\bar 02,\bar 02}$ 
& $(-1)^{\dd n_3(\hat 1)+\om_2(\bar 02,\bar 23)}$ \\
\hline
\end{tabular}
\caption{Calculations of $\mathcal O_5^\gamma[\dd n_3]$ for all possible Kitaev chain configurations in the super hexagon equation shown in Figs.~\ref{fig:hexagon} and \ref{fig:hexagon_dual}. The first column is the Majorana fermion parity change quadruple $[\dd n_3(\hat 0)+\dd n_3(\hat 1)+\dd n_3(\hat 2), \dd n_3(\hat 3), \dd n_3(\hat 4), \dd n_3(\hat 5)]$ (mod 2). There are in total 8 different cases. The second column is a simplified version of \eq{3D:O5g}. The third and forth columns are the Majorana pairing projection operators we used in calculation. And the last column is the final result of $\mathcal O_5^\gamma[\dd n_3]$, which can be summarized by \eq{3D:O5g_}.}
\label{tab:O5g}
\end{table}

Let us consider the third case (the fourth row in \tab{tab:O5g}) as an example. The Majorana fermion parity change quadruple $[\dd n_3(\hat 0)+\dd n_3(\hat 1)+\dd n_3(\hat 2), \dd n_3(\hat 3), \dd n_3(\hat 4), \dd n_3(\hat 5)]$ (mod 2) corresponds to $(1,0,1,0)$. So the dangling Majorana fermions present in \eq{3D:O5g} are $\g_{345B}^{\bar 03}$ and $\g_{235B}^{\bar 02}$. We can expand the projection operators $P_i$ to Majorana fermion operators. Since $\g_{345B}^{\bar 03}$ and $\g_{235B}^{\bar 02}$ are paired inside the tetrahedron $\langle 2345\rangle$ of the lower right figure (with projection operator $P_5$) in Figs.~\ref{fig:hexagon} and \ref{fig:hexagon_dual}, we can consider only their pairing projection operator in $P_5$ [recall \eq{proj}]
\begin{align}
P_{235B,345B}^{\bar 02,\bar 03} = \frac{1}{2} \left[1-(-1)^{\om_2(\bar 02,\bar 23)} i \g_{235B}^{\bar 02} \g_{345B}^{\bar 03}\right].
\end{align}
We can choose the second term in the above equation and all other projection operators $P_i$ with $i\neq 5$ to be 1. Then the obstruction function \eq{3D:O5g} can be expressed as
\begin{align}
\mathcal O_5^\g[\dd n_3](012345) \big|_{(1,0,1,0)}
&= \left\langle \big(\g_{345B}^{\bar 03}\big)^{\dd n_3(\hat 0)+\dd n_3(\hat 2)} \g_{235B}^{\bar 02} \big(\g_{345B}^{\bar 03}\big)^{\dd n_3(\hat 1)} P_5 \right\rangle \\\label{3D:O5g:ex1}
&= \left\langle \big(\g_{345B}^{\bar 03}\big)^{\dd n_3(\hat 0)+\dd n_3(\hat 2)} \g_{235B}^{\bar 02} \big(\g_{345B}^{\bar 03}\big)^{\dd n_3(\hat 1)} (-1)^{\om_2(\bar 02,\bar 23)} \big(-i\g_{235B}^{\bar 02} \g_{345B}^{\bar 03}\big) \right\rangle\\\label{3D:O5g:ex2}
&= (-i)(-1)^{\dd n_3(\hat 1)+\om_2(\bar 02,\bar 23)}.
\end{align}
In such way, \eq{3D:O5g:ex1} contains all Majorana fermions even times. After reordering these Majorana fermion operators, they all square to one. Finally we obtain \eq{3D:O5g:ex2} as the obstruction function for this case.

We can similarly calculate $\mathcal O_5^\g[\dd n_3]$ for all the 8 cases of Majorana fermion parity changes. The information we need in the calculation is shown in \tab{tab:O5g}. Sometimes, we need not only $P_5$ but also $P_4$. And in $P_4$ and $P_5$, we may need to many Majorana pairing projection operators \eq{proj}. They are shown in the fourth and fifth columns of \tab{tab:O5g}. The final results of $\mathcal O_5^\g$ are shown in the last column of \tab{tab:O5g}. They can be summarized into two equivalent expressions as ($\mathcal O_5^\g$ is a functional of $\dd n_3$ only):
\begin{align}\nonumber\label{3D:O5g_}
\mathcal O_5^\g[\dd n_3](012345)\big |_{\be_4=0} &= (-1)^{\dd n_3(\hat 1)\dd n_3(\hat 4) + \om_2(\bar 02,\bar 23) \left[\dd n_3(\hat 3) + \dd n_3(\hat 4) + \dd n_3(\hat 5)\right]} \\\nonumber
&\quad\times i^{\dd n_3(\hat 3) \left[1-\dd n_3(\hat 4)\right] \dd n_3(\hat 5) \text{ (mod 2)}}
\times (-i)^{\left[1-\dd n_3(\hat 3)\right] \dd n_3(\hat 4) \left[1-\dd n_3(\hat 5)\right] \text{ (mod 2)}}\\\nonumber
&= (-1)^{\dd n_3(\hat 1)\dd n_3(\hat 4) + \om_2(\bar 02,\bar 23) \left[\dd n_3(\hat 3) + \dd n_3(\hat 4) + \dd n_3(\hat 5)\right]} \\
&\quad\times i^{\dd n_3(\hat 3) \dd n_3(\hat 5) \text{ (mod 2)}}
\times (-i)^{\left[\dd n_3(\hat 0)+\dd n_3(\hat 1)+\dd n_3(\hat 2)\right] \dd n_3(\hat 4) \text{ (mod 2)}}.
\end{align}
Note that the $\dd n_3$ terms in the exponent of $(\pm i)$ should be understood as taking mod 2 values (can only be 0 or 1). They correspond to the third and sixth cases in \tab{tab:O5g}. The term $(-1)^{\dd n_3(\hat 1)}$ appears in the last column of \tab{tab:O5g} if and only if $\dd n_3(\hat 4)=1$. And the term $(-1)^{\om_2(\bar 02,\bar 23)}$ appears whenever $\dd n_3(\hat 3) + \dd n_3(\hat 4) + \dd n_3(\hat 5)=1$ (mod 2). That is the origin of the $(-1)$ terms in \eq{3D:O5g_}. Note that the exponent of $\pm i$ for the first (second) expression of $\mathcal O_5^\g[\dd n_3]$ is a cubic (quadratic) $\Z_2$-valued function of $n_3$. The expression of $\mathcal O_5^\g[\dd n_3]$ is the same as the claimed \eq{3D:O5_g}.

If we consider the unitary symmetry group $G_f=\Zf\times G_b$ ($\om_2=s_1=0$), the Majorana obstruction function \eq{3D:O5g_} will be reduced to an expression as a functional of $n_2^2$ (because of $\dd n_3=n_2\smile n_2$ in this case). Although the expression is different  from the the result in \Ref{WangGu2017} in appearance, one can show that they are exactly the same after some calculations. It is also the same as the (3+1)D spin cobordism result \cite{Kapustin2017,Morgan2018}.


\subsection{Boundary ASPT states}

In the above discussions in this section, we constructed 3D FSPT states by decorating several layers of fermion modes to the BSPT states. The decorations of Kitaev chains and complex fermions are specified by two $\Z_2$-valued cochains $n_2$ and $n_3$.

In this section, we will show that some of the $n_2$ and $n_3$ data in fact correspond to FSPT states belonging to the trivial FSPT phase. This data trivialization can be understood by investigating the boundary ASPT states on the 2D boundary of the 3D FSPT bulk.

\subsubsection{Boundary ASPT states in $\Gamma^2$ with $p+ip$ superconductors}
\label{sec:ASPT:pip1}

Similar to the 2D case, we will show in this section that the 3D FSPT data $n_2=\om_2$ is trivialized by boundary ASPT state. The 2D boundary is in fact a state with (one layer) $p+ip$ chiral superconductor decorations.

Since there is no fixed-point wave function construction for 2D $p+ip$ superconductor on discrete lattice \cite{KapustinFidkowski2018}, its decoration is very different from the Kitaev chains and complex fermions. In the boundary ASPT state construction, we put $p+ip$ superconductor domains around each vertex $i$ of the triangulation lattice, and use symmetric mass terms to gap out the edge modes and glue them together. 
In the following, we first discuss the procedure of decorating $p+ip$ superconductors. Then we propose a field theory description of symmetrically gapping out the chiral Majorana modes along the edge.

The detailed decorations are as follows. For an arbitrary triangulation of the 2D boundary of a 3D FSPT state, we put a $p+ip$ superconductors on a disk centered at each vertex $i$. 
The 2D disks of the superconductors are always gapped. And on the boundary, there are right-moving chiral Majorana modes [see the green arrows in \eq{fig:2D:012_}]. We denote these chiral Majorana modes by $\psi_{i,R}^{g_i}$. 
A picture of the chiral Majorana modes in a triangle of the 2D boundary of 3D FSPT bulk is
\begin{align}\label{fig:2D:012_}
\tikzfig{scale=2.3}{
\coordinate (g1) at (0,0);
\coordinate (g2) at (1,0);
\coordinate (g3) at (1/2,1.732/2);
\coordinate (center) at (1/2,1/2/1.732);
\pgfmathsetmacro{\s}{.11}
\coordinate (11) at (3/4-\s*1.732/2, 1.732/4-\s/2);
\coordinate (12) at (3/4+\s*1.732/2, 1.732/4+\s/2);
\coordinate (21) at (1/4+\s*1.732/2, 1.732/4-\s/2);
\coordinate (22) at (1/4-\s*1.732/2, 1.732/4+\s/2);
\coordinate (31) at (1/2, 0+\s);
\coordinate (32) at (1/2, 0-\s);
%
%
\draw [->-=.6,thick] (g1) node[scale=1,below left]{$g_0$} -- (g2);
\draw [->-=.6,thick] (g2) node[scale=1,below right]{$g_1$} -- (g3) node[scale=1,above]{$g_2$};
\draw [->-=.6,thick] (g1) -- (g3);
%
\pgfmathsetmacro{\r}{.35};
\pgfmathsetmacro{\theta}{30};
\draw[->,>=stealth',semithick,black!15!green] (-\theta:\r) arc (-\theta:{60+\theta}:\r);
\draw[->,>=stealth',semithick,black!15!green] ([shift=(0:1)]{120-\theta}:\r) arc ({120-\theta}:{180+\theta}:\r);
\draw[->,>=stealth',semithick,black!15!green] ([shift=(60:1)]{240-\theta}:\r) arc ({240-\theta}:{300+\theta}:\r);
\node[black!15!green,scale=.8,left] at ([shift=(0:0)]{60+\theta}:\r) {$\psi_{0,R}^{g_0}$};
\node[black!15!green,scale=.8,right] at ([shift=(0:1)]{180+\theta}:\r) {$\psi_{1,R}^{g_1}$};
\node[black!15!green,scale=.8,right] at ([shift=(60:1)]{-60+\theta}:\r) {$\psi_{2,R}^{g_2}$};
%
}
\end{align}

The decorated state should be symmetric under $G_b$-action. If the chiral Majorana modes around vertex $i$ with $g_i=e$ are $\psi_{i,R}^e$, then for generic $g_i\in G_b$, the decorated $p+ip$ superconductors should have chiral Majorana modes $\psi_{i,g_i(R);\al}^{g_i}$ according to the symmetry transformation rules:
\begin{align}
\label{2D:symm:R}
U(g) \psi_{i,R}^\s U(g)^\dagger &= (-1)^{\om_2(g,\s)} \psi_{i,g(R)}^{g\s},\\
\label{2D:symm:L}
U(g) \psi_{i,L}^\s U(g)^\dagger &= (-1)^{\om_2(g,\s)+s_1(g)} \psi_{i,g(L)}^{g\s}.
\end{align}
Here, we denote $g_i(R)=R$ if $s_1(g_i)=0$ ($g_i$ is unitary), and $g_i(R)=L$ if $s_1(g_i)=1$ ($g_i$ is antiunitary). The physical meaning is that the unitary symmetry $U(g)$ only changes the group element label of chiral modes from $g_i$ to $gg_i$. However, time reversal symmetry would transform $p+ip$ superconductors to $p-ip$ superconductors, and vice versa. So the right-moving and left-moving chiral Majorana modes on the edge are switched.

The next step in our construction is to gap out all the chiral Majorana modes to obtain a fully gapped boundary state. Consider the interface between two $p+ip$ domains labelled by $g_0$ and $g_1$ shown below
\begin{align}\label{fig:mass}
\tikzfig{scale=2.3}{
\coordinate (g1) at (0,0);
\coordinate (g2) at (1,0);
\draw[line width=1] (0,0)--(0,.04);
\draw[line width=1] (1,0)--(1,.04);
\draw [->-=.75,thick] (g1) node[scale=1,below]{$g_0$} -- (g2) node[scale=1,below]{$g_1$};
\draw[thick,red,densely dotted] (.5,-.4)--(.5,.4);
\node[red,above,scale=.8] at (.5,.4) {effective Kitaev chain};
\pgfmathsetmacro{\r}{.4};
\pgfmathsetmacro{\theta}{60};
\draw[->,>=stealth',semithick,black!15!green] (-\theta:\r) arc (-\theta:{0+\theta}:\r);
\draw[->,>=stealth',semithick,black!15!green] ([shift=(0:1)]{180-\theta}:\r) arc ({180-\theta}:{180+\theta}:\r);
\node[black!15!green,scale=.8,below] at ([shift=(0:0)]{-\theta}:\r) {$\psi_{0,R;\alpha}^{g_0}$};
\node[black!15!green,scale=.8,below] at ([shift=(0:1)]{180+\theta}:\r) {$\psi_{1,R;\alpha}^{g_1}$};
}.
\end{align}
There is a necessary condition for gapping out the chiral Majorana modes $\psi_{0,R/L}^{g_0}$ and $\psi_{1,R/L}^{g_1}$ shown by green arrows above: the number of net chiral modes along the direction dual to $\langle 01\rangle$ is zero. If $s_1(g_0^{-1}g_1)=1$, then one of the two right-moving chiral Majorana modes is reversed. There are two chiral modes moving in the same direction along the link dual to $\langle 01\rangle$. This is illegal for we want the boundary to be a gapped state. So we conclude the boundary ASPT state with $p+ip$ superconductors is impossible if there are some antiunitary symmetries in $G_b$. In the following discussions of ASPT state with $p+ip$ superconductors, we will always assume $G_b$ is unitary.


Let us introduce explicitly the symmetric mass terms to gap out the chiral Majorana modes. Similar to the previous discussions, we can first assume the group element labels of link $\langle 01\rangle$ are $e$ and $g_0^{-1}g_1$. The standard mass terms along the edge dual to link $\langle 01\rangle$ is
\begin{align}\label{2D:mass_std}
H_\mathrm{std}^{\langle 01\rangle}
=
im \int \dd x \psi_{0,R}^{e}(x) \psi_{1,R}^{g_0^{-1}g_1}(x).
\end{align}
The non-standard mass terms are obtained from the above standard ones by a symmetry action:
\begin{align}\label{2D:mass}
H_\mathrm{mass}^{\langle 01\rangle}
= U(g_0) H_\mathrm{std}^{\langle 01\rangle} U(g_0)^\dagger
= 
(-1)^{\om_2(g_0,g_0^{-1}g_1)} im \int \dd x \psi_{0,R}^{g_0}(x) \psi_{1,R}^{g_1}(x),
\end{align}
where we used the symmetry transformation rule \eq{2D:symm:R} of Majorana modes. Note that there is no minus sign from the imaginary unit $i$, for $G_b$ should be unitary according to the previous discussions. The mass terms constructed above are of course $G_b$-symmetric, since they are all obtained from the standard mass terms by a symmetry action.

It is well known that changing the sign of the mass $m$ of two counter-propagating chiral Majorana modes would induce a phase transition from a non-topological superconductor to a topological superconductor \cite{Kitaev2001}. If the resulting gapped edge is a 1D topological superconductor, there is effectively a Kitaev chain going along the direction of the chiral modes [see the dashed red line in \eq{fig:mass}]. We can assume the standard mass terms \eq{2D:mass_std} with $m>0$ correspond to the trivial gapped phase. Then the true mass terms \eq{2D:mass} induce effective Kitaev chains going though the link $\langle 01\rangle$. The number of effective Kitaev chains equals to the number of negative mass terms:
\begin{align}\label{eff_Maj}
\om_2(g_0,g_0^{-1}g_1).
\end{align}
If we consider the triangle $\langle 012\rangle$ with mass terms $H_\mathrm{mass}^{\langle 12\rangle}$, $H_\mathrm{mass}^{\langle 02\rangle}$ and $H_\mathrm{mass}^{\langle 01\rangle}$ on the boundary, the (mod 2) number of effective Kitaev chains going though the three links is
\begin{align}\label{2D:num_p}
\om_2(g_0^{-1}g_1,g_1^{-1}g_2) = \om_2(g_1,g_1^{-1}g_2) + \om_2(g_0,g_0^{-1}g_2) + \om_2(g_0,g_0^{-1}g_1),
\end{align}
where we have used $(\dd\om_2)(g_0,g_0^{-1}g_1,g_1^{-1}g_2)=0$ (mod 2).

If $\om_2$ is nontrivial in \eq{2D:num_p}, there may be odd number of Kitaev chains going into the triangle $\langle 012\rangle$. Since our state is on the boundary of a 3D FSPT bulk, we can connect the Kitaev chain to the 3D bulk FSPT state [see \eq{3D:n2}]. The number of Kitaev chains going though a triangle of 3D FSPT state is exactly the $n_2$ data which is discussed in detail in section~\ref{sec:3D:Maj}. So we have $n_2=\om_2$ to have a gapped state (including both the boundary and the bulk). Since we have constructed a gapped symmetric (ASPT) state without topological order on the boundary of the 3D FSPT state, we conclude that the bulk FSPT with $n_2=\om_2$ is trivial. This is the origin of the trivialization group
\begin{align}\label{Gamma2}
\Gamma^2 = \{\omega_2\smile n_0 \in H^2(G_b,\Z_2) | n_0\in H^0(G_b,\Z_T) \},
\end{align}
for 3D FSPT phases claimed in \eq{3D:Gamma}. We note that $p+ip$ superconductor is incompatible with time reversal symmetry. So there is no $\Gamma^2$ trivialization if $G_b$ contains anti-unitary symmetry. In other words, \eq{Gamma2} is equivalent to
\begin{align}
\Gamma^2 =
\begin{cases}
\{\omega_2 \in H^2(G_b,\Z_2) \}, \quad G_b\text{ is unitary}\\
0, \quad\quad\quad\quad\quad\quad\quad\quad\ \  G_b\text{ contains anti-unitary},
\end{cases}
\end{align}
because $H^0(G_b,\Z_T)=\Z$ if $G_b$ is unitary, and $H^0(G_b,\Z_T)=0$ if $G_b$ contains anti-unitary symmetries.

\subsubsection{Boundary ASPT states in $\Gamma^3$ with $p+ip$ superconductors}

The ASPT states in \eq{Gamma2} is realized as one layer of 2D $p+ip$ superconductor [$n_0=1 \in H^0(G_b,\Z_T)$] on the boundary of 3D FSPT states with $n_2=\om_2$ Majorana chain decorations. The fluctuating Majorana chains (the $n_2$ data) in the 3D bulk become the effective Majorana chains along the 1D domain walls of 2D boundary $p+ip$ superconductors. Therefore, there is a gapped symmetric boundary without topological orders and the 3D bulk is trivialized.

The ASPT state considered in this section has two layers of 2D $p+ip$ superconductors [$n_0=2\in H^0(G_b,\Z_T)$]. So according to \eq{Gamma2}, there is no $\Gamma^2$ trivialization. However, we will show below that there is a $\Gamma^3$ trivialization of 3D FSPT state with $n_3=\om_2\smile_1\om_2$ complex fermion decorations. This is related to the fact that the $F$ move of $n_0=2$ $p+ip$ superconductors on the 2D boundary changes fermion parity by $\Delta P_f^\psi(F)=(-1)^{\om_2\smile_1\om_2}$. This is similar to the $\Gamma^3$ trivialization related to $n_1$ (2D boundary Majorana chain decorations) discussed in the next section.

\begin{figure}[ht]
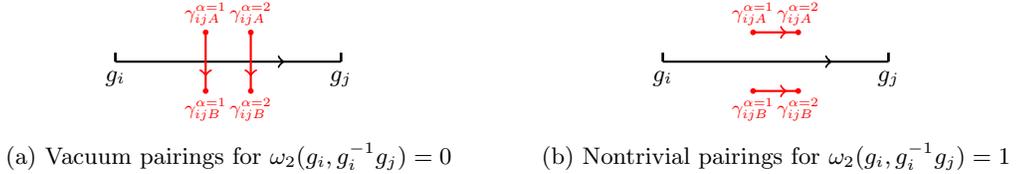

\centering
\begin{subfigure}[h]{.4\textwidth}
\centering
$
\tikzfig{scale=3}{
\coordinate (g1) at (0,0);
\coordinate (g2) at (1,0);
\draw[line width=1] (0,0)--(0,.04);
\draw[line width=1] (1,0)--(1,.04);
\draw [->-=.75,thick] (g1) node[scale=1,below]{$g_i$} -- (g2) node[scale=1,below]{$g_j$};
\coordinate (1A) at (.4,.13);
\coordinate (1B) at (.4,-.13);
\coordinate (2A) at (.6,.13);
\coordinate (2B) at (.6,-.13);
\draw[->-=.75,red,thick](1A) node[scale=.8,above]{$\g_{ijA}^{\al=1}$} -- (1B) node[scale=.8,below]{$\g_{ijB}^{\al=1}$};
\draw[->-=.75,red,thick](2A) node[scale=.8,above]{$\g_{ijA}^{\al=2}$}--(2B) node[scale=.8,below]{$\g_{ijB}^{\al=2}$};
\node[circle,draw=red,fill=red,scale=.2] at (1A) {};
\node[circle,draw=red,fill=red,scale=.2] at (1B) {};
\node[circle,draw=red,fill=red,scale=.2] at (2A) {};
\node[circle,draw=red,fill=red,scale=.2] at (2B) {};
}
$
\caption{Vacuum pairings for $\om_2(g_i,g_i^{-1}g_j)=0$}
\label{fig:n0=2_1}
\end{subfigure}
\begin{subfigure}[h]{.4\textwidth}
\centering
$
\tikzfig{scale=3}{
\coordinate (g1) at (0,0);
\coordinate (g2) at (1,0);
\draw[line width=1] (0,0)--(0,.04);
\draw[line width=1] (1,0)--(1,.04);
\draw [->-=.75,thick] (g1) node[scale=1,below]{$g_i$} -- (g2) node[scale=1,below]{$g_j$};
\coordinate (1A) at (.4,.13);
\coordinate (1B) at (.4,-.13);
\coordinate (2A) at (.6,.13);
\coordinate (2B) at (.6,-.13);
\draw[->-=.75,red,thick](1A) node[scale=.8,above]{$\g_{ijA}^{\al=1}$} -- (2A) node[scale=.8,above]{$\g_{ijA}^{\al=2}$};
\draw[->-=.75,red,thick](1B) node[scale=.8,below]{$\g_{ijB}^{\al=1}$}--(2B) node[scale=.8,below]{$\g_{ijB}^{\al=2}$};
\node[circle,draw=red,fill=red,scale=.2] at (1A) {};
\node[circle,draw=red,fill=red,scale=.2] at (1B) {};
\node[circle,draw=red,fill=red,scale=.2] at (2A) {};
\node[circle,draw=red,fill=red,scale=.2] at (2B) {};
}
$
\caption{Nontrivial pairings for $\om_2(g_i,g_i^{-1}g_j)=1$}
\label{fig:n0=2_2}
\end{subfigure}
\caption{There are $n_0=2$ effective Majorana chains ($\al=1,2$) going though link $\langle ij\rangle$ iff $\om_2(g_i,g_i^{-1}g_j)=1$. The vacuum and nontrivial pairing directions are shown in figures (a) and (b). We note that $\al=1,2$ effective Majorana chains should be understood as the stacking in the direction perpendicular to the paper. They are not related to the branching structure (black arrows).}
\label{fig:n0=2}
\end{figure}

The setup of $n_0=2$ ASPT state is similar to the state discussed in Section~\ref{sec:ASPT:pip1}. Near the vertex $i$ of the space triangulation, we put a disk of two layers of $p+ip$ superconductors labelled by $g_i$. Along the boundary, we have two chiral Majorana modes indicated by green curves in, for example, \eq{fig:2D:012_} and \eq{fig:mass}. After adding symmetric mass terms along the line dual to link $\langle ij\rangle$, we can gap out the chiral Majorana modes. If the sign of mass $(-1)^{\om_2(g_i,g_i^{-1}g_j)}$ [see the discussions above \eq{eff_Maj}] is negative, there are $n_0=2$ effective Majorana chains going though the link $\langle ij\rangle$. Since we want to analyze the fermion parities, we have to pair up the Majorana fermions in the effective Majorana chains. If $\om_2(g_i,g_i^{-1}g_j)=1$, there are $n_0=2$ effective Majorana chains, leaving two Majorana fermions on each side of link $\langle ij\rangle$. We can pair them up as shown in \fig{fig:n0=2_2} (always from $\al=1$ to $\al=2$ on each side). Note that the vacuum pairings for $\om_2(g_i,g_i^{-1}g_j)=0$ are the standard $A$ to $B$ pairings (see \fig{fig:n0=2_1}). So the fermion parity of the nontrivial pairings is always odd (compared to the vacuum pairings), as the small loop with length four is non-Kasteleyn-oriented.

\begin{figure}[ht]
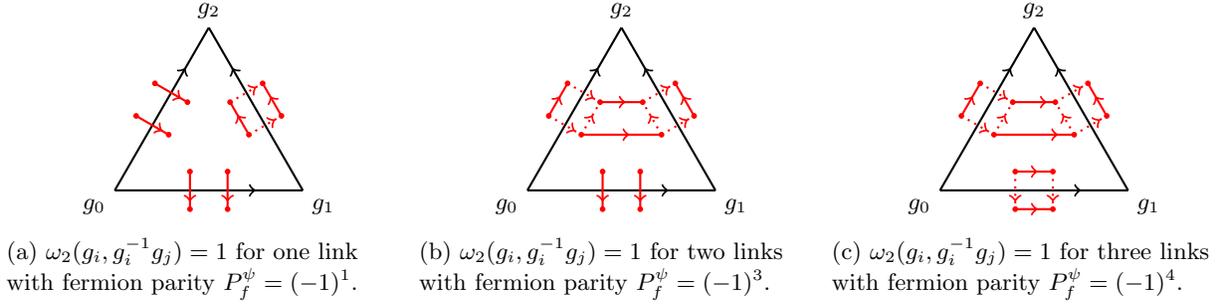

\centering
\begin{subfigure}[h]{.3\textwidth}
\centering
$
\tikzfig{scale=2.5}{
\coordinate (g1) at (-1/2,{-1/2/sqrt(3)});
\coordinate (g2) at (1/2,{-1/2/sqrt(3)});
\coordinate (g3) at (0,{sqrt(3)/2-1/2/sqrt(3)});
\draw [->-=.75,thick] (g1) node[scale=1,below left]{$g_0$} -- (g2);
\draw [->-=.75,thick] (g2) node[scale=1,below right]{$g_1$} -- (g3) node[scale=1,above]{$g_2$};
\draw [->-=.75,thick] (g1) -- (g3);
\pgfmathsetmacro{\x}{.1}
\begin{scope}[rotate=0]
\coordinate (01_A1) at (-\x,{-1/2/sqrt(3)+.1});
\coordinate (01_A2) at (\x,{-1/2/sqrt(3)+.1});
\coordinate (01_B1) at (-\x,{-1/2/sqrt(3)-.1});
\coordinate (01_B2) at (\x,{-1/2/sqrt(3)-.1});
\node[circle,draw=red,fill=red,scale=.2] at (01_A1) {};
\node[circle,draw=red,fill=red,scale=.2] at (01_A2) {};
\node[circle,draw=red,fill=red,scale=.2] at (01_B1) {};
\node[circle,draw=red,fill=red,scale=.2] at (01_B2) {};
\draw [->-=.8,thick,red] (01_A1) -- (01_B1);
\draw [->-=.8,thick,red] (01_A2) -- (01_B2);
\end{scope}
\begin{scope}[rotate=120]
\coordinate (01_A1) at (-\x,{-1/2/sqrt(3)+.1});
\coordinate (01_A2) at (\x,{-1/2/sqrt(3)+.1});
\coordinate (01_B1) at (-\x,{-1/2/sqrt(3)-.1});
\coordinate (01_B2) at (\x,{-1/2/sqrt(3)-.1});
\node[circle,draw=red,fill=red,scale=.2] at (01_A1) {};
\node[circle,draw=red,fill=red,scale=.2] at (01_A2) {};
\node[circle,draw=red,fill=red,scale=.2] at (01_B1) {};
\node[circle,draw=red,fill=red,scale=.2] at (01_B2) {};
\draw [->-=.6,thick,red] (01_A1) -- (01_A2);
\draw [->-=.6,thick,red] (01_B1) -- (01_B2);
\draw [->-=.8,thick,dotted,red] (01_A1) -- (01_B1);
\draw [->-=.8,thick,dotted,red] (01_A2) -- (01_B2);
\end{scope}
\begin{scope}[rotate=240]
\coordinate (01_A1) at (-\x,{-1/2/sqrt(3)+.1});
\coordinate (01_A2) at (\x,{-1/2/sqrt(3)+.1});
\coordinate (01_B1) at (-\x,{-1/2/sqrt(3)-.1});
\coordinate (01_B2) at (\x,{-1/2/sqrt(3)-.1});
\node[circle,draw=red,fill=red,scale=.2] at (01_A1) {};
\node[circle,draw=red,fill=red,scale=.2] at (01_A2) {};
\node[circle,draw=red,fill=red,scale=.2] at (01_B1) {};
\node[circle,draw=red,fill=red,scale=.2] at (01_B2) {};
\draw [->-=.8,thick,red] (01_B1) -- (01_A1);
\draw [->-=.8,thick,red] (01_B2) -- (01_A2);
\end{scope}
}
$
\caption{$\om_2(g_i,g_i^{-1}g_j)=1$ for one link\\with fermion parity $P_f^\psi=(-1)^1$.}
\label{fig:n0=2_ijk_1}
\end{subfigure}
\begin{subfigure}[h]{.3\textwidth}
\centering
$
\tikzfig{scale=2.5}{
\coordinate (g1) at (-1/2,{-1/2/sqrt(3)});
\coordinate (g2) at (1/2,{-1/2/sqrt(3)});
\coordinate (g3) at (0,{sqrt(3)/2-1/2/sqrt(3)});
\draw [->-=.75,thick] (g1) node[scale=1,below left]{$g_0$} -- (g2);
\draw [->-=.75,thick] (g2) node[scale=1,below right]{$g_1$} -- (g3) node[scale=1,above]{$g_2$};
\draw [->-=.75,thick] (g1) -- (g3);
\pgfmathsetmacro{\x}{.1}
\begin{scope}[rotate=0]
\coordinate (01_A1) at (-\x,{-1/2/sqrt(3)+.1});
\coordinate (01_A2) at (\x,{-1/2/sqrt(3)+.1});
\coordinate (01_B1) at (-\x,{-1/2/sqrt(3)-.1});
\coordinate (01_B2) at (\x,{-1/2/sqrt(3)-.1});
\node[circle,draw=red,fill=red,scale=.2] at (01_A1) {};
\node[circle,draw=red,fill=red,scale=.2] at (01_A2) {};
\node[circle,draw=red,fill=red,scale=.2] at (01_B1) {};
\node[circle,draw=red,fill=red,scale=.2] at (01_B2) {};
\draw [->-=.8,thick,red] (01_A1) -- (01_B1);
\draw [->-=.8,thick,red] (01_A2) -- (01_B2);
\end{scope}
\begin{scope}[rotate=120]
\coordinate (12_A1) at (-\x,{-1/2/sqrt(3)+.1});
\coordinate (12_A2) at (\x,{-1/2/sqrt(3)+.1});
\coordinate (12_B1) at (-\x,{-1/2/sqrt(3)-.1});
\coordinate (12_B2) at (\x,{-1/2/sqrt(3)-.1});
\node[circle,draw=red,fill=red,scale=.2] at (12_A1) {};
\node[circle,draw=red,fill=red,scale=.2] at (12_A2) {};
\node[circle,draw=red,fill=red,scale=.2] at (12_B1) {};
\node[circle,draw=red,fill=red,scale=.2] at (12_B2) {};
\draw [->-=.6,thick,dotted,red] (12_A1) -- (12_A2);
\draw [->-=.6,thick,red] (12_B1) -- (12_B2);
\draw [->-=.8,thick,dotted,red] (12_A1) -- (12_B1);
\draw [->-=.8,thick,dotted,red] (12_A2) -- (12_B2);
\end{scope}
\begin{scope}[rotate=240]
\coordinate (02_A1) at (-\x,{-1/2/sqrt(3)+.1});
\coordinate (02_A2) at (\x,{-1/2/sqrt(3)+.1});
\coordinate (02_B1) at (-\x,{-1/2/sqrt(3)-.1});
\coordinate (02_B2) at (\x,{-1/2/sqrt(3)-.1});
\node[circle,draw=red,fill=red,scale=.2] at (02_A1) {};
\node[circle,draw=red,fill=red,scale=.2] at (02_A2) {};
\node[circle,draw=red,fill=red,scale=.2] at (02_B1) {};
\node[circle,draw=red,fill=red,scale=.2] at (02_B2) {};
\draw [->-=.6,thick,dotted,red] (02_A2) -- (02_A1);
\draw [->-=.6,thick,red] (02_B2) -- (02_B1);
\draw [->-=.8,thick,dotted,red] (02_B1) -- (02_A1);
\draw [->-=.8,thick,dotted,red] (02_B2) -- (02_A2);
\end{scope}
\draw [->-=.6,thick,red] (02_A2) -- (12_A1);
\draw [->-=.6,thick,red] (02_A1) -- (12_A2);
}
$
\caption{$\om_2(g_i,g_i^{-1}g_j)=1$ for two links\\with fermion parity $P_f^\psi=(-1)^3$.}
\label{fig:n0=2_ijk_2}
\end{subfigure}
\begin{subfigure}[h]{.3\textwidth}
\centering
$
\tikzfig{scale=2.5}{
\coordinate (g1) at (-1/2,{-1/2/sqrt(3)});
\coordinate (g2) at (1/2,{-1/2/sqrt(3)});
\coordinate (g3) at (0,{sqrt(3)/2-1/2/sqrt(3)});
\draw [->-=.75,thick] (g1) node[scale=1,below left]{$g_0$} -- (g2);
\draw [->-=.75,thick] (g2) node[scale=1,below right]{$g_1$} -- (g3) node[scale=1,above]{$g_2$};
\draw [->-=.75,thick] (g1) -- (g3);
\pgfmathsetmacro{\x}{.1}
\begin{scope}[rotate=0]
\coordinate (01_A1) at (-\x,{-1/2/sqrt(3)+.1});
\coordinate (01_A2) at (\x,{-1/2/sqrt(3)+.1});
\coordinate (01_B1) at (-\x,{-1/2/sqrt(3)-.1});
\coordinate (01_B2) at (\x,{-1/2/sqrt(3)-.1});
\node[circle,draw=red,fill=red,scale=.2] at (01_A1) {};
\node[circle,draw=red,fill=red,scale=.2] at (01_A2) {};
\node[circle,draw=red,fill=red,scale=.2] at (01_B1) {};
\node[circle,draw=red,fill=red,scale=.2] at (01_B2) {};
\draw [->-=.8,thick,dotted,red] (01_A1) -- (01_B1);
\draw [->-=.8,thick,dotted,red] (01_A2) -- (01_B2);
\draw [->-=.6,thick,red] (01_A1) -- (01_A2);
\draw [->-=.6,thick,red] (01_B1) -- (01_B2);
\end{scope}
\begin{scope}[rotate=120]
\coordinate (12_A1) at (-\x,{-1/2/sqrt(3)+.1});
\coordinate (12_A2) at (\x,{-1/2/sqrt(3)+.1});
\coordinate (12_B1) at (-\x,{-1/2/sqrt(3)-.1});
\coordinate (12_B2) at (\x,{-1/2/sqrt(3)-.1});
\node[circle,draw=red,fill=red,scale=.2] at (12_A1) {};
\node[circle,draw=red,fill=red,scale=.2] at (12_A2) {};
\node[circle,draw=red,fill=red,scale=.2] at (12_B1) {};
\node[circle,draw=red,fill=red,scale=.2] at (12_B2) {};
\draw [->-=.6,thick,dotted,red] (12_A1) -- (12_A2);
\draw [->-=.6,thick,red] (12_B1) -- (12_B2);
\draw [->-=.8,thick,dotted,red] (12_A1) -- (12_B1);
\draw [->-=.8,thick,dotted,red] (12_A2) -- (12_B2);
\end{scope}
\begin{scope}[rotate=240]
\coordinate (02_A1) at (-\x,{-1/2/sqrt(3)+.1});
\coordinate (02_A2) at (\x,{-1/2/sqrt(3)+.1});
\coordinate (02_B1) at (-\x,{-1/2/sqrt(3)-.1});
\coordinate (02_B2) at (\x,{-1/2/sqrt(3)-.1});
\node[circle,draw=red,fill=red,scale=.2] at (02_A1) {};
\node[circle,draw=red,fill=red,scale=.2] at (02_A2) {};
\node[circle,draw=red,fill=red,scale=.2] at (02_B1) {};
\node[circle,draw=red,fill=red,scale=.2] at (02_B2) {};
\draw [->-=.6,thick,dotted,red] (02_A2) -- (02_A1);
\draw [->-=.6,thick,red] (02_B2) -- (02_B1);
\draw [->-=.8,thick,dotted,red] (02_B1) -- (02_A1);
\draw [->-=.8,thick,dotted,red] (02_B2) -- (02_A2);
\end{scope}
\draw [->-=.8,thick,red] (02_A2) -- (12_A1);
\draw [->-=.8,thick,red] (02_A1) -- (12_A2);
}
$
\caption{$\om_2(g_i,g_i^{-1}g_j)=1$ for three links\\with fermion parity $P_f^\psi=(-1)^4$.}
\label{fig:n0=2_ijk_3}
\end{subfigure}
\caption{Different effective Majorana chain configurations and their total fermion parities $P_f^\psi$. The fermion parity can be easily calculated by counting the number of length-4 (non-Kasteleyn-oriented) loops in the figures.}
\label{fig:n0=2_ijk}
\end{figure}

Now we can consider the $p+ip$ superconductor configurations in one triangle $\langle 012\rangle$. The three mass term signs $\om_2(g_i,g_i^{-1}g_j)$ for the three links are independent to each other. So we have in total $2^3=8$ configurations, which can be divided into four cases. (1) All the three $\om_2$'s are one. Then all the effective Majorana fermions near the three links are in the vacuum pairings shown in \fig{fig:n0=2_1}. And the fermion parity of this configuration is $P_f^\psi=+1$. (2) One of the three $\om_2$'s is one (see \fig{fig:n0=2_ijk_1} of $\om_2(g_1,g_1^{-1}g_2)=1$ for example). Then there is only one length-4 loop with non-Kasteleyn-orientation. And the fermion parity is $P_f^\psi=(-1)^1=-1$. (3) Two of the three $\om_2$'s are one (see \fig{fig:n0=2_ijk_2} of $\om_2(g_0,g_0^{-1}g_2)=\om_2(g_1,g_1^{-1}g_2)=1$ for example). We connect the effective Majorana chains going though the two links (according to the Kasteleyn orientation rules \fig{fig:Kasteleyn}). So the total fermion parity of this configuration is $P_f^\psi=(-1)^3=-1$. (4) All the three $\om_2$'s are one (see \fig{fig:n0=2_ijk_3} for example). We again connect the effective Majorana chains going though two (arbitrary) links, and leave the third link Majorana chain unchanged. Then the total fermion parity of this configuration is $P_f^\psi=(-1)^4=+1$. For all the configurations, one can check easily the fermion parity can be summarized as
\begin{align}\label{Pf012}
P_f^\psi(\langle 012\rangle) = (-1)^{\om_2(\bar 01,\bar 12) + \om_2(0,\bar 01)\om_2(0,\bar 02) + \om_2(0,\bar 01)\om_2(1,\bar 12) + \om_2(0,\bar 02)\om_2(1,\bar 12)},
\end{align}
where we used $(\dd \om_2)(0,\bar 01,\bar 12) = \om_2(\bar 01,\bar 12) + \om_2(1,\bar 12) + \om_2(0,\bar 02) + \om_2(0,\bar 01) = 0$ (mod 2).

With the fermion parity formula for one triangle, we can derive the fermion parity change of the standard $F$ move \eq{2D:Fmove}. Since all $\om_2(g_i,g_i^{-1}g_j)$ for link $\langle ij\rangle$ are in general independent to each other, there are in total $2^6=64$ possible configurations for the $F$ move. The fermion parity change of the $F$ move is obtained from the fermion parities of the four relevant triangles on the two sides. However, the small dimer cover loop crossing links $\langle 02\rangle$ and $\langle 13\rangle$ are counted twice by the two adjacent triangles. Therefore, the fermion parity change of the $F$ move is the product of four triangle fermion parities [see \eq{Pf012} for triangle $\langle 012\rangle$] with a modification factor $(-1)^{\om_2(0,\bar 02)+\om_2(1,\bar 13)}$. After some tedious calculations, one can show that the final result is the cup-1 product of $\om_2$:
\begin{align}\label{Pf_psi_0123}\nonumber
\Delta P_f^\psi(F) &= P_f^\psi(\langle 012\rangle) \cdot P_f^\psi(\langle 023\rangle) \cdot P_f^\psi(\langle 013\rangle) \cdot P_f^\psi(\langle 123\rangle) \cdot (-1)^{\om_2(0,\bar 02)+\om_2(1,\bar 13)}\\\nonumber
&= (-1)^{\om_2(\bar 02,\bar 23)\om_2(\bar 01,\bar 12)+\om_2(\bar 01,\bar 13)\om_2(\bar 12,\bar 23)}\\
&= (-1)^{(\om_2\smile_1\om_2)(\bar 01,\bar 12,\bar 23)}.
\end{align}

If the exponent in \eq{Pf_psi_0123} is a nontrivial cocycle in $H^3(G_b,\Z_2)$, then the 2D $F$ move necessarily breaks the fermion parity of the $p+ip$ superconductor system. However, if we introduce a 3D FSPT bulk with complex fermion decoration $n_3=\om_2\smile_1\om_2$, the total fermion parity will be preserved. This is simply because the 2D boundary $F$ move will also change the bulk complex fermion number by $n_3(g_0,g_1,g_2,g_3)$. Therefore, two layers of $p+ip$ superconductors can be viewed as a 2D ASPT state which trivializes the 3D FSPT state with complex fermion decoration $n_3=\om_2\smile_1\om_2$. So we have a trivialization group $\Gamma^3$ as
\begin{align}\label{Gamma3}
\Gamma^3 \supset \{ (\om_2\smile_1 \om_2) \floor{n_0/2} \in H^3(G_b,\Z_2) | n_0 \in H^0(G_b,\Z_T) \}.
\end{align}
Note that we introduced $n_0 \in H^0(G_b,\Z_T)$ in the above expression. Only unitary $G_b$ would be compatible with $p+ip$ chiral superconductors. And odd $n_0$ (with nontrivial $\om_2$) already leads to nontrivial trivialization group $\Gamma^2$ \eq{Gamma2} in a lower level. Only $n_0=2$ will produce nontrivial $\Gamma^3$ and a nontrivial ASPT state on the boundary of 3D FSPT with $n_3=\om_2\smile_1 \om_2$. An explicit example of this $\Gamma^3$ trivialization \eq{Gamma3} with $n_0=2$ is the FSPT state with $G_f=Q_8^f$ (see Appendix~\ref{sec:E:Q8}).

\subsubsection{Boundary ASPT states in $\Gamma^3$ with Kitaev chains}

There is another layer of boundary ASPT state for the 3D FSPT state. This boundary ASPT state has Kitaev chain decorations, and trivialize the complex fermion decoration data
\begin{align}\label{n3n1}
n_3 = \om_2\smile n_1 + s_1 \smile n_1 \smile n_1,
\end{align}
for some $n_1\in H^1(G_b,\Z_2)$. The second part $s_1 \smile n_1 \smile n_1$ is discussed in detail in \Ref{WQG2018}. It is used to construct a 2D ASPT state on the boundary of 3D FSPT state with time reversal symmetry $T^2=1$.

The generic boundary ASPT state with Kitaev chains can be constructed similar to the 2D FSPT state with Kitaev chain decorations in section~\ref{2D:decoration:Maj}. We put (at most) one Kitaev chain along the (red) dual link on the 2D boundary according to the decoration data $n_1\in H^1(G_b,\Z_2)$. The fermion parity change of the Majorana fermions under the boundary 2D $F$ move is again given by [see \eq{2D:Pf_Maj}]:
\begin{align}\label{2D:Pf_Maj_}
\Delta P_f^\g (F) = (-1)^{(\om_2\smile n_1 + s_1\smile n_1\smile n_1)(g_0^{-1}g_1,g_1^{-1}g_2,g_2^{-1}g_3)}.
\end{align}
If the exponent of $(-1)$ is not a $\Z_2$-valued coboundary, we can not preserve the boundary fermion parity even introduce complex fermion decorations on the boundary. However, since the ASPT state is on the boundary of a 3D FSPT bulk, we can simply choose the bulk complex fermion decoration data $n_3$ with \eq{n3n1}. Therefore, the total system (including both the boundary and the bulk) has definite total fermion parity.

Since we construct a gapped symmetric ASPT state without topological order on the boundary of the 3D FSPT state, we conclude that the bulk FSPT with $n_3 = \om_2\smile n_1 + s_1 \smile n_1 \smile n_1$ is trivialized. This is the origin of the $n_1$ part of the trivialization group $\Gamma^3$:
\begin{align}
\Gamma^3 \supset \{ \om_2\smile n_1 + s_1 \smile n_1 \smile n_1 \in H^3(G_b,\Z_2) | n_1 \in H^1(G_b,\Z_2) \}.
\end{align}
Combine it with \eq{Gamma3}, we obtain the trivialization group
\begin{align}
\Gamma^3 = \{ \om_2\smile n_1 + s_1 \smile n_1 \smile n_1 + (\om_2\smile_1 \om_2) \floor{n_0/2} \in H^3(G_b,\Z_2) | n_1 \in H^1(G_b,\Z_2), n_0 \in H^0(G_b,\Z_T) \},
\end{align}
which is claimed in \eq{3D:Gamma}.

\subsection{An additional layer of $p+ip$ SC decorations}
\label{sec:3D:pip}

Apart from the Kitaev chain and complex fermion layers, there is an additional layer of 2D $p+ip$ chiral superconductor (SC) decorations specified by $n_1\in H^1(G_b,\Z_T)$ for 3D FSPT states. The decoration of this layer is possible only when $G_b$ is not unitary. Since there is no fixed-point wave function constructions of chiral states on discrete lattice \cite{KapustinFidkowski2018}, we discuss it after all the other layers in this section. However, we can put a 2D continuum (infinite number of degrees of freedom) free fermion $p+ip$ SC state on the decoration plane with the bulk mass approaching positive infinity. As will be shown below, the obstruction function for the $p+ip$ chiral superconductor decoration is $\om_2\smile n_1+s_1\smile n_1\smile n_1$. For the 3D topological superconductor with time reversal symmetry $T^2=-1$, the obstruction function equals to zero identically. So this $p+ip$ layer will not twist the obstruction functions of other higher layers. In this way, we can fully classify 3D $T^2=-1$ topological superconductors.

\subsubsection{Consistency condition}

\begin{figure}[ht]
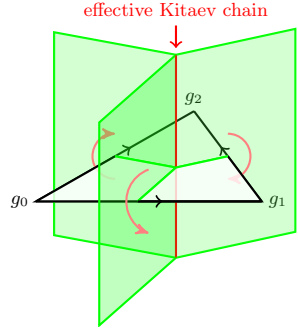

\centering
$
\tikzfig{scale=3}{
\coordinate (0) at (0,0);
\coordinate (1) at (1,0);
\coordinate (2) at (.7,.4);
\draw[thick] (0)--(1);
\draw[->-=.6,thick] (1)--(2);
\draw[->-=.6,thick] (0)--(2);
\node[left,scale=.8] at (0) {$g_0$};
\node[right,scale=.8] at (1) {$g_1$};
\node[above,scale=.8] at (2) {$g_2$};
\coordinate (01C) at ($(0)!.45!(1)$);
\coordinate (12C) at ($(1)!.5!(2)$);
\coordinate (02C) at ($(0)!.5!(2)$);
\coordinate (C) at (.62,.15);
\draw[thick,green] (C)--(01C);
\draw[thick,green] (C)--(12C);
\draw[thick,green] (C)--(02C);
\coordinate (CA) at ($(C)+(0,.5)$);
\coordinate (CB) at ($(C)-(0,.4)$);
\draw[thick,red] (C)--(CA);
\draw[thick,red] (C)--(CB);
\coordinate (02C') at ($(C)!2!(02C)$);
\coordinate (02A) at ($(02C')+(CA)-(C)$);
\coordinate (02B) at ($(02C')+(CB)-(C)$);
\draw[thick,green] (CA)--(02A)--(02B)--(CB);
\fill[green!70,nearly transparent] (CA) -- (02A) -- (02B) -- (CB) -- cycle;
\draw[->,>=stealth',thick,red!50] ($(02C)+(270:.1)$) arc (270:90:.1);
%
\fill[white,opacity=0.8] (C) -- (02C) -- (0) -- (01C) -- cycle;
\draw[thick,black] (02C)--(0)--(01C);
\draw[thick,green] (02C)--(C)--(01C);
\coordinate (12C') at ($(C)!2.3!(12C)$);
\coordinate (12A) at ($(12C')+(CA)-(C)$);
\coordinate (12B) at ($(12C')+(CB)-(C)$);
\draw[thick,green] (CA)--(12A)--(12B)--(CB);
\fill[green!70,nearly transparent] (CA) -- (12A) -- (12B) -- (CB) -- cycle;
\draw[->,>=stealth',thick,red!50] ($(12C)+(90:.1)$) arc (90:-90:.1);
\coordinate (01C') at ($(C)!2!(01C)$);
\coordinate (01A) at ($(01C')+(CA)-(C)$);
\coordinate (01B) at ($(01C')+(CB)-(C)$);
\draw[thick,green] (CA)--(01A)--(01B)--(CB);
\fill[green,nearly transparent] (CA) -- (01A) -- (01B) -- (CB) -- cycle;
\draw[->,>=stealth',thick,red!50] ($(01C)+(.1,0)+(110:.15)$) arc (110:250:.15);
\draw[red] (CA)--(CB);
%
\fill[white,opacity=0.8] (C) -- (12C) -- (1) -- (01C) -- cycle;
\draw[thick,black] (12C)--(1)--(01C);
\draw[thick,green] (12C)--(C)--(01C);
\draw[->-=.2,thick] (01C)--(1);
\node[red,above,scale=.8,yshift=15] at (CA) {effective Kitaev chain};
\draw[->,thick,red] ($(CA)+(0,.13)$) -- ($(CA)+(0,.03)$);
}
$
\caption{Decorations of $p+ip$ chiral superconductors for a triangle $\langle 012\rangle$ (see \fig{fig:3D:dof_b} for a tetrahedron). The 2D chiral superconductors are decorated on the (green) planes dual to the three links $\langle 01\rangle$, $\langle 12\rangle$ and $\langle 02\rangle$. The direction of the chiral Majorana mode along the boundary of the (green) plane dual to link $\langle ij\rangle$ is indicated by a (red) arc arrow (right-hand rule with respect to the link direction) if $n_1>0$. So there are three different kinds of chiral Majorana modes along the (red) link dual to the triangle $\langle 012\rangle$. After gapping out these chiral modes symmetrically, there may be an effective Kitaev chain along the (red) link.}
\label{fig:3D:012}
\end{figure}

We use $n_1(g_i,g_j)\in\Z$ to indicate the number of decorated $p+ip$ chiral superconductor layers on the plane dual to link $\langle ij\rangle$ (see \fig{fig:3D:012}). If $n_1(g_i,_j)<0$, we would decorate inverse $p+ip$, i.e. $p-ip$, chiral superconductors. So the number of chiral Majorana modes on the boundary of the plane is $|n_1(g_i,g_j)|$. The direction of the chiral Majorana modes (see red arc arrows in \fig{fig:3D:012}) form right-hand (left-hand) rule with respect to the oriented link $\langle ij\rangle$ if $n_1(g_i,_j)>0$ ($n_1(g_i,_j)<0$). For a triangle $\langle ijk\rangle$, there are three $p+ip$ superconductor planes intersecting at the link dual to the triangle (see the red link in \fig{fig:3D:012}). Since we are constructing a gapped state, there should be no chiral Majorana mode along this link. So we have the gappable condition:
\begin{align}\label{3D:dn1}
(\dd n_1)(g_0,g_1,g_2) = \emp^{g_0}n_1(g_1,g_2) - n_1(g_0,g_2) + n_1(g_0,g_1) = 0,
\end{align}
which merely states that the number of left-moving and right-moving chiral Majorana modes along the (red) link dual to triangle $\langle 012\rangle$ equal to each other. Note that the first term in \eq{3D:dn1} has a $g_0$ symmetry action, which adds a minus sign if $s_1(g_0)=1$. The physical meaning is that the time reversal symmetry would change $n_1$ to $-n_1$, and reverse the direction of the boundary chiral Majorana modes.

If $g\in G_b$ is unitary, then one can show $n_1(g^k)=kn_1(g)$ for all $k\in \Z$ from \eq{3D:dn1}. Since we consider only finite $G_b$, we must have $n_1(g)=0$ for all unitary $g\in G_b$. So $H^1(G_b,\Z_T)$ is trivial if $G_b$ is a finite unitary group, and there is no $p+ip$ chiral superconductor decoration layer. For symmetry group $G_b$ with antiunitary elements, one can show 
\begin{align}\label{3D:n1}
n_1(g)=
\begin{cases}
0,\quad g\text{ is unitary}\\
1,\quad g\text{ is antiunitary}\\
\end{cases}
\end{align}
by adding some coboundaries. So we have $H^1(G_b,\Z_T)=\Z_2$ for $G_b$ with antiunitary elements.

\subsubsection{Symmetry transformations}
\label{sec:3D:pip_symm}

The chiral Majorana modes along the boundary of the plane dual to link $\langle ij\rangle$ will be denoted by $\psi_{ij,R;\al}^{g_i}$ or $\psi_{ij,L;\al}^{g_i}$ ($\al=1,2,...,|n_1(ij)|$) for $n_1(ij)>0$ or $n_1(ij)<0$, respectively. As discussed in \ref{sec:3D:dof}, the symmetry transformation rules for the chiral Majorana modes are given by \eqs{3D:symm:R}{3D:symm:L}. The $\om_2$ part is the usual projective representation factor for $G_b$ (such that it is a linear representation of $G_f$). The $s_1$ part is more subtle, which needs some explanations.

To be more specific, the time reversal symmetry acts on the chiral Majorana modes as
\begin{align}
\label{3D:symm:R_}
U(T) \psi_{ij,R;\al}^{g_i} U(T)^\dagger &= (-1)^{\om_2(T,g_i)} \psi_{ij,L;\al}^{Tg_i},\\
\label{3D:symm:L_}
U(T) \psi_{ij,L;\al}^{g_i} U(T)^\dagger &= -(-1)^{\om_2(T,g_i)} \psi_{ij,R;\al}^{Tg_i}.
\end{align}
Basically, it changes $\psi_R$ to $\psi_L$, and $\psi_L$ to $-\psi_R$ (forgetting the $\om_2$ factor). So the left-moving/right-moving chiral Majorana modes form time reversal Kramers doublet with $T^2=-1$. The reason to choose this symmetry transformation convention is as follows. If we fold the three (green) planes in \fig{fig:3D:012} into one plane, we should obtain a symmetric state on the pure 2D plane. The system consists of several copies of $p+ip$ and $p-ip$ chiral superconductors. However, it is known that the we only have nontrivial 2D topological superconductors for time reversal symmetry $T^2=-1$ \cite{WangNingChen2017}. The $T^2=1$ case corresponds to the ASPT state which can only exist on the boundary of a 3D bulk \cite{WQG2018} (see also the FSPT classification examples in section~\ref{sec:E:Z2}). Therefore, we choose the chiral Majorana modes to form Kramers doublet under time reversal symmetry.

\subsubsection{Symmetric mass terms}

To obtain a symmetric gapped state, we should add symmetric mass terms to the three kinds of chiral Majorana modes along the link dual to each triangle $\langle ijk\rangle$. The strategy is again first considering the standard mass term for the standard triangle with first vertex label $e\in G_b$. The standard mass term will have the following form
\begin{align}\label{3D:std}
im \psi_\mathrm{out}^{e} \psi_\mathrm{in}^{g_0^{-1}g_i}
\quad \mathrm{or} \quad
im \psi_\mathrm{out}^{g_0^{-1}g_i} \psi_\mathrm{in}^{e},
\end{align}
where at least one of the two chiral Majorana modes are labelled by the identity element $e$ of $G_b$. The subscript ``out'' and ``in'' indicate the actual direction of the chiral Majorana modes along the (red) link dual to the triangle, i.e., going outside or inside the triangle (using right-hand rule). This direction is not locked with the sign of $n_1$, for the three links of the triangle have different right-hand rule direction (see the red arc arrows in \fig{fig:3D:012}). We always put out-going mode in front of in-going mode. This is because we have to set a rule to know whether there is an effective Kitaev chain compared to the ``trivial'' gapped edge by checking simply the sign of the mass.

Other non-standard mass terms are obtained from the standard one by a symmetry action. There will be additional minus sign for the mass from the symmetry action. The negative mass will induce effective Kitaev chain along the gapped edge (see \fig{fig:3D:012}). In the following, we will consider separately the symmetric mass terms for nontrivial $\om_2$ and $s_1$.

\emph{(c.1) Mass term signs from $\om_2$.}

We first consider the \emph{standard} mass terms for the standard triangle $\langle 012\rangle$ with group element labels $e$, $g_0^{-1}g_1$ and $g_0^{-1}g_2$ for three vertices. According to \eq{3D:dn1}, the number of left-moving and right-moving modes along the (red) link dual to the triangle should be the same. For simplicity, we assume all of $n_1(01)=n_1(e,g_0^{-1}g_1)$, $n_1(12)=n_1(g_0^{-1}g_1,g_0^{-1}g_2)$ and $n_1(02)=n_1(e,g_0^{-1}g_2)$ are positive integers, and satisfy $n_1(02)=n_1(01)+n_1(12)$. Then we need two types of standard mass terms to gap out these modes (assume $m>0$):
\begin{align}\label{3D:stdmass}
H_\mathrm{std}^{\langle 012\rangle} =
\sum_{\al=1}^{n_1(01)} im \int\dd x \psi_{01,R;\al}^{e}(x) \psi_{02,R;\al}^{e}(x)
+ \sum_{\al=1}^{n_1(12)} im \int\dd x \psi_{12,R;\al}^{g_0^{-1}g_1}(x) \psi_{02,R;\al}^{e}(x).
\end{align}
We note that the chiral modes $\psi_{01,R;\al}^{e}$ and $\psi_{12,R;\al}^{g_0^{-1}g_1}$ are going outside the triangle, while $\psi_{02,R;\al}^{e}$ is inside (see the red arc arrows in \fig{fig:3D:012}). So it satisfies the standard mass term rule \eq{3D:std}.

For the non-standard triangle with vertex label $g_0$, $g_1$ and $g_2$, we should use a symmetry action to obtain the mass terms:
\begin{align}\nonumber\label{3D:mass}
H_\mathrm{mass}^{\langle 012\rangle}
&= U(g_0) H_\mathrm{std}^{\langle 012\rangle} U(g_0)^\dag\\
&= \sum_{\al=1}^{n_1(01)} im \int\dd x \psi_{01,R;\al}^{g_0}(x) \psi_{02,R;\al}^{g_0}(x)
+ \sum_{\al=1}^{n_1(12)} (-1)^{\om_2(g_0,g_0^{-1}g_1)} im \int\dd x \psi_{12,R;\al}^{g_1}(x) \psi_{02,R;\al}^{g_0}(x),
\end{align}
where the sign $(-1)^{\om_2(g_0,g_0^{-1}g_1)}$ of the second type mass term comes from the symmetry transformation rule for $\psi_{12,R;\al}^{g_0^{-1}g_1}$. The total number of effective Kitaev chain going though the triangle $\langle 012\rangle$ is then
\begin{align}\label{3D:effK}
\om_2(g_0,g_0^{-1}g_1) n_1(g_1^{-1}g_2),
\end{align}
which is just the number of negative mass terms in \eq{3D:mass}.

If the three $n_1$ are not all positive, the mass terms are different from \eq{3D:mass}. In general, there are three types of mass terms which gap out different pairs of chiral Majorana modes. The standard form of these three types are
\begin{align}\label{3D:m1}
H_\mathrm{std}^{01,02} &= \sum_{\al=1}^{N-|n_1(12)|} im \int\dd x \psi_{01;\al}^{e}(x) \psi_{02;\al}^{e}(x),\\\label{3D:m2}
H_\mathrm{std}^{12,02} &= \sum_{\al=1}^{N-|n_1(01)|} im \int\dd x \psi_{12;\al}^{g_0^{-1}g_1}(x) \psi_{02;\al}^{e}(x),\\\label{3D:m3}
H_\mathrm{std}^{01,12} &= \sum_{\al=1}^{N-|n_1(02)|} im \int\dd x \psi_{01;\al}^{e}(x) \psi_{12;\al}^{g_0^{-1}g_1}(x),
\end{align}
where we defined $N=\max\left(|n_1(01)|,|n_1(12)|,|n_1(02)|\right)$. Note that (at most) only two types of mass terms appear for a given $n_1$ configuration. We should choose the mass terms which involve the chiral modes corresponding to the biggest $|n_1|$. For example, the two types of mass terms in \eq{3D:stdmass} both involve $\psi_{02}$, because $n_1(02)>0$ is the biggest and $N-|n_1(02)|=0$ for the summation in \eq{3D:m3}. The order of the two Majorana modes in a mass term should follow the rule \eq{3D:std} of the out-going mode appears in the front. It depends on the configuration of $n_1$. The above three equations are only one example. (We also omit the $R/L$ label of the chiral modes for the sign of $n_1$ is indefinite.)

Independent of the signs of $n_1$, the effective Kitaev chain number is always \eq{3D:effK}. The minus signs of the masses all come from the symmetry sign $(-1)^{\om_2(g_0,g_0^{-1}g_1)}$ of $\psi_{12,R/L;\al}^{g_0^{-1}g_1}$. Independent of the signs of $n_1$, there are always $|n_1(12)|$ mass terms associated with this chiral modes [see \eqs{3D:m2}{3D:m3}]. Under a $U(g_0)$-action from the standard mass term to the actual mass term, all the terms in \eqs{3D:m2}{3D:m3} have negative masses. So the (mod 2) number of effective Kitaev chains is always \eq{3D:effK}.

\emph{(c.2) Mass term signs from $s_1$.}

There are additional signs for the mass terms related to time reversal symmetry. In general, our standard mass term has the form (we omit the group element labels)
\begin{align}\label{3D:std_}
im \int \dd x \psi_{\mathrm{out},s}(x) \psi_{\mathrm{in},s'}'(x),
\end{align}
where the out-going mode (going up in \fig{fig:3D:012}) is in front of the in-going mode (going down in \fig{fig:3D:012}). The labels $s$ and $s'$ denote $R$ or $L$, depending on $n_1>0$ or $n_1<0$. Note that the out-going/in-going is not locked with $R/L$ (see the red arc arrows in \fig{fig:3D:012}).

Under a unitary symmetry action $U(g_0)$, the mass term is transformed to
\begin{align}
im \int \dd x \psi_{\mathrm{out},s}(x) \psi_{\mathrm{in},s'}'(x)
\quad\xrightarrow[U(g_0)]{\ \mathrm{unitary}\ }\quad
(-1)^{\om_2} im \int \dd x \psi_{\mathrm{out},s}(x) \psi_{\mathrm{in},s'}'(x),
\end{align}
where the sign $(-1)^{\om_2}$ is exactly the transformation sign of $\psi^{g_0^{-1}g_1}$ discussed previously. So we can obtain the effective Kitaev chain number \eq{3D:effK} in a simple way.

On the other hand, if the symmetry $U(g_0)$ is antiunitary, it will reverse the directions of all the chiral modes. The action on the standard mass term \eq{3D:std_} is
\begin{align}
im \int \dd x \psi_{\mathrm{out},s}(x) \psi_{\mathrm{in},s'}'(x)
\quad\xrightarrow[U(g_0)]{\ \mathrm{antiunitary}\ }\quad
(-1)^{\om_2}(-1)^{1+1+(1-\delta_{s,s'})} im \int \dd x \psi_{\mathrm{out},-s'}'(x) \psi_{\mathrm{in},-s}(x).
\end{align}
Apart from the $\om_2$ term, there are three additional signs. One minus sign comes from the antiunitary action on the imaginary unit $i$. The second is the sign of switching two chiral modes $\psi$ and $\psi'$. This is because time reversal changes the out-going modes to in-going modes, and vice versa. According to the rule \eq{3D:std}, we should change their orders. The third sign $(-1)^{1-\delta_{s,s'}}$ appears only when $s$ and $s'$ are different, i.e., the two modes are of different $R/L$ types. This is a consequence of the symmetry transformation rules \eqs{3D:symm:R_}{3D:symm:L_}: $R\rightarrow L$ and $L \rightarrow -R$. The only mass term between $R$ and $L$ type modes is \eq{3D:m3} for $\psi_{01,R/L}$ and $\psi_{01,L/R}$. Using the ``canonical'' 1-cocycle \eq{3D:n1}, we conclude that the only mass term with (time reversal related) minus sign is $-im\psi_{12}^{g_1}\psi_{01}^{g_0}$ with antiunitary $g_0$. This case corresponds to $s_1(g_0)=1-s_1(g_1)=s_1(g_2)=1$ and $n_1(12)=-n_1(01)=1-n_1(02)=1$. We can summarize this (time reversal related) sign for mass term by $(-1)^{s_1(g_0)n_1(\bar 01)n_1(\bar 12)}$. So the (mod 2) number of effective Kitaev chain going though the triangle related to time reversal symmetry is
\begin{align}\label{3D:effK2}
s_1(g_0)n_1(g_0^{-1}g_1)n_1(g_1^{-1}g_2)
\end{align}

In summary, we can use symmetric mass terms to gap out all the chiral Majorana modes along the (red) link dual to the triangle $\langle 012\rangle$. There are effective Kitaev chains left along the link, with (mod 2) number
\begin{align}\label{3D:effK_}
(\om_2\smile n_1+s_1\smile n_1\smile n_1)(g_0,g_0^{-1}g_1,g_1^{-1}g_2)
= \om_2(g_0,g_0^{-1}g_1) n_1(g_1^{-1}g_2) + s_1(g_0)n_1(g_0^{-1}g_1)n_1(g_1^{-1}g_2),
\end{align}
which is a combination of \eqs{3D:effK}{3D:effK2}.

\subsubsection{Obstruction function}

If we consider a tetrahedron $\langle 0123\rangle$ of the 3D triangulation lattice of the spacial manifold, there are four triangles on the boundary and four (red) links with chiral Majorana modes (see \fig{fig:3D:dof_b}). We should add mass terms $H_\mathrm{mass}^{\langle 123\rangle}$, $H_\mathrm{mass}^{\langle 023\rangle}$, $H_\mathrm{mass}^{\langle 013\rangle}$ and $H_\mathrm{mass}^{\langle 012\rangle}$ for all the triangles, following the discussions above. The total number of effective Kitaev chains crossing the boundary of the tetrahedron can be calculated as the summation of four terms similar to \eq{3D:effK_}. Using the cocycle equations for $\om_2$, $s_1$ and $n_1$, we have $\dd (\om_2\smile n_1+s_1\smile n_1\smile n_1)(g_0,g_0^{-1}g_1,g_1^{-1}g_2,g_2^{-1}g_3)=0$. So total (mod 2) number of effective Kitaev chains for the tetrahedron is
\begin{align}\label{3D:pip_K}
(\om_2\smile n_1+s_1\smile n_1\smile n_1)(g_0^{-1}g_1,g_1^{-1}g_2,g_2^{-1}g_3).
\end{align}

To make sure that there are no dangling Majorana fermion inside any tetrahedron of the lattice, the number of effective Kitaev chains should equal to the number of decorated Kitaev chains specified by $n_2$. So we have the consistency equation
\begin{align}
\dd n_2 = \om_2\smile n_1+s_1\smile n_1\smile n_1.
\end{align}
If the right-hand-side of the above equation is not a $\Z_2$-valued 3-coboundary, there is no solution for $n_2$ on the left-hand-side. We note that the above equation is the same as the 1D consistency equation \eq{2D:dn2}, although the physical meanings are totally different.

\subsection{Classification of 3D FSPT phases}

The general classification of 3D FSPT phases is as follow. We first calculate the cohomology groups $H^1(G_b,\Z_T)$, $H^2(G_b,\Z_2)$, $H^3(G_b,\Z_2)$ and $H^4(G_b,U(1)_T)$. For each $n_1\in H^1(G_b,\Z_T)$, we solve the twisted cocycle equation \eq{3D:eq} for $n_2$. For each solution $n_2$, we solve the twisted cocycle equation \eq{3D:eq} for $n_3$. And for each solution $n_3$, we solve the twisted cocycle equation \eq{3D:eq} for $\nu_4$. If $n_2$, $n_3$ and $\nu_4$ are in the trivialization subgroup $\Gamma^2$, $\Gamma^2$ and $\Gamma^3$ in \eq{3D:Gamma}, then they are trivialized by boundary ASPT states. So the obstruction-free and trivialization-free $(n_1,n_2,n_3,\nu_4)$ fully classify the 3D FSPT phases.

We can also use the 3D FSLU transformations to construct the commuting projector parent Hamiltonians. The procedure is again tedious but straightforward. Each term of the Hamiltonian is a sequence of 3D fermionic $F$ moves that changes the group element label of a vertex from $g_\ast$ to $g_\ast'$. Different terms commute with each other for the 3D FSPT wave function is at the fixed-point.

\section{Conclusions and discussions}
\label{sec:con}

In this paper, we construct gapped fermionic state with symmetry $G_f$ by decorating fermionic degrees of freesom. In $d$ spacial dimension, they are constructed using several layers of data $(...,n_{d-1},n_d,\nu_{d+1})$, which is an element in $...\times C^{d-1}(G_b,\Z_2) \times C^{d}(G_b,\Z_2) \times C^{d+1}(G_b,U(1)_T)$ (see \tab{tab:data}). There are several consistency conditions for them. Basically, the coboundary of one layer data should equal to a functional of the data of lower layers (see \tab{tab:eq}). We can summarize them as a system of twisted cocycle equations:
\begin{align}\label{cocycleeq}
\dd (...,n_{d-1},n_d,\nu_{d+1}) = (...,\mathcal O_{d},\mathcal O_{d+1},\mathcal O_{d+2}).
\end{align}
Note that the obstruction function $\mathcal O_{d+2}$ is a $U(1)_T$-valued $(d+2)$-cocycle. And all other obstruction functions $\mathcal O_i$ ($i\leq d+1$) are in $H^i(G_b,\Z_2)$ with $\Z_2$ coefficients. The data $(...,n_{d-1},n_d,\nu_{d+1})$ corresponds to a valid FSPT state if and only if all the obstruction functions are coboundaries. Otherwise there are no solutions for data of the next layer.

There are two related questions about the constructed FSPT states. The first is that whether the states with different $(...,n_{d-1},n_d,\nu_{d+1})$ data represent distinct FSPT phases. If we have a path of FSLU transformations to connect them, they are in fact in the same phase. So the FSPT classification data should quotient these cases. 
The second question is what happens to the state that is obstructed by some nontrivial cocycle $\mathcal O_i$. There are some physical inconsistencies for these states, because all the obstruction functions have physical meanings such as fermion parity conservation (see the last column of \tab{tab:eq}). But is it possible to construct such state on the boundary of a one higher dimensional state? If it is possible, we need to understand the physical properties of the bulk, such as whether it is long range entangled or short range entangled.




The answers of the the above two questions are related to the concept of anomalous SPT states \cite{WQG2018}. 
If one of the obstruction functions on the right-hand-side of \eq{cocycleeq} is not a coboundary, the state will be obstructed for it violates some physical consistency constraints (see the last column of \tab{tab:eq}). However, this state can exist as ASPT state on the boundary of an FSPT state in $(d+1)$ spacial dimension. 
The ASPT states in $(d-1)$ spacial dimensions induces new kinds of coboundaries for the classification data of FSPT phases in $d$ dimensions \cite{GuWen2014,WangGu2017,WQG2018}. Formally, we can write the new coboundaries as
\begin{align}
(...,n_{d-1},n_d,\nu_{d+1}) &\sim (...,n_{d-1},n_d,\nu_{d+1}) + \dd(...,n_{d-2},n_{d-1},\nu_{d})\\
&= (...,n_{d-1},n_d,\nu_{d+1}) + (...,\mathcal O_{d-1},\mathcal O_{d},\mathcal O_{d+1}).
\end{align}
In the first line of the above equation, we identify the FSPT state with $(...,n_{d-1},n_d,\nu_{d+1})$ with another state by stacking the coboundary of one lower dimension state with data $(...,n_{d-2},n_{d-1},\nu_{d})$. The coboundary of the latter data, by \eq{cocycleeq}, are exactly the obstruction functions for the FSPT in one lower dimensions. Therefore, the lower dimensional FSPT obstruction functions will trivialize the higher dimensional FSPT data.

Mathematically, the obstruction functions $\mathcal O_i$ for FSPT states in $(d-1)$ spacial dimensions form a subgroup of the cohomology group $H^i(G_b,\Z_2)$ [or $H^i(G_b,U(1)_T)$]:
\begin{align}
\Gamma^i = \{ \mathcal O_i[n_{i-2}] | n_{i-2} \in C^{i-2}(G_b,\Z_2) \text{ is a classification data for FSPT in $(d-1)$ dimension} \}.
\end{align}
If $\mathcal O_i[n_{i-2}]$ is a nontrivial cocycle in $\Gamma^i$, the $(d-1)$ dimensional state with classification data $n_{i-2}$ is obstructed. On the other hand, if the classification data $n_i$ (or $\nu_i$) for $d$ dimensional FSPT state belongs to the subgroup $\Gamma^i$, it will be trivialized because of the boundary ASPT state. Therefore, the distinct classification data for $d$ dimensional FSPT state is in fact $n_i \in C^i(G_b,\cdot)/B^i(G_b,\cdot)/\Gamma^i$ (the coefficient is $\Z_2$ or $U(1)_T$).

For each solution of \eq{cocycleeq}, we can use the classification data to construct an FSPT state by decorating several layers of fermionic degrees of freedom to the BSPT state. These state belong to different FSPT phases if the data are different in $C^i(G_b,\cdot)/B^i(G_b,\cdot)/\Gamma^i$. We can also use the FSPT moves to construct commuting projector parent Hamiltonians. The Hamiltonian consists of local operator which corresponds to a sequence of $F$ moves that changes one vertex label from $g_\ast$ to $g_\ast'$. Different terms commute with each other, because the $F$ moves satisfy the coherence conditions.

We conjecture that in principle our classification scheme for FSPT phases can also be applied to point/space group symmetry, so long as the crystalline principle and spin statistics relations \footnote{Meng2018} are carefully considered. For example, the mirror symmetry with $\sigma^2=1$ ($\Z_2^f\times\Z_2^P$) should be regarded as a time reversal symmetry with $T^2=P_f$ ($\Z_4^{Tf}=\Z_2^f\timesw\Z_2^T$), while the mirror symmetry with $\sigma^2=P_f$ ($\Z_4^{TPf}=\Z_2^f\timesw\Z_2^P$) should be regarded as a time reversal symmetry with $T^2=1$ ($\Z_2^f\times\Z_2^T$)\cite{Hemele2017}. The full details for the classification and construction of point/space group protected FSPT phases will be presented in our future work. Moreover, we also believe that our construction and classification scheme can be applied for continuum Lie group symmetry by using the Borel cohomology. However, it is very difficult to compute the obstruction functions for general Lie group, and we will develop special tools to handle this problem in the future. Finally, how to generalize our framework into FSPT phases protected by super symmetry (SUSY) will be an extremely interesting future direction!

\begin{acknowledgments}
We are grateful to X.-G. Wen for very enlightening discussions. This work is supported Direct Grant no. 4053300 from The Chinese University of Hong Kong and funding from Hong Kong’s Research Grants Council (GRF no.14306918, ANR/RGC Joint Research Scheme no. A-CUHK402/18). 
\end{acknowledgments}

\appendix

\section{Fixed-point wave function and classification of FSPT states in 0D}
\label{sec:0D}

In this Appendix, we discuss in this section the FSPT states in zero spacial dimension, which are classified by the one dimensional representations of $G_f$, i.e., $H^1(G_f,U(1)_T)$. As is shown below, we can choose equivalently the classification data to be $n_0\in H^0(G_b,\Z_2)$ and $\nu_1\in C^1(G_b,U(1)_T)/B^1(G_b,U(1)_T)$ with some consistency equations. Although the 0D case is rather degenerate, it shows the layer structure of the FSPT classifications, which is also true but more complicated in higher dimensions.

\subsection{Classification}

It is known that the 0D BSPT states with symmetry group $G_b$ are classified by the one dimensional linear representations of $G_b$, i.e., $H^1(G_b,U(1)_T)$ \cite{chen13}. This is because the SPT state should be both symmetric and non-degenerate. In zero spacial dimension, there are essentially no difference between bosonic and fermionic systems, except that there is an additional $\Zf$ symmetry for the fermionic system. We can treat a fermionic system with total fermionic symmetry group $G_f=\Zf\timesw G_b$ as a bosonic system with total bosonic symmetry group $G_f$. Therefore, we have the following conclusion:
\begin{itemize}
\item
0D FSPT phases with symmetry group $G_f=\Zf\timesw G_b$ are classified by the one dimensional irreducible representations of $G_f$, i.e., $H^1(G_f,U(1)_T)$.
\end{itemize}
Equivalently, we can unpack the above result and show that
\begin{itemize}
\item
0D FSPT phases with symmetry group $G_f=\Zf\timesw G_b$ are classified by a $0$-cocycle $n_0$ and a $1$-cochain $\nu_1$, with some symmetry conditions and consistency equations.
\end{itemize}
The second version of classification is more physical. The first data $n_0\in H^0(G_b,\Z_2)$ is related to the fermion parity of the state: $P_f=(-1)^{n_0}$. The second data $\nu_1\in C^1(G_b,U(1)_T)/B^1(G_b,U(1)_T)$ is the usual 0D BSPT classification.

To get a sense of the classification, we first consider the simpler case of $G_f=\Zf\times G_b$. Using the K\"unneth formula, we can split the one dimensional representation of $G_f=\Zf\times G_b$ into two parts: $H^1(G_f,U(1)_T) = H^1(\Zf,U(1)_T)\times H^1(G_b,U(1)_T) = \Z_2 \times H^1(G_b,U(1)_T)$. The first $\Z_2$ part corresponds to the one dimensional representation of $\Zf$, indicating the bosonic or fermionic nature of the state. We can use the value of fermion parity $P_f=(-1)^{n_0}$ ($n_0\in \Z_2=\{0,1\}$) to represent this $\Z_2$ classification. The second part is the same as the bosonic counterpart, which is the one dimensional irreducible representation of $G_b$.

Now let us consider the generic case $G_f=\Zf\timesw G_b$ obtained by \eq{seq}. In general, for a given one dimensional representation $\tilde U$ of $G_f$, we can always separate $\tilde U(P_f^{n}g)$ [with $g\in G_b$ and $P_f^{n}g=(P_f^n,g) \in G_f$] into three parts:
\begin{align}
\tilde U\left(P_f^{n}g\right) = P_f^{n}\nu_1(g)K^{s_1(g)},
\end{align}
where $\nu_1(g)$ is a $U(1)$ phase factor and $K$ is the complex conjugation operator depending on whether $g$ contains time reversal or not. Using the multiplication rule \eq{multi} of $G_f$, the representation condition $\tilde U\left(P_f^{n}g\right)\tilde U\left(P_f^{m}h\right) = \tilde U\left(P_f^{n}g\cdot P_f^{m}h\right)$ becomes
\begin{align}
\nu_1(g) \nu_1(h)^{1-2s_1(g)} = P_f^{\om_2(g,h)} \nu_1(gh).
\end{align}
When acting on a state with fixed fermion parity $(-1)^{n_0}$ (we can again think of $n_0\in H^0(G_b,\Z_2)=\Z_2$ as a 0-cocycle), the above equation can be summarized as
\begin{align}
(\dd \nu_1)(g,h) := \frac{\nu_1(h)^{1-2s_1(g)}\nu_1(g)}{\nu_1(gh)} = (-1)^{(\om_2\smile n_0)(g,h)},
\end{align}
which means that the cocycle equation of $\nu_1$ is twisted by $\om_2\smile n_0$. If we define the homogeneous $\nu_1$ by the inhomogeneous one as $\nu_1(g,ga)=\emp^g\nu_1(a) = \nu_1(a) \cdot (-1)^{(\omega_2\smile n_0)(g,a)}$ (we will omit the superscript $g$ of the inhomogeneous $\emp^g\nu_1$ if $g=e$ is the identity element of $G_b$), we obtain the symmetry conditions 
and consistency equations 
for $n_0$ and $\nu_1$. It is easy to see that under the condition $\om_2=0$, the classification is reduced to the previous discussed case $G_f=\Zf\times G_b$ where $n_0$ and $\nu_1$ are decoupled cocycles.

\subsection{Fixed-point wave functions}

The above discussion on 0D FSPT state is from the perspective of symmetry representation $\tilde U$. We can also construct fixed-point wave functions.

For $n_0=0$, the wave function is fermion parity even. Using the basis state $|\s\rangle$ with symmetry transformation
\begin{align}
\tilde U\left(P_f^{n}g\right) |\s\rangle = |g\s\rangle,
\end{align}
we can construct the fixed-point wave function as
\begin{align}
|\Psi\rangle = \sum_{\s\in G_b} \nu_1(\s)^{-1} |\s\rangle.
\end{align}
It is easy to check that the wave function support one dimensional representation of $G_f$:
\begin{align}
\tilde U\left(P_f^{n}g\right) |\Psi\rangle = \nu_1(g) |\Psi\rangle.
\end{align}

For $n_0=1$, the wave function is a fermionic state. The basis state is created by fermion creation operator as ($\s\in G_b$)
\begin{align}
|\s\rangle = c_{\s}^\dag |0\rangle.
\end{align}
The symmetry transformation of the basis state under $G_f$ is
\begin{align}
\tilde U\left(P_f^{n}g\right) |\s\rangle = (-1)^{n} |g\s\rangle.
\end{align}
The fixed-point wave function is also a superposition of all basis states
\begin{align}
|\Psi\rangle = \sum_{\s\in G_b} \nu_1(\s)^{-1} |\s\rangle,
\end{align}
with odd fermion parity. One can check the one dimensional representation of $G_f$ on this fixed-point wave function:
\begin{align}
\tilde U\left(P_f^{n}g\right) |\Psi\rangle = (-1)^{n} \nu_1(g) |\Psi\rangle.
\end{align}

\section{2D and 3D moves that admit a branching structure}
\label{sec:moves}

\begin{figure}
\centering
\includegraphics[scale=0.35]{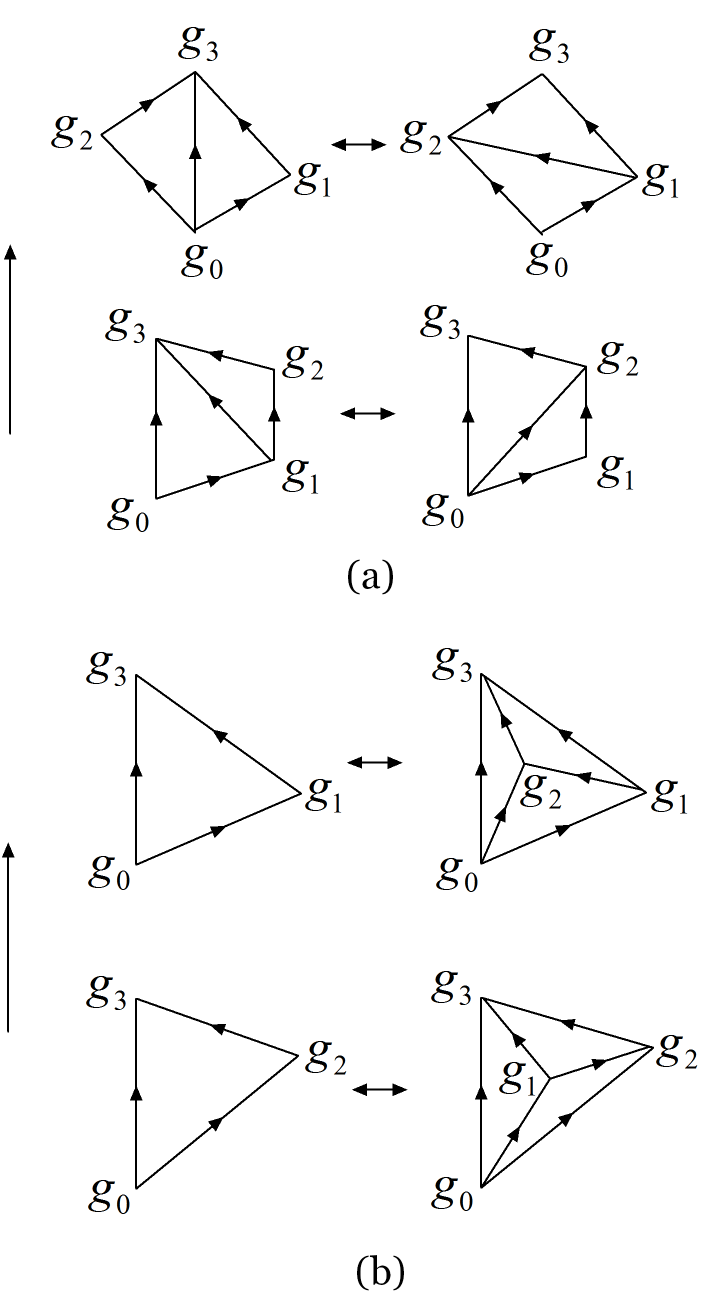}
\caption{All possible 2D (2-2) and (3-1) moves that admit a branching structure.}
\label{fig:2Dmove}
\end{figure}

\begin{figure}
\centering
\includegraphics[scale=0.3]{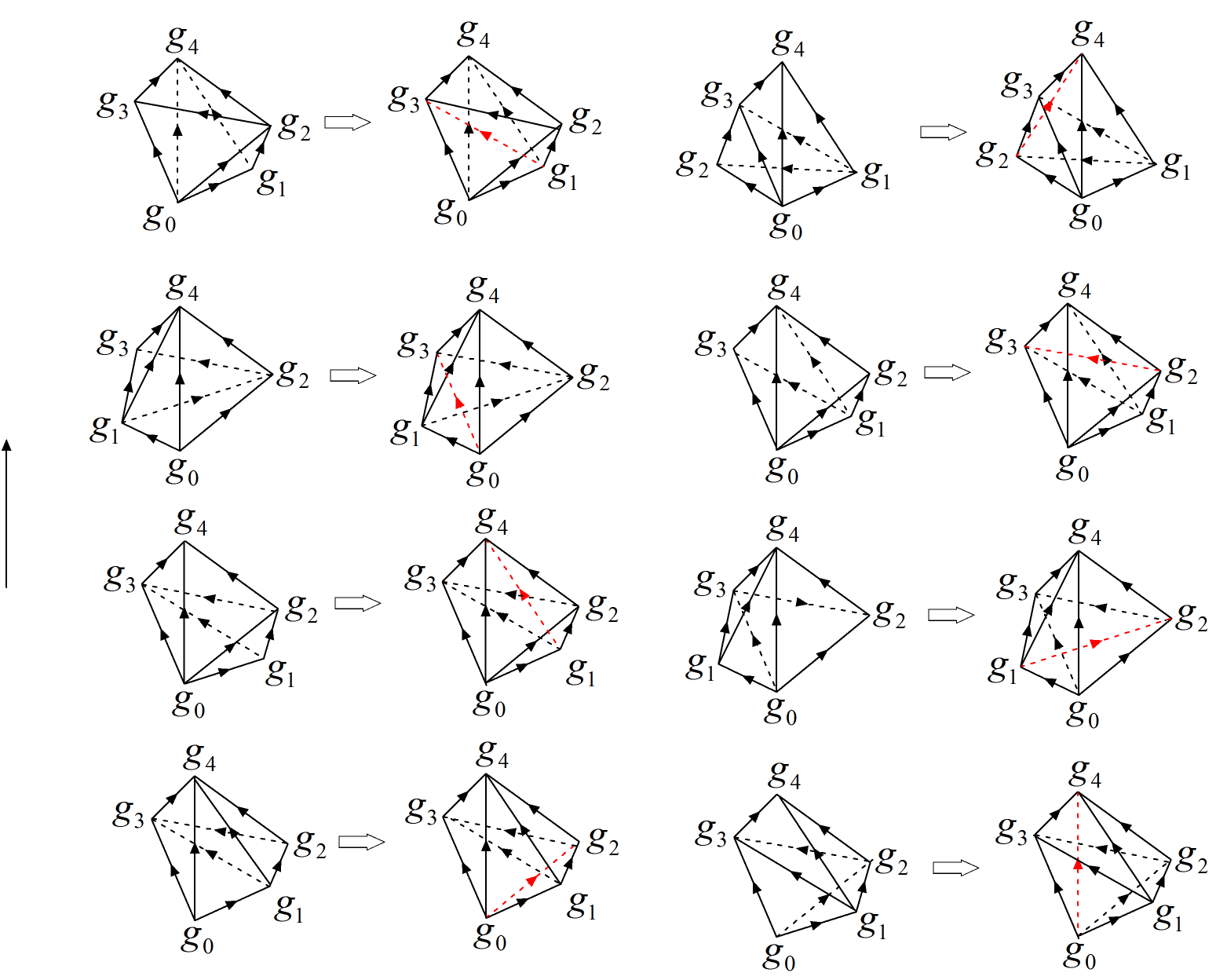}
\caption{All possible 3D (2-3) moves that admit a branching structure.}
\label{fig:3Dmovea}
\end{figure}

\begin{figure}
\centering
\includegraphics[scale=0.25]{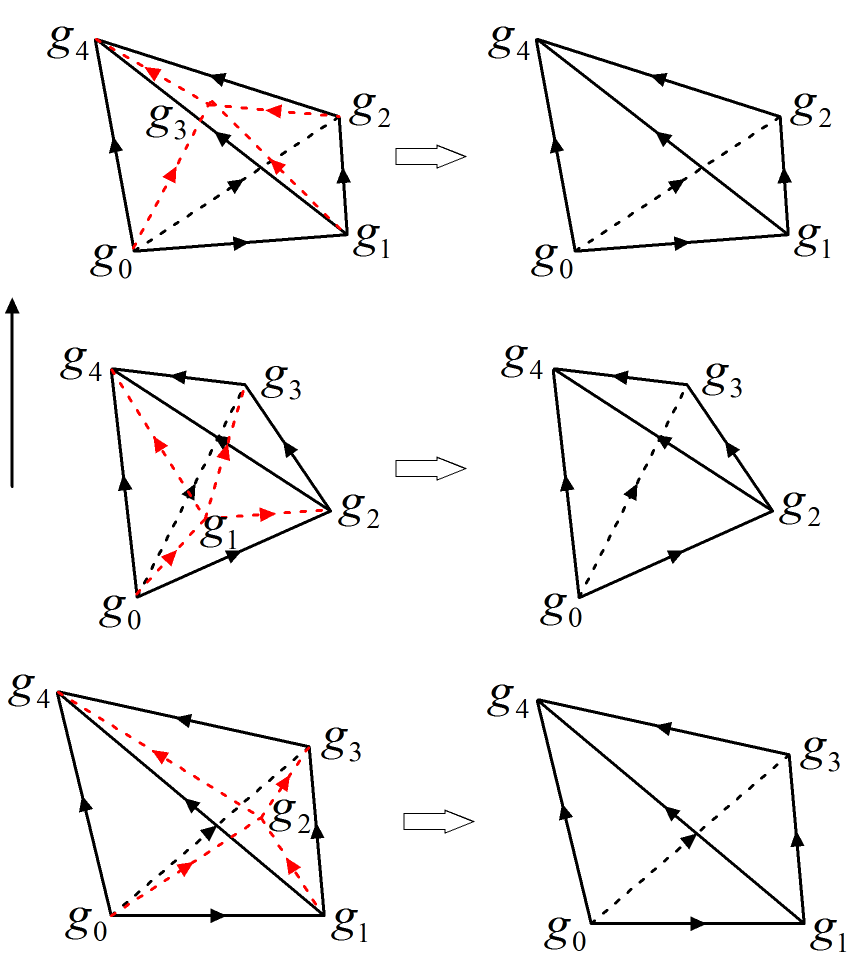}
\caption{All possible 3D (4-1) moves that admit a branching structure.}
\label{fig:3Dmoveb}
\end{figure}

In this Appendix, we list all possible 2D (2-2) and (3-1) moves that admit a branching structure (see \fig{fig:2Dmove}). The 3D (2-3) and (4-1) moves that admit a branching structure are shown in Figs.~\ref{fig:3Dmovea} and \ref{fig:3Dmoveb}. The arrow on the left hand side of the figures indicate the time direction.

\section{Kasteleyn orientations}
\label{sec:Kast}

To decorate Majorana fermions on the 2D or 3D triangulation lattices, we need (local) Kasteleyn orientations to design the pairing directions between the Majorana fermions. In the following, we will discuss briefly how to construct (local) Kasteleyn orientations from arbitrary triangulation of the spacial manifold. We refer the interested readers to \Ref{WangGu2017} for more details.

The (local) Kasteleyn orientations are constructed for arbitrary triangulation $\cT$ of a spin manifold (spacial manifold with dimension $d$) with vanishing second Stiefel-Whitney cohomology class $[w^2]=[0]$. Using the mathematical result of \Ref{Goldstein1976}, we know the representative on arbitrary triangulation of the Stiefel-Whitney homology class $w_{d-2}$, which is the Poincar\'e dual of $w^2$. Then we can construct a resolved dual lattice $\tcP$ with (local) Kasteleyn orientated links.

The procedure of constructing (local) Kasteleyn orientations is as follows \cite{WangGu2017}:
\begin{enumerate}
\item
Given a (black) triangulation lattice $\cT$ with branching structure for a $d$-dimensional spacial spin manifold;
\item
Construct the (red) resolved dual lattice $\tcP$ which is merely the lattice of Majorana fermions. The (red) link orientations are constructed rules related to the Stiefel-Whitney homology class $w_{d-2}$ in \Ref{Goldstein1976}.
\item
Find the expression of $w_{d-2}$ as a formal summation of singular $(d-2)$-simplices (i.e., non-Kasteleyn-oriented simplices) of $\cT$ by \Ref{Goldstein1976};
\item
Connect singular $(d-2)$-simplices in $\cT$ by (blue) $(d-1)$-simplices $S$ (i.e., $\partial S=w_{d-2}$);
\item
Reverse the orientations of (red) links dual to (blue) $(d-1)$-simplices belonging to $S$;
\item
Now all the $(d-2)$-simplices in $\cT$ are Kasteleyn-oriented.
\end{enumerate}
After all the above steps, the resolved dual lattice $\tcP$ now has (local) Kasteleyn orientations, meaning that the smallest (red) loop in $\tcP$ around each $(d-2)$-simplex in $\cT$ will be Kasteleyn orientated.

Using the above procedure for the special 2D and 3D triangulations, we obtain the (local) Kasteleyn orientation assignment for links inside a triangle and a tetrahedron shown in \fig{fig:Kasteleyn} and \fig{fig:3DKasteleyn}, respectively. All the Majorana fermions inside the standard triangle or tetrahedron are paired according to these (local) Kasteleyn orientations.

\section{Checking $U(1)$ coefficient obstructions by Bockstein homomorphism}
\label{App:obs}

In solving the classification equations of FSPT states, we will encounter equation of the last layer such as
\begin{align}\label{fk}
\dd \nu_{k-1} = (-1)^{f_k} = e^{2\pi i (f_k/2)},
\end{align}
where $f_k \in H^k(G_b,\Z_2)$ is a $\Z_2$-valued $k$-cocycle. This equation has no solution for $\nu_{k-1}$, iff $e^{2\pi i (f_k/2)}$ is a nontrivial $U(1)$-valued cocycle in $H^k(G_b,U(1))$.

It is well-known that $H^k(G_b,\mathbb R/\Z) \cong H^{k+1}(G_b,\Z)$. This comes from the short exact sequence $0 \rightarrow \Z \rightarrow \mathbb R \xrightarrow[]{\text{mod 1}} \mathbb R/\Z \rightarrow 0$, and the condition $H^k(G_b,\mathbb R)=0$ for finite discrete or compact Lie group $G_b$. The isomorphism is given by the connecting homomorphism of the long exact sequence, which is basically the differential operator $\dd$ acting on the $\mathbb R/\Z$-valued cocycles. According to this isomorphism, $(-1)^{f_k}=e^{2\pi i (f_k/2)}$ is a nontrivial $U(1)$-valued $k$-cocycle iff $\beta(f_k):=\dd f_k/2$ is a nontrivial $U(1)$-valued $(k+1)$-cocycle. Here, $\beta$ is the Bockstein homomorphism, which is the connecting homomorphism from the short exact sequence $0 \rightarrow \Z \xrightarrow[]{\times 2} \Z \xrightarrow[]{\text{mod 2}} \Z_2 \rightarrow 0$. It maps a $\Z_2$-valued $k$-cocycle to a $\Z$-valued $(k+1)$-cocycle.

To summarize, in order to check the last layer obstruction, we can investigate the Bockstein homomorphism of the $\Z_2$-valued obstruction functions:
\begin{align}
(-1)^{f_k} \in B^k(G_b,U(1)) \Longleftrightarrow \beta(f_k) \in B^{k+1}(G_b,\Z).
\end{align}
In many cases, the Bockstein homomorphism of $f_k$ is easy to check with the identity
\begin{align}
\be(f_k) = \frac{\dd f_k}{2}.
\end{align}
The mod 2 value of Bockstein homomorphism is also related to Steenrod square as
\begin{align}
Sq^1(f_k) = f_k\smile_{k-1}f_k = \be(f_k) \quad \text{(mod 2)}.
\end{align}
All the above discussions can be easily generalized to the obstruction functions such as $i^{f_k}=e^{2\pi i (f_k/4)}$ and the $\Z_T$ rather than $\Z$ coefficient.

The above result can be also understood from the perspective of universal coefficient theorem. We can use this theorem to obtain the $\Z_2$ coefficient cohomology groups from the $\Z$ coefficient ones:
\begin{align}\label{universal}
H^k(G_b,\Z_2) = \left[H^k(G_b,\Z)\otimes_\Z\Z_2\right] \oplus \mathrm{Tor}_1^\Z[H^{k+1}(G_b,\Z),\Z_2].
\end{align}
The above equation splits $H^k(G_b,\Z_2)$ into two types of cocycles: the first type is obtained from $\Z$-valued $k$-cocycle by mod 2; the second type is obtained from $\Z$-valued $(k+1)$-cocycle by basically the inverse of Bockstein homomorphism. So $(-1)^{f_k}$ is a trivial $U(1)$-valued cocycle iff $f_k$ is the first type $\Z_2$-valued cocycle (or equivalently $[\be(f_k)]=[0]$).

\section{FSPT examples}
\label{sec:E}

In this Appendix, we would give the classifications for FSPT phases for some symmetry groups. Some simple results are summarized in \tab{tab:examples}.

\subsection{2D FSPT phases for arbitrary unitary finite Abelian group}
\label{sec:E:Ab}

Let us consider the arbitrary unitary finite Abelian symmetry group:
\begin{align}\label{E:Gf_A}
G_f=\Z_{2N_0}^f \times \prod_{i=1}^K \Z_{N_i}.
\end{align}
We will show below that our general group super-cohomology theory gives the same classification result as \Ref{wanggu16}.

The symmetry group $G_f$ is a central extension of the bosonic unitary finite Abelian group $G_b=\prod_{i=0}^K \Z_{N_i}$ by $\Zf$ with the nontrivial ($N_0 \ge 2$) 2-cocycle
\begin{align}\label{E:A:om2}
\om_2(a,b) = \floor{\frac{a_0+b_0}{N_0}},
\end{align}
where we used $a=(a_0,a_1,...,a_K)$ with $0\leq a_i\leq N_i-1$ to denote the elements in the addictive Abelian group $G_b$. The notation $\floor{x}$ means the greatest integer less than or equal to $x$. If $N_0$ is odd, we have $[\om_2]=[0]$ in $H^2(G_b,\Z_2)$, and the fermionic symmetry group $G_f$ is merely the direct product of $\Zf$ and $G_b=\prod_{i=0}^K \Z_{N_i}$. We can treat both $N_0$ odd and $N_0$ even at one time in the following. Note that we used a different notation for $N_0$ compared to \Ref{wanggu16}, where the fermionic symmetry group is chosen to be $G_f=\Z_{N_0}^f \times \prod_{i=1}^K \Z_{N_i}$.

Before calculating the classifications, we first list the relevant cohomology groups for $G_b$ with different coefficients:
\begin{align}\label{E:A:H1}
H^1(G_b,\Z_2) &= \prod_{0\le i\le K} \Z_{\gcd(2,N_i)},\\\label{E:A:H2}
H^2(G_b,\Z_2) &= \prod_{0\le i\le K} \Z_{\gcd(2,N_i)} \prod_{0\le i<j\le K} \Z_{\gcd(2,N_{ij})},\\\label{E:A:H3}
H^3(G_b,U(1)_T) &= \prod_{0\le i\le K} \Z_{N_i} \prod_{0\le i<j\le K}\Z_{N_{ij}} \prod_{0\le i<j<k\le K}\Z_{N_{ijk}},\\\label{E:A:H4}
H^4(G_b,U(1)_T) &= \prod_{0\le i<j\le K}\Z_{N_{ij}}^2 \prod_{0\le i<j<k\le K}\Z_{N_{ijk}}^2 \prod_{0\le i<j<k<l\le K}\Z_{N_{ijkl}}.
\end{align}
Here $\gcd(x,y)$ denotes the greatest common divisor of $x$ and $y$. And $N_{ij...k}$ means the greatest common divisor of $N_i$, $N_j$, ..., and $N_k$. One can show the above results using K\"unneth formula and the universal coefficient theorem for group cohomology. Note that $N_i$ should be even (which we will assume in the following calculations), otherwise there is no nontrivial $\Z_2$-valued cocycle associated with the subgroup $\Z_{N_i}$.

There are also ``canonical'' forms for the cocycles in the cohomology groups above. To calculate the obstructions, we consider the $\Z_2$-valued cocycles:
\begin{align}
n_1^{(i)}(a) &= a_i \quad\text{(mod 2)},\\
n_2^{(i)}(a,b) &= \floor{\frac{a_i+b_i}{N_i}},\\
n_2^{(i,j)}(a,b) &= \left(n_1^{(i)}\smile n_1^{(j)}\right) (a,b)  = a_i b_j \quad\text{(mod 2)},
\end{align}
where $0\le i\le K$ for the first two cocycles and $0\le i<j\le K$ for the last cocycle. They exhaust all nontrivial cocycles in $H^1(G_b,\Z_2)$ and $H^2(G_b,\Z_2)$. We note that the 2-cocycle \eq{E:A:om2} is merely $\om_2 = n_2^{(0)}$ in terms of the above notations.

\subsubsection{Obstructions}

Different from the Abelian symmetry group $G_f=\Zf\times G_b$, not all Kitaev chain and complex fermion decorations are possible for the symmetry group \eq{E:Gf_A}. We should calculate the obstructions for each layers, using the consistency equations shown in \eq{2D:eq} (see also \tab{tab:eq}). Since we do not consider the invertible topological order $p+ip$ superconductors as FSPT states, we need only to calculate the obstructions for $n_1$ and $n_2$. We will use frequently the $\Z_2$ and $U(1)$-valued cocycle invariants in \Ref{wangcj15,ChenjieFSPT,Tantivasadakarn} to check whether a cocycle is trivial or not for finite Abelian groups.

\emph{(1) Obstructions for $n_1$.}

From \eq{2D:eq}, the obstruction function for $n_1$ is $\mathcal O_3[n_1]=\omega_2\smile n_1$ because of $s_1=0$. If we choose $n_1=n_1^{(i)}$ ($0\le i\le K$), the obstruction function is
\begin{align}
\mathcal O_3[n_1] = \omega_2\smile n_1 = n_2^{(0)} \smile n_1^{(i)}.
\end{align}
It is known that for finite Abelian groups that the above equation is always a nontrivial cocycle if both $n_2^{(0)}$ and $n_1^{(i)}$ are nontrivial cocycles.

Therefore, if $N_0$ is odd, all $n_1$ in \eq{E:A:H1} are obstruction-free and the Kitaev chain decorations are possible. If $N_0$ is even ($G_f$ is nontrivial central extension of $G_b$), all nontrivial $n_1$ in \eq{E:A:H1} are obstructed and the Kitaev chain decorations are illegal. The (cohomology class) number of obstruction-free $n_1$ is
\begin{align}\label{E:A:n1}
\#(\text{obstruction-free } n_1) = \prod_{1\le i\le K} \frac{\gcd(2,N_i)}{\gcd(2,N_{0i})}.
\end{align}

\emph{(2) Obstructions for $n_2$.} 

From \eq{2D:eq} [see also \eq{2D:O4}], The obstruction function for $n_2$ is
\begin{align}\label{E:A:O4}
\mathcal O_4[n_2]=(-1)^{n_2\smile n_2+\omega_2\smile n_2},
\end{align}
which is a cocycle in $H^4(G_b,U(1)_T)$. Note that if $N_0$ is odd, we can choose $\om_2=0$, and the equation for $n_2$ is always $\dd n_2=0$. So many terms in \eq{2D:O4} vanish, including the $(\pm i)$ terms. Since there are many different types of $n_2$ in \eq{E:A:H2}, we will discuss them seperately.

\emph{(2.1) $n_2=n_2^{(0)}$.} Since $\om_2=n_2^{(0)}$, the obstruction function \eq{E:A:O4} is always 1. So this $n_2$ is obstruction-free.

\emph{(2.2) $n_2=n_2^{(i)}\ (1\le i\le K)$.} Using the $U(1)$-valued cocycle invariants, it is easy to check both $(-1)^{n_2\smile n_2}$ and $(-1)^{\om_2\smile n_2}$ are $U(1)$-valued coboundaries. So this $n_2$ is also obstruction-free.

\emph{(2.3) $n_2=n_2^{(0,i)}\ (1\le i\le K)$.} In this case, the cocycle $(-1)^{\om_2\smile n_2} = (-1)^{n_2^{(0)}\smile n_2^{(0,i)}} = (-1)^{n_2^{(0)}\smile n_1^{(0)}\smile n_1^{(i)}}$ is nontrivial. And the other part $(-1)^{n_2\smile n_2}=(-1)^{n_2^{(0,i)}\smile n_2^{(0,i)}}$ is still a 4-coboundary. So these nontrivial $n_2$ are always obstructed.

\emph{(2.4) $n_2=n_2^{(i,j)}\ (1\le i< j\le K)$.} In this case, the obstruction function is
\begin{align}
\mathcal O_4[n_2] = (-1)^{n_2\smile n_2+\omega_2\smile n_2}
= (-1)^{n_2^{(i,j)}\smile n_2^{(i,j)}+n_2^{(0)}\smile n_2^{(i,j)}} 
= (-1)^{n_1^{(i)}\smile n_1^{(j)}\smile n_1^{(i)}\smile n_1^{(j)} + n_2^{(0)}\smile n_1^{(i)}\smile n_1^{(j)}}.
\end{align}
By calculating the $U(1)$-valued cocycle invariants, the first part $(-1)^{n_1^{(i)}\smile n_1^{(j)}\smile n_1^{(i)}\smile n_1^{(j)}}$ is always a 4-coboundary. The second part $(-1)^{n_2^{(0)}\smile n_1^{(i)}\smile n_1^{(j)}}$ is a 4-coboundary if and only if $N_{ij}/N_{0ij}$ is even. So $n_2^{(i,j)}$ ($1\le i< j\le K$) is obstruction-free if and only if $N_{ij}/N_{0ij}$ is even.

Summarizing the above results for $n_2$, the total number of obstruction-free $n_2$ is
\begin{align}\label{E:A:n2}
\#(\text{obstruction-free } n_2) = \prod_{0\le i\le K} \gcd(2,N_i) \prod_{1\le i<j\le K} \gcd\left(2,\frac{N_{ij}}{N_{0ij}}\right).
\end{align}

\subsubsection{Trivializations}

\emph{(1) Trivializations of $n_2$.}

According to the trivialized subgroups \eq{2D:Gamma} for 2D FSPT states, the 2-cocycles $n_2$ in the $\Gamma^2$ is trivialized by 1D AFSP states on the boundary. It is discussed in detail in section~\ref{sec:ASPT:1D}. For the 2-cocycle $\om_2=n_2^{(0)}$, we have $\Gamma^2 = \{\omega_2\smile n_0 \in H^2(G_b,\Z_2) | n_0\in H^0(G_b,\Z_2)\} = \langle n_2^{(0)} \rangle$. The complex fermion decoration data $n_2=n_2^{(0)}$ is trivialized. (Note that this $n_2$ is not obstructed as discussed above.) So the first subgroup $\Z_{\gcd(2,N_0)}$ of $H^2(G_b,\Z_2)$ in \eq{E:A:H2} does not correspond to nontrivial 2D FSPT state.

Combining the trivializations of $n_2$ with obstruction-free $n_2$ number \eq{E:A:n2}, the number of trivialization-free obstruction-free $n_2$ is
\begin{align}\label{E:A:n2_}
\#(\text{trivialization-free obstruction-free } n_2) = \prod_{1\le i\le K} \gcd(2,N_i) \prod_{1\le i<j\le K} \gcd\left(2,\frac{N_{ij}}{N_{0ij}}\right).
\end{align}

\emph{(2) Trivializations of $\nu_3$.}

For the bosonic $U(1)$ phase factor $\nu_3$, the trivialized subgroup in \eq{2D:Gamma} can be calculated as
\begin{align}\nonumber\label{E:A:G3}
\Gamma^3 &= \{(-1)^{\omega_2\smile n_1} \in H^3(G_b,U(1)_T) \big| n_1\in H^1(G_b,\Z_2)\}\\
&= \left\langle (-1)^{n_2^{(0)} \smile n_1^{(i)}} \Big| 0\le i \le K,\ \gcd(2,N_i)=2 \right\rangle.
\end{align}
For the cocycles in the subgroup $\Z_{N_0}\times\prod_{1\le i\le K}\Z_{0i}$ of $H^3(G_b,U(1)_T)$ in \eq{E:A:H3}, they have a ``canonical'' form expressed as lower dimensional $\Z_2$-valued cocycles $n_1^{(i)}$ ($0\le i\le K$) and $n_2^{(0)}$ as (only for even $N_i$, otherwise the cocycle $n_1^{(i)}$ is trivial)
\begin{align}\label{E:A:nu3_cano}
\nu_3=e^{2\pi i \frac{k}{N_i} n_2^{(0)} \smile n_1^{(i)}} \quad (k=0,1,...,N_{0i}-1).
\end{align}
The generating cocycle of $\Gamma^3$ in \eq{E:A:G3} can be expressed as $(-1)^{n_2^{(0)} \smile n_1^{(i)}} = e^{2\pi i \frac{N_i/2}{N_i} n_2^{(0)} \smile n_1^{(i)}}$. Comparing it with the above equation, we see that if $N_i$ is odd, or $N_i/2$ is a integral multiplier of $N_{0i}$, then the part of $\Gamma^2$ related to $\Z_{N_i}\subset G_b$ ($0\le i\le K$) is trivial in $H^3(G_b,U(1)_T)$. Otherwise, the 3-cocycle $\nu_3$ in \eq{E:A:nu3_cano} with $k=N_i/2$ (mod $N_{0i}$) is trivialized by 2D ASPT state. The results calculated above can also be obtained from calculating the cocycle invariants for $\nu_3$ \cite{wangcj15,ChenjieFSPT,Tantivasadakarn}.

In summary, (1) if $N_0$ is even, one nontrivial $\nu_3$ in $\Z_{N_0}\subset H^3(G_b,U(1)_T)$ is trivialized. Otherwise, all elements in $\Z_{N_0}\subset H^3(G_b,U(1)_T)$ are nontrivial. (2) For $1\le i\le K$, and the subgroup $\Z_{N_{0i}}\subset H^3(G_b,U(1)_T)$, we also have two possibilities. If $N_i$ is even and $N_i/N_{0i}$ is odd \footnote{This is equivalent to $\#_2(N_0)\ge \#_2(N_i) \ge 1$, where $\#_2(x)$ denotes the number of 2's in the prime factorization of integer $x$.}, then one nontrivial $\nu_3$ in $\Z_{N_{0i}}\subset H^3(G_b,U(1)_T)$ is trivialized. Otherwise, all elements in $\Z_{N_{0i}}\subset H^3(G_b,U(1)_T)$ are nontrivial. So the number of trivialization-free cocycles in the subgroup $\Z_{N_0} \times \prod_{1\le i\le K} \Z_{0i} \subset H^3(G_b,U(1)_T)$ in \eq{E:A:H3} is
\begin{align}
\frac{N_0}{\gcd(2,N_0)} \prod_{1\le i\le K} \frac{\gcd(2N_0,N_i)}{\gcd(2,N_i)}.
\end{align}
The total number of $\nu_3$ that is not trivialized in \eq{E:A:H3} is
\begin{align}\label{E:A:nu3}
\#(\text{trivialization-free }\nu_3) = 
\frac{N_0}{\gcd(2,N_0)} 
\prod_{1\le i\le K} N_i \cdot \frac{\gcd(2N_0,N_i)}{\gcd(2,N_i)} 
\prod_{1\le i<j\le K} N_{ij} 
\prod_{0\le i<j<k\le K} N_{ijk}.
\end{align}

\subsubsection{Full classification}

From the above calculations of obstructions and trivializations, we can obtain the number of 2D FSPT phases with symmetry group \eq{E:Gf_A} by combining Eqs.~(\ref{E:A:n1}), (\ref{E:A:n2_}) and (\ref{E:A:nu3}):
\begin{align}
\#(\mathrm{FSPT}) =
\frac{N_0}{\gcd(2,N_0)} 
\prod_{1\le i\le K} N_i \cdot \frac{\gcd(2,N_i)\cdot\gcd(2N_0,N_i)}{\gcd(2,N_{0i})} 
\prod_{1\le i<j\le K} N_{ij} \cdot \gcd\left(2,\frac{N_{ij}}{N_{0ij}}\right) 
\prod_{0\le i<j<k\le K} N_{ijk}.
\end{align}
If $N_0$ is even, the above equation is reduced to
\begin{align}
\#(\mathrm{FSPT})\big |_{N_0\text{ even}} =
\frac{N_0}{2} 
\prod_{1\le i\le K} N_i \cdot \gcd(2N_0,N_i)
\prod_{1\le i<j\le K} N_{ij} \cdot \gcd\left(2,\frac{N_{ij}}{N_{0ij}}\right) 
\prod_{0\le i<j<k\le K} N_{ijk}.
\end{align}
On the other hand, if $N_0$ is odd, we have
\begin{align}
\#(\mathrm{FSPT})\big |_{N_0\text{ odd}} =
N_0
\prod_{1\le i\le K} N_i \cdot \gcd(2,N_i) \cdot \gcd(2N_0,N_i)
\prod_{1\le i<j\le K} N_{ij} \cdot \gcd\left(2,N_{ij}\right) 
\prod_{0\le i<j<k\le K} N_{ijk}.
\end{align}
After some calculations \footnote{One may need some identities such as $\gcd(2N_0,N_{ij}) = N_{0ij} \cdot \gcd(2,N_{ij}/N_{0ij})$ when $N_0$ is even}, one can show that the above two equations for the number of 2D FSPT phases are exactly the same as Eqs.~(42) and (43) of \Ref{wanggu16}, which are obtained from a totally different approach of the braiding statistics data of the gauge flux. Furthermore, one can show that the group structure of general group supercohomology classification \cite{WangGu} also agrees with \Ref{wanggu16}. We note again that we use a different convention of $N_0$ definition compared to \Ref{wanggu16}.

\subsection{Four symmetry groups with $G_b\cong\Z_2$ and different $s_1,\om_2$}
\label{sec:E:Z2}

\begin{table}[ht]
\centering
\begin{tabular}{ |c|c|c|c|c| }
\hline
$G_f\ \backslash\ \mathrm{dim}$ & 0 & 1 & 2 & 3\\
\hline
\hline
$\Z_2^f\times\Z_2$ & $\Z_2\times\Z_2$ & $\Z_2$ & $\Z_8$ & $\Z_1$ \\
\hline
$\Z_4^f=\Z_2^f\timesw\Z_2$ & $\Z_4$ & $\Z_1$ & $\Z_1$ & $\Z_1$ \\
\hline
$\Z_2^f\times\Z_2^T$ & $\Z_2$ & $\Z_4$ & $\Z_1$ & $\Z_1$ \\
\hline
$\Z_4^{Tf}=\Z_2^f\timesw\Z_2^T$ & $\Z_1$ & $\Z_2$ & $\Z_2$ & $\Z_{16}$ \\
\hline
\end{tabular}
\caption{Classification of FSPT phases with $G_b\cong\Z_2$ and different choices of $s_1,\om_2$.}
\label{tab:Z2fZ2}
\end{table}

\begin{table}[ht]
\centering
\begin{tabular}{ |c|c|c|c|c| }
\hline
$G_f\ \backslash\ \mathrm{dim}$ & 0 & 1 & 2 & 3\\
\hline
\hline
$\Z_2^f\times\Z_2$ & 
$\Z_2\times\Z_2$ & 
${\color{red}\Z_2}\times\Z_2$ & 
${\color{red}\Z}\times\Z_8$ & 
$\Z_1$ \\
\hline
$\Z_4^f=\Z_2^f\timesw\Z_2$ & 
$\Z_4$ & 
$\Z_1$ & 
${\color{red}\Z}$ & 
$\Z_1$ \\
\hline
$\Z_2^f\times\Z_2^T$ &
$\Z_2$ & 
${\color{red}\Z_8}$ &
$\Z_1$ &
$\Z_1$ \\
\hline
$\Z_4^{Tf}=\Z_2^f\timesw\Z_2^T$ &
$\Z_1$ & 
$\Z_2$ & 
$\Z_2$ & 
$\Z_{16}$
\\
\hline
\end{tabular}
\caption{Classification of fermionic invertible phases with $G_b\cong\Z_2$ and different choices of $s_1,\om_2$. The differences compared to \tab{tab:Z2fZ2} (with fermionic invertible topological orders included) are emphasized by red color.
}
\label{tab:Z2fZ2_}
\end{table}

If we focus on the bosonic symmetry group $G_b$ that is isomorphic to $\Z_2=\{0,1\}$, there are in total four different $G_f$ with different choices of cocycles $s_1$ and $\om_2$. It is already discussed briefly at the end of section~\ref{sec:G}. The nontrivial cocycles in $H^1(\Z_2,\Z_2)=\Z_2$ and $H^2(\Z_2,\Z_2)=\Z_2$ are
\begin{align}
s_1(a) &= 
\begin{cases}
1, \quad a=1\\
0, \quad\text{others}
\end{cases},\\
\om_2(a,b) &= 
\begin{cases}
1, \quad a=b=1\\
0, \quad\text{others}
\end{cases}.
\end{align}
The classifications of FSPT phases for these groups are given in \tab{tab:examples}. For convenience, we resummarize them in \tab{tab:Z2fZ2}.

We note that we do not consider the invertible topological order (such as Kitaev chain in 1D) as FSPT state, since they do not need any bosonic symmetry protection ($\Zf$ can never be broken). If we include them as to consider invertible phases, the classification results are given in \tab{tab:Z2fZ2_}. There is an additional $\Z_2$ subgroup for $G_f=\Zf\times \Z_2$ and $G_f=\Zf\times \Z_2^T$ in 1D (obstructed if $\om_2$ is nontrivial). It corresponds to the $n_0$ data of 1D invertible topological order of Kitaev chain. There is an additional $\Z$ classification for symmetry group $G_f=\Zf\times \Z_2$ and $G_f=\Z_4^f$ in 2D. It corresponds to the several layers of 2D $p+ip$ chiral superconductors as fermionic invertible topological orders (the root state of the two $\Z$ classifications are different). It is not possible if there is time reversal symmetry in $G_b$ [for $H^0(G_b,\Z_T)=0$]. We note that all the results are consistent with the spin cobordism calculations \cite{kapustin14}.

\subsubsection{$G_f = \Zf \times \Z_2$}

\emph{(1) 0D.} The classification data is $(n_0,\nu_1)\in H^0(G_b,\Z_2) \times H^1(G_b,U(1)_T) = \Z_2 \times \Z_2$. There are no obstructions and trivializations for these data. In another approach, the classification of one-dimensional irreducible representations for $G_f$ is $H^1(\Zf \times \Z_2,U(1)_T)=\Z_2\times \Z_2$. So the two approaches agree to each other, and the 0D FSPT phases are classified by $\Z_2\times\Z_2$.

\emph{(2) 1D.} The classification data is $(n_1,\nu_2) \in H^1(G_b,\Z_2) \times H^2(G_b,U(1)_T) = \Z_2 \times \Z_1$. There are no obstructions or trivializations.

\emph{(3) 2D.} The classification data is $(n_1,n_2,\nu_3) \in H^1(G_b,\Z_2) \times H^2(G_b,\Z_2) \times H^3(G_b,U(1)_T) = \Z_2 \times \Z_2 \times \Z_2$. There are no obstructions or trivializations. And the group structure of the classification \cite{WangGu} can be shown to be $\Z_8$. So two copies of Kitaev chain decoration states gives the complex fermion decoration state. Two copies of complex fermion decoration states gives the nontrivial BSPT state. And finally two copies of BSPT states is trivial. This classification of 2D topological superconductors with $\Z_2$ Ising symmetry is first obtained in \Ref{gu14b}.

\emph{(4) 3D.} The classification data is $(n_1,n_2,n_3,\nu_4) \in H^1(G_b,\Z) \times H^2(G_b,\Z_2) \times H^3(G_b,\Z_2) \times H^4(G_b,U(1)_T) = \Z_1 \times \Z_2 \times \Z_2 \times \Z_1$. One can show that the nontrivial cocycles satisfy $n_2\smile n_2 \notin B^4(G_b,\Z_2)$ and $(-1)^{n_3\smile_1 n_3} \notin B^5(G_b,U(1)_T)$. According to the consistency equations in \tab{tab:eq}, all states are obstructed. There is only one trivial phase.

\subsubsection{$G_f = \Z_4^f = \Zf \timesw \Z_2$}

\emph{(1) 0D.} In terms of our classification data, we have $(n_0,\nu_1)\in H^0(G_b,\Z_2) \times H^1(G_b,U(1)_T) = \Z_2 \times \Z_2$. And both data are obstruction-free. In another way, the one-dimensional irreducible representations for $G_f$ is classified by $H^1(\Z_4^f,U(1)_T)=\Z_4$. So the classification is $\Z_4$.

\emph{(2) 1D.} The classification data is $(n_1,\nu_2) \in H^1(G_b,\Z_2) \times H^2(G_b,U(1)_T) = \Z_2 \times \Z_1$. However, the nontrivial $n_1$ is obstructed. So there is only one trivial phase.

\emph{(3) 2D.} The classification data is $(n_1,n_2,\nu_3) \in H^1(G_b,\Z_2) \times H^2(G_b,\Z_2) \times H^3(G_b,U(1)_T) = \Z_2 \times \Z_2 \times \Z_2$. The nontrivial $n_1$ is obstructed. As discussed in detail in section~\ref{sec:ASPT:1D}, the nontrivial $n_2$ is trivialized by the boundary 1D ASPT (recall the obstruction of Kitaev chain layer $n_0$ in 1D). And the nontrivial cocycle $\nu_3$ is also trivialized (recall the obstruction of $n_1$ in 1D). So the classification is $\Z_1$, which is consistent with \Ref{wangcj16,wanggu16}.

\emph{(4) 3D.} The classification data is $(n_1,n_2,n_3,\nu_4) \in H^1(G_b,\Z) \times H^2(G_b,\Z_2) \times H^3(G_b,\Z_2) \times H^4(G_b,U(1)_T) = \Z_1 \times \Z_2 \times \Z_2 \times \Z_1$. The nontrivial $n_2$ is trivialized (recall the obstruction of $n_0$ in 2D). And the nontrivial $n_3$ is also trivialized (recall the obstruction of $n_1$ in 2D). So there is only one trivial phase.

\subsubsection{$G_f = \Zf \times \Z_2^T$}

\emph{(1) 0D.} The classification data is $(n_0,\nu_1)\in H^0(G_b,\Z_2) \times H^1(G_b,U(1)_T) = \Z_2 \times \Z_1$. The cocycle $n_0$ is obstruction-free. In another approach, the classification of one-dimensional irreducible representations for $G_f$ is $H^1(\Zf \times \Z_2^T,U(1)_T)=\Z_2$. So the classification is $\Z_2$.

\emph{(2) 1D.} The classification data is $(n_1,\nu_2) \in H^1(G_b,\Z_2) \times H^2(G_b,U(1)_T) = \Z_2 \times \Z_2$. There is neither obstruction nor trivialization. The classification is $\Z_4$. If we include the invertible topological order Kitaev chain, then the classification is $\Z_8$. This classification of 1D $T^2=1$ topological superconductors is first obtained in \Ref{fidkowski10,fidkowski11}.

\emph{(3) 2D.} The classification data is $(n_1,n_2,\nu_3) \in H^1(G_b,\Z_2) \times H^2(G_b,\Z_2) \times H^3(G_b,U(1)_T) = \Z_2 \times \Z_2 \times \Z_1$. The nontrivial $n_1$ is obstructed for $s_1\smile n_1\smile n_1$ is nontrivial in $H^3(G_b,\Z_2)$. This obstruction is the fermion parity obstruction for $T^2=1$ 2D topological superconductors considered in Refs.~\cite{WangNingChen2017} and \cite{WQG2018}. The nontrivial $n_2$ is obstructed for $(-1)^{n_2\smile n_2}\notin B^4(G_b,U(1)_T)$. So there is only one trivial phase.

\emph{(4) 3D.} The classification data is $(n_1,n_2,n_3,\nu_4) \in H^1(G_b,\Z) \times H^2(G_b,\Z_2) \times H^3(G_b,\Z_2) \times H^4(G_b,U(1)_T) = \Z_2 \times \Z_2 \times \Z_2 \times \Z_2$. The nontrivial cocycle data $n_1$ of $p+ip$ chiral superconductor decoration is obstructed for $s_1\smile n_1\smile n_1 \notin B^3(G_b,\Z_2)$. The nontrivial $n_2$ is also obstructed, by calculating the cocycle invariants for the obstruction function $n_2\smile n_2+s_1\smile (n_2\smile_1 n_2) \notin B^4(G_b,\Z_2)$. The nontrivial $n_3$ is trivialized by 2D ASPT state (recall the obstruction of $n_1$ in 2D). And the nontrivial $\nu_4$ is also trivialized by another layer of 2D ASPT state (recall the obstruction of $n_2$ in 2D). In summary, there is only one trivial FSPT phase.

\subsubsection{$G_f = \Z_4^{Tf} = \Zf \timesw \Z_2^T$}

\emph{(1) 0D.} The classification data is $(n_0,\nu_1)\in H^0(G_b,\Z_2) \times H^1(G_b,U(1)_T) = \Z_2 \times \Z_1$. Since the cocycle $(-1)^{\om_2\smile n_0} \in H^2(G_b,U(1)_T)$ is nontrivial, the nontrivial $n_0$ is obstructed. In another approach, the classification of one-dimensional irreducible representations for $G_f$ is $H^1(\Zf \timesw \Z_2^T,U(1)_T)=\Z_1$. So the classification is $\Z_1$. The physical meaning is that nontrivial fermionic mode with $T^2=-1$ must be in Kramers doublet, which is two-fold degenerate. So there is only one trivial class.

\emph{(2) 1D.} The classification data is $(n_1,\nu_2) \in H^1(G_b,\Z_2) \times H^2(G_b,U(1)_T) = \Z_2 \times \Z_2$. The nontrivial $n_1$ is obstruction-free. The nontrivial $\nu_2$ is trivialized by 0D ASPT state (recall the obstruction of $n_0$ in 0D). So the classification is $\Z_2$ corresponding to the complex fermion decorations.

\emph{(3) 2D.} The classification data is $(n_1,n_2,\nu_3) \in H^1(G_b,\Z_2) \times H^2(G_b,\Z_2) \times H^3(G_b,U(1)_T) = \Z_2 \times \Z_2 \times \Z_1$. The first cocycle $n_1$ is obstruction-free, for the obstruction function is zero: $\om_2\smile n_1+s_1\smile n_1\smile n_1=0$. The nontrivial cocycle data $n_2$ is trivialized (recall the obstruction of $n_0$ in 1D). In summary, there is only one nontrivial topological superconductor with $T^2=-1$. It is exactly the Kitaev chain decoration state constructed in \Ref{WangNingChen2017}.

\emph{(4) 3D.} The classification data is $(n_1,n_2,n_3,\nu_4) \in H^1(G_b,\Z) \times H^2(G_b,\Z_2) \times H^3(G_b,\Z_2) \times H^4(G_b,U(1)_T) = \Z_2 \times \Z_2 \times \Z_2 \times \Z_2$. The nontrivial cocycle $n_1$ is obstruction-free, for the obstruction function is $\om_2\smile n_1+s_1\smile n_1\smile n_1=0$. The nontrivial cocycle $n_2$ is also obstruction-free, since the obstruction function $\om_2\smile n_2+n_2\smile n_2 + s_1\smile (n_2\smile_1 n_2)=s_1\smile (n_2\smile_1 n_2)$ is in $B^4(G_b,\Z_2)$. Since $H^5(G_b,U(1)_T)=0$, the classification data $n_3$ is always obstruction-free. In summary, all the four layers of classification data are obstruction-free and trivialization-free. So the classification of 3D $T^2=-1$ topological superconductors is $\Z_{16}$, which is first shown by Kitaev and Morgan \cite{kitaevsre}.

\subsection{FSPT states with quaternion group $G_f=Q_8^f=\Z_2^f\timesw(\Z_2\times\Z_2)$}
\label{sec:E:Q8}

The quaternion group $Q_8$ is defined as $Q_8=\langle i,j,k| i^2=j^2=k^2=ijk \rangle$ with order 8. Usually, we denote $ijk$ as $-1$. Other useful relations are $ij=k, jk=i, ki=j$, which can be easily derived from the definition of $Q_8$. Since the center of $Q_8$ is $\{\pm 1\}$, we can unambiguously identify it with the fermion parity group $\Zf$. Then the bosonic symmetry group (as a quotient group) $G_b=G_f/\Zf=Q_8^f/\Zf$ is generated by $[i]$ and $[j]$, and has relations $[i]^2=[j]^2=[i][j]=[1]$. So $G_b$ is isomorphic to $\Z_2\times\Z_2$. In terms of the short exact sequence, we have
\begin{align}
1 \rightarrow \Zf \rightarrow Q_8^f \rightarrow \Z_2\times\Z_2 \rightarrow 1.
\end{align}
The nontrivial 2-cocycle of the central extension is given by
\begin{align}\label{Q8_om2}
\om_2=n_2^{(1)}+n_2^{(2)}+n_1^{(1)}n_1^{(2)},
\end{align}
which is the most nontrivial element in $H^2(\Z_2\times\Z_2,\Z_2)=\Z_2^3=\langle n_2^{(1)},n_2^{(2)},n_1^{(1)}n_1^{(2)}\rangle$. Here, $n_1^{(i)}$ and $n_2^{(i)}=n_1^{(i)}\smile n_1^{(i)}$ are the nontrivial 1- and 2-cocycles for the $i$-th ($i=1,2$) $\Z_2$ subgroup of $G_b=\Z_2^{(1)}\times\Z_2^{(2)}$. The three terms of $\om_2$ in \eq{Q8_om2} indicate $i^2=-1$, $j^2=-1$ and $ij=-ji$ respectively in $Q_8^f$.

We will show below that there is a 3D anomaly-free FSPT state for $G_f=Q_8^f$ with complex fermion decoration $n_3 = n_2^{(1)}\smile n_1^{(2)} + n_1^{(1)}\smile n_2^{(2)}$. However, this state is trivialized by $\om_2\smile_1\om_2 \in \Gamma^3$ [see \eq{3D:Gamma}], which is related to the boundary 2D ASPT state with $n_0=2$ copies of $p+ip$ superconductors.

The relevant cohomology groups of $G_b=\Z_2^{(1)}\times\Z_2^{(2)}$ with $\Z_2$ coefficient are
\begin{align}
H^0(\Z_2\times\Z_2,\Z_2) &= \Z_2 = \langle 1 \rangle,\\
H^1(\Z_2\times\Z_2,\Z_2) &= \Z_2^2 = \langle n_1^{(1)},n_1^{(2)} \rangle,\\
H^2(\Z_2\times\Z_2,\Z_2) &= \Z_2^3 = \langle n_2^{(1)},n_2^{(2)},n_1^{(1)}n_1^{(2)} \rangle,\\
H^3(\Z_2\times\Z_2,\Z_2) &= \Z_2^4 = \langle n_3^{(1)},n_2^{(1)}n_1^{(2)},n_1^{(1)}n_2^{(2)},n_3^{(2)} \rangle,\\
H^4(\Z_2\times\Z_2,\Z_2) &= \Z_2^5 = \langle n_4^{(1)},n_3^{(1)}n_1^{(2)},n_2^{(1)}n_2^{(2)},n_1^{(1)}n_3^{(2)},n_4^{(2)} \rangle.
\end{align}
And the cohomology groups with $U(1)$ coefficient are
\begin{align}
H^1(\Z_2\times\Z_2,U(1)) &= \Z_2^2,\\
H^2(\Z_2\times\Z_2,U(1)) &= \Z_2,\\
H^3(\Z_2\times\Z_2,U(1)) &= \Z_2^3,\\
H^4(\Z_2\times\Z_2,U(1)) &= \Z_2^2,\\
H^5(\Z_2\times\Z_2,U(1)) &= \Z_2^4.
\end{align}

\subsubsection{0D}

The classification data is $(n_0,\nu_1)\in H^0(G_b,\Z_2) \times H^1(G_b,U(1)) = \Z_2 \times \Z_2^2$. From the equation $\dd \nu_1=(-1)^{\om_2n_0}$, we see that $n_0=1$ is obstructed, as $[(-1)^{\om_2}] = [(-1)^{n_1^{(1)}n_1^{(2)}}]$ is the nontrivial cocycle in $H^2(\Z_2\times\Z_2,U(1)) = \Z_2$. So the classification is $\Z_2^2$, which is the same as BSPT phases.

\subsubsection{1D}

The classification data is $(n_0,n_1,\nu_2)\in H^0(G_b,\Z_2) \times H^1(G_b,\Z_2) \times H^2(G_b,U(1)) = \Z_2 \times \Z_2^2 \times \Z_2$. The equation $\dd n_1=\om_2\smile n_0$ implies that $n_0=1$ is obstructed. From from the equation $\dd\nu_2=(-1)^{\om_2n_1}$, we see that $n_1=n_1^{(i)}$ is obstructed: $[(-1)^{\om_2n_1^{(i)}}] = [(-1)^{n_1^{(i)}n_2^{(i)}}] \notin B^3(G_b,U(1))$. The nontrivial BSPT $\nu_2=(-1)^{n_1^{(1)}n_1^{(2)}}$ is trivialized by 0D obstruction function $(-1)^{\om_2 n_0}$ with $n_0=1$. Therefore, there is only one trivial FSPT phase in 1D.

\subsubsection{2D}

The classification data is $(n_0,n_1,n_2,\nu_3)\in H^0(G_b,\Z) \times H^1(G_b,\Z_2) \times H^2(G_b,\Z_2) \times H^3(G_b,U(1)) = \Z \times \Z_2^2 \times \Z_2^3 \times \Z_2^3$. The $n_0=1$ ($p+ip$ superconductor) state is obstructed by the equation $\dd n_1=\om_2\smile n_0$. And all the $n_1$ are obstructed by the equation $\dd n_2=\om_2\smile n_1$. The data $n_2=\om_2$ is trivialized by 1D ASPT state $\om_2\smile n_0$ with $n_0=1$. So we only need to consider $n_2\in \Z_2^2=\langle n_2^{(1)},n_2^{(2)} \rangle$. Since $[(-1)^{\om_2n_2+n_2n_2}]=[(-1)^{n_1^{(1)}n_1^{(2)}n_2^{(i)}}]$ if $n_2=n_2^{(i)}$, we conclude that all $n_2$ are obstructed or trivialized. For the BSPT state $\nu_3$, two of the three root states are trivialized by 1D ASPT $(-1)^{\om_2n_1^{(i)}}$. Therefore, the 2D FSPT phases (invertible phases excluded) are classified by $\Z_2$ which is generated by one of the BSPT root phases.

\subsubsection{3D}

The classification data is $(n_1,n_2,n_3,\nu_4) \in H^1(G_b,\Z) \times H^2(G_b,\Z_2) \times H^3(G_b,\Z_2) \times H^4(G_b,U(1)) = \Z_1 \times \Z_2^3 \times \Z_2^4 \times \Z_2^2$. The data $n_2=\om_2$ is trivialized by $\om_2\smile n_0$ with $n_0=1$. And all other $n_2$ are obstructed by the equation $\dd n_3=\om_2n_2+n_2n_2$. For the $n_3$ data, the 1D ASPT with $\om_2n_1^{(i)}$ will trivialize the $n_3$ data from $\Z_2^4$ to $\Z_2^2$, which can be chosen to be $\langle n_2^{(1)}n_1^{(2)}, n_1^{(1)}n_2^{(2)} \rangle$. For the two root $n_3$, one can show that $[\dd \nu_4]=[(-1)^{\om_2n_3+n_3n_3}]=[(-1)^{n_3^{(1)}n_2^{(2)}+n_2^{(1)}n_3^{(2)}}]$. So the only obstruction-free and trivialization-free $n_3$ is $n_3=n_2^{(1)}n_1^{(2)} + n_1^{(1)}n_2^{(2)}$. However, from the $n_0=2$ trivialization \eq{Gamma3}, this $n_3$ is the same as and trivialized by $\om_2\smile_1\om_2=Sq^1(\om_2)=Sq^1[n_1^{(1)}n_1^{(2)}]$. For the BSPT $\nu_4$, they are all trivialized by 2D ASPT with $(-1)^{\om_2n_2+n_2n_2}$. In conclusion, there is only one 3D FSPT trivial phase.

%

\end{document}